\documentclass[10pt,aps,pra,floatfix,groupedaddress,showpacs]{revtex4}
\usepackage{amssymb,amsmath,amsthm,epsfig,graphicx}

\theoremstyle{plain}
\newtheorem{thm}{Theorem}[section]

\numberwithin{equation}{section}

\theoremstyle{remark}

\newtheorem{exm}[thm]{Example}

\newcommand{\sst}{\scriptstyle}
\newcommand{\intl}{\int\limits}

\newcommand{\n}{\noindent}
\newcommand{\ds}{\displaystyle}
\newcommand{\vp}{\varepsilon}

\newcommand{\bb}[1]{\mathbb{#1}}
\newcommand{\cl}[1]{\mathcal{#1}}

\newcommand{\ep}{\epsilon}

\newcommand{\forget}[1]{}
\newcommand{\be}{\begin{equation}}
\newcommand{\ee}{\end{equation}}
\newcommand{\ba}{\begin{eqnarray}}
\newcommand{\ea}{\end{eqnarray}}
\newcommand{\nn}{\nonumber}
\newcommand{\la}{\label}
\newcommand{\br}{{\bf r}}
\newcommand{\bbr}{{\bf R}}
\newcommand{\e}{{\rm e}}

\begin{document}

\title{The two electron molecular bond revisited:
from Bohr orbits to two-center orbitals
\footnote{
MOS wishes to express his appreciation and admiration for Prof. Benjamin
Bederson for his insights into and nurturing of AMO physics.
Most recently,  Ben's work focused on the experimental determination of
atomic polarizabilities.
The classic determination was in bulk material.
However, Ben developed a beautiful atomic beam method which he brought to
marvelous perfection.
It is a pleasure to dedicate this article to him.
}}

\author{Goong Chen$^{1,2}$, Siu A.~Chin$^1$, Yusheng Dou$^{1,7}$,
Kishore T.~Kapale$^{1,4,5}$, Moochan Kim$^{1,5}$, Anatoly A. Svidzinsky$^{1,5}$,
Kerim Urtekin$^1$, Han Xiong$^1$, and Marlan O.~Scully$^{1,3,5,6}$}
\affiliation{Institute for Quantum Studies and
Depts. of Physics$^1$, Mathematics$^2$, \\
Chemical and Electrical Engineering$^3$,
Texas A\&M University, College Station, TX 77843 \\
$^4$Jet Propulsion Laboratory, California Institute of Technology,
Pasadena, CA 91109 \\
$^5$Depts. of Chemistry, and Mechanical and Aerospace Engineering,
Princeton University, Princeton, NJ 08544  \\
$^6$Max-Planck-Institut f\"ur Quantenoptik, D-85748 Garching, Germany\\
$^7$Now at Dept.\ of Physical Sciences, Nicholls State University,
Thibodaux, LA 70310
}

\date{\today}
\begin{abstract}
Niels Bohr originally applied his approach to quantum mechanics
to the H atom with great success. He then went on to show in 1913
how the same ``planetary-orbit'' model can predict binding for the
H$_2$ molecule. However, he misidentified the correct dissociation
energy of his model at large internuclear separation, forcing
him to give up on a ``Bohr's model for molecules". Recently, we have
found the correct dissociation limit of Bohr's model for H$_2$
and obtained good potential energy curves at all internuclear separations.
This work is a natural extension of Bohr's original paper and
corresponds to the $D=\infty$ limit of a dimensional scaling (D-scaling)
analysis, as developed by Herschbach and coworkers.

In a separate but synergetic approach to the two-electron problem,
we summarize recent advances in constructing analytical models for
describing the two-electron bond. The emphasis here is not maximally
attainable
numerical accuracy, but beyond textbook accuracy
as informed by physical insights. We demonstrate how the interplay
of the cusp condition, the asymptotic condition, the electron-correlation,
configuration interaction, and the exact one electron two-center orbitals, can
produce energy results approaching chemical accuracy. To this end, we provide a
tutorial on using the Riccati form of the ground state wave function as a
unified way of understanding the two-electron wave function and collect a
detailed account of mathematical derivations on the exact one-electron
two-center wave functions. Reviews of more traditional calculational
approaches, such as Hartree-Fock, are also given.

The inclusion of electron correlation via
Hylleraas type functions is well known to be important, but difficult
to implement for more than two electrons. The use of the D-scaled Bohr
model offers the tantalizing possibility of obtaining electron correlation
energy in a non-traditional way.

\end{abstract}

\pacs{31.10.+z, 31.15.-p, 31.25.-v, 31.50.-x}

\maketitle

\tableofcontents

\section{Introduction}

\label{sec1}

\subsection{Overview}

We are in the midst of a revolution at the interface between chemistry and
physics, largely due to the interplay between quantum optics and
quantum chemistry. For example, the explicit control of molecules afforded
by modern femtosecond lasers and adaptive computer feedback \cite{Juds92} has
opened new frontiers in molecular science. In such studies, molecules are
controlled by sculpting the amplitude and phase of femtosecond pulses, forcing
the molecule into predetermined electronic and rotational-vibrational states.
This holds great promise for vital applications, from the trace detection of
molecular impurities, such as dipicolinic acid as it appears in anthrax
\cite{PNAS}, to the utilization of molecular excited states for quantum
information storage and retrieval \cite{Niel99}.

We are thus motivated to rethink certain aspects of molecular physics and
quantum chemistry especially with regard to the excited state dynamics and
coherent processes of molecules. The usual discussions of molecular
structure are based on solving the many-particle Schr\"odinger
equation with varying degree of sophistication, from exacting Diffusion
Monte Carlo methods, coupled cluster expansion, configuration interactions, to
density functional theory. All are intensely numerical.
Despite these successful tools of modern computational chemistry,
there remains the need for understanding electron correlations in some
relatively simple way so that we may describe excited states dynamics with
reasonable accuracy.

In this work, we propose to reexamine these questions in two complementary
ways. One approach is based on the recently resurrected Bohr's model
for molecules \cite{Svid04,Svid05}.
In particular, we show that by modifying
the original Bohr's model \cite{Bohr1} in a simple way, specially
when augmented by dimensional scaling (D-scaling), we can describe both
the singlet and triplet potential of H$_2$ with remarkable accuracy
(see Figs. \ref{f2c}, \ref{h2int}).

In another approach, following the lead of the French school
of Le Sech \cite{Le,Sech96}, we use correlated two-center orbitals of
the H$_2^{+}$ molecule to model H$_2$'s ground and excited state.
This approach worked  well, even when only a simple
electron correlation function is used, see Table~\ref{Table:BatesLaguerre}.

\begin{table}[htpb]
\caption{\label{Table:BatesLaguerre} The binding energy of H$_2$ molecule based on
``exact'' two-center  H$_2^+$ orbitals.}
\vskip0.2cm
\centerline{
\begin{tabular}{lll}
\hline
Orbital & Binding energy (eV) & \\
\cline{2-3}
 & No Free  Parameter & 1 Free (screening)\\
 &  & parameter \\
 \hline
 Jaff\'e \eqref{M13.2}& 4.50 & 4.60 \\
 Hylleraas \eqref{VI.61a}& 4.51& 4.62 \\
 \hline
\end{tabular}
}
\vskip0.2cm
{\small When we allow the $\alpha$ and $B$ parameters of Eqs. \eqref{M13.2}
and \eqref{VI.61a} to vary, we obtain a binding
energy of 4.7 eV.  The binding energy is comparable to the experimental value of 4.7
eV. }
\end{table}

The Bohr model and D-scaling technique taken together with good (uncorrelated)
molecular orbitals is especially interesting and promising. As
discussed in subsection C, the Bohr model yields a good approximation to the
electron-electron Coulomb energy, which can be used to choose a renormalized
nuclear charge and a much improved (correlated) two electron wave function.

\begin{figure}[htpb]
\bigskip
\centerline{\epsfxsize=0.35\textwidth\epsfysize=0.35\textwidth
\epsfbox{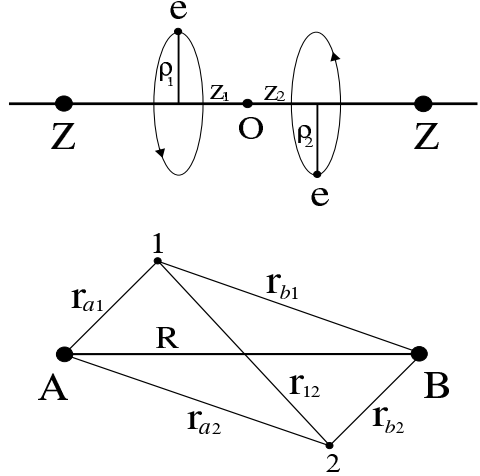}}

\caption{Cylindrical coordinates (top) and electronic distances (bottom) in
H$_2$ molecule. The nuclei $Z$ are fixed at a distance $R$ apart.
In the Bohr model, the two
electrons rotate about the internuclear axis $z$ with coordinates $\rho_1$,
$z_1$ and $\rho_2$, $z_2$ respectively; the dihedral angle $\phi$ between the
($\rho_1,z_1$) and ($\rho_2,z_2$) planes remains constant at either $\phi=\pi$
or $\phi=0$. The sketch corresponds to configuration
2 of Fig. \ref{f2c}, with $\phi=\pi$.
}
\label{f1c}
\end{figure}

\subsection{The Bohr molecule}

\begin{figure}[htpb]
\bigskip
\centerline{\epsfxsize=0.5\textwidth\epsfysize=0.6\textwidth
\epsfbox{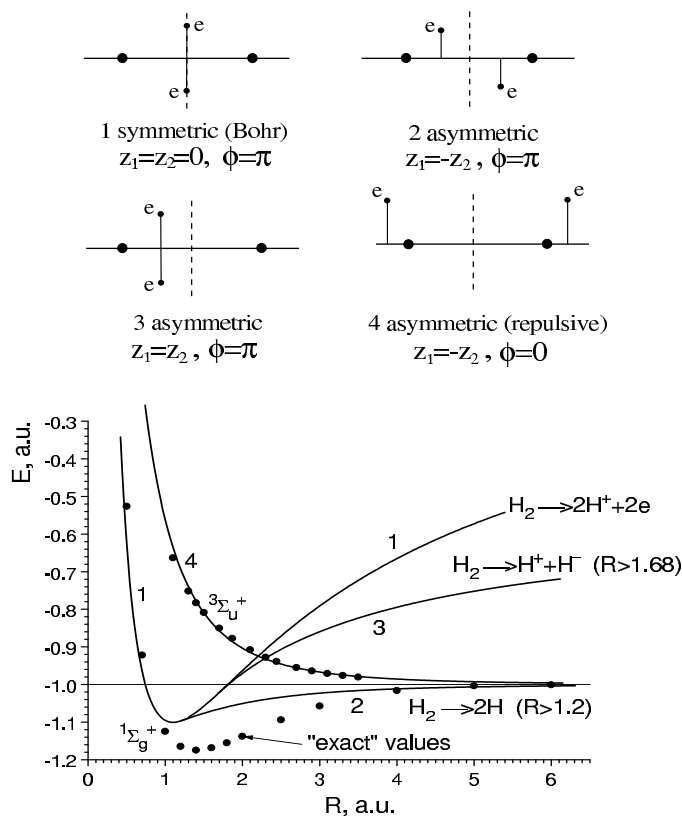}}
\vspace{0.2cm}
\caption{
Energy $E(R)$ of H$_2$ molecule
for four electron configurations (top) as a
function of internuclear distance $R$ calculated within the Bohr model
(solid lines) and the ``exact'' ground $^1\Sigma_g^+$
and first excited $^3\Sigma_u^+$
state energy of Ref. \cite{Kolo60} (dots). Unit
of energy is 1 a.u.$=27.21$ eV, and
unit of distance is the Bohr radius.
}
\label{f2c}
\end{figure}

Figure \ref{f1c} displays the Bohr model for a hydrogen molecule
\cite{Bohr1,Svid04,Svid05}, in which
two nuclei with charges $Z|e|$ are separated by a fixed distance $R$
(adiabatic approximation) and the two electrons move in the space between
them. The model assumes that the electrons move with constant speed on
circular trajectories of radii $\rho _1=\rho _2=\rho $. The circle centers
lie on the molecule axis $z$ at the coordinates $z_1=\pm z_2=z$. The
separation between the electrons is constant. The net force on each electron
consists of three contributions: attractive interaction between an electron
and the two nuclei, the Coulomb repulsion between electrons, and the
centrifugal force on the electron. We proceed by writing the energy function
$E=T+V$, where the kinetic energy $T=p_1^2/2m+p_2^2/2m$ for electrons 1 and
2 can be obtained from the quantization condition that the circumference is
equal to the integer number $n$ of the electron de Broglie wavelengths $2\pi
\rho =nh/p$, so that we have $T=p^2/2m=n^2\hbar ^2/2m\rho ^2$; the unit of
distance
is taken to be
 the Bohr radius $a_0=\hbar ^2/me^2$, and the unit of energy the atomic energy,
 $e^2/a_0$ where $m$ and $e$ are, respectively,  the
 mass and charge of the electron. The Coulomb potential
energy $V$ is given by
\begin{equation}
\label{b1}V=-\frac Z{r_{a1}}-\frac Z{r_{b1}}-\frac Z{r_{a2}}-\frac
Z{r_{b2}}+\frac 1{r_{12}}+\frac{Z^2}R,
\end{equation}
where $r_{ai}$ ($i=1,2$) and $r_{bi}$ are the distances of the $i$th
electron from nuclei A and B, as shown in Fig. \ref{f1c} (bottom), $r_{12}$
is the separation between electrons. In cylindrical coordinates the
distances are%
$$
r_{ai}=\sqrt{\rho _i^2+\left( z_i-\frac R2\right) ^2},\quad r_{bi}=\sqrt{%
\rho _i^2+\left( z_i+\frac R2\right) ^2},
$$
$$
r_{12}=\sqrt{(z_1-z_2)^2+\rho _1^2+\rho _2^2-2\rho _1\rho _2\cos \phi },
$$
here $R$ is the internuclear spacing and $\phi $ is the dihedral angle
between the planes containing the electrons and the internuclear axis. The
Bohr model energy for a homonuclear molecule having charge $Z$ is then given
by
\begin{equation}
\label{b2}E=\frac 12\left( \frac{n_1^2}{\rho _1^2}+\frac{n_2^2}{\rho _2^2}%
\right) +V(\rho _1,\rho _2,z_1,z_2,\phi ).
\end{equation}
Possible electron configurations correspond to extrema of
the energy function (\ref{b2}).
For $n_1=n_2=1$ the energy has extrema at $\rho _1=\rho _2=\rho $%
, $z_1=\pm z_2=z$ and $\phi =\pi $, $0$. These four configurations are
pictured in Fig. \ref{f2c} (upper panel). For example, for configuration 2,
with $z_1=-z_2=z$, $\phi =\pi $, the extremum equations $\partial E/\partial
z=0$ and $\partial E/\partial \rho =0$ read
$$
\frac{Z(R/2-z)}{\left[ \rho ^2+(R/2-z)^2\right] ^{3/2}}+\frac z{4[\rho
^2+z^2]^{3/2}}-
$$
\begin{equation}
\label{b3}\frac{Z(R/2+z)}{\left[ \rho ^2+(R/2+z)^2\right] ^{3/2}}=0,
\end{equation}
$$
\frac{Z\rho }{\left[ \rho ^2+(R/2-z)^2\right] ^{3/2}}+\frac{Z\rho }{\left[
\rho ^2+(R/2+z)^2\right] ^{3/2}}-
$$
\begin{equation}
\label{b4}\frac \rho {4[\rho ^2+z^2]^{3/2}}=\frac 1{\rho ^3},
\end{equation}
which are seen to be equivalent to Newton's second law applied to the motion
of each electron. Eq. (\ref{b3}) specifies that the total Coulomb force on
the electron along the $z-$axis is equal to zero; Eq. (\ref{b4}) specifies
that the projection of the Coulomb force toward the molecular axis equals
the centrifugal force. At any fixed internuclear distance $R$, these
equations determine the constant values of $\rho $ and $z$ that describe the
electron trajectories. Similar force equations pertain for the other
extremum configurations.

In Fig. \ref{f2c} (lower panel) we plot $E(R)$ for the four Bohr model
configurations (solid curves), together with ``exact'' results (dots)
obtained from extensive variational calculations for the singlet ground
state $^1\Sigma _g^{+}$, and the lowest triplet state, $^3\Sigma _u^{+}$
\cite{Kolo60}. In the model, the three configurations 1, 2, 3 with the
electrons on opposite sides of the internuclear axis ($\phi =\pi $) are seen
to correspond to the $^1\Sigma _g^{+}$
singlet ground states, whereas the other
solution 4 with the
electrons on the same side ($\phi =0$) corresponds to the first
excited, $^3\Sigma _u^{+}$ triplet state. At
small internuclear distances, the symmetric configuration 1 originally
considered by Bohr agrees well with the ``exact'' ground state quantum
energy; at larger $R$, however, this configuration's energy
rises far above that of the
ground state and ultimately dissociates to the doubly ionized limit, 2H$^{+}$%
+2e. In contrast, the solution for the asymmetric configuration 2 appears
only for $R>1.20$ and in the large $R$ limit dissociates to two H atoms. The
solution for asymmetric configuration 3 exists only for $R>1.68$ and climbs
steeply to dissociate to an ion pair, H$^{+}+$H$^{-}$. The asymmetric
solution 4 exists for all $R$ and corresponds throughout to repulsive
interaction of two H atoms.

We then extend these ``Bohr molecule'' studies in several ways. In
particular, we use a variant of the dimensional scaling (D scaling) theory as
it was originally developed in quantum chromodynamics and applied with great
success to molecular and statistical physics \cite{Witt80,Hers92}.
This is based on an analysis
in which the usual kinetic energy terms in the Schr\"{o}dinger equation are
written in D dimensions, {\it i.e.},
\begin{equation}
- \frac{\hbar^2}{2m} \sum_{i=1}^{3} \frac{\partial^2}{\partial x_i^2}
\rightarrow - \frac{\hbar^2}{2m} \sum_{i=1}^{D} \frac{\partial^2}{\partial
x_i^2}
\end{equation}
This provides another avenue into the interface between the old
(Bohr-Sommerfeld) and the new (Heisenberg-Schr\"{o}dinger) quantum
mechanics. In particular, when $D\rightarrow \infty$ the two electron
Schr\"{o}dinger equation can be scaled and sculpted into a form which is
when $D\rightarrow \infty$ identical to the Bohr theory of the H$_2$
molecule, and much better than when $1/D$ or other corrections are included.

\subsection{Simple correlation energy from the Bohr model}

The Bohr model offers an effective way to treat most of the correlation
energy absent in the conventional Hartree--Fock (HF) approximation. Here we
show how a charge renormalization method can be applied to improve the
ground state energy obtained in the HF approximation. We start from He-like
ions and consider a nucleus with charge $Z$ and two electrons moving around
it. According to the Bohr model the ground state energy is given by the
minimum of the following expression
\begin{equation}
\label{b6}E=\frac 12\left( \frac 1{\rho _1^2}+\frac 1{\rho _2^2}\right)
-\frac Z{\rho _1}-\frac Z{\rho _2}+\frac 1{\sqrt{\rho _1^2+\rho _2^2-2\rho
_1\rho _2\cos \phi }},
\end{equation}
where $\phi $ is the dihedral angle between electrons. At the minimum $\phi
=\pi $ and Eq. (\ref{b6}) reduces to
\begin{equation}
\label{b7}E=\frac 12\left( \frac 1{\rho _1^2}+\frac 1{\rho _2^2}\right)
-\frac Z{\rho _1}-\frac Z{\rho _2}+\frac 1{\rho _1+\rho _2}.
\end{equation}
Optimization with respect to $\rho _1$ and $\rho _2$ yields
\begin{equation}
\label{b8}\rho _1=\rho _2=\frac 1{Z-\frac 14},\qquad E_B=-\left( Z-\frac
14\right) ^2.
\end{equation}
The HF approximation in the framework of the Bohr model means that optimum
parameters $\rho _1$ and $\rho _2$ are determined by minimization of Eq. (%
\ref{b7}) with no electron repulsion term, i.e., omitting the
electron-electron correlation. In the HF approximation the Bohr model gives
\begin{equation}
\label{b9}\rho _1=\rho _2=\frac 1Z,\qquad E_{\text{B-HF}}=-Z^2+\frac Z2.
\end{equation}
For the He atom ($Z=2$) we obtain $E_{\text{B-HF}}=-3$ a.u., while $%
E_B=-3.0625$ a.u. Thus, the inclusion of correlation shifts the ground state
energy down by
\begin{equation}
\label{b10}E^{{\rm B}}_{\text{corr}}=-\frac 1{16}=-0.0625\text{ a.u.}
\end{equation}
The Bohr model itself is quasiclassical and, as a consequence, it predicts
the He ground state energy with only $5.4\%$ accuracy ($E_{\text{exact}%
}=-2.9037$ a.u.). However, the Bohr model provides a quantitative way to
include the correlation energy. Let us consider the He ground state energy
calculated using the HF (effective charge) variational wave function
\begin{equation}
\label{b11}\Psi (r_1,r_2)=C\exp [-\tilde Z(r_1+r_2)],
\end{equation}
where $\tilde Z$ is a variational parameter (effective charge), which is
determined by minimizing the energy $E=\tilde Z^2-2Z\tilde Z+5\tilde Z/8$, $%
\tilde Z=Z-5/16.$ The wave mechanical HF energy is
\begin{equation}
\label{b12}E^{{\rm W}}_{\text{HF}}=-\left( Z-\frac 5{16}\right) ^2.
\end{equation}
For $Z=2$ we obtain
$E^{{\rm W}}_{\text{HF}}=-2.8476$ a.u. The difference between
$E^{{\rm W}}_{\text{HF}}$
and the exact value is due to the correlation energy missing in
the HF treatment. One can notice that if we add the correlation energy (\ref
{b10}) to $E^{{\rm W}}_{\text{HF}}$ we obtain
$$
E^{{\rm W}}_{\text{HF}}+E^{{\rm B}}_{\text{corr}}=-2.9101\text{ a.u.,}
$$
which substantially improves the answer and deviates by only $0.2\%$ from
the exact value. Such an idea can be incorporated by renormalization of the
nuclear charge \cite{Kais93}. Let us define an effective charge $Z_{\text{eff%
}}$ by the condition
\begin{equation}
\label{b13}E_{\text{B-HF}}(Z_{\text{eff}})=E^{{\rm W}}_{\text{HF}}(Z),
\end{equation}
which yields
\begin{equation}
\label{b14}Z_{\text{eff}}=\frac 14+\sqrt{\frac 1{16}+\left( Z-\frac
5{16}\right) ^2}.
\end{equation}
The effective charge improves the Bohr model energy by taking into account
the difference between the quasiclassical and fully quantum mechanical
description. The effective charge is calculated from the correspondence
between the Bohr model in the HF approximation and the quantum mechanical HF
answer. Now, if we take the Bohr model energy $E_B(Z)$ (\ref{b8}) (that
includes correlation) but with $Z_{\text{eff}}$ instead of $Z$ it improves
the quantum mechanical HF answer:
\begin{equation}
\label{b15}E_B(Z_{\text{eff}})=-\left( Z-\frac 5{16}\right) ^2-\frac
1{16}=E^{{\rm W}}_{\text{HF}}(Z)-\frac 1{16}.
\end{equation}
The correction energy $-1/16$ is independent of $Z$ and coincides with Eq. (%
\ref{b10}). Table \ref{HF} compares the quantum mechanical HF answer for the
ground state energy of He-like ions and the improved value (\ref{b15}).
Depending on $Z$ Eq. (\ref{b15}) improves the accuracy $10-20$ times.

\begin{table}
\caption{\label{HF}Ground state energy of the He-like ions in the HF approximation
$E^{{\rm W}}_{\rm HF}$
and the value improved by the Bohr model $E_{B}$. The last two columns
compare the accuracy of the HF and the improved result.}
\centering
\begin{tabular}{|c|c|c|c|c|c|c|}
\hline
$Z$ & $E^{{\rm W}}_{\text{HF}}(Z)$ & $Z-Z_{\text{eff}}$ &
$E_B(Z_{\text{eff}})$ & $E_{\text{exact}}$ & $\Delta E_{\text{HF}},\text{ \%}$ &
$\Delta E_B(Z_{
\text{eff}}), \text{ \%}$
\\
\hline
2 & -2.8476 & 0.0441 & -2.9101 &
-2.9037 & 1.93 & 0.22
\\
\hline
3 & -7.2226 & 0.0508 & -7.2851 & -7.2799 & 0.79 & 0.072
\\
\hline
4 & -13.597 & 0.0540 & -13.6602 & -13.6555 & 0.42 & 0.033
\\
\hline
5 & -21.9726 & 0.0558 & -22.0351 & -22.0309 & 0.26 &
0.019
\\
\hline
6 & -32.3476 & 0.0570 & -32.4102 & -32.4062 & 0.18 & 0.012
\\
\hline
7 & -44.7226 & 0.0578 & -44.7852 & -44.7814 & 0.13 &
0.008
\\
\hline
8 & -59.0976 & 0.0584 & -59.1602 & -59.1566 & 0.10 & 0.006
\\
\hline
9 & -75.4726 & 0.0589 & -75.5352 & -75.5317 & 0.08 &
0.004
\\
\hline
\, 10 \, & \, -93.8476 \, & \, 0.0592 \, & \,
-93.9102 \, & \, -93.9068 \, & 0.06 & 0.003
\\
\hline
\end{tabular}
\end{table}

\subsection{Correlated two-center orbitals}

>From the preceding discussion it is clear that we need good (hopefully
simple) HF wave functions. There is, of course, a great deal of work on this
problem but we find the two-center orbital approach of Le Sech and coworkers
\cite{ABB,A-FL,SL,Le,Sech96}
and of Patil \cite{PTT} to be especially useful. In a previous publication
\cite{ScullyEtAl}, we attempted a first principle (semi-tutorial) presentation
employing that the exact two-center orbitals obtained from solving the
Schr\"odinger equation for the H$_2^{+}$ ion. As shown by Le Sech these are
the most useful building blocks for constructing the electronic
wave functions of the homonuclear H$_2$ molecule. One simple form of the
electronic ground state constructed with two-center orbitals is
\begin{equation}
\label{Eq:psiH212}\Psi _{{\rm H}_2}(1,2)=\Psi _{{\rm H}_2^{+},1\sigma
}(1)\Psi _{{\rm H}_2^{+},1\sigma }(2)\chi _{00}\left( 1+\frac
12r_{12}\right) ,
\end{equation}
where $\Psi _{{\rm H}_2^{+},1\sigma }$ is the solution of the Schr\"odinger
equation for the H$_2^{+}$ in prolate-spheroidal (ellipsoidal) coordinates, $%
\chi _{00}=[|\uparrow _1\downarrow _2\rangle -|\downarrow _1\uparrow
_2\rangle ]/\sqrt{2}$ is singlet spin function, and $(1+\frac 12r_{12})$ is
the Hylleraas correlation factor. See more detailed discussions of
\eqref{Eq:psiH212} in Sections IV--VI below.
This wave function yields a binding energy
of 4.5 eV for H$_2$ molecule without any variational parameters. Variation
with respect to a couple of parameters in the function~(\ref{Eq:psiH212})
shifts the binding energy to 4.7 eV, giving remarkable agreement with the
experimental value. To achieve the same result, sums over many one-centered
atomic orbitals or Hylleraas type of wave function (cf.\ \eqref{eqA.30} below)
 that explicitly include
the interelectronic distance are usually used. This has been demonstrated by
the earlier work of Kolos and Wolniewicz \cite{KW}. Kolos and Szalewicz \cite
{KS}, and that of James and Coolidge \cite{james}, respectively.

In these studies the introduction of a correlation factor, taking into
account the Coulomb interaction between the two electrons, is naturally
motivated by considering the trial wave function as broken into three parts,
we write
\begin{equation}\label{I.17}
\Psi (r_1,r_2)=\Psi (r_1)\Psi (r_2)f(r_1,r_2),
\end{equation}
where $\Psi (r_1)$ and $\Psi (r_2)$ are exact one electron solutions in the
absence of interaction between electrons. For $\Psi (r_1,r_2)$ the
Schr\"odinger equation  (in atomic units) gives
\begin{multline}
\Psi(r_2) f(r_1,r_2) \left( -\frac{\nabla_1^2}{2} - \frac{Z_a}{r_{a1}}
-\frac{Z_b}{r_{b1}}
\right)\Psi(r_1)
+\Psi(r_1) f(r_1,r_2) \left( -\frac{\nabla_2^2}{2} - \frac{Z_a}{r_{a2}}
-\frac{Z_b}{r_{b2}}
\right)\Psi(r_2) \\
+ \Psi(r_1)\Psi(r_2) \left( -\frac{\nabla_1^2}{2} - \frac{\nabla_2^2}{2}
+\frac{1}{r_{12}}
\right) f(r_1,r_2) + \mbox{cross terms} \\
= \left( E - \frac{Z_a Z_b}{R}\right)\Psi(r_1)\Psi(r_2)f(r_1,r_2)\,.
\end{multline}
where cross terms mean terms that go as $\nabla _1\Psi (r_1)\cdot\nabla
_1f(r_1,r_2)$, etc. The solution with only the first two terms is just that
for H$_2^{+}$. The functions $\Psi (r_1)$ and $\Psi (r_2)$ exponentially
decay at large distances from the nuclei. The third term corresponds to the
Schr\"odinger equation for two free electrons,
\begin{equation}
\label{Eq:f}\left( -\frac{\nabla _1^2}2-\frac{\nabla _2^2}2+\frac
1{r_{12}}\right) f(r_1,r_2)=\varepsilon f(r_1,r_2)
\end{equation}
that is the solution to Eq.~(\ref{Eq:f}) is well known \cite{Landau} and is
given in terms of Coulomb wave functions, {\it i.e.}, confluent
hypergeometric functions (see more detailed discussions
in Subsection V.B), which yields the $(1+r_{12}/2)$ Hylleraas
factor as an asymptotic at small $r_{12}$. In order to place this part of the
present review in perspective and be ready for the paradigm shift from the
old quasi quantum-mechanical Bohr model to the new fully wave-mechanical
Schr\"odinger--Born--Oppenheimer model,
we next give a brief history of the molecular
orbital concepts and computations.

\subsection{Context}

Molecular quantum chemistry is a fascinating success story in the annals of
20th Century science. In 1926, Schr\"odinger introduced the all-important
wave equation which soon bore his name. In the
following year Schr\"odinger's new theory was applied to the simplest
molecular systems of the hydrogen molecular ion H$^+_2$ by Burrau \cite{Bur}
and to the hydrogen molecule H$_2$ by Heitler and London \cite
{Heit27} and Condon \cite{Con}. In the same year, Born and Oppenheimer \cite{BO}
published their important paper dealing with molecular nuclear motion.
Further, in 1928, Hund and Mulliken \cite{HM} presented their venerable
molecular orbital (MO) theory, which provided a powerful computational tool
for chemistry and a foundation for the subsequent
development of modern molecular science.

Diatomic molecules such as H$_2$ and HeH$^+$ are the simplest of all
molecules. Their analysis, modeling and computation constitute the bedrock
of the study of chemical bonds in molecular structures. To quote a recent
insightful article by Cotton and Nocera \cite{CN}:

\begin{quote}
``It can be said without fear of contradiction that the two-electron bond is
the single most important stereoelectronic feature of chemistry.''
\end{quote}

\noindent Indeed, the description of the covalent bond in diatomics,
based on the methods of Heitler--London and Hund--Mulliken,
is one of the crowning achievements of quantum mechanics and
fundamental physics.

Computational quantum chemistry dawned in 1927 with the advent of the
Heitler--London method. However, the accuracy of these early numerical results
were not satisfactory, as can be seen from the following quotation
(Hinchliffe \cite[p.~254]{Hin}):

\begin{quote}
``The calculated bond length and dissociation energy [of MO theory] are in
poor agreement with experiment than those obtained from simple VB [valence
bond] treatment (Table 15.3), and this puzzled many people at the time. It
has also led them to believe that the VB method was the correct way forward
for the description of molecular structure $\ldots$~.''
\end{quote}

New ideas were then proposed to improve the numerical accuracy of the
Heitler--London and Hund--Mulliken method. The first idea of
{\it configuration interaction} (CI) is to incorporate
{\it excited states} into the wave function. The second idea
of {\it correlation} introduces explicit
dependence of the {\it interelectronic distance } in the
wave function. The idea of correlation was first demonstrated by Hylleraas
\cite{hylleraas} in 1929 for the helium atom and by James and
Coolidge \cite{james} in 1933 for H$_2$. The use of configuration
interaction and correlation are key evolutionary steps in
improving the original ideas of Heitler--London and Hund--Mulliken. We can
quote the following from Rychlewski \cite{Ryc}:

\begin{quote}
``$\ldots$ Very soon it has been realized that inclusion of interelectronic
distance into the wave function is a powerful way of improving the accuracy
of calculated results $\ldots$~. Today, methods based on explicitly
correlated wave functions are capable of yielding the ``spectroscopic''
accuracy in molecular energy calculations (errors less than the orders of
one $\mu$ Hartree) $\ldots$~.''
\end{quote}

\noindent
For more than two electrons, it is difficult for most numerical methods
to include electron correlations directly except in
Monte Carlo simulations. When it is possible, as in the two electron case,
excellent results can be achieved with very compact wave functions.

Molecular calculations are inherently more difficult than atomic calculations.
The fundamental difficulty is well stated by Teller and Sahlin \cite{teller}:
\begin{quote}
``The molecular problem has a greater inherent complexity than the
corresponding atomic problem.... In atoms, degeneracy due to spherical
symmetry causes many levels to have nearly the same energy.
This grouping of levels is responsible for the presence of a shell
structure in atoms, and this shell structure is in turn the primary
reason for the striking and simple behavior of atoms and the
consequent successes of the independent-electron approximation for
atomic systems. In passing from atoms to molecules the symmetry is
reduced and the amount of degeneracy for the electronic levels becomes
smaller, and, as a consequence, the power of the independent-particle
model is decreased relative to the atomic case.''
\end{quote}
Nothing illustrates this loss of symmetry, and its consequent loss
of validity of the independent-particle picture better
than the complete failure of the molecular orbital picture to account
for the correct dissociation energy of H$_2$. At large internuclear
separation, the symmetry is greatly reduced, and the independent
occupation of single-particle molecular orbitals fails
catastrophically. This failure can of course be averted by
configuration-interaction, but this extra work makes it
plain that molecular problems are inherently more complicated
than atomic problems. Fortunately, for the investigation of
ground and excited molecular states near equilibrium, one is
far from the dissociation limit; the loss of symmetry complicates
the calculation of the molecular orbital, but the independent
particle model remains a good approximation.

In the case of H$_2$, {\em a natural candidate is
the orbital of the two-center one-electron molecular ion}. Such orbitals
will be referred to as {\em diatomic orbitals} (DO) or, in more
complicated cases, {\em shielded diatomic orbitals\/} (SDO) when shielding
is a factor to be considered. The early (1930s) ansatz wave functions of
James and Coolidge \cite{james} are expressed in terms of
prolate spheroidal coordinates of the two electrons with respect to the two
centers of the diatomic nuclei.	However, their wave functions are not
DOs in that they are not expansions of the exact one electron H$_2^+$
states. Rather, their approach is  CI with a basis conveniently
chosen for numerical evaluation.
Their work is the forerunner of	the Polish quantum chemistry
group \cite{Kolo60,KW,KS} of Kolos, Wolniewicz, etc., which have
achieved the highest accuracy in numerical computation of two-electron
molecules. The high accuracy obtained by Kolos and Wolniewicz in \cite{KW}
is admirable, but as noted by Patil, Tang and
Toennies \cite{PTT},

\begin{quote}
``$\ldots$ It is, however, perhaps somewhat unfortunate that these very
impressive accomplishments have largely discouraged further fundamental
studies on novel approaches to obtain accurate wave functions more directly $%
\ldots$~.''
\end{quote}

\noindent A similar comment was made much earlier by Mulliken \cite{Mul}:

\begin{quote}
``[T]he more accurate the calculations become the more concepts tend to
vanish into thin air.''
\end{quote}

Thus the human quest for comprehension remains, and the recent research
on novel approaches to obtain accurate wave functions have indeed yielded
accurate, physically motivated, and compact two-electron wave functions.
Patil {\it et al.} \cite{Klei98,pat,K, Pati00,Pati03a,Pati03}
have advocated the construction of {\it coalescense} wave functions
by incorporating both cusp and asymptotic conditions. We have provided
a detailed review with simplfied derivations of this development
in Section III.	The other approach is the use of diatomic orbitals.
Historically, the original idea of using DOs as basis for
diatomic molecules seems to begin from the work of Wallis and Hulburt \cite
{WH}. More extensive references and history can be found in the works of
Mclean, Weiss and Yoshimine \cite{mclean},
Teller and Sahlin \cite{teller} and Shull \cite{shull}.
Wallis and Hulburt's result was not particularly successful,
because there was no explicit electron correlation and
solving the two-center wave function was difficult. According to
Aubert, Bessis and Bessis
\cite[Part I, p.~51]{ABB}:
\begin{quote}
``$\ldots$ the use of these functions, i.e., diatomic orbitals (DOs), within
the one-configuration molecular-orbital scheme has not been very successful,
owing to the difficulty of taking into account the interelectronic
interactions and, moreover, owing to the complexity of calculations.''
\end{quote}

The calculation of H$_2^+$ wave function improved over the years, culminating
in the extensive tabulations by Teller and Sahlin \cite{teller}.
In 1974--75, Aubert, Bessis and Bessis published a three-part series
\cite{ABB} detailing how to determine SDOs for diatomic molecules. These
three papers emphasize the determination of shielded DOs.
Surprisingly, the use of DO with correlation to study H$_2$ was not
undertaken until 1981 by
Aubert--Fr\'econ and Le Sech \cite{A-FL}. Le Sech, et al.\ have since then
made further refinements to the method (Siebbeles and Le Sech \cite{SL}, Le
Sech \cite{Le}).

Our study of the DO's approach was motivated by our strong interest
in the modeling and computation of molecules. We were especially attracted
by DOs as a natural and accurate description of chemical bonds.
In Scully {\it et al.}\ \cite{ScullyEtAl}, largely unaware of the prior
work done by Aubert--Fr\'econ and Le Sech of the French school, we obtained
{\em simple correlated} DOs for diatomic molecules with good accuracy.
The present paper represents part of our continued efforts in this direction.
In this work, we will first study the mathematical properties of
wave functions such as their cusp conditions, asymptotic behaviors, and
forms of correlation functions. We summarize methods, techniques
and formulas in a tutorial style, interspersed with
some unpublished results of our own. It it not our intention
to complete an exhaustive review on this vast subject, rather, only
to record developments relevant to our interest in sufficient
details. We apologize in advance for any inadvertent omissions.

\subsection{Outline}

The present paper is
organized as follows:
\begin{itemize}
\item[(i)] In Section \ref{secII}, we present some recent progress of an
interpolated
Bohr model.
\item[(ii)] In Section \ref{sec2a}, we discuss some general and fundamental
properties of atoms and molecules, including the Born--Oppenheimer separation,
the Feynman--Hellman Theorem, Riccati form of the ground state wave function,
proximal and asymptotic conditions, the coalescent construction and
the dissociation limit.
\item[(iii)]  In Section \ref{sec2}, we introduce the basics of the 1-electron
two-centered orbitals from the classic explicit solutions of Hylleraas and
Jaff\'e. The one-centered and multi-centered orbitals will also be reviewed.
\item[(iv)] In Section \ref{secV}, we present the details of the derivations of
the all-important cusp conditions of Kato, and show examples as to how to verify
them in prolate spheroidal coordinates. We also provide a glossary of various correlation
functions that satisfy the interelectronic cusp condition.
\item[(v)] In Section \ref{sec5}, we discuss numerical modeling of diatomic
molecules and compare  results with the classic methods such as the
Heitler--London, Hund--Muliken, Hartree--Fock and James--Coolidge, and the new
approach of two-centered orbitals by Le~Sech, et al.
\item[(vi)] In Section \ref{secVII} we discuss alternative methods for molecular
calculations based on the generalized Bohr model and the dimensional scaling.
\item[(vii)] Finally, in Section \ref{secVIII} we give our conclusions and present an
outlook.
\end{itemize}

\section{Recent progress based on Bohr's model}\label{secII}

The diatomic molecules in a fully quantum mechanical
treatment addressed in Subsection I.E  requires solution of the
many-particle Schr\"odinger
equation. However, such an approach also requires complicated numerical
algorithms. As a consequence, for a few electron problem the results become
less accurate and sometimes unreliable. This is pronounced for excited
electron states when the application of the variational principle is
much less involved. Therefore,
invention of simple and, at the same time, relatively accurate methods of
molecule description is quite desirable.

In this section we discuss a method which is based on the Bohr model and its
modification \cite{Svid04,Svid05}. In particular, we show that for H$_2$ a
simple extension of the original Bohr model \cite{Bohr1} describes the
potential energy curves for the lowest singlet and triplet states just about as
nicely as those from the
wave mechanical treatment.

The simplistic Bohr model provides surprisingly accurate energies for the
ground singlet state at large and small internuclear distances and for the
triplet state over the full range of $R$. Also, the model predicts the
ground state is bound with an equilibrium separation $R_e\approx 1.10$ and
gives the binding energy as $E_B\approx 0.100$ a.u.$=2.73$ eV. The
Heitler--London calculation, obtained from a two-term variational function,
yields $R_e=1.51$ and $E_B=3.14$ eV \cite{Heit27}, whereas the ``exact''
results are $R_e=1.401$ and $E_B=4.74$ eV \cite{Kolo60}. For the triplet
state, as seen in Fig. \ref{f2c}, the Bohr model gives a remarkably close
agreement with the ``exact'' potential curve and is in fact much better than
the Heitler--London result (which, e.g., is 30\% higher for $R=2$). One should
mention that in 1913, Bohr found only the symmetric configuration solution,
which fails drastically to describe the ground state dissociation limit.

The simple Bohr model offers valuable insights for the
 description of other diatomic
molecules. For $N$ electrons the model reduces to finding configurations
that deliver extrema of the energy function
\begin{equation}
\label{b5}E=\frac 12\sum_{i=1}^N\frac{n_i^2}{\rho _i^2}+V({\bf r}_1,{\bf r}%
_2,...,{\bf r}_N,R),
\end{equation}
where the first term is the electron kinetic energy, while $V$ is the
Coulomb potential energy. In such formulation of the model there is no need
to specify electron trajectories nor to incorporate nonstationary electron
motion. Eq. (\ref{b5}) assumes that only at some moment in time the electron
angular momentum equals an integer number of $\hbar $ and the energy is
minimized under this constraint. In the general case, the angular momentum of
each electron changes with time; nevertheless, the total energy remains a
conserved quantity.

\begin{figure}[htpb]
\bigskip
\centerline{\epsfxsize=0.4\textwidth\epsfysize=0.5\textwidth
\epsfbox{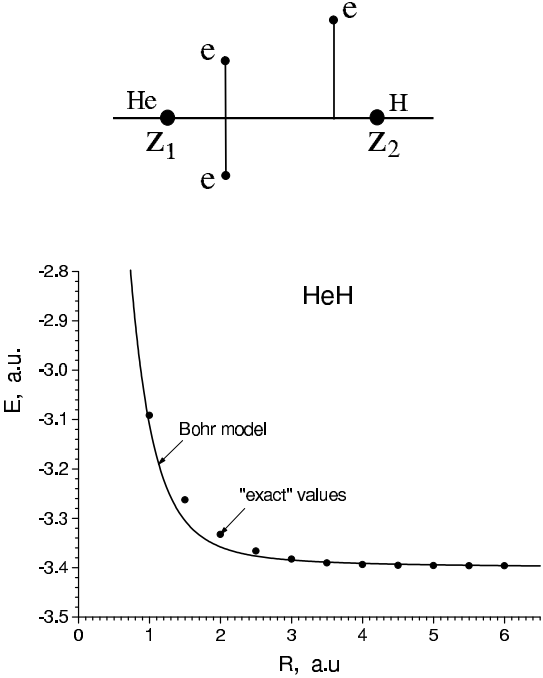}}

\caption{Energy $E(R)$ of HeH molecule
for the electron configuration shown on the upper panel as a
function of internuclear spacing $R$ calculated within the Bohr model
for $n_1=n_2=n_3=1$, $Z_{He}^{\text{eff}}=1.954$ (solid line) and the ``exact''
ground state energy of Ref. \cite{Theo84} (dots).  }
\label{hehcc}
\end{figure}

Let us now discuss the ground state potential curve of HeH. To incorporate the
Pauli exclusion principle one can use a prescription based on the sequential
filling of the electron levels. In the case of HeH the three electrons can
not occupy the same lowest level of HeH$^{++}$. As a result, we must
disregard the lowest potential energy curve $E(R)$ obtained by the minimization
of Eq. (\ref{b5}) and take the next possible electron configuration, which is
shown in Fig. \ref{hehcc} (upper panel). For this configuration,
$n_1=n_2=n_3=1$, however, the right energy corresponds to a saddle point rather
than to a global minimum. In order to obtain the correct dissociation limit
we assign the helium an effective charge $Z_{\text{He}}^{\text{eff}}=1.954$.
The charge matches the difference between the exact ground state energy of
the He atom and the Bohr model result. Fig. \ref{hehcc} shows the ground
state potential curve of HeH in the Bohr model (solid curve) and the
``exact'' result (dots) obtained from extensive variational calculations
\cite{Theo84}. The Bohr model gives a remarkably close agreement with the
``exact'' potential curve.

\subsection{Interpolated Bohr model}

\begin{figure}[htpb]
\bigskip
\centerline{\epsfxsize=0.45\textwidth\epsfysize=0.4\textwidth
\epsfbox{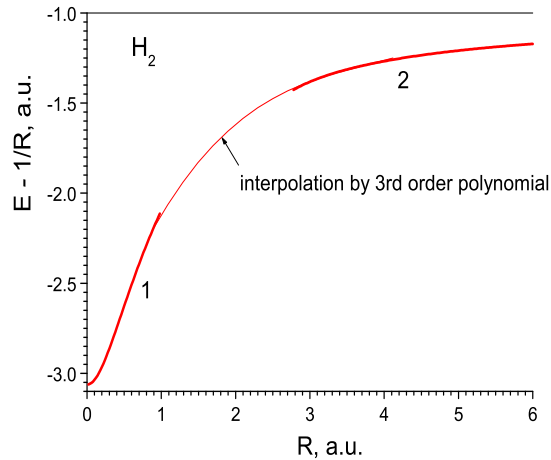}}

\caption{
The Bohr model $E(R)$ for H$_2$ molecule without $1/R$ term. Curves 1 and 2
are obtained based on the quantization relative to the molecular axis (small
$R$) and the nearest nuclei (large $R$) respectively. Thin line is the
interpolation between two regions.
}
\label{h2i}
\end{figure}

\begin{figure}[htpb]
\bigskip
\centerline{\epsfxsize=0.45\textwidth\epsfysize=0.4\textwidth
\epsfbox{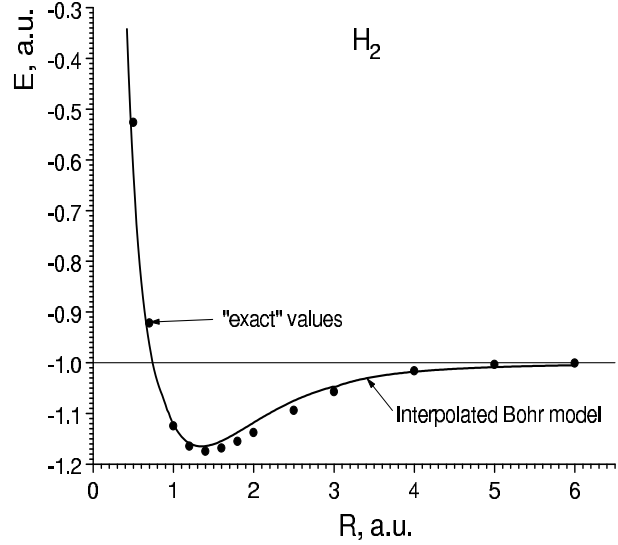}}

\caption{Ground state $E(R)$ of H$_2$ molecule as a
function of internuclear distance $R$ calculated within the interpolated Bohr
model (solid line) and the
``exact'' energy of Ref. \cite{Kolo60} (dots). }
\label{h2int}
\end{figure}

The original Bohr model assumes quantization of the electron angular
momentum relative to the molecular axis. This yields a quite accurate
description of the H$_2$ ground state $E(R)$ at small $R$. However, $E(R)$
becomes less accurate at larger internuclear separation as seen in Fig. \ref
{f2c}. To obtain a better result one can use the following observation. At
large $R$ each electron in H$_2$ feels only the nearest nuclear charge
because the remaining charges form a neutral H atom. Therefore, at large $R$
the momentum quantization relative to the nearest nuclei, rather than to the
molecular axis, must yield a better answer. This leads to the following
expression for the energy of the H$_2\,$ molecule
\begin{equation}
\label{b16}E=\frac 12\left( \frac{n_1^2}{r_{a1}^2}+\frac{n_2^2}{r_{b2}^2}%
\right) -\frac Z{r_{a1}}-\frac Z{r_{b1}}-\frac Z{r_{a2}}-\frac
Z{r_{b2}}+\frac 1{r_{12}}+\frac{Z^2}R
\end{equation}
For $n_1=n_2=1$ and $R>2.8$ the expression (\ref{b16}) has a local minimum
for the asymmetric configuration 2 of Fig. \ref{f2c}. We plot the
corresponding $E(R)$ without the $1/R$ term in Fig. \ref{h2i} (curve 2). At $%
R<2.8$ the electrons collapse into nuclei. At small $R$ we apply the
quantization condition relative to the molecular axis which yields  curve
1 in Fig. \ref{h2i}. To find $E(R)$ at intermediate separation we connect
smoothly the two regions by a third order polynomial (thin line). Addition
of the $1/R$ term yields the final potential curve, plotted in Fig. \ref
{h2int}. The simple interpolated Bohr model provides a remarkably close
agreement with the ``exact'' potential curve over the full range of $R$.

\begin{figure}[htpb]
\bigskip
\centerline{\epsfxsize=0.6\textwidth\epsfysize=0.5\textwidth
\epsfbox{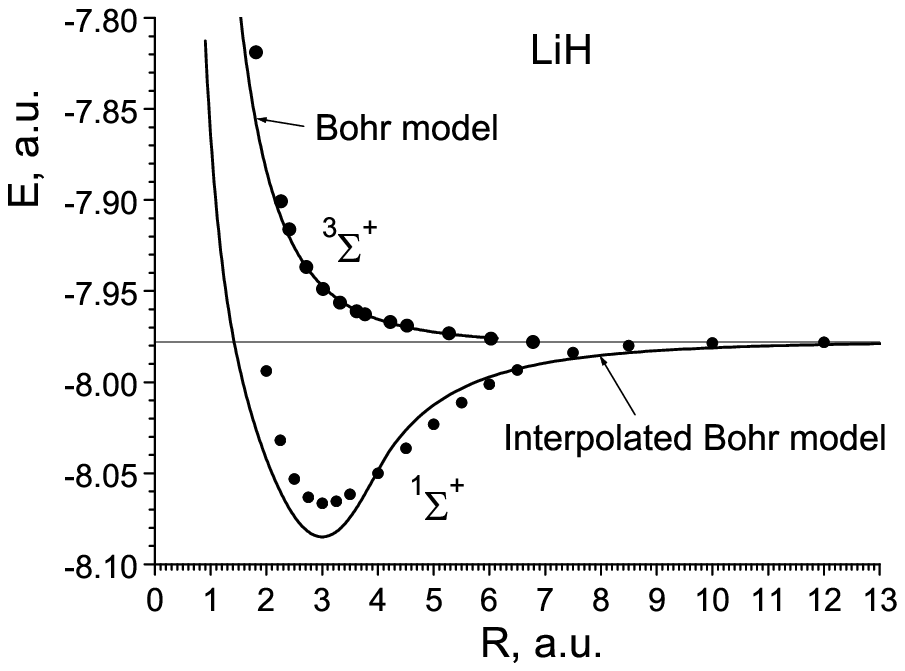}}

\caption{Ground ($^1\Sigma^+$) state energy $E(R)$ of LiH molecule as a
function of internuclear distance $R$ calculated within the interpolated Bohr
model (lower solid line) and the
``exact'' energy of  \cite{Dock72} (dots). Upper solid curve is the first
excited ($^3\Sigma^+$) state energy of LiH obtained from the Bohr model
$^3\Sigma^+_g$ $E(R)$ of H$_2$ molecule by adding
the difference between the ground state energy of Li
and H.}
\label{Lih}
\end{figure}

As an example of application of the interpolated Bohr model to other
diatomic molecules, we discuss the ground state potential curve of LiH. The
Li atom contains three electrons, two of which fill the inner shell. Only the
outer electron with the principal quantum number $n=2$ is important in
forming the molecular bond. This makes the description of LiH similar to the
excited state of H$_2$ in which two electrons possess $n_1=1$ and $n_2=2$.
So, we start from the H$_2$ excited state and apply the interpolated Bohr
model as described above. Then, to obtain $E(R)$ for LiH, we take the H$_2$
potential curve and add the difference between the ground state energy of Li
(-7.4780 a.u.) and H in the $n=2$ state, i.e., add $-7.3530$ a.u. The final
result is shown in Fig. \ref{Lih} (lower solid line), while dots are the
``exact'' numerical answer from \cite {Dock72}. One can see that the simple
interpolated Bohr model provides quite good quantitative description of the
potential curve of LiH, which is already a relatively complex four
electron system.

\section{General results and fundamental properties
of wave functions}\label{sec2a}

\indent

Having examined the quasi-quantum mechanical Bohr model in the preceding two
sections, we now attend to the fully quantum mechanical model and its approximation
and analysis.

\subsection{The Born--Oppenheimer separation}\label{sec2a.1}

\indent

The underlying theoretical basis for problems involving a few particles in
atomic and molecular physics is the Schr\"odinger equation for the electrons and
nuclei, which provides satisfactory explanations of the chemical,
electromagnetic and spectroscopic properties of the atoms and molecules. Assume
that the system under consideration has $N_1$ nuclei with masses $M_K$ and
charges $eZ_K,e$ being the electron charge, for $K=1,2,\ldots, N_1$, and $N_2 =
\sum\limits^{N_1}_{K=1}Z_K$ is the number of electrons. The position vector of
the $K$th nucleus will be denoted as $\pmb{R}_K$, while that of the $k$th
electron will be $\pmb{r}_k$, for $k=1,2,\ldots, N_2$. Let $m$ be the mass of
the electron. The Schr\"odinger equation for the overall system is given by
\begin{align}
H\Psi(\pmb{R},\pmb{r}) &= \left[-\sum^{N_1}_{K=1} \frac{\hslash^2}{2M_K}
\nabla^2_K - \sum^{N_2}_{k=1} \frac{\hslash^2}{2m} \nabla^2_k - \sum^{N_1}_{K=1}
\sum^{N_2}_{k=1} \frac{Z_Ke^2}{|\pmb{R}_K-\pmb{r}_k|} + \frac12
\sum^{N_2}_{\underset{\sst k,k'=1}{k\ne k'}}
\frac{e^2}{|\pmb{r}_k-\pmb{r}_{k'}|}\right.\nonumber\\
\label{eq2a.1}
&\quad \left. + \frac12 \sum^{N_1}_{\underset{\sst K,K'=1}{K\ne K'}}
\frac{Z_KZ_{K'}e^2}{|\pmb{R}_K-\pmb{R}_{K'}|}\right] \Psi(\pmb{R},\pmb{r}) =
E\Psi(\pmb{R},\pmb{r}),
\end{align}
where $H$ is the Hamiltonian and
\[
\pmb{R} = (\pmb{R}_1, \pmb{R}_2,\ldots, \pmb{R}_{N_1}),\quad \pmb{r}  =
(\pmb{r}_1, \pmb{r}_2,\ldots, \pmb{r}_{N_2}).
\]
The above equation is often too complex for practical purposes of studying
molecular problems. Born and Oppenheimer \cite{BO} provide a reduced order model
by approximation, permitting a particularly accurate decoupling of the motions
of the electrons and the nuclei. The main idea is to assume that $\Psi$ in
\eqref{eq2a.1} takes the form of a product
\begin{equation}\label{eq2a.2}
\Psi(\pmb{R},\pmb{r}) = G(\pmb{R}) F(\pmb{R},\pmb{r}).
\end{equation}
Substituting \eqref{eq2a.2} into \eqref{eq2a.1}, we obtain
\begin{align}
&G(\pmb{R})\left\{\left[-\sum^{N_2}_{k=1} \frac{\hslash^2}{2m} \nabla^2_k -
\sum^{N_1}_{K=1} \sum^{N_2}_{k=1} \frac{Z_Ke^2}{|\pmb{R}_K - \pmb{r}_k|} +
\frac12 \sum^{N_2}_{\underset{\sst k,k'=1}{k\ne k'}}
\frac{e^2}{|\pmb{r}_k-\pmb{r}_{k'}|}\right]
F(\pmb{R},\pmb{r})\right\}\nonumber\\
\intertext{\newpage}
&\quad + \left\{\left[- \sum^{N_1}_{K=1} \frac{\hslash^2}{2M_K} \nabla^2_K +
\frac12 \sum^{N_1}_{\underset{\sst K,K'=1}{K\ne K'}}
\frac{Z_KZ_{K'}e^2}{|\pmb{R}_K-\pmb{R}_{K'}|}\right] G(\pmb{R})\right\}
F(\pmb{R},\pmb{r})\nonumber\\
\label{eq2a.3}
&\quad + {\cl T}_1 + {\cl T}_2 = EG(\pmb{R}) F(\pmb{R},\pmb{r}),
\end{align}
where
\begin{align}
\label{eq2a.4}
{\cl T}_1 &\equiv -G(\pmb{R}) \sum^{N_1}_{K=1} \frac{\hslash^2}{M_K} \nabla_K
G(\pmb{R}) \cdot \nabla_KF(\pmb{R},\pmb{r}),\\
\label{eq2a.5}
{\cl T}_2 &\equiv -G(\pmb{R}) \sum^{N_1}_{K=1} \frac{\hslash^2}{2M_K}
\nabla^2_KF(\pmb{R},\pmb{r}).
\end{align}
The essential step in the Born--Oppenheimer separation consists in dropping
${\cl T}_1$ and ${\cl T}_2$ in \eqref{eq2a.3}. This leads to the separation of
the electronic wave function
\begin{equation}\label{eq2a.5a}
\left(-\frac{\hslash^2}{2m} \sum^{N_2}_{k=1} \nabla^2_k + \frac12
\sum^{N_2}_{\underset{\sst k,k'=1}{k\ne k'}} \frac{e^2}{|\pmb{r}_k-\pmb{r}'_k|}
- \sum^{N_1}_{K=1} \sum^{N_2}_{k=1} \frac{Z_Ke^2}{|\pmb{R}_K-\pmb{r}_k|}\right)
F(\pmb{R},\pmb{r}) = E_e(\pmb{R}) F(\pmb{R},\pmb{r}),
\end{equation}
and a second equation for the nuclear wave function
\[
\left[-\sum^{N_1}_{K=1} \frac{\hslash^2}{2M_K} \nabla^2_K + \frac12
\sum^{N_1}_{\underset{\sst K,K'=1}{K\ne K'}} \frac{Z_KZ_{K'}e^2}{|\pmb{R}_K -
\pmb{R}_{K'}|} + E_e(\pmb{R})\right] G(\pmb{R}) = EG(\pmb{R}),
\]
where $E_e(\pmb{R})$ is the constant of separation.

In ``typical" molecules, the time scale for the valence
electrons to orbit about the nuclei is about once every $10^{-15}$ s (and
that of the inner-shell electrons is even smaller), that of the molecular
vibration is about once every $10^{-14}$ s, and that of the molecule
rotation is every $10^{-12} $ s. This difference of time scale is what make $%
{\mathcal{T}}_1$ and ${\mathcal{T}}_2$ in \eqref{eq2a.4} and \eqref{eq2a.5}
negligible, as the electrons move so fast that they can instantaneously
adjust their motions with respect to the vibration and rotation movements of
the slower and much heavier nuclei.

The Born--Oppenheimer separation breaks down in several cases, chief
among them when the nuclear motion is strongly coupled to electronic
motions, e.g., when the Jahn--Teller effects \cite{JT1,JT2} are present. It
also requires corrections for loosely held electrons such as
those in Rydberg atoms.

Research work on non-Born--Oppenheimer effects, the inclusion of the
spin-orbit coupling, and on models without using the Born--Oppenheimer
separation by treating the coupled dynamical motions of the electrons and
nuclei simultaneously, may be found in Yarkony \cite{Ya1} and \"Ohrn \cite{Oh1}%
, for example.

\subsection{Variational properties:\ the virial theorem and the
Feynman--Hellman \protect\newline theorem}

\indent

The variational form of \eqref{eq2a.1} is
\begin{equation}
\label{eq2a.6}\min _{\langle \Psi |\Psi \rangle =1}\langle \Psi |H|\Psi
\rangle =E,
\end{equation}
where the trial wave functions $\Psi =\Psi (\pmb{R},\pmb{r})$ belong to a
proper function space (which is actually the Sobolev space $H^1({\mathbb{R}}
^{3(N_1+N_2)})$ in the mathematical theory of partial differential
equations (\cite[Chapter 2]{chen2})). The (unique) solution $\Psi _0$ attaining the minimum of $%
\langle \Psi |H|\Psi \rangle $ is called the ground state and the associated
value $E_0\equiv \langle \Psi _0|H|\Psi _0\rangle $ is the corresponding
ground state energy. Excited states $\Psi _k$ with successively higher
energy levels may be obtained recursively through
\begin{equation}
\label{eq2a.7}\left.
\begin{array}{l}
\min \langle \Psi |H|\Psi \rangle \equiv E_k=\langle \Psi _k|H|\Psi
_k\rangle , \\
\text{where $\Psi $ is subject to the constraints} \\ \langle \Psi |\Psi
\rangle =1,\langle \Psi |\Psi _0\rangle =\langle \Psi |\Psi _1\rangle
=\cdots =\langle \Psi |\Psi _{k-1}\rangle =0,\text{ for }k=1,2,3,\ldots ~.
\end{array}
\right\}
\end{equation}

There are two useful theorems related to the above variational formulation:\
the virial theorem and the Feynman--Hellman theorem. We discuss them below.

Let $\psi $ be any trial wave function. The expectation value of the kinetic
energy is
\begin{align}
E_{kin} &\equiv \left\langle \Psi\left| -\sum^{N_1}_{K=1} \frac{\hslash^2}{2M_K}
\nabla^2_K - \sum^{N_2}_{k=1} \frac{\hslash^2}{2m} \nabla^2_k\right|
\Psi\right\rangle
\label{eq2a.8}
\end{align}

Now, consider the scaling of all spatial variables by $1+\lambda $:
\begin{equation}
\label{eq2a.9}\pmb{\cl R}\equiv (\pmb{R},\pmb{r})\longrightarrow (1+\lambda
)(\pmb{R},\pmb{r})\equiv (1+\lambda )\pmb{\cl R}.
\end{equation}
Subject to the above transformation \eqref{eq2a.9}, we have
$$
\widetilde{E}_{kin}=\frac 1{(1+\lambda )^2}E_{kin}.
$$
By taking the variation of $\widetilde{E}_{kin}$ with respect to $\lambda $,
we see that it is the same as taking the variation of $E_{kin}$ with respect
to $\Psi $. Thus
$$
\frac d{d\lambda }\widetilde{E}_{kin}\Big|_{\lambda =0}=\delta E_{kin},
$$
\begin{equation}
\label{eq2a.10}\frac d{d\lambda }\left[ \frac 1{(1+\lambda
)^2}E_{kin}\right] \bigg|_{\lambda =0}=-2(1+\lambda )^{-3}\Big|_{\lambda
=0}\cdot E_{kin}=-2E_{kin}=\delta E_{kin}.
\end{equation}
For the potential energy, the expectation is given by
\begin{equation}
\label{eq2a.11}E_{pot}\equiv \langle \Psi |V|\Psi \rangle \equiv \langle
V\rangle ,
\end{equation}
where $V$ consists of the 3 summations of Coulomb potentials inside the
bracket of \eqref{eq2a.1} (but $V$ can be allowed to be any potential whose
negative gradient is force). The spatial scaling \eqref{eq2a.9} implies the
change of displacement $\delta {\mathcal{}}=\lambda {\mathcal{}}$, which is
now regarded as infinitesimal. Therefore,
\begin{align}
\delta E_{pot} &= \frac{d}{d\lambda} \int\limits_{{\bb R}^{3N}} \psi^*(\pmb{\cl
R})
V((1+\lambda) \pmb{\cl R}) \psi(\pmb{\cl R}) d\pmb{\cl R})\Big|_{\lambda=0}
\nonumber\\
&= \int\limits_{{\bb R}^{3N}} \psi^*(\pmb{\cl R})[\pmb{\cl R} \cdot \nabla
V(\pmb{\cl R})] \psi(\pmb{\cl R}) d\pmb{\cl R}\nonumber\\
\label{eq2a.12}
&= \langle \pmb{\cl R}\cdot \nabla V\rangle \equiv \text{Virial},
\end{align}
where Virial is the quantum form of the classical virial. For $\Psi =\Psi _0$%
, the variational form \eqref{eq2a.6} demands that
\begin{equation}
\label{eq2a.13}\delta \langle H\rangle =\delta E_{kin}+\delta E_{pot}=0.
\end{equation}
Substituting \eqref{eq2a.10} and \eqref{eq2a.12} into \eqref{eq2a.13}, we
obtain
\begin{equation}
\label{eq2a.14}-2E_{kin}+\text{Virial}=0.
\end{equation}
This is the general form of the virial theorem for an isolated system of an
atom or a molecule. In the particular case that
\[
V(\pmb{\cl R}) = -\sum^{N_1}_{K=1} \sum^{N_2}_{k=1} \frac{Z_ke^2}{|\pmb{R}_K -
\pmb{r}_k|} + \frac12 \sum^{N_2}_{\underset{\sst k,k'=1}{k\ne k'}}
\frac{e^2}{|\pmb{r}_k - \pmb{r}_{k'}|} + \frac12 \sum^{N_1}_{\underset{\sst
K,K'=1}{K\ne K'}} \frac{Z_KZ_{K'}e^2}{|\pmb{R}_K - \pmb{R}_{K'}|},
\]
which is the power law ${\cl R}^n$ (where ${\cl R}$ is the distance)
 with $n=-1$, the classical virial property
holds:
\begin{equation}\label{eq2a.15}
\text{Virial } = n E_{pot} = -E_{pot}
\end{equation}
as
\[
\pmb{\cl R} \cdot \nabla {\cl R}^n = (n{\cl R}^{n-1}) \pmb{\cl R}\cdot \nabla
{\cl R} = n{\cl R}^n.
\]
>From \eqref{eq2a.14} and \eqref{eq2a.15}, we obtain the virial theorem for an
(exact, not trial) wave function:
\begin{equation}\label{eq2a.16}
2E_{kin} + E_{pot} = 0,
\end{equation}
or
\begin{equation}\label{eq2a.17}
E_{kin} = -\frac12 E_{pot}.
\end{equation}
This property is used as a check for accuracy of calculations and the
properness of the choices of trial wave functions.

The next, Feynman--Hellman theorem, shows how the energy of a system varies
when the Hamiltonian changes.

Assume that the Hamiltonian of an atom or molecule system depends on a
parameter $\alpha $, $H=H(\alpha )$. For example, $\alpha $ may represent
the internuclear distance of the molecular ion H$_2^{+}$. The exact
wave function $\Psi =\Psi (\alpha )$ also depends on $\alpha $, so does the
energy of the system $E=E(\alpha )$. Let's see how $E(\alpha )$ changes with
respect to $\alpha $, i.e.,
\begin{align*}
dE(\alpha)/d\alpha &= \frac{d}{d\alpha} \int \Psi^*(\alpha) H(\alpha)
\Psi(\alpha)d\pmb{r}\\
&= \int \frac{\partial\Psi^*(\alpha)}{\partial\alpha} H(\alpha) \Psi(\alpha) d
\pmb{r} + \int \Psi^*(\alpha) \frac{\partial H(\alpha)}{\partial\alpha}
\Psi(\alpha) d\pmb{r} + \int \Psi^*(\alpha) H(\alpha)
\frac{\partial\Psi(\alpha)}{\partial\alpha} d\pmb{r}\\
&= E(\alpha) \int \frac{\partial\Psi^*(\alpha)}{\partial\alpha} \Psi(\alpha)
d\pmb{r} + \int \Psi^*(\alpha) \frac{\partial H(\alpha)}{\partial\alpha}
\Psi(\alpha) d\pmb{r} + E(\alpha) \int \Psi^*(\alpha)
\frac{\partial\Psi(\alpha)}{\partial\alpha} d\pmb{r}\\
&= E(\alpha) \frac{d}{d\alpha} \langle\Psi(\alpha)|\Psi(\alpha)\rangle + \int
\Psi^*(\alpha) \frac{\partial H(\alpha)}{\partial\alpha} \Psi(\alpha)d\pmb{r}\\
&= \left\langle\Psi(\alpha)\left| \frac{\partial H(\alpha)}{\partial\alpha}
\right| \Psi(\alpha)\right\rangle
\end{align*}
as $\langle \Psi (\alpha )|\Psi (\alpha )\rangle =1$ and is independent of $%
\alpha $. The above is the Feynman--Hellman theorem. Its advantage is that
oftentimes $\partial H(\alpha )/\partial \alpha $ is of a very simple form.
For example, for the H$_2^{+}$-like equation \eqref{eqA.1}, upon taking $\alpha
=R$ we have
$$
\frac{\partial H}{\partial R}=-\frac{Z_aZ_b}{R^3}{\bf R+}\frac{Z_b{\bf %
r}_b}{r_b^3}\qquad
$$
and the average force on the nucleus B is%
$$
{\bf F}=-\frac{\partial E}{\partial R}=\frac{Z_aZ_b}{R^3}{\bf R-}%
\left\langle \frac{Z_b{\bf r}_b}{r_b^3}\right\rangle .\qquad
$$
That is the force and, hence, the potential energy curve requires
calculation of $\left\langle Z_b{\bf r}_b/r_b^3\right\rangle $ only. This
substantially simplifies the problem since the matrix element from the $%
\nabla ^2$ is no longer necessary.

\subsection{Fundamental properties of one and two-electron
wave functions}

\subsubsection{Riccati form, proximal and asymptotic conditions}
\label{newsec2}

In this subsection, we introduce the Riccati form of the
ground state wave functions as a unified way of understanding
and deriving various cusp, asymptotic and correlation functions.
This forms the basis by which compact wave functions for diatomic
molecules can be derived. Consider the Schr\"odinger
equation with a spherically symmetric potential in reduced units,
\begin{equation}
-\frac1{2\mu}\nabla^2\psi(\br)+V(r)\psi(\br)=E\psi(\br),
\la{sheq}
\end{equation}
where $\mu=1$ is the central-force case, and $\mu=\frac12$ is

the equal-mass, relative coordinate case. We will be primary

interested
in
\be
V_c(r)=-\frac{Z}{r},\quad \text{(cf.\  \eqref{eq2.12} below)}
\la{ncoul}
\ee
however, the long range Coulomb potential is special in many
ways and we can best understand the Coulomb-potential wave function by
comparing and contrasting it to the short-range, Lennard-Jones
potential
\be
V_{LJ}(r)=\ep_0\left(\frac1{r^{12}}-\frac1{r^{6}}\right).
\la{lj}
\ee
Since the ground state of (\ref{sheq}) is strictly positive and
spherically symmetric, it can always be written as (unnormalized)
\be
\psi(r)=\e^{-S(r)}.
\la{re}
\ee
Substituting this into (\ref{sheq}) gives
the Riccati equation for $S(r)$
\be
\frac1{2\mu}\nabla^2S(r)
-\frac1{2\mu}\nabla S(r)\cdot\nabla S(r)
+V(r)=E_0.
\ee
Since $\nabla S(r)=S\,^\prime(r)\hat\br$ and
$\nabla\cdot\hat\br=2/r$, we have simply,
\be
\frac1{2\mu}S^{\,\prime\prime}+\frac1{\mu r}{S^{\,\prime}}
-\frac1{2\mu}(S^{\,\prime})^2
+V(r)=E_0.
\la{catti}
\ee
The advantage of this equation is that since the RHS is a constant
$E_0$, all the singularities of $V(r)$ must be cancelled by the
derivatives of $S(r)$. In particular, if we are only seeking
an approximate ground state, then a reasonable criterion
would be to require $S(r)$ to cancel the {\it most} singular
term in $V(r)$. For both the Coulomb and the Lennard-Jones case,
the potential is most singular as $r\rightarrow 0$. We will refer to
this as the {\it proximal} limit. For both cases, the singularity
of $V(r)$ is only polynomial in $r$ and therefore we can assume
\be
S(r)=ar^n,
\la{prox}
\ee
giving explicitly
\be
\frac1{2\mu}an(n+1)r^{n-2}
-\frac1{2\mu}a^2n^2r^{2n-2}
+V(r)=E_0.
\la{catti2}
\ee
Note the structure of this equation, if the $V(r)$ has a repulsive
polynomial singularity, then it can always be cancelled by
the $r^{2n-2}$ term, leaving a less singular term $r^{n-2}$ behind.
For example, in the Lennard-Jones case,
the most singular term is cancelled if we set
\be
-\frac1{2\mu}a^2n^2r^{2n-2}
+\frac{\ep_0}{r^{12}}=0,
\ee
yielding, $n=-5$, and for $\mu=\frac12$, the famous McMillan correlation
function for quantum liquid helium,
$$
S(r)=\frac15\sqrt{\ep_0}\,r^{-5}.
$$
However, if $V(r)$ has an attractive singularity, then it can only
be cancelled by the $r^{n-2}$ term, and if $n\le 0$, would
leave behind a {\it more} singular term instead. This means that
an attractive potential $V(r)$ cannot be as singular,
or more singular, than $-1/r^{2}$, otherwise, the Schr\"odinger equation
has no solutions. Fortunately, for the
the Coulomb attraction (\ref{ncoul}),
the $-Z/r$ singularity can be cancelled by setting
\be
\frac1{2\mu}an(n+1)r^{n-2}
-\frac{Z}{r}=0,
\la{cuspz}
\ee
yielding, for $\mu=1$, $n=1$,
$$S(r)=Zr.$$
For electron-electron repulsion
with $\mu=\frac12$ and $Z=-1$, we would have instead
$$S(r)=-\frac12 r.$$
In the Coulomb case, these proximal conditions are known as
{\it cusp} conditions. A more thorough treatment of the cusp condition for
the general case will be given in Subsection V.A and Appendices F and H.
 The proximal criterion of cancelling the
leading singularity of $V(r)$ can be applied generally to any
$V(r)$. The cusp conditions are just special cases for the Coulomb
potential. Note that in (\ref{cuspz}), the Coulomb singularity is
actually cancelled only by the $S\,^\prime/(\mu r)$ term
of the Riccati equation,
$$\frac1{\mu r}{S^{\,\prime}}-\frac{Z}{r}=0,$$
since requiring $S\,^\prime$ to be a constant at the singular
point forces $S\,^{\prime\prime}=0$. Thus the
cusp condition can be stated most succinctly in term of the
Riccati function $S(r)$: {\it wherever the nuclear charge
is located, the radial derivative of $S(r)$ at that point must
be equal to the nuclear charge}.

Next, we consider the {\it asymptotic} limit of $r\rightarrow\infty$.
In the case of short range potential, such that $V(r)\rightarrow 0$
faster than $1/r$, we can just completely ignore $V(r)$ in (\ref{catti}).
Substituting in
$$S(r)=\alpha r+\beta\ln(r)$$ gives
\be
-\frac1{2\mu}\frac{\beta}{r^2}
-\frac1{2\mu}\left(\alpha+\frac\beta{r}\right)^2
+\frac1{\mu r}\left(\alpha+\frac\beta{r}\right)
=E_0.
\la{asymp}
\ee
The constant terms determine
$$\alpha=\sqrt{-2\mu E_0}=\sqrt{2\mu|E_0|}.$$
Setting the sum of $1/r$ terms zero gives,
\be
\frac1{\mu r}\alpha(1-\beta)=0,
\la{oneor}
\ee
which fixes $\beta=1$. The remaining terms will decay
faster than $1/r$ and can be neglected in the
large $r$ limit. Thus the asymptotic wave function for any
short-ranged potential (decays faster than $1/r$) must
be of the form
\be
\psi(r)\rightarrow \frac1{r}\e^{-\alpha r}.
\la{aswf}
\ee
However, for the Coulomb potential $-{\tilde Z}/{r}$
(we wish to reserve the possibility that $\tilde Z$ can be
distinct from $Z$ and unrelated to $E_0$),
we must retain it among the $1/r$ terms in (\ref{oneor}),
\be
\frac1{\mu r}\alpha(1-\beta)-\frac{\tilde Z}{r}=0,
\la{twoor}
\ee
resulting in
\be
\beta=1-\frac{\mu \tilde Z}{\alpha}.
\la{genbet}
\ee
Thus the general asymptotic wave function for a Coulomb potential
has a slower decay,
\be
\psi(r)\rightarrow \frac1{r^\beta}\e^{-\alpha r}.
\la{ascwf}
\ee
For a single electron in a central
Coulomb field, $\mu=1$, $\tilde Z=Z$, $\alpha=Z$,
$\beta=0$, and
\be
\psi(r)\rightarrow\e^{-Z r},
\la{coulbe}
\ee
the decay is the slowest, very different from (\ref{aswf}).
For more than one electron, or more than one
nuclear charge, $\alpha\ne\tilde Z$, $\beta$ does not vanish and
the correct asymptotic wave function is (\ref{ascwf}).
We tend to forget this fact because we are too familiar with
the single-electron wave function, which is the exception,
rather than the norm.

The proximal and asymptotic conditions are very stringent constraints:
{\em wave functions that can satisfy both are inevitably close to
the exact wave functions.} Satisfying the proximal condition alone
is sufficient to guarantee an excellent approximate ground state
for all radial symmetric potentials such as the Lennard-Jones,
the Yukawa $(V =\frac1r e^{-\alpha r}, \alpha>0)$, and the Morse
 potential $(V =e^{-2\alpha r} -2e^{-\alpha r},\, \alpha>0)$. Needless to say,
the proximal condition alone determines the exact ground state
for the Coulomb and the harmonic oscillator
potential. The significance of the proximal condition has always
been recognized. The current interest \cite{pat} in deriving compact
wave functions for small atoms and molecules is based on a renewed
appreciation of the importance of the correct asymptotics wave functions.

\subsubsection{The coalescence wave function}

Consider the case of two electrons orbiting a central Coulomb
field,
\be
\left(-\frac1{2}\nabla^2_1
-\frac1{2}\nabla^2_2-\frac{Z}{r_1} -\frac{Z}{r_2}
+\frac{1}{r_{12}}\right)\psi(r_1,r_2)=E \psi(r_1,r_2).
\la{shtwo}
\ee
Imagine that we assemble this atom one electron at a time. When
we bring in the first electron, its energy is $E_1=-Z^2/2$, with
wave function $\psi(r_1)=\exp(-Zr_1)$ localizing it near the origin.
When electron 2 is still very far away, we can write the two-electron
wave function as
\be
\psi(r_1,r_2)=\e^{-Zr_1-S(r_2)}.
\la{wftwo}
\ee
Substituting this into (\ref{shtwo}) yields
the Riccati equation for $S(r_2)$,
\be
\frac1{2}S^{\,\prime\prime}
-\frac1{2}(S^{\,\prime})^2+\frac1{r_2}{S^{\,\prime}}
-\frac{Z}{r_2}+ \frac{1}{r_{12}}
=E_0+\frac12 Z^2.
\la{catti3}
\ee
The RHS defines the second electron's energy,
$E_2\equiv E_0+\frac12 Z^2$, whose magnitude is just the
first removal or ionization energy. Since electron 2 is far away,
and electron 1 is close to the origin, $r_{12}\approx r_{2}$.
Thus electron 2 ``sees" an effective Coulomb field $-\tilde Z/r_2$ with
$\tilde Z=Z-1$. This is the case envisioned in (\ref{twoor})
with $\alpha=\sqrt{-2E_2}$.
The asymptotic wave function for the second electron is, therefore,
\be
\psi(r_2)\rightarrow r_2^{-\beta}\e^{-\alpha r_2},
\la{coul2}
\ee
with $\beta=1-(Z-1)/\alpha$. Since $\beta$ is always less than one,
the $r_2^{-\beta}$ term is only a minor correction. Its effect can
be accounted for by slightly altering $\alpha$. The important
point here is that the second electron need not have the same
Coulomb wave function as the first electron. This coalescence
scenario would suggest, after symmetrizing
(\ref{wftwo}), the following two-electron wave function:
\be
\psi(r_1,r_2)=\e^{-Z r_1-\alpha r_2}
    +\e^{-\alpha r_1-Zr_2}.
\la{cohewf}
\ee

For the case where there are more than two electrons, one can imagine
 building up the atom
or molecule sequentially one electron at a time.
Each electron would then acquire a different Coulomb-potential wave function.
This sequential, or coalescence scenario of approximating the ground
state, in many cases, resulted in better wave functions than
considering all the electrons simultaneously, which is the traditional
Hartree--Fock point of view; see Subsection VI.C.	In the case of He,
the simple effective charge approximation
$$
\psi(r_1,r_2)=\e^{-Z_{\rm eff} r_1}\e^{-Z_{\rm eff} r_2}
$$
with
\be
Z_{\rm eff}=Z-\frac5{16}=1.6875,
\la{zeff}
\ee
gives $E_{var}=-Z_{\rm eff}^2=-2.8476$, while the ``exact" value is -2.9037.
The standard HF wave function of the form
$$
\psi(r_1,r_2)=\phi(r_1)\phi(r_2)
$$
improves\cite{parr} the energy to $E_{var}=-2.8617$. For comparison, the
coalescent wave function (\ref{cohewf}) can achieve $E_{var}=-2.8674$
at $\alpha=1.286$. Patil \cite{pat}, by restricting $\alpha$ to be consistent
with the output variational energy via $\alpha=\sqrt{-2E_{var}-4}$,
obtained $E_{var}=-2.8671$ at $\alpha=1.317$. All these values of
$\alpha$ are very close to the exact asymptotic value of
$\alpha=\sqrt{-2E_2}=\sqrt{2(0.9037)}=1.344$, lending credence to the
coalescence construction. For arbitrary $Z$, by approximating
$E_0$ by $-Z_{\rm eff}^2$, we can estimate $\alpha$ by
\be
\alpha=\sqrt{2Z_{\rm eff}^2-Z^2}\quad(\,\approx Z-5/8\,).
\la{alest}
\ee
For $Z=2$, this gives $\alpha=1.30$ (for small $Z$, we need to
use the full expression rather than the approximation),
an excellent estimate. This obviates the need for Patil's
self-consistent procedure to determine $\alpha$, and produces
even slightly better results.
The coalescent wave function (\ref{cohewf}) with this choice
for $\alpha$, defines a set of parameter-free
two-electron wave functions for all $Z$. The resulting energy
for $Z=2-10$ is given in Table \ref{coales}.

In 1930, Eckart \cite{eckart} has used wave
functions of the form
\be
\psi(r_1,r_2)=\e^{-a r_1-b r_2}
    +\e^{-b r_1-a r_2}
\la{hewf}
\ee
to compute the energy of a two-electron Z-atom. His resulting energy
functional is
\be
E_{Eck}(Z,a,b)=-Z(a+b)
+\frac1{1+C(a,b)}\left[K(a,b)+\frac12(a^2+b^2)+ ab\,C(a,b)\right],
\la{engeck}
\ee
where
\be
K(a,b)=\frac{ab}{a+b}+\frac{a^2b^2}{(a+b)^3}+\frac{20a^3b^3}{(a+b)^5}
\quad{\rm and}\quad
C(a,b)=\frac{64a^3b^3}{(a+b)^6}.
\ee
He obtained an energy
minimum  $-2.8756613$ for {\em He\/} at $a=2.1832$ and $b=1.1885$. (We have used
his energy expression to re-determine the energy minimum more accurately.)
While the improvement in energy is a welcoming contribution, it
seems difficult to interpret the resulting wave function physically.
How can an electron ``sees" a nucleus with charge greater than 2?

To gain further insight into Eckart's result, and coalescence wave function
in general, we note that (\ref{hewf}) can be rewritten as
\be
\chi(r_1,r_2)=\e^{-A(r_1+r_2)}\cosh\left[B(r_1-r_2)\right],
\la{epch}
\ee
with $A=(a+b)/2$ and $B=(a-b)/2$. This form has the HF
part $\e^{-A(r_1+r_2)}$. If we substitute Eckart's values,
we see that $A=(2.1832+1.1885)/2=1.6859$, which is nearly identical to
the effective charge value (\ref{zeff}). This is not accidental.
If we use the approximation (\ref{alest}) for $\alpha$,
then the coalescence wave function (\ref{cohewf}) would
automatically predict
\be
A=\frac12(Z+\alpha)=Z-\frac5{16}\,!
\la{preda}
\ee
Thus within the class of Eckart wave function (\ref{hewf}),
the coalescence scenario correctly predicts the path of optimal
energy as being along $A=Z_{\rm eff}$.
Moreover, this improvement in energy, which has laid dormant
in Eckart's result for three quarters of a century, can now
be understood as due to the {\it radial} correlation
cosh term in (\ref{epch}), built-in automatically
by the coalescence construction.  This term is the smallest
(=1) when $r_1=r_2$, but is large when the separation
$r_1-r_2$ is large, {\it i.e.}, it encourages
the two electrons to be separated in the radial direction.

This suggests that we should reexamine Eckart's energy functional
in terms of parameters $A$ and $B$. Expanding
(\ref{engeck}) to fourth order in $B$ yields,
\be
E_{Eck}(Z,A,B)=-Z_{\rm eff}^2+(A-Z_{\rm eff})^2
 -\frac38 y+
\frac32 y^2-\frac1{2A} y^2,
\la{eckfour}
\ee
where $y\equiv B^2/A$. If $B=0$, $A=Z_{\rm eff}$ this yields
the effective charge energy $-Z_{\rm eff}^2$. Regarding
the effect of $B$ as perturbing on this fixed choice of $A$,
the $1/A\approx 1/Z$ term can first be ignored.
Minimizing $y$ simply yields
\be
E_{Eck}(Z,A,B)=-Z_{\rm eff}-\frac3{128}-\frac1{128Z_{\rm eff}}
\la{eckrest}
\ee
at
\be
y=\frac{B^2}A=\frac18\,,
\la{bval}
\ee
where we have restored the $1/A$ term.
This remarkably simple result is the content of
Eckart's energy functional. The energy is lower from the
effective charge value by a nearly constant amount $3/128$.
In column three of Table \ref{coales}, we compare this approximate
energy (\ref{eckrest}), with the absolute minimum of the
Eckart's energy functional on the fourth column.
The agreement is uniformly excellent.  By comparison, we see that
the coalescence construction,{\em without invoking any minimization
process}, also gives a very good account of the energy minimum.

\begin{table}
\caption{\label{coales}Ground state energy of the He-like ions
as calculated from various wave functions.
}
\centering
\begin{tabular}{|c|c|c|c|c|c|c|c|}
\hline
$Z$ & $E({\rm Coales})$ & $E({\rm Approx.})$ &
$E({\rm Abs.Min.})$&
$E({\rm exact})$ & $f_{LS}$ & $f_{PJ}$ &
$1+\frac12 r_{12}$
\\
\hline
2 &\, -2.8673 &\, -2.8757 &\,-2.8757 &-2.9037 & -2.9016(3) &-2.9017(4) & -2.8898(7)
\\
\hline
3 &\, -7.2355 &\,-7.2490 &\, -7.2488&-7.2799 & -7.268(1) & -7.271(1) & -7.264(1)
\\
\hline
4 & -13.6072 & -13.6232 &-13.6230 &-13.6555 & -13.637(2) &-13.642(1) & -13.634(2)
\\
\hline
5 & -21.9802 & -21.9978 &-21.9975 &-22.0309 & -22.004(2) &-22.014(2) & -22.011(2)
\\
\hline
6 & -32.3539 & -32.3725 &-32.3723 &-32.4062 & -32.376(3) &-32.386(3) & -32.382(3)
\\
\hline
7 & -44.7280 & -44.7473 &-44.7471 &-44.7814 & -44.740(4) &-44.755(3) &-44.753(3)
\\
\hline
8 & -59.1023 & -59.1221 &-59.1220 &-59.1566 & -59.115(4) &-59.129(4) &-59.127(4)
\\
\hline
9 & -75.4768 & -75.4970 &-75.4969 &-75.5317 & -75.490(6) &-75.502(4) &-75.494(4)
\\
\hline
\, 10 \, & \, -93.8514 \, &\,-93.8719 &-93.8718& \, -93.9068 \, & \, -93.859(6) \,
&\,-93.875(5)\, &\, -93.885(5)\,
\\
\hline
\end{tabular}
\end{table}

\subsubsection{Electron correlation functions}
\label{newsec3}

\begin{figure}[htpb]
\bigskip
\centerline{\epsfxsize=0.8\textwidth\epsfysize=0.7\textwidth
\epsfbox{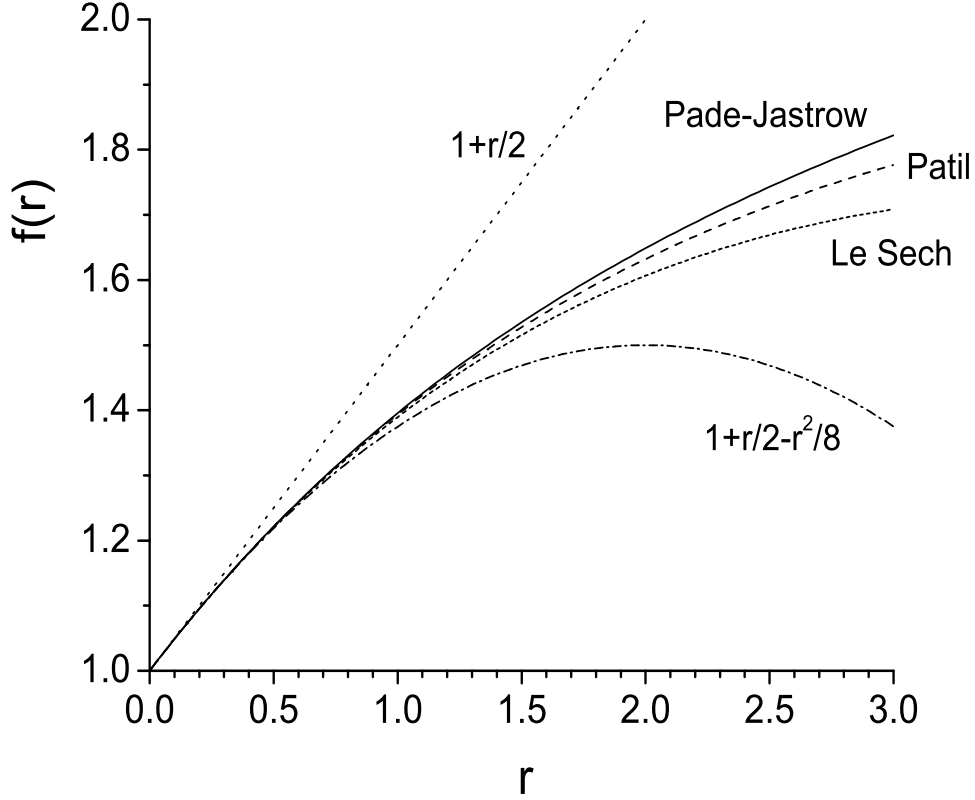}}
\caption{Comparison of three electron-electron correlation
functions.
\label{correl}
}
\end{figure}

Coalescence wave functions are better than HF wave function
because they have built-in radial correlations. To further
improve our description of He, as first realized by Hylleraas \cite{hylleraas},
one can introduce electron-electron correlation directly by forcing the
two-electron wave function to depend on $r_{12}$ explicitly. (A more detailed
discussion of correlation functions will be given in Subsection V.B below.)
 Again,
our analysis is simplified by use of the Riccati function. Let
\be
\psi(\br_1,\br_2)=\e^{-S(r_1,r_2,r_{12})},
\la{telwf}
\ee
but now consider
\be
S(r_1,r_2,r_{12})=Zr_1+Zr_2+g(r_{12}).
\la{hel2}
\ee
We have
\ba
\nabla_1S&=&Z\hat\br_1+g^\prime\hat\br_{12},
\nabla_2S=Z\hat\br_2+g^\prime\hat\br_{21},\nn\\
\nabla^2_1S&=&\frac{2Z}{r_1}
+\frac{2g^\prime}{r_{12}}+g^{\prime\prime},
\nabla^2_2S=\frac{2Z}{r_2}
+\frac{2g^\prime}{r_{12}}+g^{\prime\prime},
\nn
\ea
and (\ref{shtwo}) in terms of $S$ reads
\be
g^{\prime\prime}+\frac{2g^\prime+1}{r_{12}}-(g^\prime)^2
-Zg^\prime(\hat\br_1-\hat\br_2)\cdot\hat\br_{12}-Z^2=E_0.
\la{rcor}
\ee
In order to eliminate the $1/r_{12}$ singularity, we must have
$$\lim_{r_{12}\rightarrow0}\,g^\prime(r_{12})=-\frac12.$$
Thus, one can consider a series expansion for $g(r_{12})$ starting
out as
\be
g(r_{12})=-\frac12 r_{12}+\frac12 C r_{12}^2+...
\nn
\ee
Keeping only up to the quadratic term, in the limit of
$r_{12}\rightarrow 0$, (\ref{rcor}) reads
\be
3C-\frac14-Z^2+O(r_{12})=E_0.
\nn
\ee
If $g(r_{12})$ were exact, the LHS above would be the
constant ground state energy for {\it all} values of $r_{12}$.
Inverting the argument, we can exploit this fact to determine
$C$ at $r_{12}=0$, provided that we can estimate the ground
state energy $E_0$. The simplest estimate for a two
electron atom would be $E_0=-Z^2$, implying that
\be
C=\frac1{12}.
\la{cval}
\ee
However, since the effective charge approximation for the
energy is much better,
we should take instead, $E_0=-Z_{\rm eff}^2$,
thus fixing
\be
C=\frac1{12}+\frac13(Z^2-Z_{\rm eff}^2)
=\frac1{12}+ \frac5{24}\left(Z-\frac5{32}\right).
\la{fixc}
\ee
The determination of the quadratic term of $g(r_{12})$ was
advanced only recently by Kleinekathofer {\it et al.}
\cite{klein}. (If we also
improve the one-electron wave function from
$\exp(-Zr)\rightarrow\exp(-Z_{\rm eff}r)$, then $C$ must go
back to the value (\ref{cval}). The coefficient $C$ therefore
depends on the quality of the single electron wave function.)
The wave function (\ref{telwf}) can now be written as
\be
\psi(\br_1,\br_2)=\e^{-Zr_1}\e^{-Zr_2}f(r_{12}),
\nn
\ee
where
\be
f(r_{12})=\exp\left[\frac12 r_{12}(1-Cr_{12})\right].
\la{exform}
\ee
Since the above argument is only valid for small $r_{12}$,
the large $r_{12}$ behavior of $f(r_{12})$ is not determined.
It seems reasonable, however, barring any long range Coulomb effect,
to assume
\begin{equation}
\lim_{r_{12}\rightarrow 0}f(r_{12})\longrightarrow {\rm constant}.
\la{asbe}
\end{equation}
The form
(\ref{exform}) can have behavior (\ref{asbe}) if we just rewrite it
as
\begin{equation}
f_{PJ}(r_{12})=\exp\left[\frac{r_{12}}{2(1+Cr_{12})}\right].
\la{exform2}
\end{equation}
This {\em Pade-Jastrow form\/} has been used extensively in Monte Carlo
calculations of atomic systems \cite{hamm}. Alternatively, to achieve (\ref{asbe}),
Patil's group \cite{pat,klein} have suggested the form
\be
f_{P}(r_{12})=\frac1{2\lambda}
({1+2\lambda}-\e^{-\lambda r_{12}}); \quad \text{cf. (\ref{secV}.B.(3)) and
Fig.~\ref{ff2},}
\la{patf}
\ee
which has the small $r_{12}$ expansion
\be
f_{P}(r_{12})=
1+\frac12 r_{12}-\frac\lambda{4}r_{12}^2+\cdots~.
\la{patfex}
\ee
By comparing this with similar expansion of (\ref{exform}), we
can identify
\be
\lambda=2C-\frac12=\frac5{12}\left(Z-\frac5{32}\right)-\frac13.
\la{lval}
\ee
Le Sech's group \cite{SL} have employed the form
\be
f_{LS}(r_{12})=1+\frac12 r_{12}\e^{-a r_{12}},
\la{lesec}
\ee
with
$$a=\frac12 \lambda.$$
This function is not monotonic;
it reaches a maximum at $r_{12}=1/a$ before level off back to unity; cf.\
Fig.~\ref{ff1} in Subsection \ref{secV}.B.
However, this point may not be practically relevant,
since most electron separations do not reach beyond the maximum.
For the sake of comparison, we may use the maximum of
Le Sech's correlation function as its asymptotic limit.
This is the natural thing to do because all three functions
can now be characterized by their asymptotic value as
$r_{12}\rightarrow\infty$\,:
\ba
f_{PJ}(r_{12})\rightarrow\exp\left({\frac{1}{2C}}\right)
&\approx& 1+\frac{1}{\lambda+1/2},\nn\\
f_{P}(r_{12})&\rightarrow& 1+\frac{1}{2\lambda},\nn\\
f_{LS}(r_{12})&\rightarrow&1+{\frac{1}{\e\lambda}}.\nn
\ea
Their approaches toward unity are approximately $\lambda^{-1}$,
$\frac12\lambda^{-1}$, and $\frac13\lambda^{-1}$, respectively.
Note also that as
$\lambda$ increases with $Z$, the asymptotic value of $f(r_{12})$
decreases, this is the correct trend long observed in Monte Carlo
calculations on atomic systems \cite{hamm}.
For $Z=2$, we take $C=1/2$, $\lambda=1/2$, $a=1/4$ and
compare all three correlation functions in Fig.~\ref{correl}.
Also plotted is the simple linear and quadratic
forms. The simplest linear correlation function,
$(1+\frac12 r_{12})$, is very distinct from the other three.

We now have all the ingredients needed to construct an optimal
wave function for He. First, with the introduction of
explicit electron correlation function, there is no need for
the radial correlation introduced by the coalescence wave function,
{\it i.e.}, we are free to abandon the radial
correlation function $\cosh(B(r_1-r_2))$. Second, the
asymptotic form of the wave function $\exp(-\alpha r)$ should
be maintained. However, this asymptotic wave function, when
extended back to small $r$, violates the cusp condition.
These concerns can be simultaneously alleviated by replacing
the second electron's wave function via
$$\e^{-\alpha r_2}\rightarrow
\e^{-Zr_2}\cosh(\beta r_2).$$
At small $r_2$, the $\cosh$ function is second
order in $r_2$ and therefore will not affect the cusp condition. At large
$r_2$, $\cosh(\beta r_2)\rightarrow\exp(\beta r_2)$, and the choice
\be
\beta=Z-\alpha
\la{btdef}
\ee
will give back the correct asymptotic wave function.
Upon {\em symmetrization}, we finally arrived at the following
compact wave function for He:
\be
\psi(r_1,r_2,r_{12})=\e^{-Zr_1-Zr_2}
\left[\cosh(\beta r_1)+\cosh(\beta r_2)\right]f(r_{12}).
\la{heopt}
\ee
This wave function, first derived by
Le~Sech \cite{hesech}, satisfies all the cusp and asymptotic conditions.
We fixed $\alpha$, $\beta$ and $C$ by
(\ref{alest}), (\ref{btdef}) and (\ref{fixc}), respectively, and there are no
free parameters.
The only arbitrariness is the form the correlation function
$f(r_{12})$. Since $f_{P}(r_{12})$ is bracketed by $f_{PJ}(r_{12})$ and
$f_{LS}(r_{12})$, we only need to consider the latter two cases.
For $Z=2-10$, the resulting ground state energy for the two electron
atoms are given in Table~\ref{coales}.

\subsubsection{The one-electron homonuclear wave function}
\label{newsec4}

The one-electron, homonuclear two-center Schr\"odinger equation
\be
\left(-\frac1{2}\nabla^2
-\frac{Z}{r_{a}}
-\frac{Z}{r_{b}}\right)\psi(\br)=E\psi(\br),
\la{shellp}
\ee
where $r_a=|\br+\bbr/2|$, $r_b=|\br-\bbr/2|$, can be solved exactly,
as will be shown  in Subsection \ref{sec2}.B below.
However, its ground state wave function
can also be accurately prescribed by proximal and asymptotic conditions.
For $Z=1$, this
is hydrogen molecular ion problem.
As first pointed out by Guillemin and Zener (GZ) \cite{GZ}, when $R=0$,
the exact wave function is
\be
\psi(\br)=\e^{-2Zr}
\la{near}
\ee
and when $R\rightarrow\infty$, the exact wave function is
\be
\psi(\br)=\e^{-Zr_a}+\e^{-Zr_b}.
\la{far}
\ee
They, therefore, propose a
wave function that can interpolate between the two,
\be
\psi_{GZ}(r_a,r_a)=\e^{-Z_1 r_a-Z_2 r_b}
    +\e^{-Z_2 r_a-Z_1 r_b}.
\la{gzwf}
\ee
At $R=0$, $r_a=r_b=r$, one can take
$Z_1=Z_2=Z$. At $R\rightarrow\infty$,
one can choose
$Z_1=Z$ and $Z_2=0$. At intermediate values of $R$,
$Z_1$ and $Z_2$ can be determined variationally. This GZ
wave function gives an excellent description \cite{GZ} of the ground
state of H$_2^+$. As explained  by Patil, Tang and
Toennies \cite{PTT},
another reason why this is a good wave function is that
that (\ref{gzwf}) can satisfy both the cusp and the asymptotic condition.
We can simplify Patil {\it et al}'s discussion by rewriting the GZ wave function,
 again, in the form
\be
\psi_{GZ}(r_a,r_b)=\e^{-A(r_a+r_b)/2}\cosh\left[B(r_a-r_b)/2\right],
\la{gzwfch}
\ee
when $R=0$, $A=2Z$, and when $R\rightarrow\infty$,
$A=B=Z$. The imposition of the
cusp condition can be done most easily in terms of
the Riccati function. We, therefore, write
\be
\psi_{GZ}(r_a,r_b)=\e^{-S(r_a,r_b)},
\la{ripsi}
\ee
with
$$S(r_a,r_b)=A(r_a+r_b)/2-\ln(\,\cosh[B(r_a-r_b)/2]\,).$$
The cusp condition at $r_a=0$ is then easily computed,
\be
\frac{\partial S}{\partial r_a}\Bigl|_{r_a=0,r_b=R}=Z,\quad
A+B\tanh(BR/2)=2Z.
\la{cuspht}
\ee
One can verify that this is also the cusp condition
at $r_b=0$.	 From (\ref{cuspht}), one sees easily that at $R=0$,
$A=2Z$, and when $R\rightarrow\infty$, $A+B=2Z$.
At finite $R$, the asymptotic limit $r\rightarrow\infty$
means that $r_a=r_b=r$ and the GZ wave function
approaches
$$
\psi_{GZ}(r_a,r_b)\rightarrow\e^{-Ar}.
$$
On the other hand, the exact wave function
must be of the form (\ref{ascwf})
\be
\psi(r_a,r_b)\rightarrow \e^{-\sqrt{2|E_0|}r-\beta\ln(r)},
\la{asym}
\ee
with $\beta=1-2Z/\sqrt{2|E_0|}$. Since
$2Z\ge\sqrt{2|E_0|}\ge Z$, we can estimate that
at intermediate values of $R$, $\sqrt{2|E_0|}\approx \frac32 Z$,
suggesting a negligible $\beta\approx -\frac13$. Thus it is
suffice to take
\be
A\approx\sqrt{2|E_0|}
\la{avar}
\ee
Guillemin and Zener have allowed both $A$ and $B$ to be
variational parameters. Patil {\it et al\/}'s
estimate \cite{PTT} of $A$ is essentially that of (\ref{avar}) but with slight
improvement to incorporate the variation due to $\beta\ne 0$.
We adhere to the cusp condition (\ref{cuspht}) but allow $A$ to vary.
In practice, it is easier to just let $B$ vary and fix
A via the cusp condition (\ref{cuspht}). In all cases,
this wave function can provide an excellent description of
the hydrogen molecular ion, with energy derivation only on the
order of  $10^{-3}$ Hartree over the range of $R=0-5$.

\subsubsection{The two-electron homonuclear wave function}
\label{newsec5}

The two-electron homonuclear Schr\"odinger equation is,
\be
\left(
-\frac1{2}\nabla^2_1
-\frac1{2}\nabla^2_2
-\frac{Z}{r_{1a}}
-\frac{Z}{r_{1b}}
-\frac{Z}{r_{2a}}
-\frac{Z}{r_{2b}}+\frac{1}{r_{12}}
\right)\psi(\br_1,\br_2)=E\psi(\br_1,\br_2).
\la{shh2}
\ee
Let's denote the one-electron two-center GZ wave function as
$$
\phi(\br)=\e^{-A\sigma}\cosh(B\delta)
$$
where we have defined
$$
\sigma=\frac{r_a+r_b}2,\quad\delta=\frac{r_a-r_b}2.
$$
(The variables $\sigma$ and $\delta$ here will correspond, respectively, to $\lambda$
and $\mu$ of the prolate spheroidal coordinates in Subsection \ref{sec2}.B.)
To describe the two-electron wave function,
if one were to follow the usual approach, one would begin by
defining the Hartree-Fock like wave function
\be
\psi(\br_1,\br_2)=\phi(\br_1)\phi(\br_2)
=\e^{-A(\sigma_1+\sigma_2)}\cosh(B\delta_1)\cosh(B\delta_2).
\la{mo}
\ee
However, this molecular orbital approach is well known not
to give the correct dissociation limit of H$_2$. In the limit
of $R\rightarrow\infty$, we know that
\be
\phi(\br)\rightarrow \e^{-Zr_a}+\e^{-Zr_b}
\la{rlimit}
\ee
and therefore
\ba
\psi(\br_1,\br_2)&\rightarrow& (\e^{-Zr_{1a}}+\e^{-Zr_{1b}})
                               (\e^{-Zr_{2a}}+\e^{-Zr_{2b}})\nn\\
          &\rightarrow& (\e^{-Zr_{1a}}\e^{-Zr_{2b}}+\e^{-Zr_{1b}}\e^{-Zr_{2a}})
		       +(\e^{-Zr_{1a}}\e^{-Zr_{2a}}+\e^{-Zr_{1b}}\e^{-Zr_{2b}})
\la{lh}
\ea
Only the first parenthesis, the Heilter-London wave function, gives the
correct energy of two well separated atoms with energy
$E=2(-\frac12 Z^2)$. The remaining parenthesis describes, in the case of
H$_2$, the ionic configuration of H$^-$H$^+$, which has higher energy than
two separated neutral hydrogen atoms. Thus the molecular orbital approach
(\ref{mo}) will alway overshoot the correct dissociation limit. This is
a fundamental shortcoming of the molecular orbital approach and cannot
be cured by merely improving the one-electron wave function, {\it i.e.},
by use of the exact one-electron, two-center wave function. Even the
coalescent construction cannot overcome this fundamental problem. In the
large $R$ limit, the inner electron's wave function must be (\ref{rlimit}),
and hence no matter how one constructs the outer electron's asymptotic
wave function, one can never reproduces the Heilter-London wave function.
In both the molecular orbital and the coalescent approach, one must
resort to configuration interaction to achieve the correct dissociation
limit. Even if one were to use the exact one-electron wave function
in doing configuration intereaction, as it was done by Siebbeles and
Le~Sech \cite{SL}, the energy still overshoots the correct dissociation limit
if the correlation $(1+\frac12 r_{12})$ is used. This is because
we must have $f(r_{12})\rightarrow const$ in order to reproduce the
Heilter-London limit.

However, one can learn from Guillemin and Zener's approach,
and insist on a wave function that is correct in both
the $R=0$ and $R\rightarrow\infty$ limit. This seemed a very
stringent requirement, but surprisingly, it is possible. The wave function
is
\be
\psi(\br_1,\br_2)
=\e^{-A(\sigma_1+\sigma_2)}\cosh\left[B(\delta_1-\delta_2)\right]f(r_{12}).
\la{mo2}
\ee
For $R=0$, $\sigma_1=r_1$, $\sigma_2=r_2$, and the above function
reduces to
$$
\psi(\br_1,\br_2)
=\e^{-A(r_1+r_2)}f(r_{12}).
\la{hel3}
$$
which is not a bad description of He. In the limit of $R\rightarrow\infty$,
if we take $A=B=Z$, we have
\ba
\psi(\br_1,\br_2)
&=&(\e^{-Zr_{1a}}\e^{-Zr_{2b}}+\e^{-Zr_{1b}}\e^{-Zr_{2a}})f(r_{12}),\nn\\
&=&(\e^{-Zr_{1a}}\e^{-Zr_{2b}}+\e^{-Zr_{1b}}\e^{-Zr_{2a}}),
\ea
since $f(\infty)\rightarrow 1$. Thus wave function (\ref{mo2}) is the simplest
homonuclear two electron wave function that can describe both limits
adequately.
The wave function (\ref{mo2}) for H$_2$ without $f(r_{12})$
has been derived some time ago by Inui \cite{inui} and Nordsieck \cite{nord}.
However, they were only interested in improving the wave function and energy at
the equilibrium separation and were not concerned with whether the wave function
can yield the correct dissociation limit.

To estimate the form of the correlation function $f(r_{12})$, we repeat
our analysis as in the Helium case. The two-electron wave
function can again be written in the Riccati form (\ref{telwf}),
but now with
\be
S(r_1,r_2,r_{12})=Zr_{1a}+Zr_{2b}+g(r_{12}),
\la{hel4}
\ee
where we have assumed the unsymmetrized form of the Heilter-London
wave function. The resulting equation for $g$ is also similar to
(\ref{rcor}),
\be
g^{\prime\prime}+\frac{2g^\prime+1}{r_{12}}-(g^\prime)^2
-Zg^\prime(\hat\br_{1a}-\hat\br_{2b})\cdot\hat\br_{12}-Z^2+O
\left(\frac1 R\right)=E_0.
\la{hcor}
\ee
In the Helium case, the dot product term
vanishes in the limit of $r_{12}\rightarrow 0$, here it does not.
In the case where the two electrons meet along the molecular axis,
$\hat\br_{1a}=\hat{\bf z}$, $\hat\br_{2b}=-\hat{\bf z}$,
$\hat\br_{12}=-\hat{\bf z}$, the resulting equation
\be
g^{\prime\prime}+\frac{2g^\prime+1}{r_{12}}-(g^\prime)^2
+2Zg^\prime-Z^2+O\left(\frac1 R\right)=E_0
\la{hhcor}
\ee
can be solved by setting $E_0=-Z^2$, and expanding
\be
g(r_{12})=-\frac12 r_{12}+\frac12 C r_{12}^2+...
\nn
\ee
In the limit of
$r_{12}\rightarrow 0$, (\ref{hhcor}) reads
\be
3C-\frac14-Z+O(r_{12})=0,
\nn
\ee
giving,
\be
C=\frac{Z}3+\frac1{12}.
\la{ch2}
\ee
This agrees with Patil {\it et al}'s result \cite{PTT} of
$$\lambda=2C-\frac12=\frac13(2Z-1),$$
but without the need of consulting hypergeometric
functions. For the H$_2$ case, $C=5/12=0.42$. In our
calculation with wave function (\ref{mo2}), with
$f_{PJ}(r_{12})$ given by (\ref{exform2}),
the energy minimum at intermediate values of $R$ is
at $C=0.40$, in excellent agreement with the predicted
value. Since $C\approx 0.50$ for Helium, $C$'s variation with $R$
is very mild.

\begin{figure}[htpb]
\bigskip
\centerline{\epsfxsize=0.9\textwidth\epsfysize=0.7\textwidth
\epsfbox{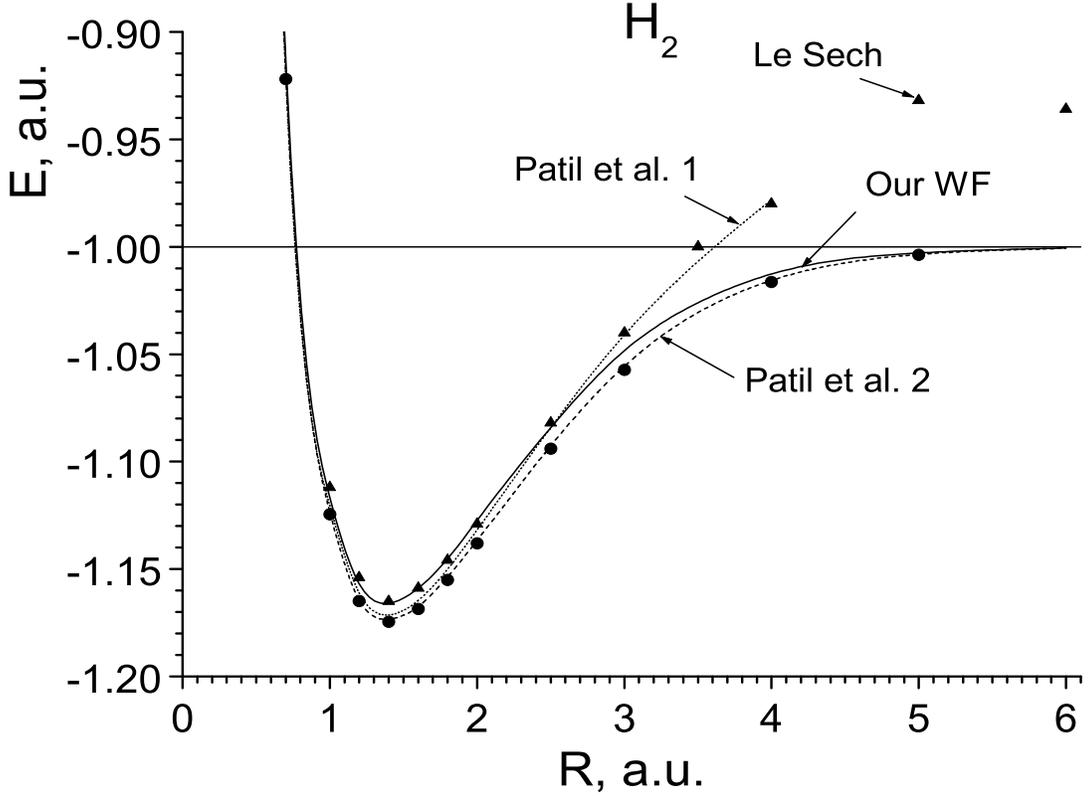}}
\caption{Ground state potential curve $E(R)$ of H$_2$ for different trial
wave functions.
Triangles correspond to ``Le Sech" molecular orbital calculation
with exact H$_2^+$ one-electron orbitals.
Our wave function (\ref{mo2})
(solid line) gives the correct dissociation limit.  Patil et al.'s results
are shown as small dot (function \eqref{pa7},
\eqref{pa9}, \eqref{pa10}) and dash (function \eqref{pa20}) lines.
Large dots are ``exact" values of Ref. \cite{Kolo60}.
}
\label{h2comp}
\end{figure}

The resulting energy for
the wave function (\ref{mo2}), is given in Fig.~\ref{h2comp} (solid line).
We vary the parameter $B$, while the other
parameters $A$ and $C$ are fixed by Eqs. (\ref{cuspht}) and (\ref{ch2}).
The parameter B is 0.8 for $R<2$, and moves graduately
toward one at larger values of $R$. The
energy at equilibrium is as good as Siebbeles and Le~Sech's
calculation \cite{SL} with unscaled
H$_2^+$ wave functions and correlation function
$(1+\frac12 r_{12})$ (triangles). Without configuration interaction, Siebbeles
and Le~Sech's energy overshot the dissociatin limit as shown. The
wave function (\ref{mo2}) can be further improved by adding a
coalescense component \`a la Patil, Tang and Toennies \cite{PTT}.
This will be detailed in the next subsection.

\subsubsection{Construction of trial wave functions by Patil and coworkers}
\label{Patilresults}

\indent

The previous discussion demonstrates that relatively simple wave functions
which incorporate the cusp conditions and the large distance asymptotics,
having no or only a few variational parameters, can be constructed to yield
fairly accurate results. Here we mention other similar trial wave functions
studied in the literature. The importance of the local properties in the
calculation of the chemical bond has been emphasized by Patil and coworkers
\cite{K,PTT,P,Pati00,Pati03a,Pati03}. Their analysis is in the spirit of our
previous discussion, however, yields more complicated trial wave functions.
Here we briefly mention the main aspects of the construction scheme and, to
be specific, consider the ground state of H$_2$ molecule described by the
Schr\"odinger equation (\ref{shh2}).

Let us assume that $r_2\gg r_1$, $R$, $1$. Then $r_{12}\approx r_{2a}\approx
r_{2b}\approx r_2$ and the Hamiltonian reads
\begin{equation}
\label{pa2}\hat H=-\frac 12\nabla _1^2-\frac Z{r_{1a}}-\frac Z{r_{1b}}+\frac{%
Z^2}R-\frac 12\nabla _2^2-\frac{(2Z-1)}{r_2}.
\end{equation}
The first four terms in \eqref{pa2} yield the H$_2^{+}$ problem, while the
last two terms correspond to motion of a particle in a Coulomb potential
with an effective charge $2Z-1$. This Hamiltonian allows us to separate
variables and write $\Psi ({\bf r}_1,{\bf r}_2)=\Psi _{\text{H}_2^{+}}({\bf r%
}_1)\varphi ({\bf r}_2)$, where the function $\varphi ({\bf r}_2)$ satisfies
the equation
\begin{equation}
\label{pa3}\left( -\frac 12\nabla _2^2-\frac{(2Z-1)}{r_2}\right) \varphi (%
{\bf r}_2)=-\varepsilon \varphi ({\bf r}_2),
\end{equation}
$\varepsilon =E_{\text{H}_2^{+}}-E>0$ is the ionization energy of the H$_2$
molecule, and $E$ is the ground state energy of H$_2$. As a result, the
asymptotic behavior of $\Psi ({\bf r}_1,{\bf r}_2)$ at $r_2\gg r_1$, $R$, $1$%
, is
\begin{equation}
\label{pa4}\Psi ({\bf r}_1,{\bf r}_2)\approx r_2^{(2Z-1)/\sqrt{2\varepsilon }%
-1}\exp (-\sqrt{2\varepsilon }r_2)\Psi _{\text{H}_2^{+}}({\bf r}_1),
\end{equation}
which is similar to the coalescence wave function \eqref{coul2} for He. Now
assume that $r_2\gg r_1\gg R$, $1$, then $\Psi _{\text{H}_2^{+}}({\bf r}_1)$
is given by Eq. \eqref{asym} and, therefore,
\begin{equation}
\label{pa6}\Psi ({\bf r}_1,{\bf r}_2)\approx r_1^{2Z/\sqrt{2\varepsilon _1}%
-1}r_2^{(2Z-1)/\sqrt{2\varepsilon }-1}\exp (-\sqrt{2\varepsilon _1}r_1-\sqrt{%
2\varepsilon }r_2),
\end{equation}
where $\varepsilon _1=Z^2/R-E_{\text{H}_2^{+}}>0$ is the separation energy
of electron in H$_2^{+}$. The power-law factor slowly varies as compared to
the exponential decaying contribution. Hence, one can assume the power-law
factor to be a constant or approximate the combination $r^a\exp (-br)$ as
$$
r^a\exp (-br)=\exp (-br+a\ln r)\approx \exp (-br+a\ln r_0+(r-r_0)a/r_0),
$$
where $r_0$ can be determined as a variational parameter or chosen to be $%
r_0=R+1/b$ \cite{PTT}.

To incorporate the cusp conditions and the large distance asymptotic the
trial wave function is separated into two parts
\begin{equation}
\label{pa7}\Psi ({\bf r}_1,{\bf r}_2)=\Phi ({\bf r}_1,{\bf r}_2)f_P(r_{12}),
\end{equation}
where $f_P(r_{12})$ is the Patil et al. electron-electron correlation
function given by Eq. \eqref{patf}. Roothaan and Weiss \cite{Root60}
have made a
very accurate numerical investigation of the desired correlation function
for the ground state of the He atom. In the vicinity of $r_{12}=0$, the
correlation function is linear and satisfies the cusp condition. It
monotonically increases and approaches a constant as $r_{12}$ becomes very
large. Clearly the function $f_P(r_{12})$ satisfies these conditions
(see Fig. \ref{ff2}).
In the united atom limit $(R=0)$ it was found that the energies
computed with the variationally determined $\lambda $ are essentially the
same as given by the analytical expression,
\begin{equation}
\label{pa81}\lambda =\frac 5{12}Z-\frac 13
\end{equation}
derived from a theory in which $1/r_{12}$ is treated as a perturbation \cite
{K}. In a molecular system, as $R$ increases, one should expect $\lambda $
to decrease monotonically and become vanishingly small for $R\rightarrow
\infty $. The small and large $R$ behavior is satisfied provided \cite
{Klei98}
\begin{equation}
\label{pa82}\lambda =\frac{5Z/6-1/3}{1+10Z^3R^2/(15Z-6)}.
\end{equation}
The electron-nucleus cusp conditions do not uniquely define the space
wave function $\Phi ({\bf r}_1,{\bf r}_2)$. If one wishes to maintain the
electronic configuration idea with an independent particle picture, one can
adopt the following form of $\Phi ({\bf r}_1,{\bf r}_2)$:
\begin{equation}
\label{pa9}\Phi ({\bf r}_1,{\bf r}_2)=\phi (r_1)\phi (r_2),
\end{equation}
with $\phi (r_j),$ $j=1,2,$ being the Guillemin-Zener \cite{GZ} trial
wave function for H$_2^{+}$
\begin{equation}
\label{pa10}\phi (r_j)=\exp (-z_1r_{ja}-z_2r_{jb})+\exp
(-z_2r_{ja}-z_1r_{jb}),
\end{equation}
where $z_1>0,$ $z_2>0$ are variational parameters. Alternatively, $z_1$ and $%
z_2$ can be determined from the cusp conditions at $r_{ja}=0$ and $r_{jb}=0$
for $j=1,2$.
The wave function (\ref{pa9}), (\ref{pa10}) is identical to  (\ref{mo}) which is
known not to give the correct dissociation limit of H$_2$. However, for the
pedagogical reason we briefly discuss it here.

As $r_{1a}$ approaches zero
\begin{equation}
\label{pa11}\Phi ({\bf r}_1,{\bf r}_2)\rightarrow \phi (r_2)\left[
(1-z_1r_{1a})\exp (-z_2R)+(1-z_2r_{1a})\exp (-z_1R)\right] .
\end{equation}
Imposing the cusp condition $\Phi ({\bf r}_1,{\bf r}_2)\rightarrow
G(r_2)(1-Zr_{1a})$ we obtain an equation for $z_1$ and $z_2$:
\begin{equation}
\label{pa12}z_1=Z+(Z-z_2)\exp [-(z_1-z_2)R].
\end{equation}
Thus, if $z_1$ and $z_2$ are related as in Eq. \eqref{pa12}, then the
electron-nucleus cusp conditions are automatically satisfied. The second
equation for $z_1$ and $z_2$ can be determined by the asymptotic condition.
For $r_2\gg r_1\gg R$, $1$ Eqs. \eqref{pa9}, \eqref{pa10} yield
\begin{equation}
\label{pa13}\Psi ({\bf r}_1,{\bf r}_2)\approx \exp
[-(z_1+z_2)r_1-(z_1+z_2)r_2].
\end{equation}
>From the other hand, according to Eq. \eqref{pa6}, the wave function must
have the following exponential behavior $\Psi ({\bf r}_1,{\bf r}_2)\sim \exp
(-\sqrt{2\varepsilon _1}r_1-\sqrt{2\varepsilon }r_2)$. The two parameters $%
z_1$ and $z_2$ do not allow to match the asymptotic exactly. However, one
can approximately choose \cite{Klei98}
\begin{equation}
\label{pa14}z_1+z_2=\varepsilon _1+\varepsilon =\frac{Z^2}R-E.
\end{equation}
Eqs. \eqref{pa12}, \eqref{pa14} determine $z_1$ and $z_2$ self-consistently
together with the ground state energy $E$.

Kleinekath\"ofer et al. \cite{Klei98} used the trial function \eqref{pa7},
\eqref{pa9}, \eqref{pa10} with $\lambda $, $z_1$ and $z_2$ determined from
Eqs. \eqref{pa82}, \eqref{pa12}, \eqref{pa14}. The wave function has no free
parameters and yields $4.661$ eV for the binding energy of the H$_2$
molecule which is very close to the exact value of $4.745$ eV. However,
$E(R)$ becomes less accurate at large $R$ and fails to describe the dissociation
limit. The corresponding $E(R)$ is shown as a small dot line (Patil et al. 1) in
Fig. \ref{h2comp}.

Both the cusp conditions and the large distance asymptotic can be satisfied
exactly provided more sophisticated trial functions are introduced. For
example, for small and intermediate $R$, Patil et al. \cite{PTT} suggested
to use a combination of ``inner'' and ``outer'' molecular orbitals which are
build from the Guillemin-Zener one-electron wave functions:
\begin{equation}
\label{pa15}\Psi _m({\bf r}_1,{\bf r}_2)=[\phi _{\text{in}}({\bf r}_1)\phi _{%
\text{out}}({\bf r}_2)+\phi _{\text{in}}({\bf r}_2)\phi _{\text{out}}({\bf r}%
_1)]f_P(r_{12}),
\end{equation}
where the ``inner'' orbital is
\begin{equation}
\label{pa16}\phi _{\text{in}}({\bf r}_j)=\exp (-z_1r_{ja}-z_2r_{jb})+\exp
(-z_2r_{ja}-z_1r_{jb}).
\end{equation}
Analogously, an ``outer'' orbital is defined as
\begin{equation}
\label{pa17}\phi _{\text{out}}({\bf r}_j)=\exp (-z_3r_{ja}-z_4r_{jb})+\exp
(-z_4r_{ja}-z_3r_{jb}).
\end{equation}
All the parameters $z_1$, $z_2$, $z_3$ and $z_4$ are determined by the cusp
and asymptotic conditions.

At large $R$, the atomic orbital wave function provides a better description
of the two electron system. The appropriate wave function is \cite{PTT}
\begin{equation}
\label{pa18}\Psi _a({\bf r}_1,{\bf r}_2)=[\Phi ({\bf r}_1,{\bf r}_2)+\Phi (%
{\bf r}_2,{\bf r}_1)]f_P(r_{12}),
\end{equation}
where
\begin{equation}
\label{pa19}\Phi ({\bf r}_1,{\bf r}_2)=\exp
[-Z(r_{1a}+r_{1b}+r_{2a}+r_{2b})]\left\{ \cosh (z_5r_{1b})\cosh
(z_6r_{2a})+\cosh (z_6r_{1b})\cosh (z_5r_{2a})\right\} .
\end{equation}
Eq. \eqref{pa18} satisfies all the electron-nucleus cusp conditions. At the
same time, it has two free parameters $z_5$ and $z_6$ which can be used to
satisfy the two asymptotic conditions.

For a description in the entire range of internuclear distances, one can use
a linear combination of the two wave functions just discussed
\begin{equation}
\label{pa20}\Psi =\Psi _m+D\Psi _a,
\end{equation}
where $D$ is a variational parameter. For H$_2$ the molecular orbital $\Psi
_m$ dominates in the region $R<1.7$, while the atomic orbital $\Psi _a$
dominates at $R>1.7$. With this complicated one parameter wave function Patil
et al. \cite{PTT} obtained $4.716$ eV for the binding energy of H$_2$
molecule and a very accurate potential curve in the entire range of $R$. The
corresponding $E(R)$ is shown as a dash line (Patil et al. 2) in Fig. \ref
{h2comp}. Similar wave functions which take full advantage of the asymptotic
and proximal boundary conditions are useful in variational
calculations of larger systems \cite{Pati03a}.

\section{Analytical wave mechanical solutions for one electron molecules}
\label{sec2}

\indent

From now on throughout the Sections \ref{sec2}, \ref{secV} and \ref{sec5},
unless otherwise noted, we
assume the Born--Oppenheimer separation, where there are $N$ nuclei,
containing $Z_k$ protons located at $\pmb{R}_k$, respectively, for $%
k=1,2,\ldots, N$, and $N_e$ electrons. Each electron's coordinates are
denoted as $\pmb{r}_j$, $j=1,2,\ldots, N_e$, where $\pmb{r}_j = (x_j,y_j,z_j)
$. The steady-state equation, in atomic units, can be written as
\begin{equation}
\label{eq2.1}H\psi = E\psi,\quad H = - \frac12 \sum^{N_e}_{j=1} \nabla^2_j +
\sum_{1\le j<k\le N_e} \frac1{r_{jk}} - \sum^{N_e}_{j=1} \sum^{N}_{k=1} \frac{%
Z_k}{|\pmb{r}_j-\pmb{R}_k|} + \sum_{1\le j<k\le N} \frac{Z_jZ_k}{|\pmb{R}_j-%
\pmb{R}_k|},
\end{equation}
where
$$
\psi =\psi(\pmb{r}_1,\pmb{r}_2,\ldots, \pmb{r}_{N_e}), \quad \nabla^2_j
\equiv \frac{\partial^2}{\partial x^2_j} + \frac{\partial^2}{\partial y^2_j}
+ \frac{\partial^2}{\partial z^2_j},\quad r_{jk} \equiv |\pmb{r}_j - \pmb{r}%
_k|,\quad j,k=1,2,\ldots,N_e.
$$
We wish to solve the eigenvalue problem \eqref{eq2.1}.

Closed-form solutions to \eqref{eq2.1} are hard to come by in general. What
is known today is the following:

\begin{itemize}
\item[(i)]  $\pmb{N_e=1, N=1}$:\newline
This is the case of hydrogen atom H (or H-like ions with a single
nucleus and electron) whose solutions are known {\em explicitly\/} in closed
form, to be briefly reviewed in Subsection \ref{sec2.1}.

\item[(ii)]  $\pmb{N_e=1, N=2}$:\newline
This is the H$_2^{+}$ (or H$_2^{+}$-like) two-centered molecular ion, whose
solutions are separable and expressible as an infinite series of products of
special functions in prolate spheroidal coordinates, where coefficients of
the series are {\em not explicitly\/} given. Such are the two renowned
classic solutions due to Hylleraas and Jaff\'e, to be discussed in
Subsection \ref{sec2.2}.

\item[(iii)]  $\pmb{N_e=1, N\ge 3}$:\newline
This one-electron multi-centered molecular ion has an {\em analytic\/}
solution due to Shibuya and Wulfman \cite{SW} in terms of integral equations
on the unit hypersphere of the 4-dimensional momentum space. This will be
reviewed in Subsection \ref{sec2.3}.
\end{itemize}

\noindent Except for Case (i) above, one must resort to numerical methods in
order to derive quantitative and qualitative information, for all cases
where $N_e\ne 1$, $N\ne 1$. Our particular interest in this paper is the
{\em diatomic\/} case, with $N_e = N = 2$, using the orbitals in Cases (i)
and (ii) above as the building blocks.

\subsection{The hydrogen atom}

\label{sec2.1}

\indent

When $N_e= N =1$, with $Z_1=1$ and $\pmb{R}_1=\pmb{0}$, equation
\eqref{eq2.1} becomes
\begin{equation}
\label{eq2.2}\left(-\frac12\nabla^2-\frac1r\right) \psi(\pmb{r}) = E\psi(%
\pmb{r}),\qquad \pmb{r}\in {\mathbb{R}}^3,
\end{equation}
the Born--Oppenheimer separation of the hydrogen atom.

We write \eqref{eq2.2} in spherical coordinates in view of the symmetry
involved:
\begin{equation}
\label{eq2.3}\frac1r \frac{\partial^2}{\partial r^2} (r\psi) + \frac1{r^2}
\Lambda^2\psi + \frac2r \psi - 2E\psi = 0,
\end{equation}
where
\begin{align}\label{eq2.4}
\Lambda^2 &\equiv \frac1{\sin^2\theta} \frac{\partial^2}{\partial\phi^2} +
\frac1{\sin \theta} \frac\partial{\partial\theta} \left(\sin
\theta\frac\partial{\partial\theta}\right),\quad \text{(the Legendrian);}\\
r &= \text{the radial variable,}\qquad 0<r<\infty;\nonumber\\
\theta &= \text{the colatitude,}\qquad 0\le\theta\le\pi;\nonumber\\
\phi &= \text{the azimuth,}\qquad 0\le\phi\le 2\pi.\nonumber
\end{align}
Equation \eqref{eq2.2} has separable solutions
\begin{equation}
\label{eq2.5}\psi(\pmb{r}) = \psi(r,\theta,\phi) = R(r) Y(\theta,\phi).
\end{equation}
The angular variables are quantized first as we know that angular functions
are the spherical harmonics
\begin{equation}
\label{eq2.6}Y(\theta,\phi) = Y_{\ell m}(\theta,\phi) = \Theta_{\ell
m}(\theta) \Phi_{m}(\phi),\qquad \ell=0,1,2,\ldots, m = \ell, \ell-1,\ldots,
-\ell,
\end{equation}
on the unit sphere $\mathcal{S}_2 \equiv\{\pmb{r}\in {\mathbb{R}}^3\mid |%
\pmb{r}| = 1\}$, satisfying
\begin{equation}
\label{eq2.7}\Lambda^2 Y_{\ell m}(\theta,\phi) = -\ell(\ell+1) Y_{\ell
m}(\theta,\phi),
\end{equation}
where in \eqref{eq2.6},
\begin{align}\label{eq2.8}
&\Phi_{m}(\phi) = (2\pi)^{-\frac12} e^{im\phi},\\
\label{eq2.9}
&\Theta_{\ell m}(\theta) = \left\{\frac{(2\ell+1)(\ell-|m|)!}{2(
\ell+|m|)!}\right\}^{\frac12} P^{|m|}_\ell(\cos\theta),\\
&(P^{|m|}_\ell \text{ is the associated Legendre function}).\nonumber
\end{align}
Using \eqref{eq2.5}--\eqref{eq2.7} in \eqref{eq2.2}, we obtain the equation
for the radial function
\begin{equation}
\label{eq2.10}\frac1r (rR)^{\prime\prime}- \frac{\ell(\ell+1)}{r^2} R +
\left(\frac2r - 2E\right)R = 0.
\end{equation}
Solutions to the eigenvalue problem \eqref{eq2.10} that are square
integrable over $0<r<\infty$ are known to be
\begin{align}\label{eq2.11}
&R(r) = R_{n\ell}(r) = (-2) \left\{\frac{(n-\ell-1)!}{2n[(n+\ell)!]^3}\right\}
(2r)^\ell L^{2\ell+1}_{n+\ell}(2r)e^{-r},\\
&E_n = -\frac12 \frac1{n^2},\quad n=1,2,\ldots, \text{ independent of } \ell,
\nonumber
\end{align}
where $L^{2\ell+1}_{n+\ell}$ are the associated Laguerre functions such that for
$m,n=0,1,2,\ldots$
\begin{align*}
&xL^m_{m+n}{}'' + (m+1-x) L^m_{m+n}{}' + (m+n) L^m_{m+n} = 0,\\
&L^m_{m+n}(x) = \frac{e^xx^{-(m+n)}}{(m+n)!} \frac{d^{m+n}}{dx^{m+n}} (e^{-x}
x^{2m+n}),
\end{align*}
(when $m=0, L^0_n(x)$ is simply denoted as $L_n(x)$).

In the subsequent sections, we will utilize mainly the ground, or {\em 1s
state\/} of the hydrogen atom, where $n=1, \ell=0$, i.e.,
\begin{equation}
\label{eq2.12}\Phi(\pmb{r}) = \frac1{\sqrt{2}} e^{-r};\quad \text{cf.\
Fig.~\ref{cuspf}.}
\end{equation}

\subsection{$\pmb{H^+_2}$-like molecular ion in prolate spheroidal coordinates}

\label{sec2.2}

\indent

We now consider the eigenvalue problem for two-centered H$^+_2$-like
molecular ion with one electron and two fixed nuclei with effective charges $%
Z_a$ and $Z_b$. Given $R$ the internuclear separation distance, we want to
find $E$ and $\Psi$ such that
\begin{equation}
\label{eqA.1}-\frac12 \nabla^2\Psi - \left(\frac{Z_a}{r_a} + \frac{Z_b}{r_b}
- \frac{Z_aZ_b}{R}\right) \Psi = E\Psi.
\end{equation}

\begin{figure}[htpb]

\includegraphics[angle=270,width=10cm]{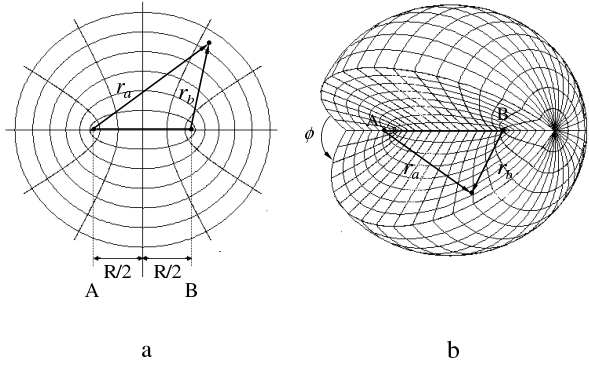}

\caption{(a) Elliptical coordinates $(\lambda ,\mu )$. (b) Prolate spheroidal
coordinates $(\lambda ,\mu ,\phi )$ with $\lambda =(r_a+r_b)/R$ and $\mu
=(r_a-r_b)/R$. The range of coordinates is $1\le \lambda \le \infty$,
$-1\le \mu \le 1$ and $0 \le \phi \le 2\pi $.
}
\label{prosph}
\end{figure}

In Appendix A we show how to separate the variables through the use of the
ellipsoidal (or, prolate spheroidal) coordinates (see Fig. \ref{prosph})
\begin{equation}
x = \frac{R}2 \sqrt{(\lambda^2-1)(1-\mu^2)}\ \cos \phi,\quad y = \frac{R}2
\sqrt{(\lambda^2-1)(1-\mu^2)}\ \sin\phi,\quad z = \frac{R}2 \lambda\mu.
\end{equation}
In such coordinates the wave function can be written as
\begin{equation}
\Psi = \Lambda(\lambda) M(\mu) e^{im\phi}.
\end{equation}
Separation of variables yields
\begin{alignat}{2}
\label{eqA.9}
\frac{d}{d\lambda} \left\{(\lambda^2-1) \frac{d\Lambda}{d\lambda}\right\} +
\left\{ A + 2R_1\lambda - p^2\lambda^2 - \frac{m^2}{\lambda^2-1}\right\} \Lambda
&= 0,  &\quad R_1 &\equiv \frac{R(Z_a+Z_b)}{2}, \ \lambda\ge 1;\\
\label{eqA.10}
\frac{d}{d\mu} \left\{(1-\mu^2) \frac{dM}{d\mu}\right\} + \left\{-A-2R_2\mu +
p^2\mu^2 - \frac{m^2}{1-\mu^2}\right\}M &= 0, &\quad R_2 &\equiv
\frac{R(Z_a-Z_b)}{2}, \ |\mu| \le 1.
\end{alignat}
Note that $A$ and $p$ are unknown and must be solved from \eqref{eqA.9} and
\eqref{eqA.10} as eigenvalues of the coupled system. Once $A$ and $p$ are
solved, then $E$ can be obtained from \eqref{eqA.8}.

In the next two subsections, we address the issues of solving \eqref{eqA.9}
and \eqref{eqA.10}, respectively.

\subsubsection{Solution of the $\pmb{\Lambda}$-equation \eqref{eqA.9}}

\label{sec2.2.1}

\indent

To solve \eqref{eqA.9}, it is important to understand the {\em asymptotics\/}
of the solution. Rewrite \eqref{eqA.9} as
\begin{equation}
\label{eqA.11}(\lambda^2-1) \Lambda^{\prime\prime}(\lambda) +
2\lambda\Lambda^{\prime}(\lambda) + \left(A+2R_1\lambda - p^2\lambda^2 -
\frac{m^2}{\lambda^2-1}\right) \Lambda(\lambda) = 0.
\end{equation}
First, consider the case $\lambda\gg 1$; we have
\begin{align}
\label{eqA.11a}
0 &= \Lambda''(\lambda) + \frac{2\lambda}{\lambda^2-1} \Lambda'(\lambda) +
\left[\frac{A+2R_1\lambda}{\lambda^2-1} - p^2 \frac{\lambda^2}{\lambda^2-1} -
\frac{m^2}{(\lambda^2-1)^2}\right] \Lambda(\lambda)\\
&\approx \Lambda''(\lambda) - p^2\Lambda(\lambda),\quad \text{for}\quad \lambda
\gg 1.\nonumber
\end{align}
This gives
\begin{equation}
\label{eqA.11b}\Lambda(\lambda) \approx a_1e^{-p\lambda} + a_2e^{p\lambda},
\text{ for } \lambda \gg 1, \text{ where } p>0.
\end{equation}
But the term $a_2e^{p\lambda}$ has exponential growth for large $\lambda$,
which is physically inadmissible and must be discarded. Thus
\begin{equation}
\label{eqA.12}\Lambda(\lambda) \approx a_1 e^{-p\lambda},\quad \text{for}%
\quad \lambda \gg 1.
\end{equation}

A finer estimate than \eqref{eqA.12} can be stated as follows
\begin{equation}
\label{eqA.12a}\Lambda(\lambda) = a_0e^{-p\lambda} \lambda^\beta
\sum^\infty_{j=0} \frac{c_j}{\lambda^j} \qquad (\lambda\gg 1)
\end{equation}
where $\beta \equiv R_1/p - 1$, $c_0 =1$, $c_1=(p^2-\beta^2-2\beta%
)/(2R_1-2p\beta-p)$, and $a_0$ is an arbitrary constant. Proof is given in
Appendix \ref{A2}.

Next, we consider the case $\lambda>1$ but $\lambda \approx 1$. In such a
limit we have (see Appendix \ref{Lg1})
\begin{equation}
\label{eqA.241}\Lambda(\lambda) \approx (\lambda-1)^{\frac{|m|}2}
\sum^\infty_{k=0} c_k(\lambda-1)^k.
\end{equation}

Our results in \eqref{eqA.12} and \eqref{eqA.241} suggest that the form
\begin{equation}
\label{eqA.25}\Lambda (\lambda )=e^{-p\lambda }(\lambda -1)^{\frac{|m|}%
2}\lambda ^\beta f(\lambda ),\qquad \text{for some function }f(\lambda )
\end{equation}
would contain the right asymptotics for both $\lambda \gg 1$ and $\lambda
\approx 1$. Here, obviously, $f(\lambda )$ must satisfy
\begin{equation}
\label{eqA.25a}f(1)\ne 0,\qquad \lim _{\lambda \to \infty }|f(\lambda )|\le
C,\text{ for some constant }C>0.
\end{equation}
Actually, in the literature (\cite{baber,jaffe,bates}), two improved or
variant forms of the substitution of \eqref{eqA.25} are found to be most
useful:\newline
{\bf (i)~~(Jaff\'e's solution \cite{jaffe})}
\begin{equation}
\label{eqA.26}\Lambda (\lambda )=e^{-p\lambda }(\lambda ^2-1)^{\frac{|m|}%
2}(\lambda +1)^\sigma \sum_{n=0}^\infty g_n\left( \frac{\lambda -1}{\lambda
+1}\right) ^n,\qquad \left( \sigma \equiv \frac{R_1}p-|m|-1\right) .
\end{equation}
This leads to a 3-term recurrence relation
\begin{equation}
\label{eqA.27}\alpha _ng_{n-1}-\beta _ng_n+\gamma _ng_{n+1}=0;\qquad
n=0,1,2,\ldots ;\quad g_{-1}\equiv 0,
\end{equation}
where
\begin{equation}
\label{eqA.28}\left.
\begin{array}{ll}
\alpha _n=(n-1-\sigma )(n-1-\sigma -m), &  \\
\beta _n=2n^2+(4p-2\sigma )n-A+p^2-2p\sigma -(m+1)(m+\sigma ), &  \\
\gamma _n=(n+1)(n+m+1), &
\end{array}
\right\}
\end{equation}
and, consequently, the continued fraction
\begin{equation}
\label{eqA.29}\frac{\beta _0}{\gamma _0}=\frac{\alpha _1}{\beta _1-\dfrac{%
\gamma _1\alpha _2}{\beta _2-\dfrac{\gamma _2\alpha _3}{\beta _3-\ldots }}}
\end{equation}
for $A$ and $p$.

\noindent {\bf (ii) (Hylleraas' solution \cite{hylleraas})}
\begin{equation}
\label{eqA.30}\Lambda (\lambda )=e^{-p(\lambda -1)}(\lambda ^2-1)^{\frac{|m|}%
2}\sum_{n=0}^\infty \frac{c_n}{(m+n)!}L_{m+n}^m(x),\qquad x\equiv 2p(\lambda
-1),
\end{equation}
where $L_{m+n}^m$ is the associated Laguerre polynomial
and $c_n$ satisfy the 3-term recurrence relation
\begin{equation}
\label{eqA.31}\alpha _nc_{n-1}-\beta _nc_n+\gamma _nc_{n+1}=0,\qquad
n=0,1,2,\ldots ;\quad c_{-1}\equiv 0,
\end{equation}
where
\begin{equation}
\label{eqA.32}\left.
\begin{array}{l}
\alpha _n=(n-m)(n-m-1-\sigma ), \\
\beta _n=2(n-m)^2+2(n-m)(2p-\sigma )-[A-p^2+2p\sigma +(m+1)(m+\sigma )], \\
\gamma _n=(n+1)(n-2m-\sigma ),
\end{array}
\right\}
\end{equation}
and the same form of continued fractions \eqref{eqA.29}.

\subsubsection{Solution of the $\pmb{M}$-equation \eqref{eqA.10}}

\label{sec2.2.2}

\indent

Equation \eqref{eqA.10} has close resemblance in form with \eqref{eqA.9}
and, thus, it can almost be expected that the way to solve \eqref{eqA.9}
will be similar to that of \eqref{eqA.9}.

First, we make the following substitution
\begin{equation}
\label{eqA.33}M(\mu )=e^{\pm p\mu }\widetilde{M}(\mu ),\qquad -1\le \mu \le
1,
\end{equation}
in order to eliminate the $p^2\mu ^2$ term in \eqref{eqA.10}. We obtain
\begin{equation}
\label{eqA.34}[(1-\mu ^2)\widetilde{M}^{\prime }]^{\prime }\pm 2p(1-\mu ^2)%
\widetilde{M}^{\prime }+\left[ (-2R_2\mp 2p)\mu +(p^2-A)-\frac{m^2}{1-\mu ^2}%
\right] \widetilde{M}=0.
\end{equation}
To simplify notation, let us just consider the case $M(\mu )=e^{-p\mu }%
\widetilde{M}(\mu )$, but note that for $M=e^{p\mu }\widetilde{M}(\mu )$, we
need only make the changes of $p\to -p$ in \eqref{eqA.35b} below.
Write
\begin{equation}
\label{eqA.35}M(\mu )=e^{-p\mu }\sum_{k=0}^\infty f_kP_{m+k}^m(\mu ),
\end{equation}
where $P_n^m(\mu)$ are the associated Legendre polynomials,
and substitute \eqref{eqA.35} into \eqref{eqA.10}. We obtain a 3-term
recurrence relation
\begin{equation}
\label{eqA.35a}\alpha _nf_{n-1}-\beta _nf_n+\gamma _nf_{n+1}=0;\qquad
n=0,1,2,\ldots ;f_{-1}\equiv 0,
\end{equation}
where
\begin{equation}
\label{eqA.35b}\left.
\begin{array}{ll}
\alpha _n=\dfrac 1{2(m+n)-1}[-2nR_2+2pn(m+n)], &  \\
\noalign{\smallskip}\beta _n=A-p^2+(m+n)(m+n+1), &  \\
\noalign{\smallskip}\gamma _n=\dfrac{2m+n+1}{2(m+n)+3}\{-2R_2-2p(m+n+1)\}, &
\end{array}
\right\}
\end{equation}
and, consequently, again the continued fractions of the same form as
\eqref{eqA.29}. The continued fractions obtained here should be {\em %
coupled\/} with the continued fraction \eqref{eqA.29} for the variable $\mu $
to solve $A$ and $p$.

In the {\em homonuclear\/} case, $R_2=R(Z_a-Z_b)/2=0$, equation
\eqref{eqA.10} reduces to
$$
[(1-\mu ^2)M^{\prime }]^{\prime }+\left( -A+p^2\mu ^2-\frac{m^2}{1-\mu ^2}%
\right) M=0.
$$
In this case, several different optional representations of $M$ can be used:
\begin{align}\label{eqA.35c}
&\begin{array}{lll}
\text{(a)}&
M(\mu) = (1-\mu^2)^{\frac{|m|}2} \sum\limits^\infty_{k=0}
c_k\mu^{2k},~~~~~~~~~~~~& M(\mu)
= (1-\mu^2)^{\frac{|m|}2} \sum\limits^\infty_{k=0}
c_k\mu^{2k+1};\hspace{.5in}\end{array}\\
\label{eqA.35d}
&\begin{array}{lll}
\text{(b)}&
M(\mu) = (1-\mu^2)^{\frac{|m|}2} \sum\limits^\infty_{k=0} c_k P^m_{m+2k} (\mu),
~~~&M(\mu) = (1-\mu^2)^{\frac{|m|}2} \sum\limits^\infty_{k=0} c_k
P^m_{m+2k+1}(\mu);\end{array}\\
&\begin{array}{lll}
\text{(c)}&
M(\mu) = e^{\pm p\mu}(1-\mu^2)^{\frac{|m|}2} \sum\limits^\infty_{k=0} c_k (1\mp
\mu)^k.&\phantom{M(\mu= (1-\mu^2)^{\frac{|m|}2} \sum\limits^\infty_{k=0}}
\end{array}\nonumber
\end{align}

In Appendix \ref{sec2.2.3} we discuss expansions of solution near
$\pmb{\lambda\approx 1}$ and $ \pmb{\lambda\gg 1}$ and their connection with the
James--Coolidge trial wave functions.

As a conclusion of this section, we note that the eigenstates of the hydrogen
atom given in the preceding subsection can also be easily represented in
terms of the prolate spheroidal coordinates. We let the nucleus of H
(i.e., a proton) sit at location $a$ where $(0,0,-R/2)$ with $Z_a=1$ while
at location $b$ where $(0,0,R/2)$ we let $Z_b=0$. Thus, the hydrogen
atom satisfies Eq.\ \eqref{eqA.1} in the form
\begin{equation}\label{IV.39b}
H\psi = \left(-\frac12\nabla^2 - \frac1{r_a}\right)\psi = E\psi.
\end{equation}
Now, in terms of the prolate spheroidal coordinates  \eqref{eqA.2a} in Appendix A,
 and
\begin{equation}\label{IV.39a}
\psi(\lambda,\mu,\phi) = \Lambda(\lambda) M(\mu) \Phi(\phi), \text{ where }
\Phi(\phi)  = e^{im\phi}
\end{equation}
in the form of separated variables, we have
\begin{align*}
&-\frac12 \frac4{R^2(\lambda^2-\mu^2)} \left\{\frac\partial{\partial\lambda}
\left[(\lambda^2-1) \frac\partial{\partial\lambda} \Lambda\right] M +
\frac\partial{\partial\mu} \left[(1-\mu^2)\frac\partial{\partial\mu} M\right]
\Lambda - \frac{(\lambda^2-\mu^2)m^2}{(\lambda^2-1)(1-\mu^2)} M\Lambda\right\}\\
&\qquad - \frac2R \frac1{\lambda-\mu} M\Lambda = EM\Lambda,
\end{align*}
which has two fewer terms than \eqref{eqA.6} does as now $Z_b=0$. Set
$p^2=-R^2E/2$, we again have \eqref{eqA.9} and \eqref{eqA.10} except that now
$R_1=R_2=R/2$ therein. The rest of the procedures follows in the same way
with some minor adjustments as noted above.

The above discussion also leads to a sequence of identities between
\eqref{eq2.6} and \eqref{IV.39a}, as
\[
\psi^{(1)}(x,y,z) = \psi^{(2)}\left(x,y,z + \frac{R}2\right),
\]
where $\psi^{(1)}(x,y,z)$ is an eigenstate of the hydrogen atom obtained
from  \eqref{eq2.2} but expressed in terms of the Cartesian coordinates
$(x,y,z)$ while $\psi^{(2)}(x,y,z)$ is that for the solution of \eqref{IV.39b}.

\subsection{The many-centered, one-electron problem}
\label{sec2.3}

\indent

When $N_e=1$ and $N\ge 3$ in \eqref{eq2.1}, we have a molecular ion with
three or more nuclei sharing one electron. A simple example is a $CO_2$-like
structure, with $N=3$. For
such a problem, separable closed-form solutions are extremely difficult to
come by from the traditional line of attack. However, we want to describe an
elegant analysis by T.\ Shibuya and C.E. Wulfman \cite{SW} (see also the
book by B.R.\ Judd \cite{Judd}) which works in momentum space and expand
electron's eigenfunction as a linear combination of 4-dimensional spherical
harmonics. This analysis may offer useful help to the modeling and
computation of complex molecules after proper numerical realization.

The model equation reads
\begin{equation}
\label{eq2.3.1}\left( -\frac 12\nabla ^2-\sum_{j=1}^N\frac{Z_j}{|\pmb{r}-%
\pmb{R}_j|}\right) \psi (\pmb{r})=E\psi (\pmb{r}),\qquad \pmb{r}=(x,y,z)\in {%
\mathbb{R}}^3,
\end{equation}
where ${\bf R}_j$ are positions of the nuclei.

Appendix \ref{many} shows how to reduce the problem to a matrix form. Here
we provide the answer for the energy; it is determined from the solution of
the eigenvalue equation
\begin{equation}
\label{ms1}
\pmb{P}{\bf c}=\sqrt{-2E}{\bf c},
\end{equation}
where ${\bf c}$ is an infinite dimensional vector, $\pmb{P}$ is an infinite
matrix with entries
\begin{gather}
\label{ms2}P_{n^{\prime }\ell ^{\prime }m^{\prime }}^{n\ell
m}=\sum_jZ_j\sum_{n^{\prime \prime }\ell ^{\prime \prime }m^{\prime \prime
}}\left[ S_{n^{\prime \prime }\ell ^{\prime \prime }m^{\prime \prime
}}^{n^{\prime }\ell ^{\prime }m^{\prime }}({\bf R}_j)\right] ^{*}\frac
1nS_{n^{\prime \prime }\ell ^{\prime \prime }m^{\prime \prime }}^{n\ell m}(%
{\bf R}_j),\\
n,n',n''=1,2,3,\ldots; \ell,\ell',\ell''=0,1,2,\ldots; m,m',m''=-\ell,
-\ell+1,\ldots, -1,0,1,\ldots, \ell-1,\ell;\nonumber
\end{gather}
the matrix $S$ is given by an integral over 4-dimensional unit hypersphere S$%
_3$ with the surface element $d\Omega =\sin ^2\chi \sin \theta d\chi d\theta
d\phi $,
\begin{equation}
\label{ms3}S_{n^{\prime }\ell ^{\prime }m^{\prime }}^{n\ell m}({\bf R}%
_j)=\int_{S_3}\exp (i{\bf R}_j\cdot
{\bf p})Y_{n\ell m}(\Omega )Y_{n^{\prime }\ell
^{\prime }m^{\prime }}(\Omega )d\Omega \text{,}
\end{equation}
$Y_{n\ell m}(\Omega )$ is a product of the spherical function $Y_{\ell
m}(\theta ,\phi )$ and the associated Gegenbauer function $C_n^\ell (\chi )$%
$$
Y_{n\ell m}(\Omega )=(-i)^\ell C_n^\ell (\chi )Y_{\ell m}(\theta ,\phi ).
$$
The 3-dimensional vector ${\bf p}$ in Eq. \eqref{ms3} has components
$$
p_x=p\sin \theta \cos \phi ,\quad p_y=p\sin \theta \sin \phi ,\quad
p_z=p\cos \theta, \quad \text{where} \quad
p=\sqrt{-2E}\tan (\chi /2)\text{.}
$$

In practice, the infinite matrix $\pmb{P}$ in \eqref{ms2} is truncated to a
finite size square matrix according to the quantum numbers $(n\ell m)$ for
which the restriction $n\le n_0$ is specified for some positive integer $n_0$%
.

In the derivation, if we restrict $N=1$, $Z_1=1$ and set $\pmb{R}_1=0$, then
the matrix $\pmb{P}$ is {\em diagonal\/} and we recover the
hydrogen atom as derived in Subsection \ref{sec2.1}. Obviously, if $N=2$, by
setting $\pmb{R}_1=(0,0,-R/2)$ and $\pmb{R}_2=(0,0,R/2)$, we should also be
able to recover those H$_2^{+}$-like solutions given in Subsection
\ref{sec2}.B.

The 1-electron one-centered or two-centered orbitals derived in Subsection
\ref{sec2}.A and \ref{sec2}.B will be utilized frequently in the rest of the
paper. At the present time, there is very limited knowledge about the
1-electron many-centered orbitals as discussed in Subsection \ref{sec2}.C.
There seems to be abundant space for their exploitation in molecular
modeling and computation in the future.

\section{Two electron molecules: cusp conditions and correlation functions}
\label{secV}

\subsection{The cusp conditions}

\label{sec3.1}

\indent

In the study of any linear partial differential equations with singular
coefficients, it is well known to the theorists that solutions will have
important peculiar behavior at and near the locations of the singularities.
We have first encountered such singularities in Subsection \ref{newsec2}.
Here we give singularities of the Coulomb type a more systematic treatment.
The critical mathematical analysis was first made by Kato \cite{Ka} in the
form of {\em cusp conditions\/} for the Born--Oppenheimer separation.

Consider the following slightly more general form of the Schr\"odinger
equation for a 2-particle system
\begin{equation}
\label{C0}(\hat H-E)\psi =0,
\end{equation}
where
\begin{equation}
\label{C1}\hat H=-\frac 1{2m_1}\nabla _1^2-\frac 1{2m_2}\nabla _2^2-\frac{Z_a%
}{r_{1a}}-\frac{Z_b}{r_{1b}}-\frac{Z_a}{r_{2a}}-\frac{Z_b}{r_{2b}}+\frac{%
q_1q_2}{r_{12}}+\frac{Z_aZ_b}R.
\end{equation}
The operator $\hat H$ has five sets of singularities, at
\begin{equation}
\label{C2}r_{1a}=0,\quad r_{1b}=0,\quad r_{2a}=0,\quad r_{2b}=0\quad \text{%
and}\quad r_{12}=0.
\end{equation}
It has been proved by Kato \cite{Ka} that the wave function $\psi $ is
H\"older continuous, with bounded first order partial derivatives. However,
these first order partial derivatives $\partial \psi /\partial x_i$, etc., $%
i=1,2,\ldots ,6$, are discontinuous at \eqref{C2}. In the terminology of the
mathematical theory of partial differential equations, \eqref{C1} is said to
have a nontrivial solution in the Sobolev space $H^1({\mathbb{R}}^6)$.

We now discuss the {\em cusp conditions\/} at these singularities. {\em What
is a cusp condition}? It can be simply explained in the following paragraph.
Let us elucidate it for the two particle Hamiltonian \eqref{C1}; for a
multi-particle Hamiltonian the idea is the same.

In order for the wave function $\psi $ to satisfy the eigenvalue problem
\eqref{C0} at the singularities \eqref{C2}, the kinetic energy operators $%
-\nabla _1^2/2m_1$ and $-\nabla _2^2/2m_2$, after acting on $\psi $, must
produce terms that {\em exactly cancel\/} those singularity terms in the
potential in order to give us back just a constant $E$ times $\psi $,
because the wave function $\psi $ is {\em bounded\/} everywhere in space,
including the points where the nuclei are located, without exception. One
can see that, if the cusp conditions are not satisfied, then there is some
{\em unboundedness\/} at the singularities \eqref{C2} which can
affect the accuracy in numerical computations. Conversely, if the cusp
conditions are satisfied, this normally improves the numerical accuracy.

In case we don't know the exact eigenstate, but only a certain {\em trial
wave function}, say $\phi $, then $(\hat H-E)\phi =0$ will not be satisfied
in general. Rather, we have
$$
(\hat H-E)\phi (\pmb{r}_1,\pmb{r}_2)=f(\pmb{r}_1,\pmb{r}_2)
$$
for some function $f$ depending on the spatial variables $\pmb{r}_1$ and $%
\pmb{r}_2$. However, we can insist on choosing parameters in $\phi $ such
that the residual $f(\pmb{r}_1,\pmb{r}_2)$ is a {\em bounded function
everywhere}; in particular, $f(\pmb{r}_1,\pmb{r}_2)$ cannot contain any
singularity at \eqref{C2}. We say that the trial wave function $\phi $
satisfies

\begin{itemize}
\item[(i)]  the {\em electron-nucleus cusp condition\/} at $a$ (resp.\ $b$)
if $f$ is not singular at $a$ (resp.\ $b$);

\item[(ii)]  the {\em interelectronic (or electron-electron) cusp condition\/%
} if $f$ is not singular when $r_{12}=0$.
\end{itemize}

\begin{figure}[htpb]
\bigskip
\centerline{\epsfxsize=0.4\textwidth\epsfysize=0.25\textwidth
\epsfbox{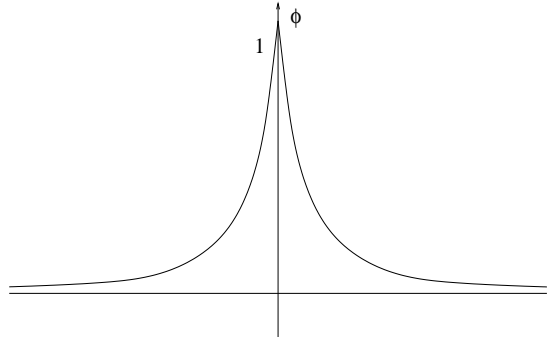}}

\caption{A 1-dimensional cross section of $\pmb{\phi(r)= e^{-r}}$
showing a cusp.}
\label{cuspf}
\end{figure}

For example, in the simple case of a hydrogen atom,
$$
\hat H=-\frac 12\nabla ^2-\frac 1r,
$$
let $\phi (\pmb{r})=Ce^{-\alpha r}$ be a trial wave function. Then for any $E$%
,
$$
(\hat H-E)\phi =C\left\{
\frac{(\alpha-1)e^{-\alpha r}}r
-(E+\alpha^2/2)e^{-\alpha r}\right\} .
$$
The singularity $1/r$ can be eliminated only by choosing $\alpha =1$. This
is the cusp condition, which actually forces $\phi $ to be the ground state
(with $E=-1/2$). The profile of $\phi $, as shown in Fig. \ref{cuspf}
illustrates the appearance of a cusp at the origin.

In Appendix \ref{cuspd} we derive the cusp conditions for the two particle
electron wave function $\psi $ of \eqref{C0}:
\begin{equation}
\label{cu1}\left. \frac{\partial \psi }{\partial r_{1a}}\right|
_{r_{1a}=0}=-m_1Z_a\psi (r_{1a}=0),\quad \left. \frac{\partial \psi }{%
\partial r_{1b}}\right| _{r_{1b}=0}=-m_1Z_b\psi (r_{1b}=0),
\end{equation}
\begin{equation}
\label{cu2}\left. \frac{\partial \psi }{\partial r_{2a}}\right|
_{r_{2a}=0}=-m_2Z_a\psi (r_{2a}=0),\quad \left. \frac{\partial \psi }{%
\partial r_{2b}}\right| _{r_{2b}=0}=-m_2Z_b\psi (r_{2b}=0),
\end{equation}
\begin{equation}
\label{cu3}\left. \frac{\partial \psi }{\partial r_{12}}\right| _{r_{12}=0}=%
\frac{m_1m_2}{m_1+m_2}q_1q_2\psi (r_{12}=0).
\end{equation}
Eqs. \eqref{cu1}, \eqref{cu2} are the {\em electron-nucleus cusp condition, }%
while Eq. \eqref{cu3} is the {\em interelectronic} condition.

In forming
trial wave functions from one-centered or two-centered orbitals for a
homonuclear diatomic molecule a commonly used wave function is
\begin{equation}
\label{C12a}\psi (\pmb{r}_1,\pmb{r}_2)=\phi (\pmb{r}_1)\phi (\pmb{r}%
_2)f(r_{12}),\quad f(r_{12})=1+\frac 12r_{12},\qquad \text{(cf.\ \eqref{I.17})}
\end{equation}
where $\phi (\pmb{r}_i)$, $i=1,2$, is an orbital for the molecular ion. In
Appendix \ref{cuspd} we show that \eqref{C12a} satisfies the interelectronic
cusp condition. If, however, $\phi _1\ne \phi _2$, then the trial
wave function
$$
\psi (\pmb{r}_1,\pmb{r}_2)=\phi _1(\pmb{r}_1)\phi _2(\pmb{r}_2)\left(
1+\frac 12r_{12}\right)
$$
satisfies the interelectronic cusp condition if and only if
$$
\phi _2(\pmb{r})\nabla \phi _1(\pmb{r})-\phi _1(\pmb{r})\nabla \phi _2(%
\pmb{r})=\pmb{0}.
$$

The actual verifications of cusp conditions for {\em specifically given\/}
examples of trial wave functions in the cases of one-centered orbitals or their
products are not difficult.
But such work is nontrivial when the trial wave functions are expressed in terms
of prolate spheroidal coordinates. In Appendix \ref{cuspprol} we
illustrate through concrete examples how to carry out this task.

\subsection{Various forms of the correlation function $\pmb{f(r_{12})}$}

\label{sec4}

\indent

We have learned the importance of the interelectronic cusp condition in
Section \ref{sec3.1}. But there are, in addition, three important constructs
that are crucial for diatomic calculations:\ {\em orbitals, configurations
and electronic correlation}. In this section, we compile a list of often
cited correlation functions $f$ which help the satisfaction of the
interelectronic cusp conditions in the context of Eq. \eqref{cu3}. The study
of the other two, i.e., orbitals and configurations, will be addressed in
the next section.

\vspace{0.1cm}
\noindent
{\large {\bf (1)}}~$\displaystyle%
\pmb{f(r_{12}) = 1+\frac12 r_{12}}$.
\vspace{0.1cm}

\noindent This is the simplest possible interelectronic
configuration function. The specific form is due to the {\em correlation
cusp condition only}. We derive it as follows (see Patil, Tang \& Toennies
\cite{PTT}). Consider two charged particles which are described by the
Schr\"odinger equation
\begin{equation}
\label{F1.1}\left( -\frac 1{2m_1}\nabla _1^2-\frac 1{2m_2}\nabla _2^2+\frac{%
q_1q_2}{r_{12}}\right) \psi =\epsilon \psi .
\end{equation}
First, transform the above equation to the center-of-mass coordinates (cf.
Appendix \ref{CMC})
\begin{equation}
\label{F1.2}\left( -\frac 1{2M}\nabla _S^2-\frac 1{2\mu }\nabla _{r_{12}}^2+%
\frac{q_1q_2}{r_{12}}\right) \psi =\epsilon \psi ,
\end{equation}
where
\begin{equation}
\label{F1.3}M=m_1+m_2,\quad \mu =\frac{m_1m_2}{m_1+m_2},\quad \pmb{S}=\frac{%
m_1\pmb{r}_1+m_2\pmb{r}_2}{m_1+m_2},\qquad \pmb{r}_{12}=\pmb{r}_1-\pmb{
r}_2.
\end{equation}
In the ground state $\psi $ is independent of $\pmb{S}$. Near the
singularity point $r_{12}=0$, the wave function $\psi $ has a local
representation as a power series
\begin{equation}
\label{F1.4}\psi =C_0+C_1r_{12}+{\cl O}(r_{12}^2).
\end{equation}
Substituting \eqref{F1.4} into \eqref{F1.2} and using $\nabla _r^2=\frac
1{r^2}\frac \partial {\partial r}\left( r^2\frac \partial {\partial
r}\right) $ we obtain

\begin{equation}
\frac1{r_{12}} \left(-\frac{C_1}\mu + q_1q_2C_0\right) + \left(-\frac{3C_2}%
\mu + C_1q_1q_2\right) +\cdots = \epsilon(C_0+C_1r_{12} + {\cl O}%
(r^2_{12}))
\end{equation}
The above mandates that the coefficient of $1/r_{12}$ must vanish, that is
$$
C_1=\mu q_1q_2C_0,
$$
which for $\mu =1/2$ (i.e., $m_1=m_2$), and $q_1=q_2=1$, yields
$$
C_1=\frac 12C_0.
$$
This gives us the small $r_{12}$ behavior
\begin{equation}
\label{cfr12}\psi =C_0\left( 1+\frac 12r_{12}\right) ,
\end{equation}
where we have dropped all the ${\cl O}(r_{12}^2)$ terms. The asymptotic
expression \eqref{cfr12} motivates the choice of $f(r_{12})$ in such
particular form.

This simple $f(r_{12})$ captures {\em short distance\/} interelectronic interaction
very well. It offers elegant representations of molecular orbitals and great
facility to computation. Nevertheless, its asymptotic behavior of linear
growth for large $r_{12}$ is not physically correct.

When we write molecular orbitals as
$$
\psi(\pmb{r}_1,\pmb{r}_2,r_{12}) = \phi(\pmb{r}_1,\pmb{r}_2) f(r_{12}),
$$
if the function $\phi(\pmb{r}_1,\pmb{r}_2)$ is already quite small in the
region where $r_{12}$ becomes large compared to 1, then this simple $%
f(r_{12}) = 1 + \frac12 r_{12}$ can work quite well (Kleinekath\"ofer et
al.\ \cite[pp.~2841--2842]{K}).

\noindent
{\large {\bf (2)}}~~$\displaystyle%
\pmb{f(r_{12}) = 1 + \frac{r_{12}}2 e^{-r_{12}/d},\quad
(d>0)}$

\vspace{0.1cm}

\begin{figure}[htpb]
\bigskip
\centerline{\epsfxsize=0.3\textwidth\epsfysize=0.25\textwidth
\epsfbox{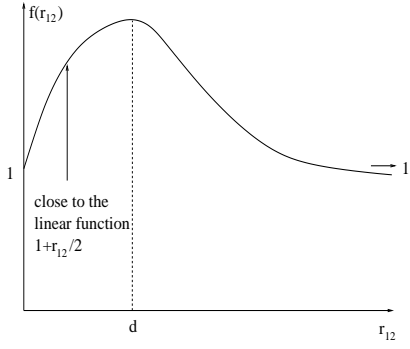}}

\caption{
Graph of $\pmb{f(r_{12}) = 1 + \frac{r_{12}}{2}
e^{-r_{12}/d}}$, where the maximum happens at $\pmb{r_{12}=d}$.
}
\label{ff1}
\end{figure}

This function was proposed by Hirschfelder
\cite{Hir} where $d$ is a variational parameter. Its profiles is shown in
Fig. \ref{ff1}.
It satisfies the cusp condition near $r_{12}=0$. This function was
used by Siebbles, Marshall and Le Sech \cite{SML}. Nevertheless, Le Sech et
al. \cite{SML,Sech96} reported that in performing variational calculations
by writing $f(r_{12}) = 1 + \frac{r_{12}}2 e^{-\alpha r_{12}}$, they found
that $\alpha$ is computed to be very close to 0 for small and intermediate $R$.
That is, it virtually degenerates into $f(r_{12}) = 1 + \frac12 r_{12}$.
However, at large $R$ one should use $\alpha\ne 0$ in order to obtain the
correct dissociation limit.

>From physical considerations, there is no reason to believe why $f(r_{12})$
should have a local maximum as shown in Fig. \ref{ff1}. Even though the
choice of
this $f(r_{12})$ seems satisfactory asymptotically for both $r_{12}$ small
and large, it may not be satisfactory for medium values of $r_{12}$.\medskip

\vspace{0.1cm}
\noindent
{\large {\bf (3)}}~$\displaystyle%
\pmb{f(r_{12}) =1-\frac1{1+2\lambda}e^{-\lambda r_{12}},
(\lambda>0)}$
\vspace{0.1cm}

\noindent This correlation function is partly motivated by Hirschfelder's
work \cite{Hir}, and partly by Hylleraas study of the helium atom \cite
{hylleraas}. It was introduced by Kleinekath\"ofer et al.\ \cite{K}.

\begin{figure}[htpb]
\bigskip
\centerline{\epsfxsize=0.4\textwidth\epsfysize=0.25\textwidth
\epsfbox{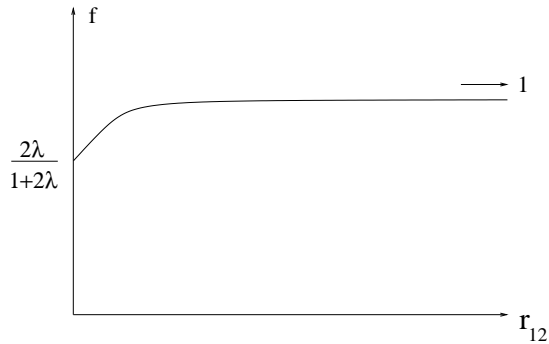}}

\caption{
Graph of $\pmb{f(r_{12}) = 1 - \frac1{1+2\lambda}e^{-\lambda
r_{12}}}$.
}
\label{ff2}
\end{figure}

At small $r_{12}$ we have the expansion

$$
f(r_{12})=\frac{2\lambda }{1+2\lambda }\left( 1+\frac 12r_{12}-\frac \lambda
4r_{12}^2\pm \cdots \right) .
$$
Therefore the cusp condition is satisfied for any $\lambda >0$. This $%
\lambda $ can be either used as a variational parameter, or be determined
from a given Hamiltonian. For example, for a helium-like 2-electron atom
with nuclear charge $Z$,
using a perturbation argument, Kleinekath\"ofer et al.\ \cite{K} analyzed
that the best value for $\lambda $ is
$$
\lambda =\frac 5{12}Z-\frac 13.
$$
This $f(r_{12})$ has a monotone profile and correct asymptotics for both $%
r_{12}$ small and large. See Fig. \ref{ff2}.

\vspace{0.1cm}
\noindent
{\large {\bf (4)}}~$\displaystyle%
\pmb{f(r_{12}) = e^{\frac12 r_{12}}}$.
\vspace{0.1cm}

\noindent This $f(r_{12})$ satisfies the correlation-wave equation
$$
\left( -\frac 12\nabla _1^2-\frac 12\nabla _2^2+\frac 1{r_{12}}\right)
e^{\frac 12r_{12}}=-\frac 14e^{\frac 12r_{12}}
$$
with a negative energy $-1/4$. For small $r_{12}$, its expansion is
$$
e^{\frac 12r_{12}}=1+\frac 12r_{12}+{\cl O}(r_{12}^2).
$$
Therefore, it satisfies the correlation-cusp condition and its asymptotics
for small $r_{12}$ is good. However, this $f(r_{12})$ has exponential growth
for large $r_{12}$ and is thus physically incorrect.\medskip

\vspace{0.1cm}
\noindent
{\large {\bf (5)}}~$\displaystyle%
\pmb{f(r,r_{12})=\frac{\sinh(tr)}{tr}\cdot
\frac{F_0(\frac1{2k},kr_{12})}{r_{12}}}$
\vspace{0.1cm}

\noindent where $F_j(\eta ,\rho )$ is the Coulomb wave function regular at
the origin, $j$ is an integer, $t,k$ are separation of variables constants
and $r=|{\bf r}_1+{\bf r}_2|$.

\begin{figure}[htpb]
\bigskip
\centerline{\epsfxsize=0.4\textwidth\epsfysize=0.25\textwidth
\epsfbox{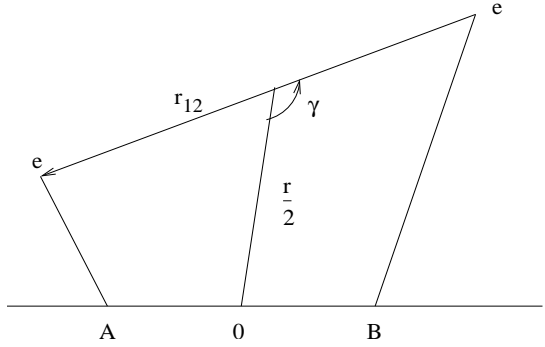}}

\caption{
The variables $\pmb{r,r_{12}}$ and angle
$\pmb{\gamma}$ in the center-of-mass coordinates.
}

\label{ff3}
\end{figure}

This is perhaps the most complex form of the correlation function in the
literature, given by Aubert--Fr\'econ and Le Sech \cite{A-FL}. It comes from
solving
\begin{equation}
\label{F6.1}\left\{ -\frac 12\nabla _1^2-\frac 12\nabla _2^2+\frac
1{r_{12}}\right\} f=\epsilon f.
\end{equation}
Rewrite the above in center-of-mass coordinates:
\begin{equation}
\label{F6.2}\left( -\frac 14\nabla _S^2-\nabla _{r_{12}}^2+\frac
1{r_{12}}\right) f=\epsilon f.
\end{equation}
Assume that $f$ {\em depends only on $r,r_{12}$ and\/} $\gamma $, where $%
\gamma $ is the angle between $\pmb{S}$ and $\pmb{r}_{12}$ (see Fig. \ref{ff3}).

Then equation \eqref{F6.2} can be written as
\begin{align}
&-\frac1{r^2_{12}} \frac\partial{\partial r_{12}} \left(r^2_{12} \frac{\partial
f}{\partial r_{12}}\right) - \frac1{r^2} \frac\partial{\partial r} \left(r^2
\frac{\partial f}{\partial r}\right) - \left[\frac1{r^2_{12}}
\frac\partial{\partial q} (1-q^2) \frac\partial{\partial q}f\right]\nonumber\\
\label{F6.3}
&\quad - \left[\frac1{r^2} \frac\partial{\partial q} (1-q^2) \frac{\partial
f}{\partial q}\right] + \frac1{r_{12}} f = \ep f, \qquad q\equiv \cos\gamma.
\end{align}
Equation \eqref{F6.3} can be separated by writing
$$
f=P_j(q)g^j(r)u^j(r_{12}),
$$
where
\begin{equation}
\label{F7.1}\frac \partial {\partial q}(1-q^2)\frac{\partial P_j}{\partial q}%
=-j(j+1)P_j,
\end{equation}
$P_j$ is the Legendre polynomial of degree $j$ and $g^j(r)$ and $u^j(r_{12})$
satisfy, respectively,
\begin{align}
\label{F7.2}
&\left\{\frac{d^2}{dr^2} + \frac2r \frac{d}{dr} - \frac{j(j+1)}{r^2} +
\ep_g\right\} g^j(r) = 0,\\
\label{F7.3}
&\left\{\frac{d^2}{dr^2_{12}}  + \frac2{r_{12}} \frac{d}{dr_{12}} -
\frac{j(j+1)}{r^2_{12}} - \frac1{r_{12}} + \ep_u\right\} u^j(r_{12}) =0,
\end{align}
with
$$
\epsilon =\epsilon _g+\epsilon _u.
$$
Take
\begin{equation}
\label{F7.4}\epsilon _g=-t^2,\quad \epsilon _u=k^2,\quad j=0.
\end{equation}
Then
$$
P_0(q)= 1, \quad g^0_t(r) = \frac{\sinh(tr)}{tr}, \quad u^0_k = \frac{%
F_0(\frac1{2k},kr_{12})}{r_{12}},
$$
where the function $F_0(\eta ,\rho )$ belongs to the class of regular
Coulomb wave function $F_L(\eta ,\rho )$, satisfying
$$
\frac{d^2}{d\rho ^2}F_L(\eta ,\rho )+\left[ 1-\frac{2\eta }\rho -\frac{L(L+1)%
}{\rho ^2}\right] F_L(\eta ,\rho )=0
$$
for the two parameters $\eta $ and $L$.

Note that $t$ and $k$ in \eqref{F7.4} can be used as {\em variational
parameters}. For small $r_{12}$,
$$
F_0\left( \frac 1{2k},kr_{12}\right) =C\cdot kr_{12}\left[ 1+\frac 12r_{12}+
{\cl O}((kr_{12})^2)\right]
$$
and, thus,
$$
u_k^0(r_{12})=\frac{F_0(\frac 1{2k},kr_{12})}{r_{12}}=C\cdot k\left[ 1+\frac
12r_{12}+{\cl O}((kr_{12})^2)\right]
$$
satisfies the correlation-cusp condition for {\em any given\/} $k$.

For large $r_{12}$, the Coulomb wave function $F_L(\eta,\rho)$ has an
asymptotic expansion
\begin{align*}
F_L &= g\cos\theta_L + f\sin \theta_L,\\
\theta_L &\equiv \rho-\eta \ln z \rho - L\frac\pi2 + \sigma_L,\\
\sigma_L &\equiv \arg \Gamma (L+1+i\eta),\\
f &\sim \sum^\infty_{k=0} f_k,\quad g\sim \sum^\infty_{k=0} g_k,\\
f_0 &= 1, \quad g_0 =1,\\
f_{k+1} &= a_kf_k-b_kg_k,\quad g_{k+1} = a_kg_k+b_kf_k,\\
a_k &= \frac{(2k+1)\eta}{(2k+2)\rho},\quad b_k =
\frac{L(L+1)-k(k+1)+\eta^2}{(2k+2)\rho}.
\end{align*}
Thus
$$
\lim_{r_{12}\to \infty} u^0_k(r_{12}) = \lim_{r_{12}\to\infty} \frac{%
F_0(\frac1{2k},kr_{12})}{r_{12}} = 0.
$$

\begin{figure}[htpb]
\bigskip
\centerline{\epsfxsize=0.5\textwidth\epsfysize=0.15\textwidth
\epsfbox{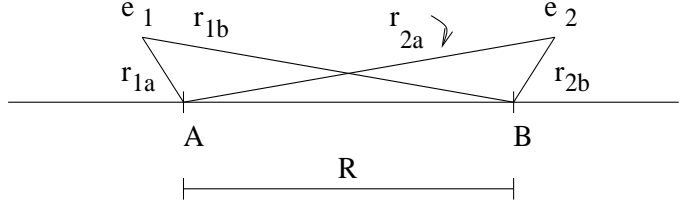}}

\caption{
For large $\pmb{R}$, electron $\pmb{e_1}$ is
localized near $\pmb{A}$, and electron $\pmb{e_2}$ is localized near $\pmb{B}$.
}

\label{ff4}
\end{figure}

\vspace{0.1cm}
\noindent
{\large {\bf (6)}}~$\displaystyle%
\pmb{f(r_{12})={}_1F_1\left(-\frac1{2Z},2,-2Zr_{12}%
\right)}$
\vspace{0.1cm}

\noindent
where ${}_1F_1(a;b;x)$ is the {\em confluent hypergeometric function\/}
satisfying the differential equation
\begin{equation}
\label{F9.1}x \frac{d^2w}{dx^2} + (b-x) \frac{dw}{dx} - aw = 0.
\end{equation}
This $f(r_{12})$ was given by Patil, Tang and Toennies \cite{PTT} for the
case {\em when the internuclear separation $R$ is large}:
\begin{equation}
\label{F9.2}R \gg \frac3Z.
\end{equation}
Its derivation can be motivated as follows. Consider
\begin{equation}
\label{F9.3}H = - \frac12 \nabla^2_1 - \frac12 \nabla^2_2 - \left(\frac{Z}{%
r_{1a}} + \frac{Z}{r_{1b}} + \frac{Z}{r_{2a}} + \frac{Z}{r_{2b}}\right) +
\frac1{r_{12}} + \frac{Z^2}{R}
\end{equation}
for an H$_2$-like molecule. Guillemin--Zener-type one-electron wave functions
\cite{GZ} suggest the molecular orbital for large $R$:
\begin{equation}
\label{F9.4}\Psi = (e^{-Zr_{1a}-Zr_{2b}} + e^{-Zr_{1b}-Zr_{2a}}) f(r_{12}).
\end{equation}
When $R$ is large, electron 1 is localized around $A$ and electron 2 is
localized around $B$, as shown in Fig. \ref{ff4}.

We have
\begin{equation}
\label{F9.5}r_{1b}\gg r_{1a},\quad r_{2a}\gg r_{2b}.
\end{equation}
Because of \eqref{F9.5}, we have, from \eqref{F9.4},
\begin{align}
\Psi &= (e^{-Zr_{1a}-Zr_{2b}} + e^{-Zr_{1b}-Zr_{2a}}) f(r_{12})\nonumber\\
\label{F9.6}
&\approx e^{-Zr_{1a}-Zr_{2b}} f(r_{12}).
\end{align}
Substituting \eqref{F9.6} into $\hat H\psi =E\psi $, we obtain
\begin{equation}
\label{F9.7}-\frac 12(\nabla _1^2+\nabla _2^2)f+Z(\nabla _1f)\cdot
(\hat{\pmb{r}}_{1a} + \hat{\pmb{r}}_{2b}) +\frac
1{r_{12}}f={\mathcal{}} \mathcal{O}\left( \frac 1R\right) .
\end{equation}
Note that Eq.\ \eqref{F9.7} {\em contains the effect of cross terms. For
large $R$, the electron-electron correlation is most significant when the
two electrons are colinear and in between the two nuclei}. In this
situation, either $\pmb{r}_{12}$ is antiparallel to $\pmb{r}_1-\pmb{r}_a$,
or $\pmb{r}_{12}$ is parallel to $\pmb{r}_2-\pmb{r}_b$.

Now, using the center-of-mass coordinates and dropping all the ${\cl O}%
\left( 1/R\right) $ terms, we obtain
\begin{equation}
\label{F9.8}-\left( \frac{\partial ^2}{\partial r_{12}^2}+\frac
2{r_{12}}\frac \partial {\partial r_{12}}\right) f+\frac 1{r_{12}}f-2Z\frac{%
\partial f}{\partial r_{12}}=0.
\end{equation}
The solution to \eqref{F9.8}, after setting it into the form of \eqref{F9.1}
is
$$
f(r_{12})={}_1F_1\left( -\frac 1{2Z},2,-2Zr_{12}\right) .
$$
For small $r_{12}$, the expansion is
$$
f(r_{12})=1+\frac 12r_{12}-\frac{(2Z-1)}{12}r_{12}^2+\cdots ~.
$$
Therefore, the correlation-cusp condition is satisfied. For large $r_{12}$,
the asymptotics is
$$
f(r_{12})\sim C(-2Zr_{12})^{\frac 1{2Z}}
$$
(cf.\ Abramowitz and Stegun \cite[p.~508, 13.5.1]{AS}).

\section{Modeling of diatomic molecules}\label{sec5}

\indent

In this section, we give a survey of major existing methods for the numerical
modeling
of diatomic molecules. These methods provide approximations of wave functions
either in explicit form through properly selected ansatzs, or in implicit form
through iterations as numerical solutions of integro-partial differential
equations.

The methods and ansatzs to be described below are

\begin{itemize}
\item[(1)] {\em The Heitler--London method};
\item[(2)] {\em The Hund--Mulliken method};
\item[(3)] {\em The Hartree--Fock (self-consistent) method};
\item[(4)] {\em The James--Coolidge wave function};
\item[(5)] {\em Two-centered orbitals};
\end{itemize}

Items (1)$\sim$(3) above historically are associated with one-centered
orbitals, while (4)$\sim$(5) are based on two-centered orbitals. But this
dichotomy is not inflexible.
An example is a {\em hybrid\/} type
containing both one-centered and two-centered orbitals considerd
in Subsection \ref{sec2a}.C.6.
We now discuss them in
sequential order below. Each approach has a set of modeling parameters which
can be optimized through calculus of variations. In particular, we will point
out what these parameters are.

\subsection{The Heitler--London method}

\indent

This method has the longest history. It was developed by Heitler and
London during the 1920s soon after Heisenberg laid the quantum mechanical
foundation of ferromagnetism. The method is usually called the valence-bond (or
atomic orbital) method. In this method, each molecule is thought of as composed
of atoms, and the electronic structure is described using atomic orbitals of
these atoms.

Here we present a version of refined
Heitler--London approach due to Slater \cite{Slat63}. In the method, electron {\em
spin-orbitals\/} are
taken from a determinant
\begin{equation}\label{M2.1}
\left|\begin{matrix}
u_1(1)&u_1(2)&\cdots&u_1(n)\\
u_2(1)&u_2(2)&\cdots&u_2(n)\\
\cdots\\
u_n(1)&u_n(2)&&u_n(n)\end{matrix}\right|,
\end{equation}
which satisfy the Fermionic property of the Pauli exclusion principle. The
orbitals in \eqref{M2.1} are called the Slater orbitals. Let us consider the
2-electron case, i.e., $n=2$ in \eqref{M2.1}. The spin-orbital $u_i(j)$,
$i,j=1,2$, consists of

\n{\bf (i)~~the electron-atomic orbital part}

\begin{alignat}{2}
\label{M2.2}
a(1) &= \sqrt{\frac{\alpha^3}\pi} e^{-\alpha r_{1a}},&\qquad a(2) &=
\sqrt{\frac{\alpha^3}\pi} e^{-\alpha r_{2a}},\\
\label{M2.3}
b(1) &= \sqrt{\frac{\alpha^3}\pi} e^{-\alpha r_{1b}},&\qquad b(2) &=
\sqrt{\frac{\alpha^3}\pi} e^{-\alpha r_{2b}},
\end{alignat}
where in \eqref{M2.2} and \eqref{M2.3}, the atomic electron wave functions are
centered at, respectively, $a$ and $b$.\medskip

\n {\bf (ii)~~the spin part}

\[
\begin{matrix}
\text{spin} &\alpha(1),& \alpha(2),& \alpha(j) = |s,m_s\rangle,& \uparrow,
&\text{for } j=1,2,\\
\text{spin}&\beta(1),&\beta(2),&\beta(j)=|s,m_s\rangle&\downarrow,&\text{for }
j=1,2.
\end{matrix}
\]
The linear combinations of the total spin-orbital wave function that are
antisymmetric are tabulated below:
\begin{equation}\label{M3.1}
\begin{array}{|c|l|c|}
\hline
&\multicolumn{1}{c|}{\text{spin-orbital wave functions}}&M_s \text{ (total
spin)}\\
\hline
\text{singlet state}&[a(1)b(2) + b(1)a(2)] [\alpha(1)\beta(2)-\beta(1)
\alpha(2)]&0\\
\hline
&[a(1)b(2)-b(1)a(2)][\alpha(1)\alpha(2)]&1\\
\cline{2-3}
\text{triplet
states}&[a(1)b(2)-b(1)a(2)][\alpha(1)\beta(2)+\beta(1)\alpha(2)]&0\\
\cline{2-3}&[a(1)b(2)-b(1)a(2)][\beta(1)\beta(2)]&-1\\
\hline
\end{array}
\end{equation}

\n {\bf Table 1. Spin-orbital wave functions of singlet and triplet
states}\bigskip

\n The singlet state has lower energy than the triplets. For example, if we aim
to calculate the ground state of H$_2$, we use
\eqref{M3.1} as the trial wave function to minimize the total energy
\begin{equation}\label{M3.3}
\langle\Psi|
{\hat H}|\Psi\rangle, \text{ subject to } \langle\Psi|\Psi\rangle = 1,
\end{equation}
where
\begin{align}\label{M3.3a}
{\hat H}
&= -\frac12 \nabla^2_1 - \frac12 \nabla^2_2 - \frac1{r_{1a}} - \frac1{r_{1b}}
- \frac1{r_{2a}} - \frac1{r_{2b}} + \frac1{r_{12}} + \frac1R\\
\label{M3.3b}
\Psi &= {\cl N}[a(1)b(2) + b(1)a(2)],\quad {\cl N} = \text{ a normalization
factor.}
\end{align}
Heisenberg, and Heitler--London's basic point of view is that ${\hat H}$ in
\eqref{M3.3a} can be written as
\[
{\hat H} = \left(-\frac12 \nabla^2_1 - \frac1{r_{1a}}\right) + \left(-\frac12
\nabla^2_2 - \frac1{r_{2b}}\right) + \left(-\frac1{r_{1b}} - \frac1{r_{2a}} +
\frac1{r_{12}} + \frac1R\right),
\]
where the terms inside the third pair of parentheses above can be viewed as a
``perturbation.''

Since the Heitler--London method is quite fundamental in molecular chemistry,
let us give some details about the calculation of \eqref{M3.3} given
\eqref{M3.3a} and \eqref{M3.3b}.

Define
\begin{align*}
S &\equiv \text{the overlap } \equiv \langle a(1)|b(1)\rangle =
\int\limits_{{\bb R}^3} e^{-\alpha r_{1a}} e^{-\alpha r_{1b}}\ dx\cdot
\frac{\alpha^3}\pi \qquad (dx = dx_1dx_2dx_3)\\
&= e^{-\alpha R} \left[1 + \alpha R + \frac{(\alpha R)^2}3\right].
\end{align*}
Then $\langle a(1)b(2)| a(2)b(1)\rangle =\langle a(1)|b(1)\rangle \langle
b(2)|a(2)\rangle = S^2$, and the normalized state for the singlet or the
triplets, without spin, is
\[
\Psi = \frac{a(1)b(2) \pm a(2)b(1)}{\sqrt{2(1\pm S^2)}}, \text{ such that }
\langle\Psi|\Psi\rangle = 1.
\]

The total energy of the singlet and triplet states can now be written as
\begin{equation}
E_\pm = \frac1{2(1\pm S^2)} \langle a(1)b(2)\pm a(2)b(1)| - \frac12 \nabla^2_1 -
\frac12 \nabla^2_2 - \frac1{r_{1a}} - \frac1{r_{1b}} - \frac1{r_{2a}} -
\frac1{r_{2b}} + \frac1{r_{12}} + \frac1R |a(1)b(2) \pm a(2)b(1)\rangle.
\label{HLINT}
\end{equation}
The integrals involved are given in Appendix \ref{integ}.
Using these integrals, we are able to write down the
total energy $E=KE + PE$ as follows:
\begin{align*}
KE_\pm &= \text{kinetic energy}\\
&= \frac1{2(1\pm S^2)} \langle a(1)b(2) \pm a(2)b(1)| -\frac12 \nabla^2_1 -
\frac12 \nabla^2_2 |a(1)b(2) \pm a(2)b(1)\rangle\\
&= \frac{\alpha^2}{1\pm S^2} (1\mp 2KS\mp S^2);
\end{align*}
\begin{align*}
PE_\pm &= \text{potential energy}\\
&= \frac1{2(1\pm S^2)} \langle a(1)b(2) \pm a(2)b(1)| -\frac1{r_{1a}} -
\frac1{r_{1b}} - \frac1{r_{2a}} - \frac1{r_{2b}} + \frac1{r_{12}} + \frac1R
|a(1)b(2)\pm a(2)b(1)\rangle\\
&= \frac\alpha{(1\pm S^2)} (-2 + 2J + J' \pm 4KS \pm K') + \frac\alpha{w},
\end{align*}
where $w=\alpha R$, the other symbols are defined in Appendix \ref{integ}.
The parameter $\alpha$ can be used as the {\em variational parameter\/} to
minimize \eqref{M3.3} \cite{Rose31}.

Consider the special case when $\alpha=1$. Then
\begin{align*}
E_\pm &= -1 + \frac{H_0\pm H_1}{(1\pm S^2)},
\end{align*}
where
\begin{align*}
H_0
&= \iint\limits_{{\bb R}^6} a^2(1)b^2(1) \left(-\frac1{r_{1b}} - \frac1{r_{2a}}
+ \frac1{r_{12}} + \frac1R\right) \ dx dy\\
&= 2J + J' + \frac1R
\end{align*}
is the Coulomb integral,
and
\begin{align*}
H_1
&= \iint\limits_{{\bb R}^6} a(1)b(1)a(2)b(2) \left(-\frac1{r_{1b}} -
\frac1{r_{1a}} + \frac1{r_{12}} + \frac1R\right) \ dxdy\\
&= 2KS + K' + \frac{S^2}R
\end{align*}
is the exchange integral.
According to numerical values computed in Slater \cite[Table 3.2]{Slat63}, e.g., it
is
known that $H_1$ is usually many times larger than $H_0$, and is largely
responsible for the attraction between atoms in forming a molecule.

\begin{figure}[htpb]
\bigskip
\centerline{\epsfxsize=0.5\textwidth\epsfysize=0.5\textwidth
\epsfbox{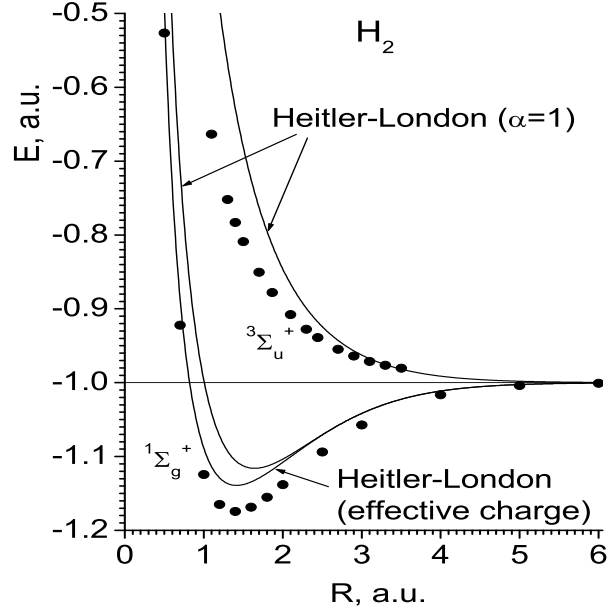}}

\caption{
Ground and first excited state energy $E(R)$ of H$_2$ molecule
for the Heitler--London wave function (solid lines)
and the ``exact'' energy of Ref. \cite{Kolo60} (dots).
}
\label{HLF}
\end{figure}

In Fig. \ref{HLF} we plot the ground $^1\Sigma_g^+$ and first excited
$^3\Sigma_u^+$ state potential energy curves $E(R)$ of the H$_{2}$
molecule.
When $\alpha =1$ the ground state curve yields the binding energy
of 0.116 a.u.=3.16 eV; the value must be compared with 4.748 eV obtained by
Kolos  and Roothaan \cite{Kolo60}. When $\alpha$ (effective charge) is treated as a
variational parameter \cite{Wang28}
the calculation yields the binding energy of 0.139
a.u.=3.78 eV and the bond length of 1.41 Bohr radii.
For the $^3\Sigma_u^+$ state the effective charge and the $\alpha =1$ curves
are practically indistinguishable.

\subsection{The Hund--Mulliken method}

\indent

In 1927, Robert Mulliken worked with Friedrich Hund and developed the
Hund--Mulliken molecular orbital theory in which electrons are assigned to
states over an entire molecule. Hund--Mulliken's molecular orbital method was
more flexible and applicable than the traditional Valence-Bond theory that had
previously prevailed. Because of this, Mulliken received the Nobel Prize in
Chemistry in 1966.

The approach has some similarity to the Heitler--London's, so we can inherit
the notation from there. Its special feature is that a linear combination of
the molecular
{\em gerade\/} (g) and {\em ungerade\/} (u) states are used:
\begin{equation}\label{M4.1}
\left.\begin{matrix}
{}^1\Sigma_{g}^+ : (g^+g^-):&\frac{a(1)+b(1)}{\sqrt{2(1+S)}}
\frac{a(2)+b(2)}{\sqrt{2(1+S)}} \frac{\uparrow_1\downarrow_2 - \downarrow_1
\uparrow_2}{\sqrt 2},\\
{}^1\Sigma_{g}^- : (u^+u^-) :&\frac{a(1)-b(1)}{\sqrt{2(1-S)}}
\frac{a(2)-b(2)}{\sqrt{2(1-S)}} \frac{\uparrow_1 \downarrow_2 - \downarrow_1
\uparrow_2}{\sqrt 2},\\
{}^1\Sigma_u : (g^+u^- - g^-u^+)&\frac{a(1)a(2) - b(1)b(2)}{\sqrt{2(1-S^2)}}
\frac{\uparrow_1\downarrow_2 - \downarrow_1\uparrow_2}{\sqrt 2},\\
{}^3\Sigma^+_u : (g^+u^+)&\frac{a(1)b(2) - b(1)a(2)}{\sqrt{2(1-S^2)}} \uparrow_1
\uparrow_2,&(M_s=1)\\
{}^3\Sigma_u : (g^+u^- + g^-u^+)&\frac{a(1)b(2) - b(1)a(2)}{\sqrt{2(1-S^2)}}
\frac{\uparrow_1\downarrow_2 + \downarrow_1\uparrow_2}{\sqrt 2},&(M_s=0)\\
{}^3\Sigma^-_u : (g^-u^-)&\frac{a(1)b(2) - b(1)a(2)}{\sqrt{2(1-S^2)}}
\downarrow_1 \downarrow_2,&(M_s=-1)
\end{matrix}\right\}
\end{equation}
with
\begin{gather*}
a(i) = \sqrt{\frac{\alpha^3}\pi} e^{-\alpha r_{ia}},\quad b(i) =
\sqrt{\frac{\alpha^3}\pi} e^{-\alpha r_{ib}},\qquad i=1,2,\\
g\colon \ \frac{a(i) + b(i)}{\sqrt{2(1+S)}}\qquad u\colon \ \frac{a(i) -
b(i)}{\sqrt{2(1-S)}},\qquad i=1,2.
\end{gather*}
Especially, note that the first three states in \eqref{M4.1}, i.e.,
${}^1\Sigma_{g^+}$, ${}^1\Sigma_{g^-}$ and ${}^1\Sigma_{g^-}$ signify the
possibility of {\em double occupancy\/} of the two electrons at a single nucleus
as products of $a(1)a(2)$ and $b(1)b(2)$ appear in the wave functions. Such
states, chemically, represent ionic bonds. On the other hand, the last three
states in \eqref{M4.1} i.e., ${}^3\Sigma^+_u$, ${}^3\Sigma^0_u$,
${}^3\Sigma^-_u$, agree with the triplet states in \eqref{M3.1}.

We denote the six molecular orbitals in \eqref{M4.1} in sequential order as
$\Psi_1,\Psi_2,\ldots, \Psi_6$. Then the energy of any linear combination
$\Sigma^6_{j=1} c_j\Psi_j$ corresponds to a quadratic form:
\[
\left\langle \sum^6_{j=1} c_j\Psi_j\bigg| H \bigg| \sum^6_{k=1}
c_k\Psi_k\right\rangle = \sum^6_{j,k=1} H_{jk} \bar c_jc_k,
\]
where $H_{jk}$ are the  $(j,k)$-entry of the following symmetric matrix
\begin{equation}\label{M4.2}
{\cl H} = \left[\begin{matrix}
\begin{array}{cc|} H_{11}&H_{12}\\ H_{12}&H_{22}\\ \hline
\end{array}&&\bigcirc\\
&H_{{}^1\Sigma_u}\\
&&H_{{}^3\Sigma_u}\\
\bigcirc&&&H_{{}^3\Sigma_u}\\
&&&&H_{{}^3\Sigma_u}\end{matrix}\right],
\end{equation}
\begin{align*}
{\cl H}_{11} &= \left\langle{}^1\Sigma_{g^+}\Big| {\cl H}
\Big|{}^1\Sigma_{g^+}\right\rangle, \qquad
{\cl H}_{22}  = \left\langle{}^1\Sigma_{g^-}\Big| {\cl H}
\Big|{}^1\Sigma_{g^-}\right\rangle\\
{\cl H}_{{}^1\Sigma_u} &= \left\langle{}^1\Sigma_u \Big| {\cl H}
\Big|{}^1\Sigma_u\right\rangle,  \qquad
{\cl H}_{{}^3\Sigma_u} = \left\langle{}^3\Sigma^j_u \Big| {\cl H}
\Big|{}^3\Sigma^j_u\right\rangle,\qquad j= +,0,-.
\end{align*}
Specifically,
\begin{align*}
{\cl H}_{11}, {\cl H}_{22} &= \alpha^2 \left[\frac{1\mp S \mp 2K}{1\pm S}\right]
+ \alpha \left[\frac{-2 + 2J\pm 4K}{1\pm S} + \frac{\frac58 + J' + 2K'\pm
4L}{2(1\pm S)^2} + \frac1w\right],\\
{\cl H}_{12} &= \left\langle{}^1\Sigma_{g^+}\Big| {\cl H}\Big|
\Sigma_{g^-}\right\rangle = \alpha \left(\frac{\frac58 -
J'}{2(1-S^2)}\right),\\
{\cl H}_{{}^1\Sigma_u} &= \alpha^ 2 \left[\frac{1+2KS+S^2}{1-S^2}\right] +
\alpha \left[\frac{-2+2J - 4KS}{1-S^2} + \frac{\frac54 - 2K'}{2(1-S^2)} +
\frac1w\right],\\
{\cl H}_{{}^3\Sigma_u} &= \alpha^2
\left[\frac{1+2KS+S^2}{1-S^2}\right] + \alpha
\left[\frac{-2+2J + J' - 4KS - K'}{(1-S^2)} + \frac1w\right],
\end{align*}
where
$$
L=(1/\alpha)\int a^2(1)a(2)b(2)\frac{1}{r_{12}}dxdy=
e^{-w}\left[w+\frac18+\frac{5}{16w}-e^{-2w}\left(\frac18+\frac{5}{16w}\right)
\right] .
$$
The other symbols are defined in Appendix \ref{integ}.

In the calculation of the ground state energy, the sub-block of $2\times 2$
matrix in the upper left corner of \eqref{M4.2} plays the exclusive role as the
remaining diagonal $4\times 4$ block in \eqref{M4.2} contributes no effect. We
thus determine the ground state energy $E$ by the determinant
\begin{gather*}
\left|\begin{matrix} H_{11}-E&H_{12}\\ H_{12}&H_{22}-E\end{matrix}\right| = 0,\\
E_\pm = \frac12\left(H_{11}+H_{22} \pm \sqrt{(H_{11} -H_{22})^2 +4
H^2_{12}}\right).
\end{gather*}
The value $E_-$ will correspond to the ground state energy.

\begin{figure}[htpb]
\bigskip
\centerline{\epsfxsize=0.5\textwidth\epsfysize=0.5\textwidth
\epsfbox{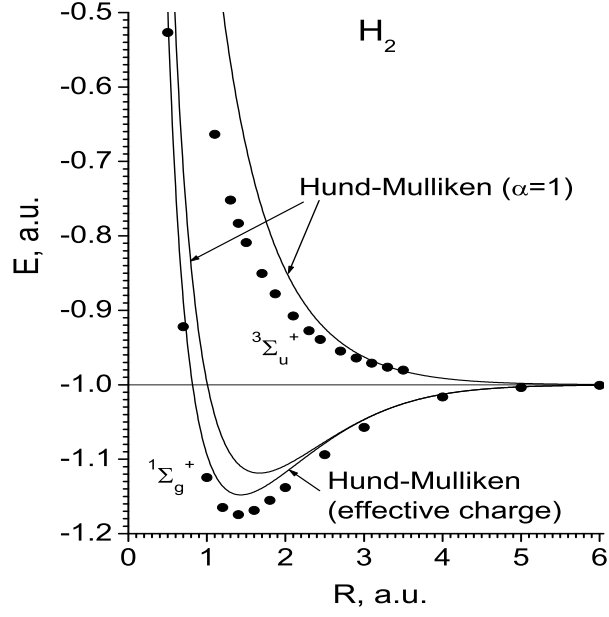}}

\caption{Ground and first excited state
$E(R)$ of H$_2$ molecule
for the Hund--Mulliken wave function (solid lines)
and the ``exact'' energy of Ref. \cite{Kolo60} (dots).
}
\label{HMa}
\end{figure}

For the Hund--Mulliken method discussed above, again $\alpha$ is the variational
parameter.
In Fig. \ref{HMa} we plot the ground $^1\Sigma_g^+$ and first excited
$^3\Sigma_u^+$ state potential energy curves $E(R)$ of the H$_{2}$
molecule. When $\alpha =1$ the ground state curve yields the binding energy
of 0.119 a.u.=3.23 eV. The effective charge calculation yields the binding
energy of 0.148 a.u.=4.03 eV and the bond length of 1.43 Bohr radii.
The $^3\Sigma_u^+$ curve is identical to the Heitler--London $E(R)$ (see Fig.
\ref{HLF}).

\subsection{The Hartree--Fock self-consistent method}

\indent

This is perhaps the best known method in molecular quantum chemistry and it
works for multi-electron and multi-center cases.
In computational physics, the Hartree--Fock calculation scheme is a
self-consistent iterative procedure to calculate the optimal
single-particle determinant solution to the time-independent Schr\"odinger
equation.  As a consequence to this, while it calculates the exchange energy
exactly, it does not calculate the effect of electron correlation at all. The
name is for Douglas Hartree, who devised the self consistent field method, and
Vladimir Fock who reformulated it into the matrix form used today and
introduced the exchange energy.

The starting point for the Hartree--Fock method is a set of approximate
orbitals. For an atomic calculation, these are typically hydrogenic orbitals.
For a molecular calculation, the initial
approximate wave functions are typically a linear combination of atomic
orbitals. This gives a collections of one electron orbitals, which due to the
Fermionic nature of electrons must be anti-symmetric; the antisymmetry is
achieved through the use of a Slater determinant.

Once an initial wave function is constructed, an electron is selected. The
effect of all the other electrons is summed up, and used to generate a
potential. This is why the procedure is sometimes called a mean-field
procedure. This gives a single electron in a defined potential, for which the
Schr\"odinger equation can be solved, giving a slightly different wave function
for that electron. This process is then repeated for all the other electrons,
which complete one iteration of the procedure. The whole procedure is then repeated
until the self-consistent solution is obtained.

Here we discuss the method in detail. Consider the Hamiltonian of a
multi-electron and multi-center molecular system (with $n$ electrons and
$N$ nuclei, respectively) under the Born--Oppenheimer separation in the
following form
\begin{equation}\label{eq4.1}
{\hat H} = - \frac12 \sum^{n}_{j=1} \nabla^2_j + \sum_{1\le
j<k\le n} \frac1{r_{jk}} - \sum^{n}_{j=1} \sum^{N}_{k=1}
\frac{Z_k}{|\pmb{r}_j-\pmb{R}_k|}.
\end{equation}
For a closed-shell system, all electrons are paired up and $n$ must be an
even number.
The trial wave function is chosen in the form of a Slater determinant $\psi =
\|\phi_1(\pmb{r}_1)\bar \phi_1(\pmb{r}_2) \phi_2(\pmb{r}_3)\bar
\phi_2(\pmb{r}_4)\ldots \phi_{n/2}(\pmb{r}_{n-1})\bar \phi_{n/2}
(\pmb{r}_{n})\|$, where $\| \ldots \|$ denotes determinant and each
$\phi_i$ corresponding to a $u_i$ in (6.1) is a molecular orbital (MO) and the
one without a ``bar'' on top denotes a spin-orbital
with $\alpha$-spin (spin-up), and the one with a ``bar'' on top denotes a
$\beta$-spin (spin-down) orbital. We assume that the MOs (the spatial part) are
orthonormal:
\[
\int_{R^3} \phi_i^*(\pmb{r})\phi_j(\pmb{r})d\pmb{r}=\delta_{ij},
\]
and we stipulate this orthonormality in our subsequent calculations.
 For the ground state calculation, we use the Ritz variational method.
Substituting the trial wave function into
$\langle\psi|{\hat H}|\psi\rangle$, we have
\begin{equation}\label{eq4.2}
\langle\psi|
{\hat H}|\psi\rangle = 2 \sum_{i=1}^{n/2}
h_{ii} + \sum_i\sum_j(2{\cl J}_{ij} - {\cl
K}_{ij}),
\end{equation}
where,
\begin{align*}
h_{ij} &= \int_{R^3} \phi^*_i(\pmb{r}) \left\{-\frac12 \nabla^2-\sum_{k=1}^{
N} \frac{Z_k}{|\pmb{r}-\pmb{R}_k|}\right\}\phi_j(\pmb{r}) d\pmb{r},\\
{\cl J}_{ij} &= \iint_{R^6} \phi^*_i(\pmb{r})\phi_i(\pmb{r})
\frac{1}{|\pmb{r}-\pmb{r}\prime|} \phi_j(\pmb{r}\prime) \phi^*_j(\pmb{r}\prime)
d\pmb{r}d\pmb{r}\prime=(ii|jj),\\
{\cl K}_{ij} &= \iint_{R^6} \phi^*_i(\pmb{r}) \phi_j(\pmb{r})
\frac{1}{|\pmb{r}-\pmb{r}\prime|} \phi_i(\pmb{r}\prime)
\phi^*_j(\pmb{r}\prime)d\pmb{r} d\pmb{r}\prime=(ij|ij).
\end{align*}
Minimizing (\ref{eq4.2}) subject to $\langle\psi|\psi\rangle = 1$ leads to a set
of $n/2$ Hartree--Fock equations for the spatial molecular orbital
$\phi_j(\pmb{r})$:
\begin{equation}\label{M8.1}
F\phi_j(\pmb{r})=\left\{h+\sum_{r=1}^{
n/2}[2J_r-K_r]\right\}\phi_j(\pmb{r})=\vp_j\phi_j(\pmb{r}), j=1,2, \ldots,
n/2,
\end{equation}
with $h$ being the single electron Hamiltonian
\[
h=-\frac12 \nabla^2-\sum_{k=1}^{N} \frac{Z_k}{|\pmb{r}-\pmb{R}_k|},
\]
$J_r$ being the Coulomb interaction operator defined by
\[
J_r\phi_j(\pmb{r}) = \left\{\int_{R^3}\phi^*_r(\pmb{r}\prime)
\phi_r(\pmb{r}\prime)\frac{1}{|\pmb{r}-\pmb{r}\prime|} d\pmb{r}\prime\right\}
\phi_j(\pmb{r})
\]
and $K_r$ being the energy exchange operator defined by
\[
K_r\phi_j(\pmb{r}) =
\left\{\int_{R^3}\phi^*_r(\pmb{r}\prime)\phi_j(\pmb{r}\prime)\frac1{|\pmb{r}-
\pmb{r}\prime|}d\pmb{r}\prime\right\}
\phi_r(\pmb{r}).
\]
$F$, called the Fock operator, involves those unknown MOs ($\phi_j(\pmb{r})$'s)
complicatedly (note that summations in $F$ are over all occupied orbitals,
including $\phi_j(\pmb{r})$ itself); this is not a simple eigenvalue problem. A
self-consistent field (SCF) method is needed to solve the Hartree--Fock equation.
For that, we take all MOs in $F$ to be known by guessing a set of MOs and
plugging them
into $F$. Then, the Hartree--Fock equation becomes a normal eigenvalue
problem. We solve this problem and compare the resulting MOs (eigenvectors)  to
those MOs we substitute into $F$. If they are different, we put these new MOs
back to $F$ and do the same calculation again. This is done iteratively until
all $\phi_j$'s converge.

The problem is then how do we guess (or express) those MOs. An often used method
is what we called the linear combination of atomic orbitals (LCAO) method. In
this method, we expand the unknown MOs linearly in a fixed basis set, such as
\begin{equation}\label{eq4.3}
\phi_i(\pmb{r})=\sum_{\mu}C_{\mu i}\chi_{\mu}(\pmb{r}), \quad C_{\mu i}\in
{\bb C}; {\bb C} \text{ is the complex number field.}
\end{equation}
Here, $\chi_{\mu}(\pmb{r})$ can be any functions deemed appropriately (not
required to be orthonormal), and they are often approximations to atomic
orbitals with respect to the individual centers. We still refer them as atomic
orbitals (AOs) in the following. By multiplying $\phi_i^*(\pmb{r})$ on equations
(\ref{M8.1}) and integrating with respect to $\pmb{r}$, we can rewrite those HF
equations (\ref{M8.1}) as
\begin{equation}\label{4.4}
F_{ij}=\vp_{i}\delta_{ij},
\end{equation}
with
\[
F_{ij}=h_{ij}+\sum_r[2(ij|rr)-(ir|jr)].
\]
If we then substitute expansions (\ref{eq4.3}) back into the new HF equations
(\ref{4.4}), we obtain
\begin{equation}\label{eq4.5}
\sum_{\nu}F_{\mu\nu}C_{\nu i}=\sum_{\nu}S_{\mu\nu}C_{\nu i}\vp_i,
\end{equation}
where
\begin{align*}
S_{\mu\nu}&=\int\chi_{\mu}(\pmb{r})\chi_{\nu}(\pmb{r})d^3
\pmb{r}\ne\delta_{\mu\nu}, \\
F_{\mu\nu}&=h_{\mu\nu}+\sum_{\lambda\sigma}D_{\lambda\sigma}\left\{(\mu\nu|
\lambda\sigma)-\frac14(\mu\lambda|\nu\sigma)-\frac14(\mu\sigma|\nu\lambda)
\right\}, \\
D_{\lambda\sigma}&=2\sum_iC_{\lambda i}C_{\sigma i}.
\end{align*}

The whole procedure for the Hartree--Fock Self-consistent Field (HF-SCF)
calculation can be summarized as follows:
\begin{itemize}
\item[(1)] Choose a basis set \{$\chi_{\mu}$\}.
\item[(2)] Calculate integrals over this basis set (and store in memory).
\item[(3)] Guess the MO coefficients $C_{\mu i}$.
\item[(4)] Construct $\pmb{D}$ and then $\pmb{F}$.
\item[(5)] Solve equations (\ref{eq4.5}) for new $C_{\mu i}$, and iterate until
convergence results.
\end{itemize}
Note here the total electronic energy can be calculated by use of equation
(\ref{eq4.2}).

Good choice of the basis set is very important to the HF-SCF method. Otherwise,
basis set can fail the whole calculation, produce bad or wrong results or make
calculations become very expensive. A natural choice of the basis
are hydrogenic wave functions, or
Slater-type orbitals (STOs)
\[
\chi_{l,m}^{STO}=exp(-\zeta r_A)r_A^lY_{lm}(\theta_A,\phi_A),
\]
where $Y_{lm}$ are the spherical harmonics and the STO is centered on the
nucleus A. This basis set works nicely for atoms as well as for diatomic
molecules (some numerical integrations are required) and linear polyatomics (a
large number of numerical integrations need to be done). No algorithm for
nonlinear molecules has been developed till now. Another basis set is the
Gaussian-type orbitals (GTOs). This basis set is proposed by S.F.\ Boys
\cite{Boy} in the following form
\[
\chi_{l,m}^{GTO}=exp(-\alpha r_A^2)r_A^lY_{lm}(\theta_A,\phi_A),
\]
or in Cartesian form,
\[
\chi_{l,m}^{GTO}=exp(-\alpha r_A^2)x_A^ly_A^mz_A^n.
\]
This basis set has been widely used since those multicenter integrals in
$\pmb{F}$ can be easily done \cite{Boy} over this basis set. In principle, it
works for all molecules. However, there is a trade-off. GTOs are totally
conjectural functions which have no concrete physical meanings. Because of this,
GTOs are seldomly used to form basis sets directly. They are often used to
estimate STOs (by expressing one STO into linear combination of several GTOs)
and then form the basis set, such as the STO-3G (3 GTOs are used to estimate one
STO) basis set.

So far we have discussed the HF-SCF method for closed-shell systems. For the
open-shell systems, there are unpaired electrons, orbitals could be doubly
occupied or singly occupied. This makes calculations more complicated. The
simplest way to deal with this problem is to treat one doubly occupied orbital
as two independent orbitals. For example, orbitals $\phi_i$ and $\bar {\phi_i}$
are treated independently and they are not required to have the same spatial
part. With a few modifications, the closed-shell HF-SCF method can be migrated
to this case, which is called the unrestricted open-shell HF-SCF method. If we
do put restrictions on $\phi_i$ and $\bar {\phi_i}$  requiring them to have the
same spatial part, then we need to treat the doubly occupied and the singly
occupied orbitals differently. Calculations can be done still in a more
complicated way called the restricted open-shell HF-SCF method \cite{Kobu96}.

Many strategies have been developed to improve the HF-SCF calculations such as
introducing unoccupied molecular orbitals (virtual orbitals), or using two
(double-zeta basis set) or three (triple-zeta basis set) STOs to describe one
molecular orbital (note here that STOs could be linear combinations of GTOs).

\begin{figure}[htpb]

\bigskip
\centerline{\epsfxsize=0.5\textwidth\epsfysize=0.4\textwidth
\epsfbox{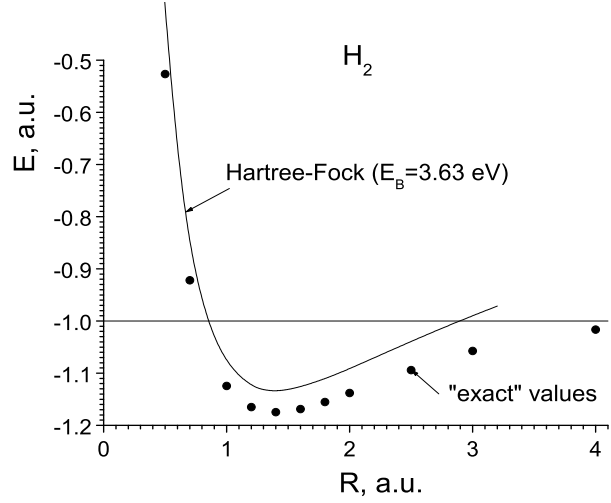}}

\caption{Ground state $E(R)$ of H$_2$ molecule
calculated by the self-consistent Hartree--Fock method (solid line)
and the ``exact'' energy (dots).
}
\label{SHF1}
\end{figure}

The HF-SCF method succeeds in many purposes of calculations such as the ground
state energy, chemical bond lengths, angles calculations, etc. It yields
3.636 eV for the binding energy of H$_2$ molecule \cite{Kolo60}
(to be compared against the ``exact value of $4.745$~eV).
Nevertheless, this method fails to describe the long
distance behavior of a chemical bond (see Fig. \ref{SHF1}). More subtly, the HF
method views that each electron interacts with a mean potential field generated by
the other electrons. It does not take into account the electron correlation.
Researchers have been trying to mend these by combining other methods, such as
the multiconfigurational SCF, configuration interaction and interelectronic
correlation terms in their calculations.

\subsection{The James--Coolidge wave functions}

\indent

James and Coolidge \cite{james} suggested a
wave function of the form
\begin{equation}\label{M9.1}
\psi = \frac1{2\pi} e^{-\alpha(\lambda_1+\lambda_2)}
\sum^\infty_{\overset{m,n,j,k,p}{=1}} C_{mnjkp} (\lambda^m_1 \lambda^n_2 \mu^j_1
\mu^k_2 + \lambda^n_1 \lambda^m_2 \mu^k_1 \mu^j_2)\rho^p
\end{equation}
for the H$_2$-molecule, where $\psi$ is a function of five variables
\begin{equation}\label{M9.2}
\left.\begin{array}{lll}
\lambda_j = \dfrac{r_{ja} + r_{jb}}{R};&\mu_j = \dfrac{r_{ja}-r_{jb}}{R},&
j=1,2,\\
\noalign{\medskip}
\rho = \dfrac{2r_{12}}{R}.\end{array}\right\}
\end{equation}
>From homonuclear (such as H$_2$) symmetry considerations, $j+k$ in \eqref{M9.1}
must be even. But this restriction can be removed for heteronuclear cases. For
given $R>0$, the coefficients $\alpha$, and $C_{mnjkp}$ for finitely many
indices $m,n,j,k$ and $p$, can be used as variational parameters to minimize the
energy $\langle\psi|H|\psi\rangle$, subject to the normalization condition
\begin{equation}\label{M9.3}
\langle\psi|\psi\rangle = 1.
\end{equation}

For example, for fixed $\alpha$ and $R$ in \eqref{M9.1} and \eqref{M9.2},
introduce a Lagrange multiplier $\lambda$ for the constraint \eqref{M9.3} and
choose only a total of $s$ terms in \eqref{M9.1} by truncation, then the
variational problem
\begin{equation}\label{M10.1}
\min_\psi [\langle\psi|{\hat H}
|\psi\rangle + \lambda(\langle\psi|\psi\rangle-1)]
\end{equation}
leads to a set of $s$ linear equations (\cite{james}, p.~826]) for the
coefficients $C_j$, $j=1,2,\ldots, s$,
\begin{equation}\label{M10.2}
\left\{\begin{array}{l}
(H_{11}-\lambda S_{11}) C_1 + (H_{12}-\lambda S_{12})C_2 +\cdots+
(H_{1s}-\lambda S_{1s})C_s=0,\\
(H_{12}-\lambda S_{12}) C_1 + (H_{22}-\lambda S_{22})C_2 +\cdots+
(H_{2s}-\lambda S_{2s})C_2 = 0,\\
~~~\vdots\\
(H_{1s}-\lambda  S_{1s})C_1 + (H_{2s}-\lambda S_{2s})C_2 +\cdots+ (H_{ss} -
\lambda S_{ss})C_s = 0\end{array}\right.
\end{equation}
where
\begin{align*}
&H_{ij} = \langle\phi_i|{\hat H}|\phi_j\rangle,\quad S_{ij} =
\langle\phi_i|\phi_j\rangle,\\
&\phi_i \text{ is a typical summand term in \eqref{M9.1} without the coefficient
} C_j,\\
&\text{for } i,j=1,2,\ldots, s.
\end{align*}
Solving \eqref{M10.2} then leads to the ground state $\psi_0$ and its energy
$E_0$.

The trial wave function \eqref{M9.1} may be further adapted to
\begin{equation}\label{eq6.17a}
\psi = \frac1{2\pi} e^{-\alpha_1\lambda_1-\alpha_2\lambda_2}
e^{-\beta_1\mu_1-\beta_2\mu_2}
\sum^\infty_{\overset{m,n,j,k,p}{=1}} C_{mnjkp} (\lambda^m_1 \lambda^n_2
\mu^j_1\mu^k_2 \pm
\lambda^n_1 \lambda^m_2 \mu^k_1 \mu^j_2)\rho^p
\end{equation}
for the calculation of heteronuclear cases and excited states.

Next, we address the analytic treatment and provide a complete compendium
for the integrals given in \eqref{M9.1}--\eqref{eq6.17a}, which constitutes
the keystone in the two-centered variational treatment.

The model Hamiltonian that we will be using here is the (homonuclear case) H$_2$
as given in \eqref{M3.3a}.
Let us rewrite the James--Coolidge wave function as
\begin{align}
\label{eq7a.5}
\Psi(1,2) &= \sum_r C_r\Phi_r(1,2),\\
\intertext{where}
\label{eq7a.6}
\Phi_r(1,2) &= \frac1{2\pi} e^{-\alpha(\lambda_1+\lambda_2)} \lambda^{m_r}_1
\lambda^{n_r}_2 \mu^{j_r}_1 \mu^{k_r}_2 r^{\ell_r}_{12}.
\end{align}
We remark that for the heteronuclear case and for excited states, the
trial wave function \eqref{eq7a.5}--\eqref{eq7a.6} can be easily re-adjusted to
the form \eqref{eq6.17a} with relative ease.

In the evaluation of the energy for given two-electron trial
wave functions, a typical term is of the form
\begin{equation}\label{eq7a.8}
 Z^\nu(m,n,j,k;\ell) = \frac{1}{4\pi^2} \int \cdots \int
e^{-\alpha (\lambda_1 + \lambda_2)} \lambda_1^m \lambda_2^n
\mu_1^j \mu_2^k r_{12}^\ell M^\nu \cos^\nu(\phi_1 - \phi_2) \,
d\lambda_1 d\lambda_2 d\mu_1 d\mu_2 d\phi_1 d\phi_2
\end{equation}
where
\begin{equation}\label{eq7a.9}
 M = [( \lambda_1^2 - 1 ) ( \lambda_2^2 -1) (1-\mu_1^2)
(1-\mu_2^2)]^{1/2},
\end{equation}
and $\nu$, $m$, $n$, $j$, $k$ and $\ell$ are integers ($\ell \geq -1 $ and
others are nonnegative integers).

Using the above function, we can write every integration in terms
of $Z^\nu(m,n,j,k;\ell)$. For example, the Coulomb interaction
energy between the
nuclei and electrons may be easily written as follows:
\begin{align}
\label{eq7a.10}
\frac{1}{r_{a1}} &= \frac{2}{R(\lambda_1 + \mu_1)} \\
\label{eq7a.11}
\frac{1}{r_{b1}} &= \frac{2}{R(\lambda_1 - \mu_1)} \\
\label{eq7a.12}
\frac{1}{r_{a2}} &= \frac{2}{R(\lambda_2 + \mu_2)} \\
\label{eq7a.13}
\frac{1}{r_{b2}} &= \frac{2}{R(\lambda_2 - \mu_2)}
\end{align}
The evaluation of the Coulomb interactions can be done in terms of
$Z$. The evaluation of the Laplacian matrix element,
however, somewhat more involved. But it also can be expressed via $Z^{\nu}$.
We provide details of the derivation in Appendix \ref{Lapder}.

Since we can write every term in the energy in
terms of $Z^\nu$, let us express $Z^\nu$ via simpler
functions defined by recurrence relations. Such relations
are given in Appendix \ref{recrel}.
For $\ell \geq 1$ and $\nu = 0$, one can reduce $\ell$ to $0$ or $-1$
using the following identity:
\begin{equation}\label{eq7a.15}
r_{12}^2 = \frac{R^2}{4} ( \lambda_1^2 + \lambda_2^2 + \mu_1^2 +
\mu_2^2 - 2 - 2\lambda_1 \lambda_2 \mu_1 \mu_2 - 2 M \cos(\phi_1 -
\phi_2) ).
\end{equation}

For $\nu=0$ and $\ell = 0$, the integration can be evaluated as follows. Given
\begin{equation}\label{eq7a.16}
Z^0(m,n,j,k;0)\equiv Z(m,n,j,k;0) = \frac{1}{4\pi^2} \int \cdots \int
{\rm e}^{-2\alpha (\lambda_1 + \lambda_2)} \lambda_1^m \lambda_2^n
\mu_1^j \mu_2^k \, d\lambda_1\, d\lambda_2\,d\mu_1\,d\mu_2\,
d\phi_1\,d\phi_2,
\end{equation}
note that the $\phi$ integrals are trivial to evaluate and the
$\mu$ integrals survive only for even integers $j$ and $k$, i.e.,
\begin{equation}\label{eq7a.17}
\int_{-1}^{+1} \mu^j d \mu = \frac{2}{j+1}, \quad \mbox{for even
$j$}\,;
\end{equation}
we arrive at
\begin{equation}\label{eq7a.18}
 Z(m,n,j,k;0) = \begin{cases}
  A(m;\alpha) A(n;\alpha) \left[\frac{4}{(j+1)(k+1)}\right],&\text{for even $j$
and $k$,}\\
 0,&\text{for odd $j$ or $k$,}\end{cases}
\end{equation}
where we have introduced the function
\begin{equation}\label{eq7a.19}
A(m;\alpha) = \int^\infty_1 e^{-\alpha \lambda} \lambda^m \,
d\lambda\,,
\end{equation}
for the $\lambda$ integration. Using integration by parts one can
show that the $A(m;\alpha)$ satisfies the recursion relation
\begin{equation}\label{eq7a.20}
A(m;\alpha) = \frac{1}{\alpha} \left[ e^{-\alpha} + m
A(m-1;\alpha) \right] \,,
\end{equation}
with
\begin{equation}\label{eq7a.21}
 A(0;\alpha) = \frac{e^{-\alpha}}{\alpha}\,.
\end{equation}
Thus we have given a recipe for the evaluation of $Z(m,n,j,k;0)$
for arbitrary values of the power parameters $m,n,j,$ and $k$.

Next, consider the case where $\ell = -1$, i.e.,
\begin{equation}\label{eq7a.22}
 Z(m,n,j,k;-1) = \frac{1}{4\pi^2} \int \cdots \int
{\rm e}^{-2\alpha (\lambda_1 + \lambda_2)} \lambda_1^m \lambda_2^n
\mu_1^j \mu_2^k \frac{1}{r_{12}}\,
d\lambda_1\,d\lambda_2\,d\mu_1\,d\mu_2\, d\phi_1\,d\phi_2\,.
\end{equation}
Recalling the Neumann expansion for $1/r_{12}$:
\begin{equation}\label{eq7a.22a}
\frac{1}{r_{12}} = \frac{2}{R} \sum_{\tau = 0}^\infty
\sum_{\nu=-\tau}^\tau (-1)^\nu (2\tau+1) \left[ \frac{ (\tau -
|\nu|)!}{(\tau +|\nu|)!} \right]^2 P_\tau^\nu (\lambda_<)
Q_\tau^\nu (\lambda_>) P_\tau^\nu (\mu_1) P_\tau^\nu (\mu_2)
e^{i\nu(\phi_1-\phi_2)}\, ,
\end{equation}
where $\lambda_<=min(\lambda_1,\lambda_2)$,
$\lambda_>=max(\lambda_1,\lambda_2)$ and $P_\tau^\nu$, $Q_\tau^\nu$
are the associated Legendre functions of the 1st and 2nd kind respectively.
After the angular
integration, only terms corresponding to $\nu=0$ survive, as long as
the wave function has no angular or $r_{12}$ dependence.
Separating the $\lambda$ and $\mu$ integrals, we arrive at
\begin{equation}\label{eq7a.23}
Z (m,n,j,k;-1) = \frac{2}{R} \sum_{\tau=0}^\infty (2 \tau + 1)
R_\tau(j) R_\tau(k) H_\tau(m,n;\alpha) \,,
\end{equation}
where  $R_\tau$ and
$H_\tau$ are defined by
\begin{align}\label{eq7a.25}
R_\tau(j) &\equiv \int^1_{-1} \mu^j P_\tau(\mu) \, d\mu,
\\
\label{eq7a.26}
 H_\tau(m,n;\alpha) &\equiv \int^\infty_1 \int^\infty_1
e^{-\alpha (\lambda_1 + \lambda_2)} \lambda_1^m \lambda_2^n
 P_\tau
(\lambda_<) Q_\tau (\lambda_>) \, d\lambda_1 d\lambda_2.
\end{align}
In the discussion to follow we give recursion relations for the
evaluation of the various auxiliary functions. For $\tau = 0$,
\begin{equation}\label{eq7a.27}
H_0(m,n;\alpha) = A(m;\alpha)F(n;\alpha) + A(n;\alpha) F(m;\alpha)
- T(m,n;\alpha) - T(n,m;\alpha).
\end{equation}
Here, $F(m;\alpha)$ can be evaluated for arbitrary $m$ by noting
its recursion relation
\begin{align*}
F(m;\alpha) &= \int^\infty_1 e^{-\alpha \lambda} \lambda^m
Q_0(\lambda) \, d\lambda \\
&= F(m-2;\alpha) + \frac{1}{\alpha} \left[ m F(m-1;\alpha) -
(m-2) F(m-3;\alpha) - A(m-2;\alpha) \right]
\end{align*}
and the initial values
\begin{equation}\label{eq7a.28}
F(0;\alpha) = \frac{1}{2} \left[ (\ln 2\alpha + \gamma)
\frac{e^{-\alpha}}{\alpha} - \text{Ei}[-2\alpha]
\frac{e^\alpha}{\alpha} \right],
\end{equation}
and
\begin{equation}\label{eq7a.29}
F(1;\alpha) = \frac{1}{2} \left[ (\ln 2\alpha + \gamma)
e^{-\alpha} \left( \frac{1}{\alpha} + \frac{1}{\alpha^2} \right) -
\text{Ei}[-2\alpha] e^{\alpha} \left( - \frac{1}{\alpha} +
\frac{1}{\alpha^2} \right) \right].
\end{equation}
where
\begin{equation}\label{eq7a.30}
\text{Ei}(-x) = - \int^\infty_x \frac{e^{-t}}{t} \, dt
\end{equation}
and $\gamma = 0.577216\cdots$ is the Euler constant; see Appendix I.

Similarly the quantity $T(m,n;\alpha)$ can be determined for
arbitrary values of $m$, $n$ through
\begin{align*}
T(m, n;\alpha) &\equiv \frac{m!}{\alpha^{m+1}} \sum^m_{\nu=0}
\frac{\alpha^\nu}{\nu!} F(n+\nu;2\alpha) \\
&= \frac{1}{\alpha} \left[ m T(m-1,n;\alpha) + F(m+n;2\alpha)
\right]
\end{align*}
with the initial value
\begin{equation}\label{eq7a.31}
T(0,n;\alpha) = \frac{1}{\alpha} F(n;2\alpha)\,.
\end{equation}
Note that we have so far considered only a special case where $\tau=0$.
We turn to the case where $\tau = 1$. Once again we note the
recursion relation for $H_1(m,n;\alpha)$:
\begin{equation}\label{eq7a.32}
 H_1(m,n;\alpha) = H_0(m+1,n+1;\alpha) - S(m,n+1;\alpha) -
S(n,m+1;\alpha)
\end{equation}
where $S(m, n;\alpha)$ can be determined according to
\begin{align*}
S(m, n;\alpha) &\equiv \frac{m!}{\alpha^{m+1}} \sum^m_{\nu=0}
\frac{\alpha^\nu}{\nu!} A(n+\nu;2\alpha) \\
&= \frac{1}{\alpha} \left[ m S(m-1,n;\alpha) + A(m+n;2\alpha)
\right]
\end{align*}
with  initial value
\begin{equation}\label{eq7a.33}
S(0,n;\alpha) = \frac{1}{\alpha} A(n;2\alpha).
\end{equation}

We now have shown relations needed to evaluate
$H_\tau(m,n;\alpha)$ for particular values of $\tau=0,1$. Here we
summarize the evaluation for $\tau>1$ through the following
recursion relations:
\begin{align*}
H_\tau(m,n;\alpha) &= \frac{1}{\tau^2} \left[ (2 \tau-1)^2
H_{\tau-1}(m+1,n+1;\alpha) + (\tau-1)^2 H_{\tau-2}(m,n;\alpha) \right. \\
& - (2\tau-1)(2\tau-3) \left\{ H_{\tau-2}(m+2,n;\alpha) +
H_{\tau-2} (m,n+2;\alpha) \right\} \\
& \qquad \qquad \qquad \qquad \qquad + 2 (2\tau-1)(2\tau-5)
H_{\tau-3}(m+1,n+1;\alpha) \\
& - (2\tau-1)(2\tau-7) \left\{ H_{\tau-4}(m+2,n;\alpha) +
H_{\tau-4} (m,n+2;\alpha) \right\} \\
& \qquad \qquad \qquad \qquad \qquad + 2 (2\tau-1)(2\tau-9)
H_{\tau-5}(m+1,n+1;\alpha) \\
&+ \text{one of the following:}
\end{align*}
\[
\left\{\begin{array}{l}
(\text{even } \tau)\qquad  - \left. (2\tau-1) \left\{
H_0(m+2,n;\alpha) + H_0(m,n+2;\alpha) - S(m+1,n;\alpha) -
S(n+1,m;\alpha) \right\}
\right], \\
(\text{odd } \tau)\qquad  + \left. (2\tau-1) \left\{ 2
H_0(m+1,n+1;\alpha) - S(m,n+1;\alpha) - S(n,m+1;\alpha) \right\}
\right],\end{array}\right.
\]
where  the starting values
$H_0(m,n;\alpha)$ and $H_1(m,n;\alpha)$ are already given in \eqref{eq7a.27} and
\eqref{eq7a.32}.

For nonzero $\nu$, the integral $Z^\nu$ can be written in terms of
various simple function as discussed below. It is fairly
straightforward to obtain the recursion relation for the $Z^{\nu}$
function, just by inspection of the equation. For $\ell \geq 1$,
\begin{align}
Z^\nu (m,n,j,k;\ell) &=
 Z^\nu (m+2,n,j,k;\ell-2)
+ Z^\nu (m,n+2,j,k;\ell-2)   \nonumber\\
&\quad + Z^\nu (m,n,j+2,k;\ell-2) + Z^\nu (m,n,j,k+2;\ell-2) - 2 Z^\nu
(m,n,j,k;\ell-2)\nonumber \\
\label{eq7a.35}
& \quad - 2 Z^\nu (m+1,n+1,j+1,k+1;\ell-2) - 2
Z^{\nu+1}(m,n,j,k;\ell-2).
\end{align}
The above recursion relation can be proved in a straightforward
manner by using the expansion of $r^2_{12}$ in the elliptical
coordinates introduced in \eqref{eq7a.15}.

Higher order terms of $Z^{\nu}$ are given by

\begin{equation}\label{eq7a.36}
Z^1 (m,n,j,k,-1) = -\frac{2}{R} \sum_{\tau=1}^\infty
\frac{(2\tau+1)}{\tau^2(\tau+1)^2} R^1_\tau(j) R^1_\tau(k)
H^1_\tau(m,n;\alpha)
\end{equation}
and
\begin{align*}
Z^2 (m,n,j,k,-1) &= \frac{2}{R} \sum_{\tau=2}^\infty
\frac{(2\tau+1)}{(\tau-1)^2\tau^2(\tau+1)^2(\tau+2)^2} R^2_\tau(j)
R^2_\tau(k) H^2_\tau(m,n) \\
&\quad + \frac{1}{R} \sum_{\tau=0}^\infty (2\tau+1) ( R_\tau(j) -
R_\tau(j+2))( R_\tau(k) - R_\tau(k+2)) \\
& \quad \times ( H_\tau(m+2,n+2;\alpha) -
H_\tau(m+2,n;\alpha) \\
& \quad  - H_\tau(m,n+2;\alpha) + H_\tau(m,n;\alpha)).
\end{align*}
For the $\ell = 0$ case,
\[ Z^1 (m,n,j,k;0) = 0, \]
\[ Z^2 (m,n,j,k;0) = ( A(m+2;\alpha) - A(m;\alpha)) ( A(n+2;\alpha) -
A(n;\alpha) ) \frac{8}{(j+1)(j+3)(k+1)(k+3)}, \]
when $j$ and $k$ are even; otherwise $Z^2(m,n,j,k;0)$  vanishes.

The definitions and recurrence relations for $R^\nu_\tau$ and
$H^\nu_\tau$ are given below:
\[ R^\nu_\tau(j) \equiv \int^1_{-1} (1-\mu^2)^{\nu/2} \mu^j P^\nu_\tau(\mu) \,
d\mu, \]
\[ H^\nu_\tau(m,n;\alpha) \equiv \int^\infty_1 \int^\infty_1 e^{-\alpha
(\lambda_1 + \lambda_2)} \lambda_1^m \lambda_2^n
(\lambda_1^2-1)^{\nu/2}(\lambda_2^2-1)^{\nu/2}   P_\tau^\nu
(\lambda_<) Q_\tau^\nu (\lambda_>) \, d\lambda_1 d\lambda_2. \]

The higher order recursions for
$H^\nu_\tau(m,n;\alpha)$ for $\nu=1$ and $\nu=2$ are listed below:
\[
H^1_\tau(m,n;\alpha) =\frac{\tau(\tau+1)^2}{(2\tau+1)}
H_{\tau+1}(m,n;\alpha) - \tau(\tau+1) H_\tau(m+1,n+1;\alpha)+
\frac{\tau^2(\tau+1)}{2\tau+1} H_{\tau-1}(m,n;\alpha), \]
 and
\begin{align*}
 H^2_\tau(m,n;\alpha) &= \frac{\tau^2(\tau-1)^2}{(2\tau+1)}
H^1_{\tau+1}(m,n;\alpha)
- (\tau+2)(\tau-1) H^1_\tau(m+1,n+1;\alpha)\\
&\quad +
\frac{(\tau+2)(\tau+1)^2}{2\tau+1} H^1_{\tau-1}(m,n;\alpha).
\end{align*}

The James--Coolidge wave functions may be said to have the most ``brawny'' power
as thousands of coefficients $C_{mnjkp}$ have been calculated with automated
computer programs, yielding very accurate values for the binding energy of
diatomic molecules. Nevertheless, there are certain associated shortcomings:
\begin{itemize}
\item[(i)] The physical insights about molecular bonding seem to be lost;
\item[(ii)] The wave functions in general do not satisfy the correlation cusp
condition;
\item[(iii)] For large $\lambda_1$ and $\lambda_2$, the asymptotic conditions
are violated (see Appendix \ref{sec2.2.3}).
\end{itemize}

\subsection{Two-centered orbitals}

\indent

Historically, the first use of two-centered orbitals for molecular calculations
is attributable to Wallis and Hulbert's paper \cite{WH} published in 1954. For
a historical summary, see \cite{teller,shull}.
New push and progress have been made by a school
of French researchers \cite{ABB,A-FL,SL,Le,lesechetal,lesechetal1} since 1976.
Their work has made the two-center orbital approach to diatomic modeling and
computation an admirable success.

The contributions by the French school are manifold, generalizing most of
the aspects of one-centered orbitals to two-centered ones. To introduce its
basic elements, let us utilize some of the semi-tutorial material from Scully,
et al. \cite{ScullyEtAl}.

\begin{figure}[htpb]
\epsfig{file=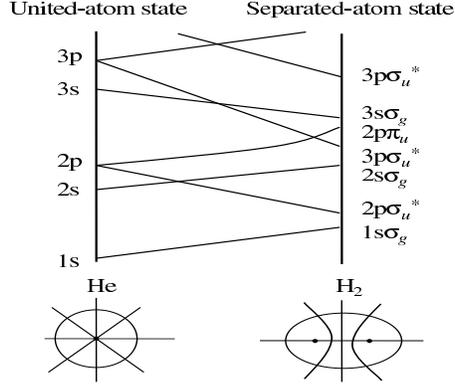,height=3in,width=3.2in}

\caption{Schematic correlation diagram for He, the united atom limit, and
H$_2$, the separated-atom case.}
\label{h2schem}
\end{figure}

Recall the correlation diagram for H$_2$ molecule
shown in Fig.~\ref{h2schem} \cite{ScullyEtAl}.
A typical form of the trial wave function
for H$_2$, e.g., may be represented as
\begin{equation}\label{M12.1}
\psi = [c_1\widetilde \psi_{H_2,1s\sigma} + c_2\widetilde\psi_{H_2,2p\sigma}]
f(r_{12}),
\end{equation}
where $\widetilde \psi_{H_2,1s\sigma}$ and $\widetilde\psi_{H_2,2p\sigma}$ are
chosen to be, respectively,
\begin{equation}\label{M12.2}
\left.\begin{array}{l}
\widetilde \psi_{H_2,1s\sigma} = \psi_{H^+_2,1\sigma}(1)
\psi_{H^+_2,1\sigma}(2),\\
\widetilde\psi_{H_2,2p\sigma} = \psi_{H^+_2,2p}(1)\psi_{H^+_2,2p}(2).
\end{array}\right\}
\end{equation}
>From \eqref{M12.1} and \eqref{M12.2}, we see
 that the trial wave function \eqref{M12.1}

\begin{itemize}
\item[\bf (i)] is {\em uncorrelated\/} if $f(r_{12})\equiv 1$. If $f(r_{12})$
has explicit dependence on $r_{12}$, then  \eqref{M12.1} is
correlated. A simple choice of $f(r_{12})$ was
\begin{equation}\label{M13.1}
f(r_{12}) = 1 + \frac12 r_{12},
\end{equation}
in Scully, et al. \cite{ScullyEtAl}, but many such correlation functions $f$
 satisfying the interelectronic cusp condition as given in
 Subsection \ref{secV}.B may be used also;
\item[\bf (ii)] has {\em configuration interaction\/} if $c_1c_2\ne 0$ in
\eqref{M12.1}.
\end{itemize}

It is found by \cite{ScullyEtAl} that
\begin{itemize}
\item[\bf (a)] for uncorrelated orbitals and without configuration interaction,
i.e., $f(r_{12})\equiv 1$ and $c_2=0$ in \eqref{M12.1}, the choice of
\begin{equation}\label{M13.2}
\psi_{H^+_2,1\sigma}(j) = {\cl N}e^{-\alpha_1\lambda_j}[1+B_2P_2(\mu_j)],\qquad
j=1,2,
\end{equation}
in \eqref{M12.2}$_1$, where $P_2$ is a Legendre polynomial,
gives the binding energy $E_B=0.132$ a.u.$=3.59$ eV
for H$_2$. Here, ${\cl
N}$ is a normalization  constant, and $\alpha_1$ and $B_2$ are two variational
parameters.
\item[\bf (b)] for correlated orbitals but without configuration interaction,
i.e., $f(r_{12}) = 1 + \frac12 r_{12}$ and $c_2=0$ in \eqref{M12.1}, the choice
of \eqref{M13.2} yields the binding energy $E_B=0.1710$ a.u.$=4.653$ eV
for H$_2$. The choice $f(r_{12}) = 1 + \kappa r_{12}$, where $\kappa$ is a
variational parameter, slightly improves the answer and gives
$E_B=0.1713$ a.u.$=4.661$ eV.
\item[\bf (c)] for correlated orbitals with configuration interaction, i.e.,
$f(r_{12}) = 1 + \frac12 r_{12}$ and $c_1c_2\ne 0$ in \eqref{M12.1}, the choice
of \eqref{M13.2} along with
\begin{equation}\label{eq6.27c}
\psi_{H^+_2,2p}(j) = {\cl N}e^{-\alpha_2\lambda_j} [P_1(\mu_j) + B_3P_3(\mu_j)]
\end{equation}
in \eqref{M12.2} ($\alpha_1$, $B_2$, $\alpha_2$ and $B_3$ are variational
parameters)
yields the binding energy $E_B=0.1712$ a.u.$=4.658$ eV. For the correlation
factor $f(r_{12})$ with the variational parameter $\kappa$ we obtain
$E_B=0.1721$ a.u.$=4.682$ eV.
\end{itemize}

What is quite striking here is that if, instead of \eqref{M13.2}, we choose a
finer two-centered orbital
\begin{equation}\label{eq6.27d}
\psi_{H^+_2,1\sigma}(j) = {\cl N}e^{-\alpha_1\lambda_j} [1+B_2P_2(\mu_j) +
B_4P_4(\mu_j)],\qquad j=1,2,
\end{equation}
and perform calculations using {\em three\/} variational parameters $\alpha_1$,
$B_2$ and $B_4$, then numerical results manifest that $B_2$ totally dominates
$B_4$, with a ratio $|B_4/B_2|\approx 2\times 10^{-3}$. This shows that {\em the
simple orbital \eqref{M12.1} with \eqref{M13.2} is able to capture the essence of
chemical bonding\/} of H$_2$ with very good accuracy.

For a heteronuclear molecule such as HeH$^+$, a simple adaptation of the above
scheme works equally well. Nevertheless, the authors have found that for
large $R$, more terms involving both $\lambda$ and $\mu$ variables should be
included in \eqref{M13.2}, such as
\begin{equation}\label{VI.61a}
\psi_{H^+_2,1\sigma}(j) = {\cl N}e^{-\alpha_1
\lambda_j} [1+B_2P_2(\mu_j) +\cdots+
B_{2m}P_m(\mu_j)] [1+A_1L_1(x_j) +\cdots+ A_nL_n(x_j)],\quad
(x_j=2\alpha_1(\lambda_j-1)),
\end{equation}
where $L_n(x)$ are Laguerre polynomials,
so that good accuracy can be maintained in the calculation of $E$.

Another choice of simple two-centered wave functions, in the same spirit of this
subsection, was given by Patil \cite{P}, where he has generalized the one-centered
Guillemin--Zener type one-centered molecular orbitals (see Item (3)) by
considering the gerade state
\begin{equation}\label{eq6.27a}
\psi_g = N(1+b\lambda)^\beta e^{-a\lambda} \cosh(a\mu)
\end{equation}
and ungerade state
\begin{equation}\label{eq6.27b}
\psi_u = N(1+b\lambda)^\beta e^{-a\lambda} \sinh(a\mu)
\end{equation}
for H$^+_2$ (or any homonuclear) ionic orbitals. Asymptotic behavior of
H$^+_2$-like orbitals can be built in through $\beta$, which plays the same role
as $\beta$ in the Jaff\'e solution \eqref{eqA.12a}.

\subsection*{(1)~~Le Sech's simplification of integrals involving cross-terms
of correlated wave functions}

\indent

Siebbeles and Le Sech \cite{SL} developed an ingenious approach to the evaluation of
energy by applying integration by parts (or, Green's Theorem) to avoid the
quadratures of cross terms of the type
\begin{equation}\label{M15.1}
\nabla_i(\Phi_i\Phi_j)\cdot \nabla_if,\qquad i,j=1,2,
\end{equation}
where $f$ is a correlation function, and $\Phi_i\Phi_j$ is a product of
two-centered orbitals. This is a fine feature of their approach.

They choose wave functions of the form
\begin{equation}\label{M15.2}
\psi(\pmb{r}_1,\pmb{r}_2) = \phi(1,2) \Omega(1,2),
\end{equation}
where $\Omega(1,2)$ plays the role of the correlation function (but may be more
general than the) $f$ discussed in \eqref{M12.1}, while
\begin{equation}\label{M15.3}
\phi(1,2) = \Phi(1) \Phi(2)
\end{equation}
where $\Phi(i)$, $i=1,2$, are the Hylleraas-type two-centered orbitals;
cf.\ \eqref{eq6.47a} below. Note here that each of $\Phi_1$ and $\Phi_2$
can have higher order molecular configurations such as $2s\sigma_g$,
$3p\sigma_u$,
$3d\sigma_g$ and $4f\sigma_u$ (in the computation of He, e.g., in \cite{Le}).

Let the diatomic molecule's Hamiltonian be
\begin{equation}\label{M16.1}
H = -\frac12(\nabla^2_1+\nabla^2_2) + V(1,2),
\end{equation}
where
\begin{equation}\label{M16.2}
V(1,2) = \widetilde V(1,2) + \frac1{r_{12}}
\end{equation}
with
\begin{equation}\label{M16.3}
\widetilde V(1,2) = -\left(\frac{Z_a}{r_{1a}} + \frac{Z_b}{r_{1b}} +
\frac{Z_a}{r_{2a}} + \frac{Z_b}{r_{2b}}\right).
\end{equation}
Let $\Phi_n(i)$, $i=1,2$, be eigenstates of the two center problem
\begin{equation}\label{M16.4}
\left(-\frac12 \nabla^2_i - \frac{Z_a}{r_{ia}} - \frac{Z_b}{r_{ib}}\right)
\Phi_n(i) = \epsilon_n\Phi_n(i),\qquad i=1,2.
\end{equation}
Denote $\phi_j(1,2)$ to be solutions of
\begin{equation}\label{M16.5}
\left[-\frac12\left(\nabla^2_1+\nabla^2_2\right) + \widetilde V(1,2)\right]
\phi_j(1,2) = E_j\phi_j(1,2).
\end{equation}
Then
\begin{equation}\label{M16.6}
\phi_j(1,2) \equiv \Phi_{n_1}(1) \Phi_{n_2}(2) \cdot \frac1{\sqrt 2} [\alpha(1)
\beta(2) - \alpha(2)\beta(1)]
\end{equation}
satisfies \eqref{M16.5} with
\begin{equation}\label{M16.7}
E_j = \epsilon_{n_1} + \epsilon_{n_2}.
\end{equation}
Now consider a trial wave function for \eqref{M16.1} in the form
\[
\phi(1,2) \Omega(1,2)
\]
where $\phi(1,2)$ is of the form \eqref{M16.6} while $\Omega(1,2)$ is intended
to model the Coulombic repulsive effect from the $1/r_{12}$ term and, therefore,
plays a similar role as (but may be more general than) the correlation function
$f(r_{12})$. Without loss of generality, $\phi(1,2)$ and $\Omega(1,2)$ are
assumed to be real.

Denote the 6-dimensional Laplacian
\[
\nabla^2_6 \equiv \nabla^2_1 + \nabla^2_2.
\]
Consider the matrix element
\begin{align}
H_{ij} &\equiv \langle\phi_i\Omega| - \frac12 \nabla^2_6 + V(1,2)|\phi_j\Omega
\rangle
= \int\limits_{{\bb R}^3} \int\limits_{{\bb R}^3} d\pmb{r}_1 d\pmb{r}_2 \left[-
\frac12 \phi_i\Omega \nabla^2_6(\phi_j\Omega) + \phi_i\phi_j
\Omega^2V(1,2)\right]\nonumber\\
\label{M17.1}
&= \iint d\pmb{r}_1 d\pmb{r}_2 \left[-\frac12 (\Omega^2\phi_i\nabla^2_6 \phi_j +
\phi_i\phi_j \Omega \nabla^2_6 \Omega + \underbrace{2 \phi_i\Omega \nabla_6
\phi_j\cdot \nabla_6\Omega}_{{\cl T}_1}) + \phi_i\phi_j\Omega^2V(1,2)\right].
\end{align}
To treat ${\cl T}_1$, write
\[
{\cl T}_1 = 2\phi_i\Omega \nabla_6\phi_j\cdot \nabla_6\Omega = \phi_i
\nabla_6\phi_j\cdot \nabla_6(\Omega^2),
\]
and note through an easy verification the following:
\begin{align}
2\phi_i\nabla\phi_j\cdot\nabla(\Omega^2) &= -[\phi_i\phi_j\nabla^2(\Omega^2) +
\Omega^2\nabla\cdot (\phi_i\nabla\phi_j - \phi_j\nabla\phi_i)]\nonumber\\
\label{M18.3}
&\quad + \nabla\cdot [\phi_i\phi_j\nabla(\Omega^2) + (\phi_i\nabla\phi_j -
\phi_j\nabla\phi_i)\Omega^2].
\end{align}
But the LHS of \eqref{M18.3} is equal to twice of ${\cl T}_1$. So we now
substitute one-half of the RHS of \eqref{M18.3} for ${\cl T}_1$ in
\eqref{M17.1}, obtaining
\begin{align}
H_{ij} &= \iint d\pmb{r}_1 d\pmb{r}_2\Big[\underbrace{-\frac12 (\Omega^2\phi_i
\nabla^2_6\phi_j}_{{\cl T}_2} + \phi_i\phi_j\Omega\nabla^2_6
\Omega)\nonumber\\
\label{M18.4}
&\quad + \underbrace{\frac14(\phi_i\phi_j\nabla^2_6(\Omega^2) + \Omega^2
\nabla_6\cdot (\phi_i\nabla_6\phi_j}_{{\cl T}_3} - \phi_j\nabla_6 \phi_i)) +
\phi_i\phi_j\Omega^2V(1,2)\Big],
\end{align}
where the divergence term $\nabla\cdot[\cdots]$ in \eqref{M18.3} disappears
after integration in the 6-dimensional space provided that $\phi_i,\phi_j,
\nabla(\Omega^2)$ and $\Omega^2$ decay fast enough as $|\pmb{r}_1|$,
$|\pmb{r}_2|\to \infty$. The RHS of \eqref{M18.4} is further simplified by
utilizing
\[
\frac14 \nabla^2(\Omega^2) =  \frac12 |\nabla\Omega|^2 +
\frac12\Omega(\nabla^2\Omega),
\]
and by combining ${\cl T}_2$ with ${\cl T}_3$ terms therein, leading to
\begin{equation}\label{M19.1}
H_{ij} = \iint d\pmb{r}_1d\pmb{r}_2 \left\{\frac12 \phi_i\phi_j
|\nabla_6\Omega|^2 - \frac14 \Omega^2[\phi_i\nabla^2_6\phi_j +
\phi_j\nabla^2_6\phi_i]  + \phi_i\phi_j\Omega^2V(1,2)\right\}.
\end{equation}
Now, with $\phi$ satisfying \eqref{M16.5} we obtain
\begin{equation}\label{M19.3}
H_{ij} = \iint d\pmb{r}_1d\pmb{r}_2 \left\{\phi_i\phi_j\Omega^2 \left[\frac12
(E_i+E_j)  + \frac1{r_{12}}\right] + \frac12\phi_i\phi_j
|\nabla_6\Omega|^2\right\}.
\end{equation}
Comparing \eqref{M19.3} with \eqref{M17.1}, we see that all the cross-derivative
terms $\nabla_6\phi_j\cdot\nabla_6\Omega$ have been eliminated, and the only
``burden of differentiation'' has been placed only on $|\nabla_6\Omega|^2$ in
\eqref{M19.3}.

If we choose $\Omega = 1+\frac12r_{12}$, then
$|\nabla_6\Omega |^2=|\nabla_1\Omega |^2+|\nabla_2\Omega |^2=1/2$
and \eqref{M19.3} becomes
\begin{equation}\label{M19.5}
H_{ij} = \iint d\pmb{r}_1d\pmb{r}_2 \left\{\phi_i\phi_j\Omega^2
\left[\frac12(E_i + E_j) + \frac1{r_{12}}\right] + \frac14 \phi_i\phi_j\right\}.
\end{equation}
But $\Omega$ doesn't have to be chosen as $\Omega = 1 + \frac12 r_{12}$.
Le Sech has preferred a special form of $\Omega$ as given in (\ref{secV}B.(2)).
 Note that
$\epsilon_g$ and $\epsilon_u$ can be used as {\em variational parameters},
so can be $d$ in (\ref{secV}B.(2)). In a personal communication from Le~Sech
to Scully (9/6/2004), Le~Sech pointed out that the trial wave functions in the
form of \eqref{M15.2} with $\Omega(1,2)$ chosen as (\ref{secV}B.(2))
is particularly suitable for
the diffusion Monte Carlo method \cite{lesechetal1}.

Regarding the two-centered orbitals $\Phi_{n_1}(1)$ and $\Phi_{n_2}(2)$ in
\eqref{M16.4} and \eqref{M16.6}, Aubert--Fr\'econ and Le Sech \cite[(4)]{A-FL}
 used a
truncated summation from the Hylleraas series solution, cf.\ \eqref{eqA.30}:
\begin{equation}\label{eq6.47a}
\Phi = \Phi(\lambda,\mu,\phi) = \sum^K_{k=m} f^m_kY^m_k(\mu,\phi) \sum^N_{n=0}
c_ne^{-p(\lambda-1)} [2p(\lambda-1)]^{m/2} L^m_n(2p(\lambda-1)),
\end{equation}
where $p$ is used as a {\em variational parameter}, and $f^m_k$ and $c_n$ are
coefficients which can be determined from the Killingbeck procedures \cite{Ki}.

\subsection*{(2)~~Generalized correlated or uncorrelated two-centered
wave functions}

\indent

The two-centered orbitals we have been using in this section to build up the
molecular orbitals, whether they be correlated, uncorrelated, with or without
configuration interaction, have been heavily influenced
by the classic solutions due to Hylleraas and Jaff\'e, cf.\ \eqref{eqA.30} and
\eqref{eqA.26}.
These two famous solutions were derived during the 1920s and 1930s with great
ingenuity, their greatest advantages being that a 3-term recurrent relation of
the series coefficients. The 3-term recurrences further lead to continued
fractions which have enabled mathematicians to perform asymptotic analysis. That
was during a time when no electronic computers were available and it
was perhaps the only way to perform any theoretical analysis at all. Nowadays,
fast computers are readily available so we don't have to rely overly on 3-term
recurrence relations. Five-term recurrence relations are just as good and can be
treated with relative ease also by the Killingbeck algorithm \cite{Ki}, for example.

Recall that for the single-electron, two-centered heteronuclear problem,
separation of variables leads to
\begin{alignat}{2}
\label{M21.1}
&[(\lambda^2-1)\Lambda']' + \left[A+2R_1\lambda - p^2\lambda^2 -
\frac{m^2}{\lambda^2-1}\right] \Lambda = 0, &\qquad &(R_1 \equiv R(Z_a+Z_b)/2)\\
\label{M21.2}
&[(1-\mu^2)M']' + \left[-A-2R_2\mu + p^2\mu^2- \frac{m^2}{1-\mu^2}\right]M = 0,
&\qquad &(R_2\equiv R(Z_a-Z_b)/2).
\end{alignat}
The appearances of the above two equations suggest the ansatz
\begin{equation}\label{M21.3}
\Lambda(\lambda) = \sum^\infty_{k=0} f_{k,1} P^m_k(\lambda),\quad M(\mu) =
\sum^\infty_{k=0} f_{k,2}P^m_k(\mu),
\end{equation}
where $P_k^m(\lambda)$ are the associated Legendre polynomials.
Substituting \eqref{M21.3} into \eqref{M21.1} and \eqref{M21.2} and equate
coefficients of $P^m_k(\lambda)$ and $P^m_k(\mu)$ to zero, see Appendix L, we
obtain {\em 5-term recurrence relations\/}
\begin{equation}\label{M21.4}
A_{kj}f_{k-2,j} + B_{kj} f_{k-1,j} + C_{kj}f_{k,j} + D_{kj}f_{k+1,j} + E_{kj}
f_{k+2,j} = 0,\qquad k=1,2; j=0,1,2,\ldots,
\end{equation}
where
\begin{equation}\label{M22.1}
\left.\begin{array}{l}
A_{k1} = A_{k2} = \dfrac{p^2(k-m-1)(k-m)}{(2k-3)(2k-1)},\\
\noalign{\medskip}
B_{kj} = -\dfrac{2R_j(k-m)}{2k-1},\qquad j=1,2,\\
\noalign{\medskip}
C_{k1} = C_{k2} = -k(k+1)-A + \dfrac{p^2(k-m+1)(k+m+1)}{(2k+1)(2k+3)} +
\dfrac{p^2(k-m)(k+m)}{(2k+1)(2k-1)}\\
\noalign{\medskip}
D_{kj} = -\dfrac{2R_j(k+m+1)}{2k+3},\qquad j=1,2,\\
\noalign{\medskip}
E_{k1} = E_{k2} = \dfrac{p^2(k+m+1)(k+m+2)}{(2k+3)(2k+5)}.
\end{array}\right\}
\end{equation}
The two-centered orbitals derived above differ from those of Hylleraas and
Jaff\'e. The ansatz \eqref{M21.3} is not the only way to obtain two-centered
orbitals. It is possible to derive several other solutions in different forms
using different orthogonal polynomials.
Numerical results indicate that these two-centered orbitals here give accuracy
compatible to that corresponding to Hylleraas and Jaff\'e type solutions.

\subsection*{(3)~~Numerical algorithm}

\indent

In contrast to the explicit analytic formulas for the quadratures of
James--Coolidge wave functions given in \ref{sec5}.D and in the affiliated Appendices
J and K, here the integrals will be computed using the Gaussian quadrature
routines. (The analytic formulas given earlier in \ref{sec5}.D can then be
compared with those obtained here as a good check.)
For the two-electron problem,
 the number of coordinates are six, so that the
energy calculation requires the 6-dimensional integration. Using
the cylindrical symmetry, we can reduce two angular variables to one.

The next simplification is the choice of the wave functions. If we only
restrict the wave functions to be the James--Coolidge type, the
5-dimensional integration can be divided as one two-dimensional
integration and two three-dimensional integration. However, if we
include the exponential function of inter-electron distance, that
is $e^{ - \kappa r_{12}}$, this 5-dimensional integration should
be evaluated as multidimensional integration, and the
numerical accuracy should be carefully investigated.

Here, we only restrict the wave functions to be the James--Coolidge type.
The usual integration has the following form.
\begin{equation}\label{paperVeq1}
\int\cdots\int \Lambda_1(\lambda_1) \Lambda_2(\lambda_2)
M_1(\mu_1) M_2(\mu_2) \Phi(\phi_1 - \phi_2) r_{12}^n d\lambda_1
d\lambda_2 d\mu_1 d\mu_2 d\phi_1 d\phi_2
\end{equation}
Using the $r_{12}^2$-identity, cf.\ \eqref{eq7a.15},
the exponent of $r_{12}$ can be reduced into 0 or -1. For $n=0$,
the whole integration is divided into only 5 one-dimensional
integrations. However, for $1/r_{12}$, by the Neumann expansion
the form of wave function includes sum of Legendre Polynomials with
proper coefficients; cf.\ \eqref{eq7a.22a}.
 Especially, the form of the $\lambda$-part is
\begin{equation}\label{paperVeq2}
P(\lambda_<) Q(\lambda_>) = \left\{ \begin{array}{lll}
P(\lambda_1) Q(\lambda_2), & \qquad & \lambda_1 < \lambda_2, \\
P(\lambda_2) Q(\lambda_1), & \qquad & \lambda_2 > \lambda_1.
\end{array} \right.
\end{equation}
This makes the form of integration over the $\lambda$ variable as
\begin{equation}\label{paperVeq3}
\int^\infty_0 \int^\infty_0 \cdots = \int^\infty_0
\int^{\lambda_1}_0 \cdots + \int^\infty_0 \int^\infty_{\lambda_1}
\cdots
\end{equation}

The  implementation of these numerical integration is
automated by standard computer software. Let us describe
the numerics
for a homonuclear dimer H$_2$ molecule and a heteronuclear
dimer HeH$^+$ molecular ion.

The approximation of wave function of two-electron system into a
multiplication of one-electron system can be easily made in the
natural orbit expansion
\begin{equation}\label{paperVeq4}
\Psi(\mathbf{r}_1, \mathbf{r}_2) = \sum c_k \chi_k(\mathbf{r}_1)
\chi_k( \mathbf{r}_2)
\end{equation}
where $\chi_k(\mathbf{r}_1)$ is the wave function of one-electron
two-center problem.

For the ground state, it's well-known that $c_1 \sim 1$, or
\begin{equation}\label{paperVeq5}
\Psi(\mathbf{r}_1, \mathbf{r}_2) \simeq \chi_1(\mathbf{r}_1)
\chi_1( \mathbf{r}_2).
\end{equation}

There are a few alternatives to approximate the $\chi_1(\mathbf{r}_1)$. One
is using the exact solution of one-electron two-center problem,
$1s\sigma_g$ state. The Jaffe's form is
\begin{equation}\label{paperVeq6}
\chi_1(\mathbf{r}_1) = e^{-\alpha \lambda} (1+\lambda)^\sigma
\sum_n g_n \left( \frac{\lambda-1}{\lambda+1} \right)^n \sum_m
f_{2m} P_{2m}(\mu),
\end{equation}
cf. \eqref{eqA.26}.

And the other is Patil's wave function
\begin{equation}\label{paperVeq7}
\chi_1(\mathbf{r}_1) = (1 + b \lambda)^\beta e^{-\alpha_P \lambda}
\cosh(a\mu);\quad \text{cf.\ \eqref{eq6.27a}.}
\end{equation}

We may also include configuration interaction, such as
\begin{equation}\label{paperVeq8}
\Psi(\mathbf{r}_1, \mathbf{r}_2) = c_1 \chi_1(\mathbf{r}_1)
\chi_1( \mathbf{r}_2) + c_2 \chi_2(\mathbf{r}_1) \chi_2(
\mathbf{r}_2)
\end{equation}
where $\chi_1$ is $1s\sigma_g$ state and $\chi_2$ is $2p\sigma_u$
state.

\begin{figure}[htpb]
\centering \epsfig{file=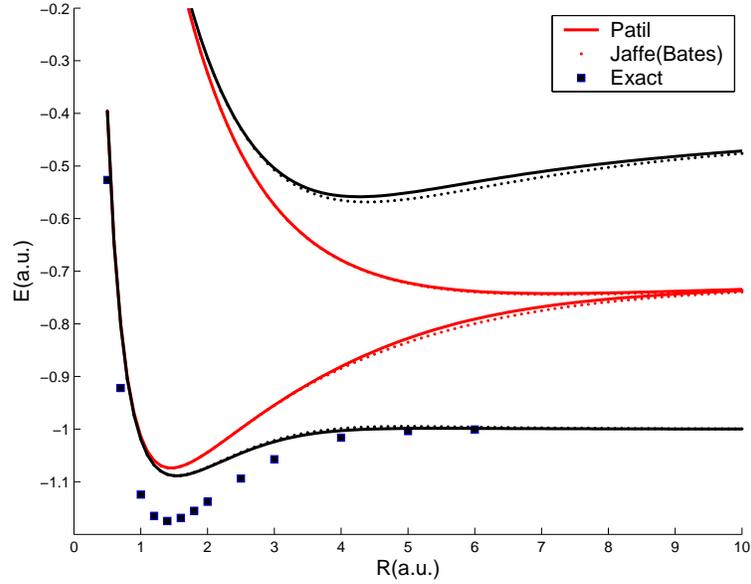,
width=10cm, angle=0} \caption{Comparison of potential curves of the H$_2$
molecule
computed with no correlation factor using the exact solution of the one-electron
wave function \eqref{paperVeq6} (small dots) and the
Patil's wave function \eqref{paperVeq7} (solid line).
The inner (outer) curves are the result without (with) configuration
interaction.
Squares are the ``exact" ground state $E(R)$ from
Ref. \cite{Kolo60}.
Upper curves correspond to ``excited states''.}\label{fig19}
\end{figure}

The result is shown in Fig.~\ref{fig19}.
Before the diagonalization (that is, CI),  the asymptotic behavior
of the ground state $E(R)$ is monotonically increasing. By diagonalizing,
the behavior at large $R$ becomes almost flat, however, $E(R)$ is
slightly above the exact asymptotic value, -1 (htr).

\begin{figure}[htpb]
\centering \epsfig{file=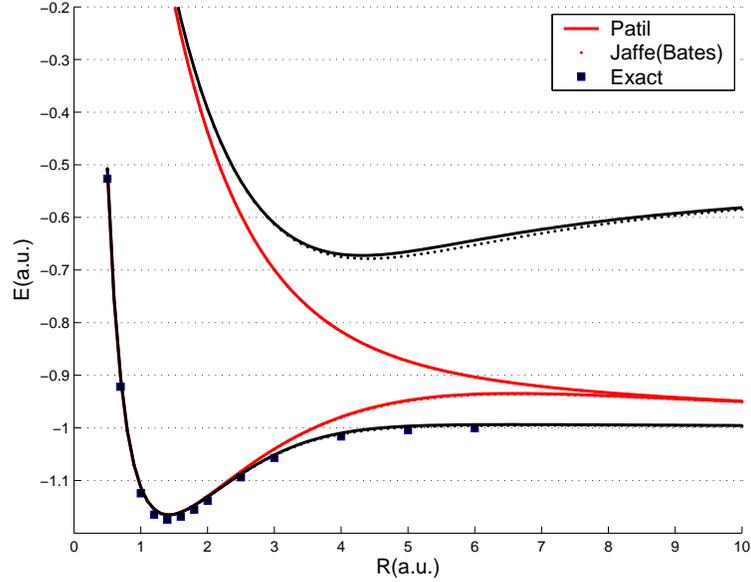, width=10cm,
angle=0} \caption{Same computations as with Fig.~\ref{fig19}, but
with the correlation factor.}\label{fig20}
\end{figure}

Next, we perform computations  with correlation
by using $f(r_{12}) = 1 + \frac{1}{2} r_{12}$. Improvement can be readily
seen in Fig.~\ref{fig20}
which is much closer to the
exact calculation done by Kolos \cite{Kolo60}.

\begin{figure}[htpb]
\centering
\epsfig{file=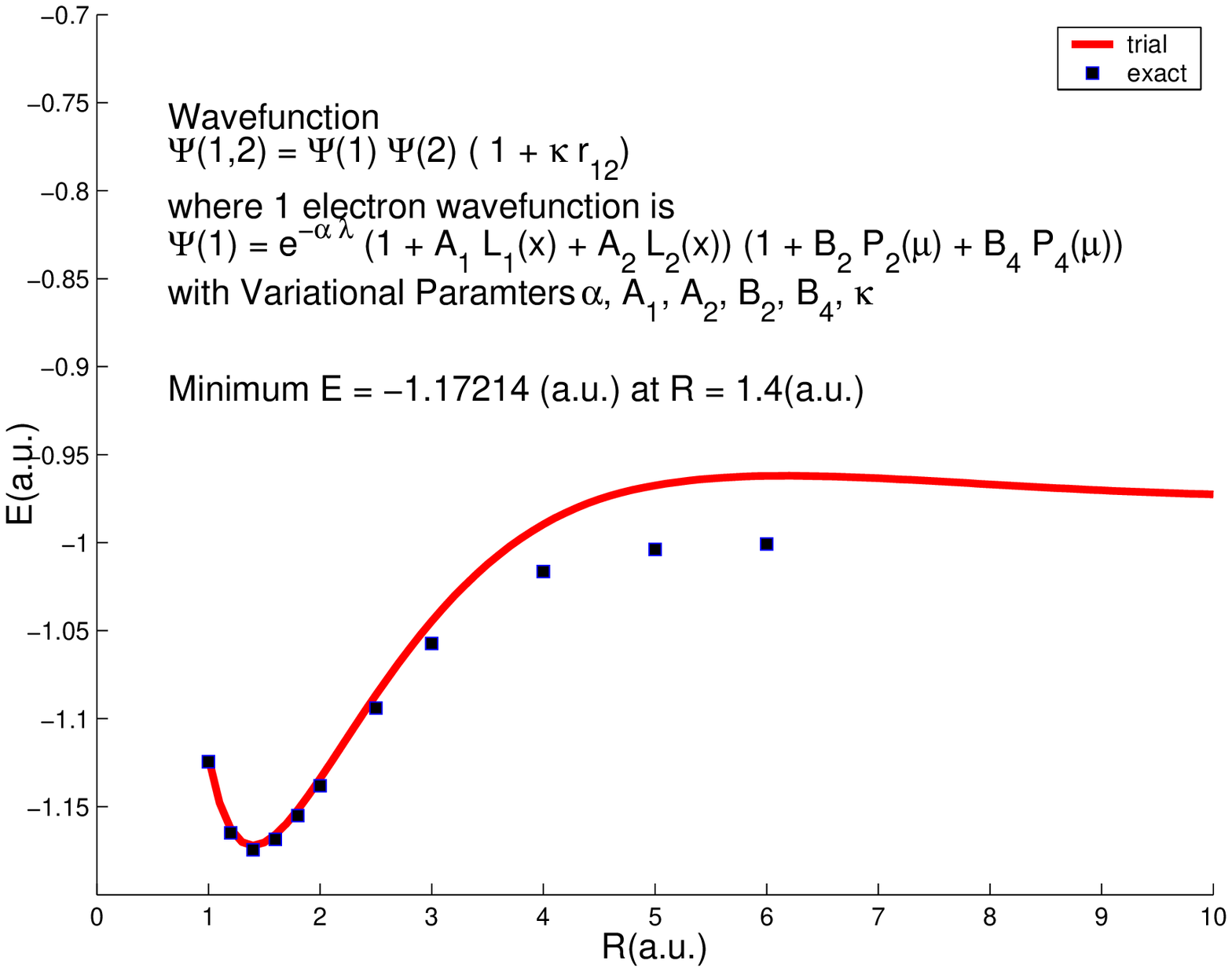,
width=12cm} \caption{
Ground state potential energy curve of the H$_2$
obtained using the truncated exact
wave functions of one-electron system with the correlation factor and
several variational parameters. The curve yields the binding energy of
$E_B =-0.1721$ a.u.$=4.684$ eV.
Squares are the ``exact" ground state $E(R)$.}
\label{fig21}
\end{figure}

\begin{figure}[htpb]
\centering
\epsfig{file=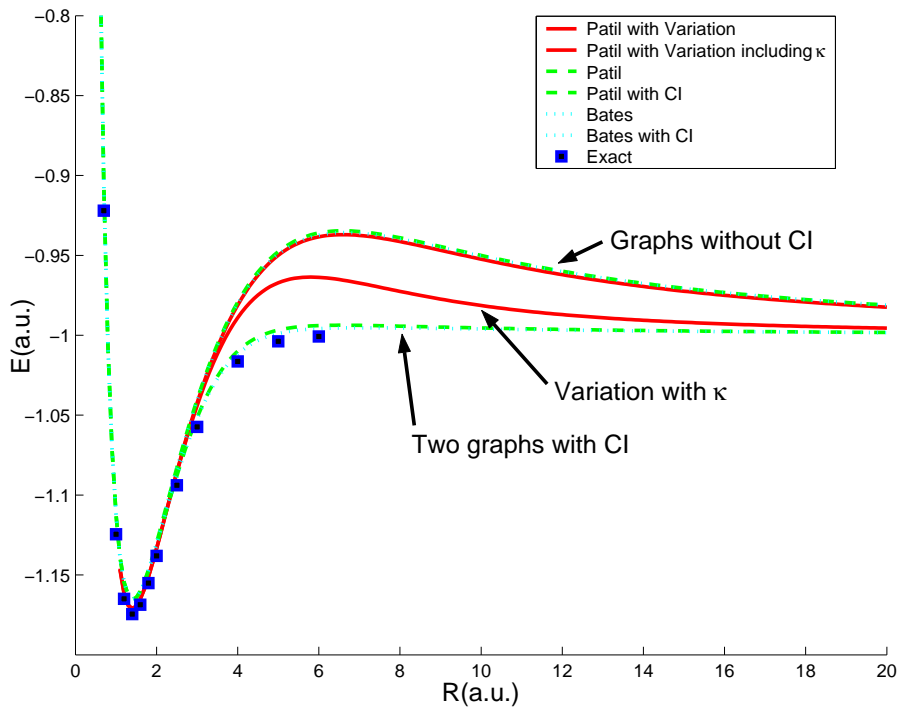,width=12cm}
\caption{Ground state potential curve of H$_2$ computed
from the Patil's wave function with the correlation factor
and variational parameters.
The curve yields the binding energy of
$E_B =-0.1713$ a.u.$=4.662$ eV.
Squares are the ``exact" ground state $E(R)$.}
\label{fig22}
\end{figure}

In the  calculations below (Figs.~\ref{fig19} and \ref{fig20}), ``exact''
solutions of H$^+_2$-solutions were used (whose coefficients are obtained through
truncated Killingbeck \cite{Ki} procedures). Instead, one can use
\eqref{M13.2}, \eqref{eq6.27c}, \eqref{eq6.27d} by optimizing
the coefficients $\alpha$ and $B$ therein, or, for the Patil's wave function,
by optimizing the coefficients $a$ and $\beta$ in \eqref{eq6.27a} and
\eqref{eq6.27b}. By doing so, we obtain the energy curve of the ground
state as shown in Figs.~\ref{fig21} and \ref{fig22}, respectively. The
energy curves are improved over a wide range of $R$ values, and we obtain the
binding energy of $0.171$ a.u.$=4.65$ eV, close to the experimental value.
However, $E(R)$ in Fig. \ref{fig21} overshoots the dissociation limit.

\section{Alternative approaches}\label{secVII}

\subsection{Improvement of Hartree--Fock results using the Bohr model}

\begin{figure}[htpb]
\bigskip
\centerline{\epsfxsize=0.6\textwidth\epsfysize=0.5\textwidth
\epsfbox{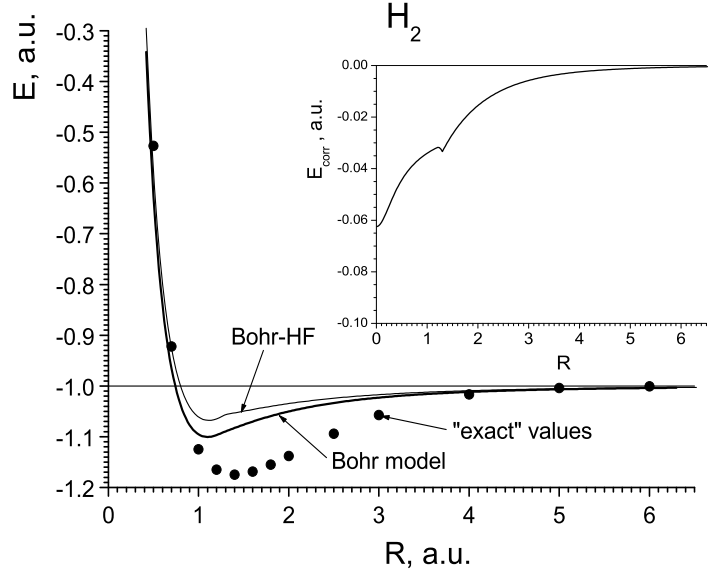}}

\caption{Ground state $E(R)$ for the H$_2$ molecule in the Bohr and Bohr-HF
models. Insert shows the correlation energy as a function of $R$.
}
\label{h2BHF}
\end{figure}

The Bohr model can also be applied to calculate the correlation energy for
molecules and then improve the HF treatment. Fig. \ref{h2BHF} shows the
ground state potential curve for H$_2$ molecule calculated in the Bohr-HF
approximation. Such an approximation omits the electron repulsion term $%
1/r_{12}$ in finding the electron configuration from Eq. (\ref{b2}). The
difference between the Bohr and Bohr-HF potential curves yields the
correlation energy plotted in the insert of Fig. \ref{h2BHF}.

\begin{figure}[htpb]
\bigskip
\centerline{\epsfxsize=0.5\textwidth\epsfysize=0.5\textwidth
\epsfbox{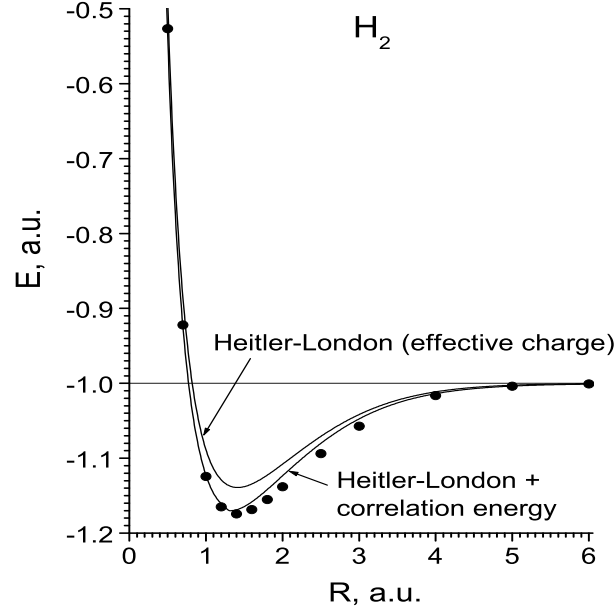}}

\caption{Ground state energy $E(R)$ of the H$_2$ molecule in the Heitler--London
method and the improved $E(R)$ after the addition of the correlation energy.
Dots are the ``exact'' result from \cite{Kolo60}.
}
\label{h2HLCE}
\end{figure}

In Fig. \ref{h2HLCE} we draw the ground state $E(R)$ for the H$_2$ molecule
obtained with the Heitler--London trial function that has the form of the
combination of the atomic orbitals \cite{Slat63}:
$$
\Psi =C\left\{
\exp [-\alpha (r_{a1}+r_{b2})]+\exp [-\alpha (r_{b1}+r_{a2})]\right\} ,
$$
where $\alpha$ is a variational parameter. Addition of the correlation energy
from Fig.  \ref{h2BHF} improves the Heitler--London result and shifts $E(R)$
close to the ``exact'' values. The improved potential curve yields the binding
energy of 4.63 eV.

\subsection{Dimensional scaling}

Dimensional scaling offers promising new computational strategies for the
study of few electron systems. This is exemplified by recent applications to
atoms, as well as H$_2^{+}$ and H$_2$ molecules \cite{Hers92,Svid04}. D-scaling
emulates a standard method of quantum chromodynamics \cite{Witt80}, by
generalizing the Schr\"odinger equation to D dimensions and rescaling
coordinates \cite{Hers92}. The $D\rightarrow \infty $ limit corresponds to
infinitely heavy electrons and reduces to a classic electrostatic problem
of finding an electron configuration that minimizes a known effective
potential.

\begin{figure}[htpb]
\bigskip
\centerline{\epsfxsize=0.45\textwidth\epsfysize=0.4\textwidth
\epsfbox{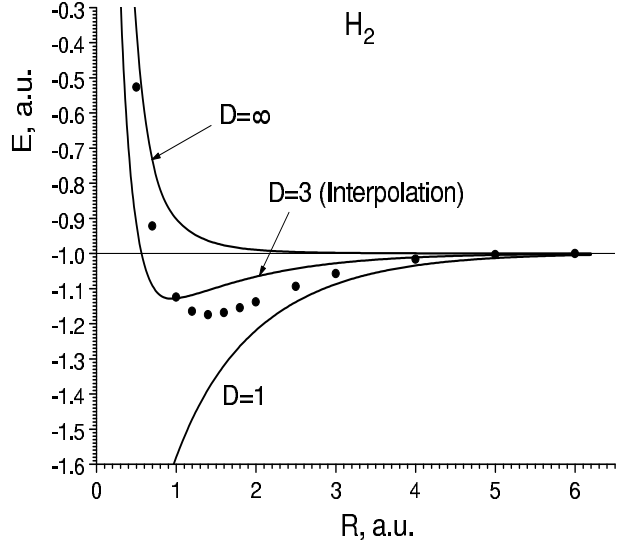}}

\caption{
Ground state energy $E(R)$ of the H$_2$ molecule in $D=\infty $, $D=1$ and three
dimensional interpolation (solid lines). Dots show the ``exact'' D=3 energy
from \cite{Kolo60}.
}
\label{h2i13}
\end{figure}

We start from the Schr\"odinger equation $\hat H\Psi =E\Psi $ for a two
particle wave function $\Psi ({\bf r}_1,{\bf r}_2)$. The H$_2$ Hamiltonian
in atomic units reads
$$
\hat H=-\frac 12\nabla _1^2-\frac 12\nabla _2^2+V(\rho _1,\rho
_2,z_1,z_2,\phi ,R),
$$
where the Coulomb potential energy $V$ is given by Eq. (\ref{b1}). In a
traditional dimensional scaling approach the Laplacian is treated in
D-dimension and the wave function is transformed by incorporating the square
root of the Jacobian via $\Psi \rightarrow J^{-1/2}\Phi $, where $J=(\rho
_1\rho _2)^{D-2}(\sin \phi )^{D-3}$ in cylindrical coordinates, and $\phi $
is the dihedral angle specifying relative azimuthal orientation of electrons
about the molecular axis \cite{Fran88}.
D-scaling in spherical coordinates is discussed in Appendix \ref{dimspher}.
On scaling the coordinates by $f^2$
and the energy by $1/f^2$, with $f=(D-1)/2$, the Schr\"odinger equation in
the limit $D\rightarrow \infty $ leads to minimization of the effective
potential \cite{Fran88}
\begin{equation}
\label{d1}E=\frac 12\left( \frac 1{\rho _1^2}+\frac 1{\rho _2^2}\right)
\frac 1{\sin ^2\phi }+V(\rho _1,\rho _2,z_1,z_2,\phi ,R).
\end{equation}
The obtained electron configuration is sometimes called the Lewis structure
because it provides a rigorous version of the familiar electron-dot formulas
introduced by Lewis in 1916 \cite{Lewi16}. Fig. \ref{h2i13} (upper curve)
displays the $D\rightarrow \infty $ potential curve $E(R)$ that exhibits no
binding and substantially deviates from the $D=3$ ``exact'' energy (dots).
In the limit $D\rightarrow 1$ the dimensional scaling reduces the
Hamiltonian to the delta function model \cite{Dore87}
\begin{equation}
\label{d2}\hat H=-\frac 12\left( \frac{d^2}{dx_1^2}+\frac{d^2}{dx_2^2}%
\right) -\delta \left( x_1+\frac R2\right) -\delta \left( x_1-\frac
R2\right) -\delta \left( x_2+\frac R2\right) -\delta \left( x_2-\frac
R2\right) +\delta \left( x_1-x_2\right) .
\end{equation}
A variational solution of the one dimensional wave equation \cite{Lapi75} is
pictured in Fig. \ref{h2i13} (lower curve). Also shown is an approximation
to $E(R)$ for $D=3$ obtained by interpolating linearly in $1/D$ between the
dimensional limits:
\begin{equation}
\label{d3}E_3(R)=\frac 23E_\infty (R)+\frac 13E_1(R).
\end{equation}
The interpolated $D=3$ curve exhibits binding which indicates the
feasibility of extending dimensional interpolation to molecules.

A proper choice of scaling can improve the method. Here we discuss a
dimensional scaling transformation of the Schr\"odinger equation that yields
the Bohr model of H$_2$ in the limit of infinite dimensionality \cite{Svid04}.
For such a transformation, the large-D limit provides a link between the old
(Bohr-Sommerfeld) and the new (Heisenberg-Schr\"odinger) quantum mechanics.
The first-order correction in 1/D substantially improves the agreement with
the exact ground state $E(R)$.

In the modified scaling the Laplacian depends on a continuous parameter $D$
as follows
\begin{equation}
\label{d4}\nabla ^2=\frac 1{\rho ^{D-2}}\frac \partial {\partial \rho
}\left( \rho ^{D-2}\frac \partial {\partial \rho }\right) +\frac 1{\rho ^2}%
\frac{\partial ^2}{\partial \phi ^2}+\frac{\partial ^2}{\partial z^2}.
\end{equation}
For $D=3$, Eq. (\ref{d4}) reproduces the 3D Laplacian. On scaling the
coordinates by $f^2$, the energy by $1/f^2$ (recalling $f=(D-1)/2$)
 and transforming the electronic
wave function $\Psi $ by
\begin{equation}
\label{d5}\Psi =(\rho _1\rho _2)^{-(D-2)/2}\Phi ,
\end{equation}
the Schr\"odinger equation is recast as
\begin{equation}
\label{d6}(K+U+V)\Phi =E\Phi ,
\end{equation}
where
$$
K=-\frac 2{(D-1)^2}\left\{ \frac{\partial ^2}{\partial \rho _1^2}+\frac{%
\partial ^2}{\partial \rho _2^2}+\frac{\partial ^2}{\partial z_1^2}+\frac{%
\partial ^2}{\partial z_2^2}+\left( \frac 1{\rho _1^2}+\frac 1{\rho
_2^2}\right) \frac{\partial ^2}{\partial \phi ^2}\right\} ,
$$
\begin{equation}
\label{d7}U=\frac{(D-2)(D-4)}{2(D-1)^2}\left( \frac 1{\rho _1^2}+\frac
1{\rho _2^2}\right) .
\end{equation}
In the $D\rightarrow \infty$ limit the derivative terms in $K$ are
quenched. The corresponding energy $E_\infty $ for any given internuclear
distance $R$ is then obtained simply as the minimum of an effective
potential
\begin{equation}
\label{d8}E=\frac 12\left( \frac 1{\rho _1^2}+\frac 1{\rho _2^2}\right)
+V(\rho _1,\rho _2,z_1,z_2,\phi ,R).
\end{equation}
This is identical in form to that for the Bohr model, Eq. (\ref{b2}), and we
thus obtain for $E_\infty (R)$ the same solutions depicted in Fig. \ref{f2c}%
. This result for $E_\infty (R)$ differs in an interesting way from that
obtained in a traditional study of the $D\rightarrow \infty $ limit for the H%
$_2$ molecule \cite{Fran88}. Here, in order to connect with the Bohr model,
it is necessary to incorporate only the radial portion of the Jacobian in
transforming the electronic wave function via Eq. (\ref{d5}). The customary
practice, which employs the full Jacobian, introduces a factor of $1/(\sin
^2\phi)$ into the centrifugal potential, as seen from Eq. (\ref{d1}). The
modified procedure yields directly a good zeroth-order approximation.

The ground state $E(R)$ can be substantially improved by use of a
perturbation expansion in powers of $1/D$, developed by expanding the
effective potential of Eq. (\ref{d8}) in powers of the displacement from the
minimum \cite{Hers92}; for He and H$_2^{+}$ this has yielded highly accurate
results \cite{Good92}. Terms quadratic in the displacement describe harmonic
oscillations about the minimum and give a $1/D$ correction to the energy.
Anharmonic cubic and quartic terms give a $1/D^2$ contribution. For the He
ground state energy (the $R=0$ limit for H$_2$) a first-order approximation
yields \cite{Svid04,ff}
\begin{equation}
\label{d9}E(0)=\frac{4E_\infty }{(D-1)^2}\left( 1-\frac{0.1532}D\right) ,
\end{equation}
where $E_\infty =-3.062$ a.u. is the Bohr model He result \cite{Bohr1,Vlec22}.
 For $D=3$ Eq. (\ref{d9}) improves the ground state energy of He
  to $E(0)=-2.906$ a.u., which
differs by $0.07\%$ only from the ``exact'' value of $-2.9037$ a.u.
\cite{Drak02}.

\begin{figure}[htpb]
\bigskip
\centerline{\epsfxsize=0.45\textwidth\epsfysize=0.4\textwidth
\epsfbox{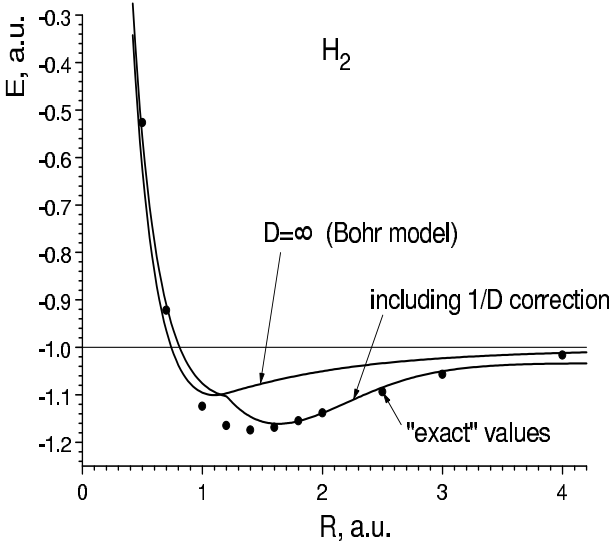}}

\caption{
Ground state energy $E$ of H$_2$ molecule as a
function of internuclear distance $R$ calculated within the dimensional
scaling approach
(solid lines) and the ``exact'' energy (dots).
}
\label{f3c}
\end{figure}

To evaluate the $1/D$ correction for arbitrary $R$, it is convenient to
introduce new coordinates%
$$
\tilde z_1=\frac 1{\sqrt{2}}(z_1-z_2),\qquad \tilde z_2=\frac 1{\sqrt{2}%
}(z_1+z_2).\qquad
$$
The effective potential of Eq. (\ref{d8}) then has a minimum at $\rho
_1=\rho _2=\rho _0,$ $\phi =\pi ,$ $\tilde z_2=0$. Along the coordinates $%
\rho _1$, $\rho _2$, $\tilde z_2$ and $\phi $ the potential has a single
well structure no matter what the internuclear spacing $R$ is. However, along
the $\tilde z_1$ direction at $R=1.2$ the potential changes shape from a
single to a double well; such symmetry breaking is a typical feature
exhibited at large D \cite{Shi01}. To avoid the inaccuracy of approximation
by a single quadratic form one can solve the Schr\"odinger equation
numerically along the $\tilde z_1$ direction for the exact potential and add
contributions from the harmonic motion associated with the other coordinates
$\rho _1$, $\rho _2$, $\tilde z_2$ and $\phi $. The result is shown in Fig.
\ref{f3c} \cite{Svid04}. The $1/D$ correction improves $E(R)$ and predicts
the equilibrium separation to be $R_e=1.62$ with binding energy $E_B=4.38$
eV.

\section{Conclusions and outlook}\label{secVIII}

\indent

Many major topics on diatomic molecules, and some other atoms and molecules in
general, have been addressed in this article, giving it a very ``locally
diverse'' or perhaps a somewhat disjoint look. But a simple,
 ``global'' picture may
be viewed and understood in/from the following diagram:

\begin{center}
\epsfig{file=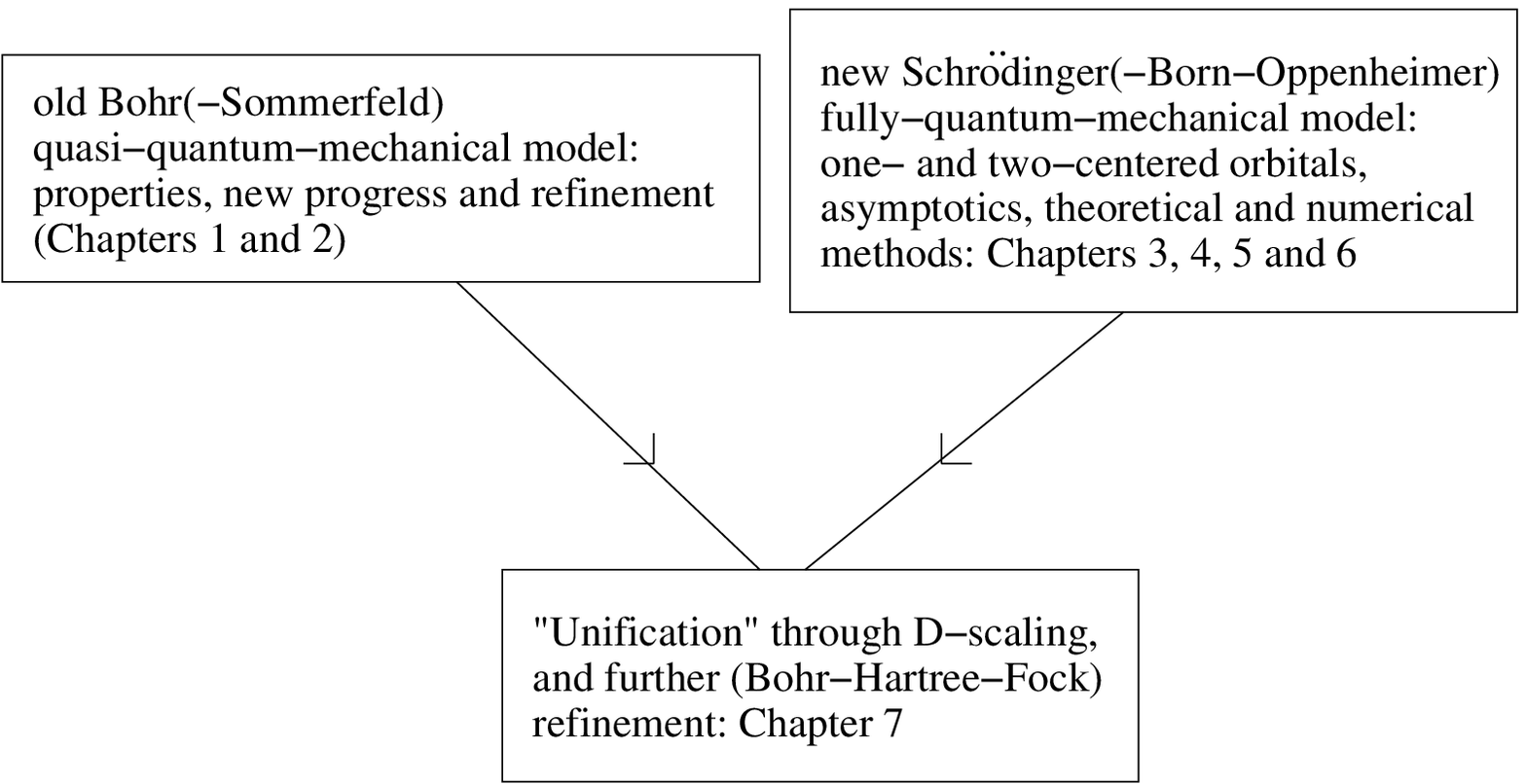,height=3in,width=5in}
\end{center}

\n This ``unification'' is by no means an easy task. Nevertheless, it was
a goal we somehow envisioned to achieve when this research was started and we
hope we have made at least some success.

At present, in the study of ultrafast laser applications to chemical physics
and molecular chemistry problems, there is a pressing need to understand
the quantum-{\em dynamical\/} behavior of molecules and the associated properties of
excited states and their computations. Many challenges are lying ahead and
awaiting elegant solutions.

\appendix

\section*{Appendices}

\section{Separation of variables for the H$_2^+$-like Schr\"odinger equation.}

\label{A1}

\indent
Let us consider \eqref{eqA.1}. Here we show how to separate
the variables through the use of the  ellipsoidal (or, prolate spheroidal)
coordinates (see Fig. \ref{prosph})
\begin{equation}
\label{eqA.2}x = \frac{R}2 \sqrt{(\lambda^2-1)(1-\mu^2)}\ \cos \phi,\quad y
= \frac{R}2 \sqrt{(\lambda^2-1)(1-\mu^2)}\ \sin\phi,\quad z = \frac{R}2
\lambda\mu .
\end{equation}
Note that the coordinates $\lambda,\mu$ and $\phi$ are
orthogonal, and we have the first fundamental form
$$
ds^2 = dx^2 + dy^2 + dz^2 = h^2_\lambda\ d\lambda^2 + h^2_\mu\ d\mu^2 +
h^2_\phi \ d\phi^2
$$
where
\begin{align*}
h^2_\lambda &= \left(\frac{\partial x}{\partial\lambda}\right)^2 +
\left(\frac{\partial y}{\partial\lambda}\right)^2 +  \left(\frac{\partial
z}{\partial\lambda}\right)^2 = \frac{R^2}4
\frac{1-\mu^2}{\lambda^2-1},\\
h^2_\mu &= \left(\frac{\partial x}{\partial\mu}\right)^2 + \left(\frac{\partial
y}{\partial\mu}\right)^2 + \left(\frac{\partial z}{\partial \mu}\right)^2  =
\frac{R^2}4 \frac{\lambda^2-1}{1-\mu^2},\\
h^2_\phi &= \left(\frac{\partial x}{\partial\phi}\right)^2 +
\left(\frac{\partial y}{\partial\phi}\right)^2 + \left(\frac{\partial
z}{\partial\phi}\right)^2 = \frac{R^2}4 (\lambda^2-1)(1-\mu^2).
\end{align*}
Thus
\begin{align}
\nabla^2\Psi &= \frac1{h_\lambda h_\mu h_\phi}
\left[\frac\partial{\partial\lambda} \left(\frac{h_\mu h_\phi}{h_\lambda}
\frac\partial{\partial\lambda} \Psi\right) + \frac\partial{\partial\mu}
\left(\frac{h_\lambda h_\phi}{h_\mu} \frac\partial{\partial\mu}\Psi\right) +
\frac\partial{\partial\phi} \left(\frac{h_\lambda h_\mu}{h_\phi}
\frac\partial{\partial\phi}\Psi\right) \right]\nonumber\\
\label{eqA.3}
&= \frac4{R^2(\lambda^2-\mu^2)} \left\{\frac\partial{\partial\lambda}
\left[(\lambda^2-1) \frac\partial{\partial\lambda}\right] +
\frac\partial{\partial\mu} \left[(1-\mu^2) \frac{\partial}{\partial\mu}\right] +
\frac{\lambda^2-\mu^2}{(\lambda^2-1)(1-\mu^2)} \frac{\partial^2}{\partial\phi^2}
\right\}\Psi.
\end{align}


\noindent Note that through the coordinate transformation \eqref{eqA.2}, we
have
\begin{equation}
\label{eqA.2a}\left\{
\begin{array}{l}
\lambda =
\dfrac{r_a+r_b}{R}, \\ \noalign{\smallskip} \mu = \dfrac{r_a-r_b}{R},
\end{array}
\right. \qquad \text{equivalently,}\qquad
\begin{array}{l}
r_a =
\dfrac{R}{2} (\lambda+\mu), \\ \noalign{\smallskip} r_b = \dfrac{R}{2}
(\lambda-\mu).
\end{array}
\end{equation}
Also, we have $\lambda \ge 1, - 1\le \mu \le 1$.

Write
\begin{equation}
\label{eqA.4}\Psi = \Lambda(\lambda) M(\mu) \Phi(\phi).
\end{equation}
$\Phi(\phi)$ must be periodic with period $2\pi$. Therefore
\begin{equation}
\label{eqA.5}\Phi(\phi) = e^{im\phi},\qquad m = 0,\pm 1, \pm 2,\ldots~.
\end{equation}
Substitute \eqref{eqA.3}, \eqref{eqA.4} and \eqref{eqA.5} into \eqref{eqA.1}%
, and then divide by $e^{im\phi}$:
\begin{align}
&- \frac12 \frac4{R^2(\lambda^2-\mu^2)} \left\{\frac\partial{\partial\lambda}
\left[(\lambda^2-1) \frac\partial{\partial\lambda} \Lambda\right] M +
\frac\partial{\partial\mu} \left[(1-\mu^2)
\frac\partial{\partial\mu}M\right]\Lambda -
\frac{(\lambda^2-\mu^2)m^2}{(\lambda^2-1)(1-\mu^2)} M\Lambda\right\}\nonumber\\
\label{eqA.6}
&- \frac2R \frac{Z_b}{\lambda+\mu} M\Lambda - \frac2R \frac{Z_a}{\lambda-\mu}
M\Lambda + \frac{Z_aZ_b}R M\Lambda = EM\Lambda.
\end{align}
Further multiplying every term by $-\frac{R^2}2 (\lambda^2-\mu^2)$, we
obtain
\begin{align}
&\frac\partial{\partial\lambda} \left[(\lambda^2-1)
\frac\partial{\partial\lambda}\Lambda\right] M + \frac\partial{\partial\mu}
\left[(1-\mu^2) \frac\partial{\partial\mu} M\right]\Lambda -
\frac{\lambda^2-\mu^2}{(\lambda^2-1)(1-\mu^2)} m^2M\Lambda\nonumber\\
\label{eqA.7}
&+ \underbrace{\left[RZ_b(\lambda-\mu) + RZ_a(\lambda+\mu) -
\left(\frac{RZ_aZ_b}{2} - \frac{R^2E}2\right)
(\lambda^2-\mu^2)\right]}_{\left(\frac{R^2E}2 - \frac{RZ_aZ_b}2\right)
(\lambda^2-\mu^2) + R[(Z_a+Z_b)\lambda + (Z_a-Z_b)\mu]} M\Lambda =
0.
\end{align}
Set
\begin{equation}
\label{eqA.8}p^2 = \frac12 (-R^2E + RZ_aZ_b) > 0.
\end{equation}
We have $p^2>0$ here due to the fact that we are mainly interested in the
electronic states that are {\em bound states}, i.e., not ionized.

Let the constant of separation of variables be $A$. Then from \eqref{eqA.7}
and \eqref{eqA.8} we obtain \eqref{eqA.9} and \eqref{eqA.10} in Subsection
\ref{sec2}.B.

\section{The asymptotic expansion of $\pmb{\Lambda(\lambda)}$
for large $\pmb{\lambda}$}

\label{A2}

\indent

Here, we provide a quick proof of Eq. \eqref{eqA.12a}. From \eqref{eqA.11a}%
, we have
\begin{equation}
\label{eqA2.1}\Lambda^{\prime\prime}(\lambda) -p^2\Lambda(\lambda) = -\left\{%
\frac{2\lambda}{\lambda^2-1} \Lambda^{\prime}(\lambda) + \left[\frac{%
A+2R_1\lambda}{\lambda^2-1} - p^2 \left(\frac{\lambda^2}{\lambda^2-1}
-1\right) - \frac{m^2}{(\lambda^2-1)^2}\right] \Lambda(\lambda)\right\}.
\end{equation}
We now find the Laurent expansions of the coefficient functions on the
right-hand side of \eqref{eqA2.1} as follows:
\begin{align*}
\frac{2\lambda}{\lambda^2-1} =
\frac2\lambda + \frac0{\lambda^2} + \frac2{\lambda^3} + \frac0{\lambda^4} +
\frac2{\lambda^5} +\cdots,
\end{align*}
\begin{align*}
\frac{A+2R_1\lambda}{\lambda^2-1} &= \frac1{\lambda^2} (A+2R_1\lambda) \left(1 +
\frac1{\lambda^2} + \frac1{\lambda^4} +\cdots\right)\\
&= \frac{2R_1}{\lambda} + \frac{A}{\lambda^2} + \frac{2R_1}{\lambda^3} +
\frac{A}{\lambda^4} +\cdots,
\end{align*}
\begin{align*}
-p^2\left(\frac{\lambda^2}{\lambda^2-1}-1\right) &=
\frac0\lambda - \frac{p^2}{\lambda^2} + \frac0{\lambda^3} -
\frac{p^2}{\lambda^4} \pm\cdots\\
-\frac{m^2}{(\lambda^2-1)^2} &=
\frac0\lambda + \frac0{\lambda^2} + \frac0{\lambda^3} - \frac{m^2}{\lambda^4}
+ \frac0{\lambda^5} - \frac{2m^2}{\lambda^6} \pm\cdots~.
\end{align*}
>From these expansions above, we now use the ansatz
$$
\Lambda(\lambda) = a_0e^{-p\lambda} \lambda^\beta\left(1 + \frac{c_1}\lambda
+ \frac{c_2}{\lambda^2} +\cdots\right)
$$
by substituting it into \eqref{eqA2.1} and equating all the coefficients of $%
\lambda^{-n}$ to zero for $n=0,1,2,\ldots$~. We easily obtain
$$
\beta = \frac{R_1}p-1,\quad c_1 = \frac{p^2-\beta^2-2\beta}{2R_1-2p\beta-p}.
$$
All the other coefficients $c_n$ can be determined in a straightforward way.
Note that the asymptotic expansion \eqref{eqA.12a} also gives
\begin{equation}
\label{eqA2.2}\Lambda(\lambda) - a_1e^{-p\lambda} \lambda^\beta\sum^n_{j=0}
\frac{c_j}{\lambda^j} = \mathcal{O}(e^{-p\lambda} \lambda^{\beta(n+1)}),
\text{ for } \lambda\gg 1.
\end{equation}

\section{The asymptotic expansion of $\pmb{\Lambda(\lambda)}$ as $
\pmb{\lambda}\rightarrow 1$}

\label{Lg1}

Here we provide a proof of Eq. \eqref{eqA.241}.
Multiply \eqref{eqA.11} by $(\lambda-1)/(\lambda+1)$ and rewrite it as
\begin{align}
0 &= (\lambda-1)^2 \Lambda''(\lambda) + \frac{2\lambda}{\lambda+1} (\lambda-1)
\Lambda'(\lambda) + \left[ \frac{(A+2R_1\lambda -p^2\lambda^2)}{\lambda+1}
(\lambda-1) - \frac{m^2}{(\lambda+1)^2}\right] \Lambda(\lambda)\nonumber\\
\label{eqA.13}
&\approx (\lambda-1)^2 \Lambda''(\lambda) + (\lambda-1) \Lambda'(\lambda) -
\frac{m^2}4 \Lambda(\lambda), \text{ for } \lambda\approx 1.
\end{align}
A differential equation (such as \eqref{eqA.13}) set in the form
\begin{equation}
\label{eqA.14}(x-1)^2 y^{\prime\prime}(x) + (x-1) q(x) y^{\prime}(x) +
r(x)y(x) = 0,
\end{equation}
near $x=1$, where $q(x)$ and $r(x)$ are analytic functions at $x=1$, is said
to have a {\em regular singular point\/} at $x=1$. The solution's behavior
near $x=1$ hinges largely on the roots $\nu$ of the {\em indicial equation\/}
\begin{equation}
\label{eqA.15}\nu(\nu-1) + q(1)\nu + r(1) = 0
\end{equation}
because the solution $y(x)$ of \eqref{eqA.14} is expressible as
$$
y(x) = b_1(x-1)^{\nu_1} \sum^\infty_{k=0} c_k(x-1)^k + b_2(x-1)^{\nu_2}
\sum^\infty_{k=0} d_k(x-1)^k,\qquad (c_0=d_0=1),
$$
where $\nu_1$ and $\nu_2$ are the two roots of the indicial equation
\eqref{eqA.15}, under the assumptions that
\begin{equation}
\label{eqA.16}\nu_1>\nu_2,\quad \nu_1-\nu_2 \quad \text{is not a positive
integer}.
\end{equation}
However, if \eqref{eqA.16} is violated, then there are two possibilities and
two different forms of solutions arise:

\noindent {\bf (a)~~$\pmb{\nu_1=\nu_2}$}. Then
\begin{align}\label{eqA.17}
y(x) &= b_1y_1(x) + b_2y_2(x),\\
\intertext{where}
\label{eqA.18}
y_1(x) &= (x-1)^{\nu_1} \sum^\infty_{k=0} c_k(x-1)^k,\quad
(c_0=1)\\
\intertext{and}
\label{eqA.19}
y_2(x) &= (x-1)^{\nu_1} \sum^\infty_{k=1} d_k(x-1)^k + [\ln (x-1)] y_1(x);
\qquad (d_1=1).
\end{align}
Solution $y_2$ in \eqref{eqA.19} should be discarded because it becomes
unbounded at $x=1$.\newline
{\bf (b)~~$\pmb{\nu_1-\nu_2=}$ a positive integer}. Then case (a) holds
except with the modification that
\begin{align}
y_2(x)  = (x-1)^{\nu_2} \sum^\infty_{k=0} d_k(x-1)^k &+ c[\ln (x-1)]
y_1(x),\nonumber\\
\label{eqA.20}
&\quad\quad (d_0=1, \, c \text{ is a fixed constant but may be
0).}
\end{align}
Applying the above and \eqref{eqA.15} to equation \eqref{eqA.13}:
\begin{equation}
\label{eqA.21}(\lambda-1)^2 \Lambda^{\prime\prime}(\lambda) + (\lambda-1)
\Lambda^{\prime}(\lambda) - \frac{m^2}4 \Lambda(\lambda) \approx 0,
\end{equation}
we obtain the indicial equation
\begin{equation}
\label{eqA.22}\nu(\nu-1) + \nu - \frac{m^2}4 = 0,
\end{equation}
with roots
\begin{equation}
\label{eqA.23}\nu_1 = |m|/2,\quad \nu_2 = -|m|/2, \quad \text{($m$ can be
either a positive or a negative integer).}
\end{equation}
Thus either
\begin{align*}
&\nu_1 = \nu_2\quad \text{(when $m=0$)}\\
\intertext{or}
&\nu_1-\nu_2 = |m|= \text{ a positive integer, where } m\ne 0.
\end{align*}
Again, we see that solution $y_2$ in \eqref{eqA.20} must be discarded
because it becomes unbounded at $x=1$. Thus, from \eqref{eqA.18} and
\eqref{eqA.21}, we have
\begin{equation}
\label{eqA.24}\Lambda(\lambda) \approx (\lambda-1)^{\frac{|m|}2}
\sum^\infty_{k=0} c_k(\lambda-1)^k.
\end{equation}

\section{Expansions of solution near $\pmb{\lambda\approx 1}$ and $
\pmb{\lambda\gg 1}$: trial wave functions of James and Coolidge}

\label{sec2.2.3}

\indent

In the pioneering work of James and Coolidge \cite{james}, the two-centered
trial wave functions for H$_2$ are chosen to be
\begin{align}
\intertext{(ground state)}
\label{eqA1}
&\psi(\lambda_1,\lambda_2,\mu_1,\mu_2,\rho) = \frac1{2\pi}
e^{-\alpha(\lambda_1+\lambda_2)} \sum_{m,n,j,k,p} C_{mnjkp} (\lambda^m_1
\lambda^n_2 \mu^j_1 \mu^k_2 + \lambda^n_1 \lambda^m_2 \mu^k_1
\mu^j_2)\rho^p,\\
\intertext{and}
\intertext{(excited state)}
\label{eqA2}
&\psi(\lambda_1,\lambda_2,\mu_1,\mu_2,\rho) = e^{-\alpha(\lambda_1+\lambda_2)}
\sum_{m,n,j,k,p} C_{mnjkp} (\lambda^m_1 \lambda^n_2 \mu^j_1\mu^k_2 -
\lambda^n_1\lambda^m_2 \mu^k_1\mu^j_2)\rho^p,
\end{align}
Where $\rho=r_{12}$ is the distance between the two electrons of H$_2$, the
coefficients $C_{mnjkp}$ are computed from the minimization of the energy,
plus possible orthogonality conditions.

We examine the asymptotics of the trial solutions \eqref{eqA1} or
\eqref{eqA2} based on the discussions in Subsection \ref{sec2.2}. We see
that the exponent $e^{-\alpha\lambda_1}$ in $e^{-\alpha(\lambda_1+\lambda_2)}
$ would correspond to the factor $e^{-p\lambda}$ in \eqref{eqA.12} or
\eqref{eqA.26}. This is excellent as it reflects the exponential decay in
the radial variable of the (first) electron. However, the other non-constant
polynomial terms of the form
\begin{equation}
\label{eqA3}\lambda^{m_1}_1\lambda^{n_1}_2 \mu^j_1\mu^k_2,\qquad
\lambda^{n_2}_1\lambda^{m_2}_2 \mu^k_1\mu^j_2,\quad \text{with}\quad |m_\ell| +
|n_\ell| \ge 1,\qquad \ell=1,2,
\end{equation}
possess polynomial growth rates in either $\lambda_1$ or $\lambda_2$, which
are at odds with the asymptotics in \eqref{eqA.12a} since for large $%
\lambda$ (which may be either $\lambda_1$ or $\lambda_2$), there should not
be any polynomial growth $\lambda^{m_1}_1 \lambda^{n_1}_2$ or $%
\lambda^{n_2}_1 \lambda^{m_2}_2$ in \eqref{eqA3} besides the exponential
decay factor $e^{-p\lambda_1}\cdot e^{-p\lambda_2}$. (One might argue that
the polynomial growth $\lambda^{m_1}_1\lambda^{n_1}_2$ or $%
\lambda^{n_2}_1\lambda^{m_2}_2$ would be killed by the exponential decay $%
e^{-p\lambda_1}\cdot e^{-p\lambda_2}$. This can be true, however, only by
increasing $p$ and thus it causes the loss of accuracy.) One might still
argue that the typical term in the series \eqref{eqA.26}
for $m=0$ (ground state) satisfies
\begin{align}
(\lambda+1)^\sigma \left(\frac{\lambda-1}{\lambda+1}\right)^k
= \lambda^\sigma \left[1 + (\sigma-k) \left(\frac1\lambda\right) +
\frac{(\sigma-1)(\sigma-k-1)}{1\cdot 2} \left(\frac1\lambda\right)^2
+\cdots\right]\nonumber\\
&= \cl O(\lambda^\sigma), \quad \text{for}\quad \lambda \gg 1.\nonumber
\end{align}
This means that either $m_\ell$
or $n_\ell$ in \eqref{eqA3} {\em should never exceed\/}
$\sigma$. Can't we, at least, use terms $\lambda^{m_1}_1\lambda^{n_1}_2$ or $%
\lambda^{n_2}_1\lambda^{m_2}_2$ with some restrictions such as
$$
0\le m_\ell,
n_\ell\le\sigma;\quad \left(\sigma \equiv \frac{R_1}p-1, \text{ for }
m=0\right), \qquad \ell=1,2~?
$$

The most important behavior of $\Psi$ happens near $\lambda = 1$. For $%
\lambda \approx 1$, the typical term in \eqref{eqA.26}
satisfies
\begin{align}
&~~~~(\lambda^2-1)^{\frac{|m|}2} (\lambda+1)^\sigma
\left(\frac{\lambda-1}{\lambda+1}\right)^k\nonumber\\
&= 2^{\frac{|m|}2+ \sigma-k} (\lambda-1)^{\frac{|m|}2+k} \bigg[1 +
\left(\frac{|m|}2 +\sigma-k\right) \left(\frac{\lambda-1}2\right)\nonumber\\
&\qquad + \frac{\left(\frac{|m|}2 + \sigma-k\right) \left(\frac{|m|}2
+\sigma-k-1\right)}{1\cdot 2} \left(\frac{\lambda-1}2\right)^2 +
\cdots\bigg]\nonumber\\
\label{eqA5}
&= (\lambda-1)^{\frac{|m|}2} [2^{\frac{|m|}2 +\sigma-k} (\lambda-1)^k + \cl
O((\lambda-1)^{k+1})], \text{ for } \lambda\approx 1,  \lambda>1.
\end{align}
When $m=0$, the case of the ground state, the above expansion in terms of
powers of $\lambda$ is consistent with the terms involving powers of $%
\lambda_1$ or $\lambda_2$ in \eqref{eqA3} since those powers in \eqref{eqA3}
can always be re-expanded in terms of powers of $\lambda_1-1$ or $\lambda_2-1
$. However, when $m=1$ (or $m = \pm$ odd integer, for that matter) for the
excited state, the function in \eqref{eqA3} is no longer consistent with
those in \eqref{eqA5} as far as the $\lambda$-variable is concerned because
the factor $(\lambda-1)^{\frac{|m|}2}$ in \eqref{eqA5} is unaccountable for
those in \eqref{eqA3}. Indeed, we have indicated in \eqref{eqA.241} through
asymptotic analysis that {\em the factor $(\lambda-1)^{\frac{|m|}2}$ is
inherent in the solution and, thus, must be properly taken into account}.

The discussion in this section points out some weaknesses in the choice of
basis functions \eqref{eqA1} or \eqref{eqA2} based on the asymptotic
arguments for $\lambda\gg 1$ and $\lambda\approx 1$. Such weaknesses may
have contributed to the fact that {\em many\/} terms are required in
\eqref{eqA1} in order to calculate or to do the variational analysis of the
energy $E$ accurately for H$_2$ by James and Coolidge \cite{james}.

Our conclusion for this subsection is:\ because of the vastly different
asymptotic behaviors of $\Lambda(\lambda)$ for $\lambda\approx 1$ and $%
\lambda\gg 1$, the best strategy for numerical computation is to use {\em %
two different representations\/} for $\Lambda(\lambda)$, one for $%
\lambda\approx 1$ and another for $\lambda\gg 1$ and match them, say, at a
medium-size value such as $\lambda=5$ or 10. This, however, will invoke more
computational work and is beyond the interest of the authors for the time
being.

\section{The many-centered, one electron problem in momentum space}\label{many}

Here we derive the eigenvalue equation \eqref{ms1}.
We start from the model equation
\begin{equation}
\label{eq2.3.1a}\left(-\frac12 \nabla^2 - \sum^{N}_{j=1} \frac{Z_j}{|\pmb{r}-%
\pmb{R}_j|} \right)\psi(\pmb{r}) = E\psi(\pmb{r}),\qquad \pmb{r} =
(x,y,z)\in {\mathbb{R}}^3.
\end{equation}
Recall the Fourier transform
\begin{equation}
\label{eq2.3.2}\Psi(\pmb{p}) = \frac1{(2\pi)^{3/2}} \int\limits_{{\mathbb{R}}%
^3} e^{-i\pmb{r}\cdot \pmb{p}} \psi(\pmb{r})d\pmb{r}.
\end{equation}
Note that the Fourier transform of the potential term
\begin{equation}
\label{eq2.3.3}\frac1{(2\pi)^{3/2}} \int\limits_{{\mathbb{R}}^3} e^{-i\pmb{r}%
\cdot\pmb{p}} \frac1{|\pmb{r}-\pmb{R}_j|} d\pmb{r}
\end{equation}
is a {\em divergent integral\/} in the classical sense. However, the modern
mathematical theory of the ``regularization of divergent integrals''
\cite[Chap.~3]{chen2} makes \eqref{eq2.3.3} well defined:
\begin{align}
\eqref{eq2.3.3} &= (2\pi)^{-3/2} \int\limits_{{\bb R}^3} e^{-i(\pmb{r}' +
\pmb{R}_j)\cdot\pmb{p}} \frac1{|\pmb{r}'|} d\pmb{r}'\qquad (\pmb{r}' \equiv
\pmb{r} - \pmb{R}_j)\nonumber\\
&\equiv (2\pi)^{-3/2} e^{-i\pmb{R}_j\cdot\pmb{p}} \lim_{\alpha\rightarrow 0}
\int\limits_{{\bb R}^3} e^{-i\pmb{r}'\cdot\pmb{p}} \frac{e^{-\alpha r'}}{r'}
d\pmb{r}'\nonumber\\
&= (2\pi)^{-3/2} e^{-i\pmb{R}_j\cdot\pmb{p}} \lim_{\alpha\rightarrow 0}
\intl^\infty_0 \intl^\pi_0 \intl^{2\pi}_0 e^{-ir'p\cos \theta-\alpha r'} r'{}^2\
dr'\ \sin \theta\ d\theta d\phi\nonumber\\
\label{eq2.3.3a}
&= (2\pi)^{-3/2} e^{-i\pmb{R}_j\cdot\pmb{p}} \lim_{\alpha\rightarrow 0}
\frac{4\pi}{p^2+\alpha^2} = 2\cdot (2\pi)^{-1/2} e^{-i\pmb{R}_j\cdot
\pmb{p}}/p^2.
\end{align}
Applying the Fourier transform to \eqref{eq2.3.1a} by utilizing
\eqref{eq2.3.3a} and other well known properties such as convolution, we
obtain
\begin{equation}
\label{eq2.3.4}\left(\frac12 p^2-E\right) \Psi(p) - (2\pi)^{-3/2}\cdot
2\cdot (2\pi)^{-1/2} \sum^{N}_{j=1} Z_j \cdot \int\limits_{{\mathbb{R}}^3}
\frac{e^{-i\pmb{R}_j\cdot (\pmb{p}-\pmb{p}^{\prime})}}{|\pmb{p}-\pmb{p}%
^{\prime}|^2} \Psi(\pmb{p}^{\prime}) d\pmb{p}^{\prime}= 0.
\end{equation}
Define
\begin{equation}
\label{eq2.3.5}p^2_0 = -2E,
\end{equation}
then we obtain the integral equation
\begin{equation}
\label{eq2.3.6}(p^2+p^2_0) \Psi(p) = \sum^{N}_{j=1} \frac{Z_j}{\pi^2}
\int\limits_{{\mathbb{R}}^3} \frac{e^{-i\pmb{R}_j\cdot (\pmb{p}-\pmb{p}%
^{\prime})}}{|\pmb{p}-\pmb{p}^{\prime}|^2} \Psi(p^{\prime})dp^{\prime}.
\end{equation}
Now, we project the 3-dimensional momentum vector $\pmb{p}$ onto the surface
of the unit sphere, ${\mathcal{S}}_3$, of the 4-dimensional space, the 1-1
correspondence $\pmb{\xi} \leftrightarrow \pmb{p}$ through
\begin{equation}
\label{eq2.3.7}\left.
\begin{array}{l}
\xi_1 = 2p_0p_x(p^2+p^2_0)^{-1} = \sin \chi \sin \theta \cos \phi, \\
\xi_2 = 2p_0p_y(p^2+p^2_0)^{-1} = \sin \chi \sin\theta \sin\phi, \\
\xi_3 = 2p_0p_z(p^2+p^2_0)^{-1} = \sin \chi \cos \theta, \\
\xi_4 = (p^2_0-p^2) (p^2+p^2_0)^{-1} = \cos\chi \\
0\le \chi \le \pi, 0\le \theta\le 2\pi, 0\le \phi\le\pi,
\end{array}
\right\}\qquad \pmb{\xi} \in {\mathbb{R}}^4, \xi = |\pmb{\xi}|=1,
\end{equation}
while keeping in mind that
\begin{equation}
\label{eq2.3.8}p_x=p\sin\theta \cos\phi,\quad p_y = p\sin
\theta\sin\phi,\quad p_z = p\cos\theta.
\end{equation}
Then for $\pmb{\xi} \leftrightarrow \pmb{p}, \, \pmb{\xi}^{\prime}%
\leftrightarrow \pmb{p}^{\prime}$, and $\kappa$ be the angle between $%
\pmb{\xi}$ and $\pmb{\xi}^{\prime}$, we have
\begin{align}
&\cos \kappa = \pmb{\xi}\cdot \pmb{\xi}',\nonumber\\
\label{eq2.3.8a}
&|\pmb{\xi}-\pmb{\xi}'|^2 = \xi^2+\xi'{}^2 - 2\pmb{\xi}\cdot \pmb{\xi}' = 2-2
\cos \kappa = 4\sin^2(\kappa/2).
\end{align}
Also, it is straightforward to verify that
$$
|\pmb{\xi}-
\pmb{\xi}^{\prime}|^2 = \frac{4p^2_0|\pmb{p}-\pmb{p}^{\prime}|^2}{%
(p^2+p^2_0)(p^{\prime}{}^2 +p^2_0)}.
$$
Hence
\begin{equation}
\label{eq2.3.9}\frac1{|\pmb{p}-\pmb{p}^{\prime}|^2} = \frac{p^2_0}{%
(p^2+p^2_0) (p^{\prime}{}^2 + p^2_0) \sin^2(\kappa/2)}.
\end{equation}
Let $d\Omega$ be the infinitesimal surface area of the 4-dimensional
hypersphere ${\cl S}_3$. Then from \eqref{eq2.3.7},
\begin{equation}
\label{eq2.3.10}d\Omega = \sin^2\chi \sin\theta \ d\chi d\theta d\phi =
-\sin^2 \chi \ d\chi d(\cos\theta)d\phi
\end{equation}
while \eqref{eq2.3.8} gives the standard
\begin{equation}
\label{eq2.3.11}d\pmb{p} = p^2\sin\theta \ dp d\theta d\phi = -p^2\
dpd(\cos\theta)d\phi.
\end{equation}
>From \eqref{eq2.3.7}$_4$, we further obtain
\begin{equation}
\label{eq2.3.12}\frac{dp}{d\chi} = \frac{p^2+p^2_0}{2p_0}.
\end{equation}
The equations \eqref{eq2.3.7} and \eqref{eq2.3.10}--\eqref{eq2.3.12} now
give
\begin{equation}
\label{eq2.3.13}d\pmb{p} = \frac{p^2}{\sin^2\chi} \frac{dp}{d\chi} d\Omega =
\left(\frac{p^2+p^2_0}{2p_0}\right)^3 d\Omega.
\end{equation}
We can now use \eqref{eq2.3.9} to eliminate the denominator inside the
Fourier integral of \eqref{eq2.3.6}. Moreover, define
\begin{equation}
\label{eq2.3.14}\varphi(\Omega) = \frac{(p^2+p^2_0)^2}{4p^{5/2}_0} \Psi(%
\pmb{p}),
\end{equation}
where $\Omega\in {\cl S}_3$ is the point $\pmb{\xi}\leftrightarrow
\pmb{p}$. Then the integral equation \eqref{eq2.3.6} simplifies to
\begin{equation}
\label{eq2.3.15}p(\Omega) = \frac1{p_0} \sum^n_{j=1} \int\limits_{
{\cl S}_3}
\frac{Z_j}{8\pi^2} \frac{e^{-i\pmb{R}_j(\pmb{p}-\pmb{p}^{\prime})}}{%
\sin^2(\kappa/2)} \varphi(\Omega^{\prime}) d\Omega^{\prime}.
\end{equation}
At this point, we need to introduce the hyperspherical harmonics on ${%
\mathcal{O}}_3$, which constitute an orthonormal basis for square summable
functions on ${\cl S}_3$ and are given by
\begin{equation}
\label{eq2.3.16}\left.
\begin{array}{l}
Y_{n\ell m}(\Omega) = (-i)^\ell C^\ell_n(\chi) Y_{\ell m}(\theta,\phi), \\
n = 1,2,\ldots, \ell=0,1,2,\ldots, m=-\ell,-\ell+1, \ldots, 0,1,\ldots,
\ell, \\
\text{cf.\ \eqref{eq2.6}--\eqref{eq2.9} for } Y_{\ell m}(\theta,\phi),
\end{array}
\right\}
\end{equation}
where
\begin{equation}
\label{eq2.3.17}C^\ell_n(\chi) \equiv \left[\frac{2n(n-\ell-1)!}{\pi(n+\ell)!%
}\right]^{1/2} (\sin^\ell \chi) \left\{\left[\frac{d}{d(\cos\chi)}%
\right]^\ell C_{n-1}(\cos \chi)\right\},
\end{equation}
are the associated Gegenbauer functions, and the $C_k$'s are the Gegenbauer
functions, with the generating function
\begin{equation}
\label{eq2.3.18}\frac1{1-2\mu h+h^2} \equiv \sum^\infty_{j=0}
h^jC_j(\mu),\qquad |h|<1.
\end{equation}
For $\pmb{\xi}, \pmb{\xi}^{\prime}\in {\mathbb{R}}^4$,
\begin{equation}
\label{eq2.3.19}\frac1{|\pmb{\xi}-\pmb{\xi}^{\prime}|^2} =
\begin{cases}
\dfrac1{\xi^2} \dfrac1{1-2\cos \kappa(\xi'/\xi) + (\xi'/\xi)^2}&\text{if
$\xi'/\xi<1$,}\\
\noalign{\medskip}
\dfrac1{\xi'{}^2}\dfrac1{1-2\cos \kappa(\xi/\xi') + (\xi/\xi')^2}&\text{if
$\xi/\xi' < 1$.}
\end{cases}
\end{equation}
Denote $\xi_> = \max(\xi,\xi^{\prime})$, $\xi_< = \min(\xi,\xi^{\prime})$.
Then \eqref{eq2.3.18} and \eqref{eq2.3.19} give the Neumann expansion
\begin{align*}
\frac1{|\pmb{\xi}-\pmb{\xi}'|^2} &= \frac1{\xi^2_>} \sum^\infty_{n=0}
\left(\frac{\xi_<}{\xi_>}\right)^n C_n(\cos \kappa)\\
&= \sum^\infty_{n=1} \frac{\xi^{n-1}_<}{\xi^{n+1}_>} C_{n-1}(\cos\kappa),
\end{align*}
which, in the limit as $\xi,\xi^{\prime}\to 1$, yields
\begin{align}
\frac1{|\pmb{\xi} - \pmb{\xi}'|^2} &= \frac1{4\sin^2(\kappa/2)}\qquad \text{(by
\eqref{eq2.3.8a}))}\nonumber\\
\label{eq2.3.20}
&= \sum^\infty_{n=1} C_{n-1} (\cos\kappa).
\end{align}
But by the addition theorem of angles, we have
\begin{align}
\label{eq2.3.21}
&C_{n-1}(\cos\kappa) = \frac{2\pi^2}n \sum_{\ell m} Y^*_{n\ell m}(\Omega')
Y_{n\ell m}(\Omega),\\
\label{eq2.3.22}
&n\ge 1, \ell=0,1,2,\ldots, m=-\ell, -\ell+1,\ldots, 0,1,\ldots,\ell,
\end{align}
so from \eqref{eq2.3.20} we obtain
$$
\frac1{4\sin^2(\kappa/2)} = 2\pi^2 \sum^\infty_{n=1} \sum_{\ell,m} Y_{n\ell
m}(\Omega) \frac1n Y^*_{n\ell m}(\Omega^{\prime}),
$$
and, thus, the kernel of the integral equation \eqref{eq2.3.15} can be
written as
\begin{equation}
\label{eq2.3.19a}Z_j e^{-i\pmb{R}_j\cdot(\pmb{p}-\pmb{p}^{\prime})}/[8\pi^2
\sin^2(\kappa/2)] = Z_j \sum_t [e^{-i\pmb{R}_j\cdot\pmb{p}}
Y_t(\Omega)] \frac1n
[e^{-i\pmb{R}_j\cdot \pmb{p}^{\prime}} Y_t(\Omega^{\prime})]^*,
\end{equation}
where $t=(n\ell m)$ runs triple summation indices according to
\eqref{eq2.3.22}. We now use the orthonormal basis functions \eqref{eq2.3.16}
to make a re-expansion
$$
e^{i\pmb{R}_j\cdot\pmb{p}} Y_t(\Omega) = \sum_\tau S^+_\tau(\pmb{R}_j)
Y_\tau(\Omega),\qquad j=1,2,\ldots, N,
$$
where $\tau = (n^{\prime}\ell^{\prime}m^{\prime})$ runs triple summation
indices similarly to $t$. Then
\begin{equation}
\label{eq2.3.20a}S^t_\tau (\pmb{R}_j) = \int\limits_{{\cl S}_3} e^{i%
\pmb{R}_j\cdot \pmb{p}} Y_t(\Omega) Y_\tau(\Omega)d\Omega
\end{equation}
and \eqref{eq2.3.19a} becomes
\begin{equation}
\label{eq2.3.21a}Z_j e^{-i\pmb{R}_j\cdot(\pmb{p}-\pmb{p}^{\prime})}/[8\pi^2
\sin^2(\kappa
/2)] = Z_j \sum_{t\tau\tau^{\prime}} [S^\tau_t(\pmb{R}_j)]^* \frac1n
S^{\tau^{\prime}}_t (\pmb{R}_j) Y_\tau(\Omega)
Y^*_{\tau^{\prime}}(\Omega^{\prime}).
\end{equation}
The orthonormal expansion of $\varphi(\Omega)$ in \eqref{eq2.3.15} is
denoted as
\begin{equation}
\label{eq2.3.22a}\varphi(\Omega) = \sum_{t^{\prime}}
c_{t^{\prime}}Y_{t^{\prime}}(\Omega).
\end{equation}
Substituting \eqref{eq2.3.21a} and \eqref{eq2.3.22a} into \eqref{eq2.3.15}
and equating coefficients, we obtain
$$
c_{t^{\prime}} = \frac1{p_0} \sum_j Z_j \sum_{t\tau} [S^{t^{\prime}}_t(%
\pmb{R}_j)]^* \frac1n S^\tau_t(\pmb{R}_j)c_\tau.
$$
This is an eigenvalue problem
\begin{equation}
\label{eq2.3.23}P\pmb{c} = p_0\pmb{c}
\end{equation}
where $\pmb{P}$ is an infinite matrix with entries
$$
P^t_{t^{\prime}} = \sum_j Z_j \sum_\tau [S^{t^{\prime}}_\tau(\pmb{R}_j)]^*
\frac1n S^t_\tau(\pmb{R}_j)
$$
where $\pmb{c}$ is an infinite-dimensional vector with entries $%
c_{t^{\prime}}$.

The value of $p_0$ will yield the energy $E$ from \eqref{eq2.3.5}. One can
obtain the wave function $\psi(\pmb{r})$ from applying the inverse Fourier
transform to $\Psi(\pmb{p})$ through \eqref{eq2.3.14} and \eqref{eq2.3.7}.

\section{Derivation of the cusp conditions}\label{cuspd}

Here we derive the cusp conditions at the singularities of Eq. \eqref{C2}.
Since the idea is the same for each of the five sets of singularities, we will
only treat the most complicated case, $r_{12}=0$.

Use the center-of-mass coordinate system, see Appendix \ref{CMC}, we can
transform \eqref{C1} into the form
\begin{align}
\hat H &= -\frac1{2M} \nabla^2_S - \frac1{2\mu} \nabla^2_{r_{12}} -
\frac{2Z_a}{|2\pmb{R}_a+\pmb{r}_{12}|} - \frac{2Z_b}{|2\pmb{R}_b +
\pmb{r}_{12}|}\nonumber\\
\label{C3}
&\quad - \frac{2Z_a}{|2\pmb{R}_a-\pmb{r}_{12}|} - \frac{2Z_b}{|2\pmb{R}_b-
\pmb{r}_{12}|} + \frac{q_1q_2}{r_{12}} + \frac{Z_aZ_b}{R},
\end{align}

where {\bf S} is the center of mass coordinate of two electrons,
see Appendix \ref{CMC} for the details of the notation.

We now define the {\em spherical means\/} of a function. Given a point
$\pmb{r}_0\in {\bb R}^3$, the spherical means of a function $u(\pmb{r})$ at
$\pmb{r}_0$ on the sphere with radius $\rho$ is defined to be
\begin{equation}\label{C4}
u_{av,\rho}(\pmb{r}_0) \equiv \frac1{4\pi} \iint\limits_{{\cl S}_1} u(\pmb{r}_0
+\rho\pmb{\nu})d\omega ,
\end{equation}
where
${\cl S}_\rho $ is the sphere with radius $\rho$ centered at $\pmb{r}_0$,
here $\rho=1$;
$d\omega = \sin\theta \ d\theta d\phi $, where $ \omega $
represents all the angular variables;
$ \pmb{\nu} = (\nu_1,\nu_2,\nu_3)$, $\nu^2_1+\nu^2_2 + \nu^2_3 = 1$;
$\pmb{\nu}$ is the unit outward pointing normal vector on ${\cl S}_1$.
It is easy to see that if $u(\pmb{r})$ is continuous in a neighborhood of
$\pmb{r}_0$, then the spherical means just converge to the pointwise value:
\[
\lim_{\rho\to 0} u_{av,\rho}(\pmb{r}_0) = u(\pmb{r}_0).
\]

We now integrate $\hat H\Psi $
over a small 3-dimensional ball $B_{\rho_0}$ with
radius $\rho_0$ centered at $r_{12}=0$, for any $\pmb{S}\in {\bb R}^3,
\pmb{S}\ne \pmb{0}$:
\begin{align*}
&\iiint\limits_{B_{\rho_0}} \left\{-\frac1{2M} \nabla^2_S\psi (\pmb{r}_{12},
\pmb{S}) - \frac1{2\mu} \nabla^2_{12} \psi(\pmb{r}_{12}, \pmb{S}) -
\frac{2Z_a}{|2\pmb{R}_a+\pmb{r}_{12}|} \psi(\pmb{r}_{12},\pmb{S})\right.\\
&\quad - \frac{2Z_b}{|2\pmb{R}_b +\pmb{r}_{12}|} \psi(\pmb{r}_{12}, \pmb{S}) -
\frac{2Z_a}{|2\pmb{R}_a - \pmb{r}_{12}|} \psi(\pmb{r}_{12},\pmb{S}) -
\frac{2Z_b}{|2\pmb{R}_b-\pmb{r}_{12}|} \psi(\pmb{r}_{12}, \pmb{S})\\
&\quad \left.+ \frac{q_1q_2}{r_{12}} \psi(\pmb{r}_{12},\pmb{S}) +
\frac{Z_aZ_b}{R} \psi(\pmb{r}_{12},\pmb{S})\right\} d\pmb{r}_{12} = 0.
\end{align*}
Note that as $\rho_0\to 0$, the integrals of all the terms above vanish (by
their continuity at $r_{12}=0$), except possibly those of
\[
-\frac1{2\mu} \nabla^2_{12}\psi (\pmb{r}_{12},\pmb{S}),\quad
\frac{q_1q_2}{r_{12}} \psi(\pmb{r}_{12},\pmb{S}).
\]
Thus, we need only consider
\begin{equation}\label{C5}
\lim_{\rho_0\to 0} \iiint\limits_{B_{\rho_0}} \left[-\frac1{2\mu} \nabla^2_{12}
\psi(\pmb{r}_{12},\pmb{S}) + \frac{q_1q_2}{r_{12}}
\psi(\pmb{r}_{12},\pmb{S})\right] d\pmb{r}_{12} = 0.
\end{equation}
Apply the Gauss Divergence Theorem to the first term of the integral to get
\begin{align}
&\iiint\limits_{B_{\rho_0}} \left(-\frac1{2\mu}\right) \nabla^2_{12}
\psi(\pmb{r}_{12},\pmb{S}) d\pmb{r}_{12} = -\frac1{2\mu} \iint\limits_{{\cl
S}_{\rho_0}} \frac{\partial\psi}{\partial r_{12}} (\pmb{r}_{12},\pmb{S}) d{\cl
S}_{\rho_0}\nonumber\\
\label{C6}
&\quad = -\iiint\limits_{B_{\rho_0}} \frac{q_1q_2}{r_{12}}
\psi(\pmb{r}_{12},\pmb{S}) d\pmb{r}_{12}
\end{align}
But $d{\cl S}_{\rho_0} = \rho^2_0\ d\omega$; thus
\begin{equation}\label{C7}
\iint\limits_{{\cl S}_{\rho_0}} \frac{\partial\psi}{\partial r_{12}} d{\cl
S}_{\rho_0} = \iint\limits_{{\cl S}_1} \frac{\partial\psi}{\partial r_{12}}
d\omega_1\cdot \rho^2_0 = \rho^2_0 \iint\limits_{{\cl S}_1}
\frac{\partial\psi(\pmb{r}_{12},\pmb{S})}{\partial r_{12}}d\omega
= 4\pi\rho^2_0 \left[\frac{\partial\psi(\pmb{0},\pmb{S})}{\partial
r_{12}}\right]_{av,\rho_0}
\end{equation}
where ${\cl S}_1 = {\cl S}_{\rho_0}|_{\rho_0=1}$. By using spherical
coordinates, $d\pmb{r}_{12} =r^2_{12}\ d\omega
dr_{12}$,
we have
\begin{align}
&\iiint\limits_{B_{\rho_0}} \frac{q_1q_2}{r_{12}} \psi(\pmb{r}_{12}, \pmb{S})
d\pmb{r}_{12} = \int^{\rho_0}_0 \left[\iint\limits_{{\cl S}_\rho}
\frac{q_1q_2}{r_{12}} \psi(\pmb{r}_{12},\pmb{S})d\omega\right] r^2_{12} dr_{12}
\nonumber\\
&\quad = q_1q_2\int^{\rho_0}_0 \left\{\iint\limits_{{\cl S}_\rho}
[\psi(\pmb{0},\pmb{S}) + \vp(\pmb{r}_{12},\pmb{S})]d\omega\right\}
r_{12}dr_{12},\nonumber\\
\intertext{where $\vp(\pmb{r}_{12},\pmb{S}) \to 0$ as $\pmb{r}_{12}\to \pmb{0}$.
This follows from the fact that $\psi(\pmb{r}_{12},\pmb{S})$ is continuous
at $\pmb{r}_{12}=\pmb{0}$
for any $\pmb{S}$.}
\text{(Continuing from the above)} &\quad \equiv q_1q_2
\left[\psi(\pmb{0},\pmb{S}) \int^{\rho_0}_0 r_{12} dr_{12}
+
\tilde\vp(\rho_0,\pmb{S}) 4\pi \frac{\rho^2_0}2\right]\nonumber\\
\label{C8}
&\quad = q_1q_2 \psi(\pmb{0},\pmb{S})4\pi\cdot \frac{\rho^2_0}2 +
2\pi\rho^2_0\tilde\vp(\rho_0,\pmb{S}).
\end{align}
>From \eqref{C5}, \eqref{C6}, \eqref{C7} and \eqref{C8}, we now get
\[
-\frac1{2\mu} \rho^2_0 \left[\frac{\partial\psi(\pmb{0}, \pmb{S})}{\partial
r_{12}}\right]_{av,\rho_0} = -\frac12 \rho^2_0 q_1q_2 \psi(\pmb{0},\pmb{S}) +
\frac{\rho^2_0}2 \tilde\vp(\rho_0,\pmb{S}).
\]
Dividing all the terms above by $\rho^2_0$ and let $\rho_0\to 0$, we have
$\tilde\vp(\rho_0,\pmb{S})\to 0$ and, therefore,
\begin{equation}\label{C9}
\lim_{\rho_0\to 0} \left[\frac{\partial\psi(\pmb{0},\pmb{S})}{\partial
r_{12}}\right]_{av,\rho_0} = \mu q_1q_2\psi(\pmb{0},\pmb{S}), \text{ for all }
\pmb{S}\in {\bb R}^3.
\end{equation}
(This is equation (2.11) in Patil, Tang and Toennies \cite{PTT}.)
Similarly, one can derive the electron-nucleus cusp condition at, e.g.,
$r_{1a}=0$, for \eqref{C0} to be
\begin{equation}\label{C9a}
\lim_{\rho_0\to 0} \left[\frac{\partial\psi(\pmb{r}_1,\pmb{r}_2)}{\partial
r_{1a}}\right]_{av,\rho_0} = -m_1 Z_a\psi(\pmb{r}_1,\pmb{r}_2)|_{r_{1a}=0},
\text{ for  all } \pmb{r}_2.
\end{equation}

\begin{thm}\label{thmC1}
Assume that $m_1=m_2$ and $q_1=q_2=-1$ in \eqref{C0}. Let
$(\pmb{S},\pmb{r}_{12})$
denote the $CM$ and relative coordinates
and $\omega_{12}$ denote the angular variables of
the vector $\pmb{r}_{12}$. Let $\psi$ be a nontrivial solution of \eqref{C0}
such that its local Taylor expansion near $r_{12}=0$ is of the form
\begin{equation}\label{C10}
\psi(\pmb{r}_1,\pmb{r}_2) = C_0(\pmb{S}) + C_1(\pmb{S})r_{12} + {\cl
O}(r^2_{12}), \quad \text{for $r_{12}$ small,}
\end{equation}
where $C_0(\pmb{S})$ and $C_1(\pmb{S})$ are independent of $\omega_{12}$ while
the remainder satisfies ${\cl O}(r^2_{12}) = C_3(\pmb{r}_{12},\pmb{S})r^2_{12}$
for some bounded function $C_3$ which depends on $\pmb{r}_{12}$ and $\pmb{S}$.
Then $C_1(\pmb{S}) = \frac12 C_0(\pmb{S})$ for all $\pmb{S}$ and, consequently,
\begin{equation}\label{C11}
\psi(\pmb{r}_1,\pmb{r}_2) = C_0(\pmb{S}) \left(1 + \frac12 r_{12}\right) + {\cl
O}(r^2_{12}).
\end{equation}
\end{thm}

\begin{proof}
We substitute the RHS of \eqref{C10} into \eqref{C9}. There is no need to take
the spherical mean on the left of \eqref{C9} anymore as the dominant term on the
RHS of \eqref{C11}
does not depend on the angular variables of $r_{12}$. We therefore obtain
\[
C_1(\pmb{S}) = \mu q_1q_2C_0(\pmb{S}) = \frac12 C_0(\pmb{S}), \text{ as } \mu =
\frac{m_1m_2}{m_1+m_2} = \frac12, \quad q_1=q_2 = -1.
\]
Hence
\begin{align*}
\psi(\pmb{r}_1,\pmb{r}_2) &= C_0(\pmb{S}) + \frac12C_0(\pmb{S}) r_{12} + {\cl
O}(r^2_{12})\\
&= C_0(\pmb{S}) \left[1 + \frac12 r_{12}\right] + {\cl O}(r^2_{12}),
\end{align*}
which is \eqref{C10}.
\end{proof}

It looks as though the condition \eqref{C10} is somewhat contrived. However,
useful application can be seen shortly, in Theorem \ref{thmC3}.

Similarly, we can obtain the cusp conditions at $r_{1a}=0$, $r_{1b}=0$,
$r_{2a}=0$ and $r_{2b}=0$, as given in the following.

\begin{thm}\label{thmC2}
Let $\psi$ be either a nontrivial solution of \eqref{C0} or a trial wave function
for \eqref{C0}. Let $r_{1a}$ and $\omega_{1a}$ denote, respectively,
the radial and angular variables of the vector $\pmb{r}_{1a}$. Assume that for
$r_{1a}$ sufficiently small, $\psi$ satisfies the Taylor expansion
\begin{equation}\label{C12}
\psi(\pmb{r}_1,\pmb{r}_2) = C_0(\pmb{r}_2) + C_1(\pmb{r}_2)r_{1a} + {\cl
O}(r^2_{1a}), \quad r_{1a} \text{ small,}
\end{equation}
for some functions $C_0$ and $C_1$ depending on $\pmb{r}_2$ only, where  the
remainder satisfies ${\cl O}(r^2_{1a}) = C_3(\pmb{r}_1,\pmb{r}_2) r^2_{1a}$ for
some bounded function $C_3$ depending on $\pmb{r}_1$ and $\pmb{r}_2$. Then
\begin{equation}\label{C13}
\psi(\pmb{r}_1,\pmb{r}_2) = C_0(\pmb{r}_2) (1-m_1Z_ar_{1a}) + {\cl O}(r^2_{1a}),
\quad r_{1a} \text{ small,}
\end{equation}
where ${\cl O}(r^2_{1a}) = C_3(\pmb{r}_1,\pmb{r}_2) r^2_{1a}$.
\end{thm}

\begin{figure}[htpb]
\bigskip
\centerline{\epsfxsize=0.4\textwidth\epsfysize=0.25\textwidth
\epsfbox{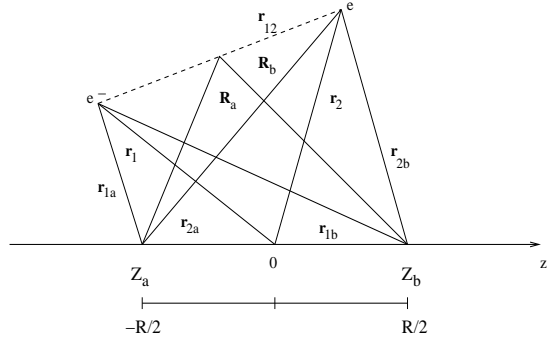}}

\caption{Various vectors are defined in this diagram.}
\label{vectors}
\end{figure}

\begin{proof}
The kinetic energy term $-\frac1{2m_1}\nabla^2_1$, after the translation
\[
z_1 \to z_1 + \frac{R}2,\quad \text{cf.\ \eqref{eqC.1} and Fig. \ref{vectors}
for notation,}
\]
becomes $-\frac1{2m_1} \nabla^2_{\pmb{r}_{1a}}$ which is centered at
$(x_1,y_1,z_1) = \left(0,0-\frac{R}2\right)$. Now apply the Hamiltonian
\begin{equation}\label{C14}
\hat H=-\frac1{2m_1} \nabla^2_{\pmb{r}_{1a}} - \frac{Z_a}{r_{1a}} +
\left[-\frac1{2m_2} \nabla^2_2 - \frac{Z_b}{r_{1b}} - \frac{Z_a}{r_{2a}} -
\frac{Z_b}{r_{2b}} +
\frac{q_1q_2}{r_{12}} + \frac{Z_aZ_b}{R}\right]
\end{equation}
to \eqref{C12}. We need only focus our attention locally near $r_{1a}=0$ and
ignore the terms in the bracket of \eqref{C14} as they have no effect on the
singularity at $r_{1a}=0$. We obtain
\begin{align*}
(\hat H-E)\psi &= - \frac1{2m_1} \frac1{r^2_{1a}} \frac\partial{\partial r_{1a}}
[r^2_{1a}\cdot C_1(\pmb{r}_2)] -
\frac{Z_a}{r_{1a}} [C_0(\pmb{r}_2) + C_1(\pmb{r}_2)r_{1a}]\\
&\quad + \text{higher order terms in  } r_{1a}\\
&= -\frac1{2m_1} \frac{2C_1(\pmb{r}_2)}{r_{1a}} -
\frac{Z_a\cdot C_0(\pmb{r}_2)}{r_{1a}} + \text{higher order terms in } r_{1a}.
\end{align*}
To eliminate the singularity at $r_{1a}=0$ above, it is necessary that
\[
\frac{C_1(\pmb{r})}{m_1} + Z_aC_0(\pmb{r}_2) = 0.
\]
The above gives \eqref{C13}, as desired.
\end{proof}

\begin{exm}\label{exmC2}
The trial wave function $\psi(\pmb{r}_1,\pmb{r}_2) = \phi(\pmb{r}_{1a})
\phi(\pmb{r}_{2b})$, where $\phi(\pmb{r}) = e^{-\alpha r}$ is a one-centered
orbital, satisfies condition \eqref{C12} of Theorem \ref{thmC2}.$\hfill\square$
\end{exm}

\n One can easily apply Theorem \ref{thmC2} to other singularities at $r_{1b} =
0$, $r_{2a}=0$ and $r_{2b} = 0$.

\begin{thm}\label{thmC3}
Let $\phi(\pmb{r})$ be a sufficiently smooth function. Let the Hamiltonian
represents a homonuclear case:
\[
\hat H = -\frac12 \nabla^2_1 - \frac12 \nabla^2_2 - \frac{Z}{r_{1a}} -
\frac{Z}{r_{1b}} - \frac{Z}{r_{2a}} - \frac{Z}{r_{2b}} + \frac1{r_{12}} +
\frac{Z^2}{R}.
\]
Then the product function
\[
\psi(\pmb{r}_1,\pmb{r}_2)  = \phi(\pmb{r}_1) \phi(\pmb{r}_2) \left(1 + \frac12
r_{12}\right)
\]
satisfies the interelectronic cusp condition as given in \eqref{C10}.
\end{thm}

\begin{proof}
We represent the variables $\pmb{r}_1$ and $\pmb{r}_2$ in terms of the CM
coordinates
\begin{align*}
\pmb{r}_1 &= \pmb{S} + \frac12(\pmb{r}_1-\pmb{r}_2) = \pmb{S} + \frac12
\pmb{r}_{12},\\
\pmb{r}_2 &= \pmb{S} - \frac12(\pmb{r}_1-\pmb{r}_2) = \pmb{S} - \frac12
\pmb{r}_{12},
\end{align*}
where $\pmb{S} = \frac12(\pmb{r}_1+\pmb{r}_2)$.

Then for $r_{12}$ small, by Taylor's expansion,
\begin{align*}
\phi(\pmb{r}_1) &= \phi\left(\pmb{S} + \frac12 \pmb{r}_{12}\right) =
\phi(\pmb{S}) + \frac12 \nabla\phi(\pmb{S})\cdot \pmb{r}_{12}
 + \frac1{2!} \frac14 \pmb{r}^T_{12} \cdot D^2\phi(\pmb{S})\cdot
\pmb{r}_{12} +\cdots,\\
\phi(\pmb{r}_2) &= \phi\left(\pmb{S}- \frac12 \pmb{r}_{12}\right) =
\phi(\pmb{S}) - \frac12 \nabla\phi(\pmb{S}) \cdot \pmb{r}_{12}
 + \frac1{2!} \frac14 \pmb{r}^T_{12}\cdot D^2\phi(\pmb{S})\cdot
\pmb{r}_{12}-\cdots~.
\end{align*}
Therefore
\begin{align*}
\psi &= \phi(\pmb{r}_1) \phi(\pmb{r}_2) \left(1+\frac12 r_{12}\right)\\
&= \left[\phi(\pmb{S}) + \frac12 \nabla\phi(\pmb{S})\cdot \pmb{r}_{12} +
\frac1{2!} \frac14 \pmb{r}^T_{12} \cdot D^2\phi(\pmb{S})\cdot \pmb{r}_{12}
+\cdots\right]\cdot\\
&\phantom{=} \left[\phi(\pmb{S}) - \frac12 \nabla\phi(\pmb{S}) \cdot
\pmb{r}_{12} + \frac1{2!} \frac14 \pmb{r}^T_{12}\cdot D^2\phi(\pmb{S}) \cdot
\pmb{r}_{12} \pm \cdots\right]\cdot \left(1+\frac12 r_{12}\right)\\
&= \phi^2(\pmb{S}) \left(1+\frac12 r_{12}\right) + \text{quadratic or higher
order terms involving } r_{12}.
\end{align*}
Hence condition \eqref{C10} is satisfied.
\end{proof}

\section{Center of mass coordinates for the kinetic energy
$\pmb{-\frac1{2m_1} \nabla^2_1-\frac1{2m_2}\nabla^2_2}$}\label{CMC}

The coordinates for electron 1 and 2 are, respectively,
\begin{equation}\label{eqC.1}
\pmb{r}_1 = (x_1,y_1,z_1), \quad \pmb{r}_2 = (x_2,y_2,z_2).
\end{equation}
The kinetic energy operator is
\[
\widetilde H = -\frac1{2m_1} \nabla^2_1 - \frac1{2m_2} \nabla^2_2,
\]
where
\[
\nabla^2_j = \frac{\partial^2}{\partial x^2_j} + \frac{\partial^2}{\partial
y^2_j} + \frac{\partial^2}{\partial z^2_j},\qquad j=1,2.
\]
Define the CM (center-of-mass) coordinate $\pmb{S}$:
\[
\pmb{S} = \frac{m_1\pmb{r}_1 + m_2\pmb{r}_2}{m_1+m_2}
\]
and (relative coordinate)
\[
\pmb{r}_{12} = \pmb{r}_1-\pmb{r}_2.
\]
Here we derive the kinetic energy term in coordinates ${\bf S}$, $r_{12}$.
For any differentiable scalar function $f(\pmb{r}_1,\pmb{r}_2)$, we have
\[
df  = \nabla_1f\cdot d\pmb{r}_1 + \nabla_2f \cdot d\pmb{r}_2 =
\nabla_{\pmb{S}}f\cdot d\pmb{S} + \nabla_{12}f \cdot d\pmb{r}_{12},
\]
where $\nabla_{\pmb{S}}$ and $\nabla_{12}$ are the gradient operators with
respect to the variables of $\pmb{S}$ and $\pmb{r}_{12}$. Then
\begin{equation}\label{eqC.2}
[\nabla_1f,\nabla_2f] \left[\begin{matrix} d\pmb{r}_1\\
d\pmb{r}_2\end{matrix}\right] = [\nabla_Sf, \nabla_{12}f] \left[\begin{matrix}
d\pmb{S}\\ d\pmb{r}_{12}\end{matrix}\right].
\end{equation}
But
\begin{equation}\label{eqC.3}
\left[\begin{matrix} d\pmb{S}\\ d\pmb{r}_{12}\end{matrix}\right] = \left[
\begin{matrix} \frac{m_1}{m_1+m_2}&\frac{m_2}{m_1+m_2}\\ 1&-1\end{matrix}\right]
\left[\begin{matrix} d\pmb{r}_1\\ d\pmb{r}_2\end{matrix}\right].
\end{equation}
Hence from \eqref{eqC.2} and \eqref{eqC.3},
\[
[\nabla_1f,~~\nabla_2f] = [\nabla_Sf,~~\nabla_{12}f] \left[\begin{matrix}
\frac{m_1}{m_1+m_2}&\frac{m_2}{m_1+m_2}\\ 1&-1\end{matrix}\right];
\]
\begin{align*}
\widetilde H &= -\frac1{2m_1} \nabla^2_1 - \frac1{2m_2} \nabla^2_2 =
[\nabla_1,~~\nabla_2] \left[\begin{matrix} -\frac1{2m_1}&0\\ 0&-\frac1{2m_2}
\end{matrix}\right] \left[\begin{matrix} \nabla_1\\
\nabla_2\end{matrix}\right]\\
&= [\nabla_S,~~\nabla_{12}] \left[\begin{matrix}
\frac{m_1}{m_1+m_2}&\frac{m_2}{m_1+m_2}\\ 1&-1\end{matrix}\right]
\left[\begin{matrix}
-\frac1{2m_1}&0\\ 0&-\frac1{2m_2}\end{matrix}\right]
\left[\begin{matrix}
\frac{m_1}{m_1+m_2}&1\\ \frac{m_2}{m_1+m_2}&-1\end{matrix}\right]
\left[\begin{matrix} \nabla_S\\ \nabla_{12}\end{matrix}\right]\\
&= [\nabla_S,~~\nabla_{12}] \left[\begin{matrix}
-\frac1{2(m_1+m_2)}&0\\ 0&-\frac12\left(\frac1{m_1}+\frac1{m_2}\right)
\end{matrix}\right] \left[\begin{matrix} \nabla_S\\
\nabla_{12}\end{matrix}\right];
\end{align*}
or
\begin{align*}
\widetilde H &= -\frac1{2(m_1+m_2)} \nabla^2_S - \frac12 \left(\frac1{m_1} +
\frac1{m_2}\right)\nabla^2_{12}\\
&= -\frac1{2M} \nabla^2_S - \frac1{2\mu} \nabla^2_{12},
\end{align*}
where $M \equiv m_1+m_2$ is the total mass of electrons and $\mu \equiv
\frac{m_1m_2}{m_1+m_2}$ is the reduced mass.

\section{Verifications of the cusp conditions for two-centered orbitals
in prolate spheroidal coordinates}
\label{cuspprol}

\begin{figure}[htpb]
\bigskip
\centerline{\epsfxsize=0.4\textwidth\epsfysize=0.20\textwidth
\epsfbox{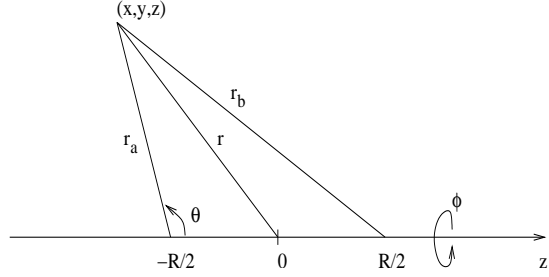}}

\caption{A local spherical coordinate system
$\pmb{(r_a,\theta,\phi)}$.}
\label{h2pt}
\end{figure}

In the work of Patil (see Eq. (2.15) in \cite{P}), he
indicated that for the ground state $\psi$ (i.e., with azimuth quantum number
$m=0$) for the molecular ion with Hamiltonian
\begin{equation}\label{C13a}
\hat H = -\frac12 \nabla^2_1 - \frac{Z_a}{r_a} - \frac{Z_b}{r_b},
\end{equation}
the ``coelescense'' condition at $r_b=0$ can be expressed as
\begin{equation}\label{C14a}
\frac12\left(\frac1\psi \frac{\partial\psi}{\partial r_b}\Big|_{\theta=0} +
\frac1\psi \frac{\partial\psi}{\partial r_b}\Big|_{\theta=\pi}\right)_{r_b=0} =
-Z_b,
\end{equation}
the angle $\theta$ is introduced in Fig. \ref{h2pt}.
He then indicates that the above is ``essentially the same'' as Kato's cusp
condition.

Actually, at least two ways are viable, which are going to be described below
through some concrete examples. The first way takes a limiting approach in a
similar spirit as Patil's \eqref{C14a}. The second way utilizes an {\em
explicit\/} representation.

\begin{exm}\label{exmC3}
For the Hamiltonian \eqref{C13a}, motivated by the Jaff\'e solution
\eqref{eqA.26}, let
us
consider a trial wave function for the ground state:
\begin{equation}\label{C15}
\psi(\lambda,\mu) = e^{-\alpha\lambda}[1+B_2P_2(\mu)].
\end{equation}
We want to consider the cusp condition in terms of the undetermined coefficients
$\alpha$ and $B_2$ in \eqref{C15}.
\end{exm}

\n Recall from \eqref{eqA.2a} that
\begin{align}
\label{C16}
&\text{~(i)}~~r_a\to 0 \text{ is equivalent to } \lambda\to 1, \mu\to -1;\\
\label{C17}
&\text{(ii)}~~r_b\to 0 \text{ is equivalent to } \lambda\to 1, \mu\to 1.
\end{align}
The singularities in the Laplacian $\nabla^2$ in prolate spheroidal coordinates,
according to \eqref{eqA.2a}, are discerned to be contained in
\[
\frac1{\lambda^2-\mu^2} = \frac1{\lambda+\mu} \cdot \frac1{\lambda-\mu}
\]
Thus, we deduce that for \eqref{C13a}, after substituting \eqref{C15} into
$\hat H$, the following:

\n (i)
\begin{equation}\label{C18}
\hat H\psi\sim \frac1{\lambda+\mu} F_1(\lambda,\mu), \text{ for } r_a\approx 0,
\text{ where } \lambda \gtrsim 1, \mu\approx -1,
\end{equation}
where we have dropped terms not containing the singularity $(\lambda+\mu)^{-1}$
and collected the dominant terms corresponding to singularity
$(\lambda+\mu)^{-1}$ in
\begin{align*}
&\quad F_1(\lambda, \mu)
= - \frac{2}{R^2 (\lambda -
\mu)} \left[ - 2 \alpha \lambda e^{-\alpha \lambda} ( 1 + B_2
P_2(\mu)) + (\lambda^2 - 1) \alpha^2 e^{-\alpha \lambda} ( 1 + B_2
P_2(\mu)) +
e^{-\alpha \lambda} 3 B_2 ( 1 - 3 \mu^2) \right] \\
& - \frac{2Z_a}{R} e^{-\alpha \lambda} [ 1 + B_2 P_2(\mu)]
\end{align*}
(ii)
\begin{equation}\label{C19}
H\psi\sim \frac1{\lambda-\mu} G_1(\lambda,\mu), \text{ for $r_b\approx 0$, where
} \lambda \gtrsim 0, \mu\approx 1,
\end{equation}
where, similarly,
\begin{align*}
&\quad G_1(\lambda, \mu)
= - \frac{2}{R^2 (\lambda +
\mu)} \left( - 2 \alpha \lambda e^{-\alpha \lambda} ( 1 + B_2
P_2(\mu)) + (\lambda^2 - 1) \alpha^2 e^{-\alpha \lambda} ( 1 + B_2
P_2(\mu)) +
e^{-\alpha \lambda} 3 B_2 ( 1 - 3 \mu^2) \right) \\
& - \frac{2Z_b}{R} e^{-\alpha \lambda} ( 1 + B_2 P_2(\mu))
\end{align*}
For \eqref{C18} and \eqref{C19} to stay bounded, it is {\em necessary\/} that
$F_1(1,-1) =0$ and $G_1(1,1)=0$.
\[
F_1(1,-1) = - \frac{1}{R^2} \left[ -2 \alpha e^{-\alpha} ( 1 + B_2)
- 6 B_2 e^{-\alpha} \right] - \frac{2Z_a}{R} e^{-\alpha} ( 1 +
B_2)=0,
\]
i.e.,
\begin{equation}\label{C20}
\alpha = R Z_a - \frac{3B_2}{1 + B_2}.
\end{equation}
\[
G_1(1,1) = - \frac{1}{R^2} \left[ -2 \alpha e^{-\alpha} ( 1 + B_2) -
6 B_2 e^{-\alpha} \right] - \frac{2Z_b}{R} e^{-\alpha} ( 1 + B_2) =0,
\]
i.e.,
\begin{equation}\label{C21}
\alpha = R Z_b - \frac{3B_2}{1 + B_2}.
\end{equation}
These are the cusp conditions at $r_a=0$ and $r_b=0$.

However, we need to remark that $F_1(1,-1)=0$ and $G_1(1,1)=0$ are only {\em
necessary conditions\/} for the desired boundedness because the limits in
\eqref{C18} and \eqref{C19} {\em do not exist\/} in general as $\lambda$ and
$\mu$ may be related in infinitely many different ways to yield totally
different limits as $\lambda\to 1$ and $\mu\to \pm 1$. Thus, the above
estimation approach has an {\em ad hoc nature}. Nevertheless, \eqref{C20} and
\eqref{C21} do provide correct answers as to be cross-validified with
\eqref{C28} and \eqref{C29} in Example \ref{exmC4}.

In inspecting \eqref{C20} and \eqref{C21}, we see that they are consistent when
and only when $Z_a=Z_b$, i.e., the {\em homonuclear\/} case. Therefore,
\eqref{C15} would not be a good choice of a trial wave function in the
heteronuclear case.
For the heteronuclear case, taking hints from the exact solutions \eqref{eqA.30}
and \eqref{eqA.35}, we
choose the trial wave function
\[
\Psi(\lambda,\mu) = e^{-\alpha \lambda} e^{\beta \mu} ( 1 + B_1 P_1(\mu) +
B_2 P_2(\mu))
\]
Here, $\alpha$ may be different to $\beta$ even though the exact
solution says $\alpha = \beta$. We have
\[
\hat  H \Psi \sim \frac{1}{\lambda + \mu}
F_2(\lambda, \mu),
\]
where the singular terms are collected in
\begin{align*}
&F_2(\lambda, \mu)
=- \frac{2}{R^2( \lambda -
\mu)} \left\{ - 2 \alpha \lambda e^{-\alpha \lambda} e^{\beta\mu} [
1
+ B_1 P_1(\mu) + B_2 P_2(\mu)) \right. \\
& + (\lambda^2 - 1) \alpha^2 e^{-\alpha \lambda} e^{\beta\mu} ( 1
+ B_1 P_1(\mu) + B_2 P_2(\mu)] \\
& + e^{-\alpha \lambda} e^{\beta \mu} [ ( \beta^2 - \beta^2 \mu^2
- 2 \beta \mu ) ( 1 + B_1 P_1(\mu) + B_2 P_2(\mu)) + ( 2 \beta - 2
\beta \mu^2 - 2 \mu) ( B_1 + 3 B_2 \mu) \\
&  + ( 1 - \mu^2) 3 B_2)]
- \frac{2Z_a}{R} e^{-\alpha \lambda} e^{\beta \mu} [ 1 + B_1
P_1(\mu) + B_2 P_2(\mu)],
\end{align*}
with the terms corresponding to the dominant singularity in
\begin{align*}
F_2(1,-1) =& - \frac{1}{R^2} e^{-\alpha - \beta} \left[ -2 \alpha (
1 - B_1 + B_2) + (2\beta)(1 - B_1 + B_2) + 2 (B_1 - 3 B_2) \right)
\\
& - \frac{2Z_a}{R} e^{-\alpha - \beta} ( 1 - B_1 + B_2),
\end{align*}
i.e.,
\[
\alpha - \beta = R Z_a + \frac{( B_1 - 3 B_2)}{1 - B_1 + B_2}.
\]
Similarly, the behavior near $r_b = 0$ gives
\[
\hat  H \Psi \sim \frac{1}{\lambda - \mu}
G_2(\lambda, \mu),
\]
where
\begin{align*}
& G_2(\lambda, \mu)
= - \frac{2}{R^2( \lambda +
\mu)} \Big\{ - 2 \alpha \lambda e^{-\alpha \lambda} e^{\beta\mu} [
1
+ B_1 P_1(\mu) + B_2 P_2(\mu)]  \\
& + (\lambda^2 - 1) \alpha^2 e^{-\alpha \lambda} e^{\beta\mu} [ 1
+ B_1 P_1(\mu) + B_2 P_2(\mu)] \\
&  + e^{-\alpha \lambda} e^{\beta \mu} [ ( \beta^2 - \beta^2 \mu^2
- 2 \beta \mu ) ( 1 + B_1 P_1(\mu) + B_2 P_2(\mu)] + ( 2 \beta - 2
\beta \mu^2 - 2 \mu) ( B_1 + 3 B_2 \mu) \\
&  + ( 1 - \mu^2) 3 B_2]
\Big\}
- \frac{2Z_b}{R} e^{-\alpha \lambda} e^{\beta \mu} [ 1 + B_1
P_1(\mu) + B_2 P_2(\mu)];
\end{align*}
\begin{align*}
G_2(1,1) = &- \frac{1}{R^2} e^{-\alpha + \beta} \left( -2 \alpha (
1 + B_1 + B_2) - (2\beta)(1 + B_1 + B_2) - 2 (B_1 + 3 B_2) \right)
\\
& - \frac{2Z_b}{R} e^{-\alpha + \beta} ( 1 + B_1 + B_2),
\end{align*}
i.e.,
\[
\alpha + \beta = R Z_b - \frac{( B_1 +3 B_2)}{1 + B_1 + B_2}.
\]

So, the cusp conditions give
\begin{align*}
B_1 &= - \frac{ 3(R Z_a+2 \beta-R Z_b)}{ 2(3+4\alpha-2RZ_a-2RZ_b-R Z_a
\alpha-RZ_a\beta+R^2Z_aZ_b-\alpha R Z_b+\beta R Z_b+\alpha^2-\beta^2) },\\
B_2 &= - \frac{ (2\alpha^2+2\alpha-2RZ_a\alpha-2\alpha R Z_b-R
Z_a+2 \beta R Z_b-2 \beta^2+2 R^2 Z_a Z_b-R Z_b-2 R Z_a \beta)
}{2(3+4 \alpha-2 R Z_a-2 R Z_b-R Z_a \alpha-R Z_a \beta+R^2 Z_a
Z_b-\alpha R Z_b+\beta R Z_b+ \alpha^2-\beta^2) }.
\end{align*}
If, in addition, we set $\alpha=\beta$, then
\begin{align*}
B_1 &= \frac{ 3(R Z_a+2 \alpha-R Z_b)}{ 2(2 R Z_a \alpha-R^2 Z_a Z_b+2 R Z_a-4
\alpha+2 R Z_b-3) },\\
B_2 &= - \frac{ (-2 R^2 Z_a Z_b+R Z_a+4 R Z_a \alpha+R Z_b-2
\alpha ) }{ 2(2 R Z_a \alpha-R^2 Z_a Z_b+2 R Z_a-4 \alpha+2 R
Z_b-3) }.\qquad \square
\end{align*}

\begin{exm}[Verification of the cusp conditions via explicit
calculations]\label{exmC4}
Consider the same Hamiltonian as given in \eqref{C13a}.
\end{exm}

We translate the origin from (0,0,0) to $(0,0,-R/2)$ and set up spherical
coordinates $(r_a,\theta,\phi)$ as shown in Fig. \ref{h2pt}.
>From the cosine law, cf. Fig. \ref{h2pt},
\begin{align}
r_b &= (r^2_a + R^2 - 2Rr_a\cos \theta)^{1/2}\nonumber\\
&= R\left(1-\frac{2\cos\theta}R r_a + \frac1{R^2} r^2_a\right)^{1/2}
= R\left[1-\frac{\cos\theta}R r_a + {\cl O} (r^2_a)\right].
\label{C22}
\end{align}
As we are exploring the singularity behavior of
$\hat H\psi$ near $r_a=0$, we need
only concentrate on the dominant terms of $r_a$ and, thus, we drop ${\cl
O}(r^2_a)$ in \eqref{C22} and approximate $r_b$ by
\[
r_b=R\left(1- \frac{\cos\theta}R r_a\right).
\]
Using \eqref{eqA.2a}, we therefore obtain
\begin{equation}\label{C23}
\left\{\begin{array}{l}
\lambda = \dfrac{r_a}R (1-\cos\theta)+1,\\
\noalign{\medskip}
\mu = \dfrac{r_a}R (1+\sin\theta)-1.\end{array}\right.
\end{equation}
The transformation \eqref{C23} will greatly facilitate our calculations.

Consider the trial wave function \eqref{C15} again here. We have
\begin{equation}\label{C24}
\frac{\partial\psi}{\partial r_a} = e^{-\alpha\lambda} \left\{-\alpha
\frac{\partial\lambda}{\partial r_a} \left[1 - \frac{B_2}2 + \frac32
B_2\mu^2\right] + 3B_2\mu \frac{\partial\mu}{\partial r_a}\right\}.
\end{equation}
>From \eqref{C23}, we have
\begin{equation}\label{C25}
\left\{\begin{array}{l}
\dfrac{\partial\lambda}{\partial r_a} = \dfrac1R (1-\cos\theta) =
\dfrac{\lambda-1}{r_a},\\
\noalign{\medskip}
\dfrac{\partial\mu}{\partial r_a} = \dfrac1R (1+\cos\theta) =
\dfrac{\mu+1}{r_a},
\end{array}\right.
\end{equation}
so we can use \eqref{C23} and \eqref{C25} to rewrite \eqref{C24} as
\begin{align}
\frac{\partial\psi}{\partial r_a} &=  e^{-\alpha[(\lambda-1)+1]} \left\{-\alpha
\left(\frac{\lambda-1}{r_a}\right) \left[\left(1-\frac{B_2}2\right) + \frac32
B_2 ((\mu+1)^2-1)^2\right]\right.\nonumber\\
&\quad \left. +3B_2((\mu+1)-1) \cdot \frac{\mu+1}{r_a}\right\}\nonumber\\
&= \frac{e^{-\alpha}}{r_a} \cdot e^{-\alpha(\lambda-1)}
\left\{-\underbrace{[\alpha(1+B_2) (\lambda-1) + 3B_2(\mu+1)]}_{{\cl
O}(r_a)}\right.\nonumber\\
\label{C26}
&\quad \left. + \underbrace{\left[\frac32 \alpha\cdot B_2\cdot 2 (\lambda-1)
(\mu+1) + 3B_2 (\mu+1)^2\right]}_{{\cl O}(r^2_a)} -\underbrace{\left[\frac32
B_2\cdot \alpha(\lambda-1)(\mu+1)^2\right]}_{{\cl O}(r^3_a)}\right\}
\end{align}
If we set $r_a=\rho$ and then substitute \eqref{C26} into the left-hand side
(LHS) of \eqref{C9a}, the two rightmost brackets on the RHS of \eqref{C26} will
become ${\cl O}(r_a) = {\cl O}(\rho)$ and ${\cl O}(r^2_a)  = {\cl O}(\rho^2)$,
and then vanish after $\rho\to 0$. So we only need to consider
the spherical mean of
$\frac{\partial\psi}{\partial r_a}$ over a sphere with radius $\rho$
which is given by
\begin{align}
&= \frac1{4\pi\rho^2} \intl_{{\cl S}_\rho} \frac{e^{-\alpha}}\rho
e^{-\alpha
\frac\rho{R}(1-\cos\theta)}
\left[-\alpha(1+B_2) \cdot \frac\rho{R}
(1-\cos\theta) - 3B_2 \frac\rho{R} (1+\cos\theta)\right]\rho^2\
d\omega\nonumber\\
&\quad= -\frac1{4\pi} e^{-\alpha} e^{-\frac{\alpha\rho}R} \int^{2\pi}_0
\int^\pi_0
e^{\frac{\alpha\rho}R\cos\theta} \left\{\left[\alpha(1+B_2)\cdot \frac1R +
3B_2\cdot \frac1R\right]\right.\nonumber\\
&\qquad \left. + \left[-\alpha(1+B_2)\cdot \frac1R + 3B_2\cdot \frac1R\right]
\cos \theta\right\} \sin \theta\ d\theta d\phi\nonumber\\
&\quad= -\frac12 e^{-\alpha-\frac{\alpha\rho}R} \left\{\frac1k (e^k-e^{-k})
[\alpha(1+B_2)+3B_2]\cdot \frac1R\right.\nonumber\\
\label{C27}
&\qquad \left. + \left[-\frac1{k^2} (e^k-e^{-k}) + \frac1k (e^k+e^{-k})\right]
[-\alpha (1+B_2) +3B_2]\cdot \frac1R\right\},
\end{align}
with $k\equiv\alpha\rho/R$. Letting $\rho\to 0$, we have $k\to 0$, and
\begin{align*}
\text{(RHS) of \eqref{C27}} &= - e^{-\alpha}\cdot [\alpha(1+B_2) +
3B_2] \cdot \frac1R\\
&\equiv \text{(RHS) of \eqref{C9a} (modified for the case of molecular ion)}\\
&= -Z_ae^{-\alpha} (1+B_2),
\end{align*}
i.e.,
\begin{equation}\label{C28}
\alpha(1+B_2) + 3B_2 =R Z_a(1+B_2),
\end{equation}
which is exactly \eqref{C20}. Similarly, at $r_b=0$, we can obtain
\begin{equation}\label{C29}
\alpha(1+B_2) + 3B_2 =R Z_b(1+B_2),
\end{equation}
which is exactly \eqref{C21}.

\section{Integrals with the Heitler--London wave functions}\label{integ}

The integrals involved in Eq. \eqref{HLINT} can be separated into
{\em eight\/} types of elementary integrals:

\begin{itemize}
\item[(i)] $\ds \frac12 \int\limits_{{\bb R}^3} a(1)(-\nabla^2_1) a(1) \ dx
\bigg(= \frac12 \int\limits_{{\bb R}^3} b(2) (-\nabla^2_2) b(2)\ dy = \frac12
\int\limits_{{\bb R}^3} a(2) (-\nabla^2_2) a(2) \ dy $\newline $=\ds\frac12
\int\limits_{{\bb R}^3} b(1) (-\nabla^2_1) b(1)\ dx\bigg) = \frac{\alpha^2}2$;
\item[(ii)] $\ds\frac12 \int\limits_{{\bb R}^3} a(1)(-\nabla^2_1) b(1) \ dx
\bigg(= \frac12 \int\limits_{{\bb R}^3} a(2)(-\nabla^2_1)b(2)dx\bigg)
= -\frac{\alpha^2}2\left(1 +
w - \frac12 w^2\right)$;
\item[(iii)] $\ds \int\limits_{{\bb R}^3} a^2(1) \left(-\frac1{r_{1a}}\right) \
dx= -\alpha$;
\item[(iv)] $S=\ds\int\limits_{{\bb R}^3} a(1)b(1)\ dx
= e^{-w} \left(1 + w
+\dfrac{w^2}3\right)$;
\item[(v)] $\alpha J =
\ds\int\limits_{{\bb R}^3} a^2(1) \left(-\frac1{r_{1b}}\right)\ dx
= \alpha \left[-\frac1w + e^{-2w}\left(1 +
\frac1w\right)\right]$;
\item[(vi)] $\alpha K =
\ds\int\limits_{{\bb R}^3} a(1)b(1) \left(-\frac1{r_{1b}}\right)\
dx =  -\alpha e^{-w}(1+w)$;
\item[(vii)] $\alpha J' = \ds\iint\limits_{{\bb R}^6} a^2(1)b^2(2)
\left(\frac1{r_{12}}\right) \ dxdy =  \alpha \left[\frac1w -
e^{-2w} \left(\frac1w + \frac{11}8 + \frac34 w + \frac16 w^2\right)\right]$;
\item[(viii)] $\ds
\alpha K'=\int\limits_{{\bb R}^6} a(1)b(1)a(2)b(2) \frac1{r_{12}} \ dxdy
=
\frac15 \alpha\bigg\{-e^{-2w} \left(-\frac{25}8 + \frac{23}4 w + 3w^2
+\frac13 w^3\right) + \frac6w [S^2(\gamma+\ln w)$\newline $+ S'{}^2E_i(-4w) -
2SS'E_i(-2w)]\bigg\}$,
\end{itemize}
where
\begin{itemize}
\item[] $w \equiv \alpha R$;\newline
$\ds S' \equiv e^w\left(1 - w + \frac13 w^2\right)$;\newline
$\gamma =$ Euler's constant $\ds = \int^1_0 \frac{1-e^{-t}}t \ dt - \int^\infty_1
\frac{e^{-t}}t \ dt = 0.57722\cdots$;\newline
$E_i(x) =$ integral logarithm $\ds= -(\text{P.V.}) \int\limits^\infty_{-x}
\frac{e^{-t}}t\ dt$\quad (for $x>0$),
\end{itemize}
here P.V. means ``principal value'' of a singular integral.

\section{Derivations related to the Laplacian for Subsection \ref{sec5}.D}
\label{Lapder}

Evaluation of the Laplacian in the prolate spheroidal coordinates
for the general form of the wave function which includes
electron-electron correlations explicitly is a demanding task.
Here we provide the details of the calculations.

The general form of each term in the wave function is
\begin{equation}\label{eqE.40}
\Phi_s(1,2) = \frac{1}{2\pi} \Phi_s(1) \Phi_s(2) P_s(r_{12})
\end{equation}
with
\begin{equation}\label{eqE.41}
\Phi_s(1)= F_s(\lambda_1) G_s(\mu_1)
\end{equation}
and
\begin{equation}\label{eqE.42}
\Phi_s(2)= \tilde{F}_s(\lambda_2) \tilde{M}_s(\mu_2).
\end{equation}
It can readily be seen that the Laplacian $\nabla_1^2$ operates only on
$\Phi_s(1)$.

According to \eqref{eqA.6}:
\begin{align}
\nabla^2_1 \Phi_s(1) &= \frac{4}{R^2 (\lambda_1^2 - \mu_1^2)}
\left[ \frac{(\lambda_1^2 - \mu_1^2)}{(\lambda_1^2 -
1)(1-\mu_1^2)} \frac{\partial^2}{\partial \phi_1^2} +
\frac{\partial}{\partial \lambda_1} (\lambda_1^2 -1)
\frac{\partial }{\partial \lambda_1} + \frac{\partial}{\partial
\mu_1} ( 1 - \mu_1^2 ) \frac{\partial }{\partial \mu_1} \right]
\Phi_s(1) \nonumber\\
 &= \frac{4}{ R^2 (\lambda_1^2 - \mu_1^2)}
\left[ \frac{(\lambda_1^2 - \mu_1^2)F_s G_s}{(\lambda_1^2 -
1)(1-\mu_1^2)} \frac{\partial^2 P_s}{\partial \phi_1^2} + G_s
\frac{\partial}{\partial \lambda_1} (\lambda_1^2 -1)
\frac{\partial (F_s P_s)}{\partial \lambda_1} \right. \nonumber \\
\label{eqE.43}
& \quad  \left. + F_s \frac{\partial}{\partial \mu_1} ( 1
- \mu_1^2 ) \frac{\partial (G_s P_s)}{\partial \mu_1} \right]\,.
\end{align}

We first single out part of the second term in the square bracket above for
further evaluation to obtain
\begin{align}
\frac{\partial}{\partial \lambda_1} (\lambda_1^2 -1)
\frac{\partial (F_s P_s) }{\partial \lambda_1} &=
\frac{\partial}{\partial \lambda_1} (\lambda_1^2 -1) \left\{P_s
\frac{\partial F_s }{\partial \lambda_1} + F_s
\frac{\partial P_s}{\partial \lambda_1} \right\}\nonumber \\
\label{eqE.44}
&= P_s \frac{\partial}{\partial \lambda_1} (\lambda_1^2 -1)
\frac{\partial F_s}{\partial \lambda_1} + 2 (\lambda_1^2-1)
\frac{\partial F_s}{\partial \lambda_1} \frac{\partial
P_s}{\partial \lambda_1} + F_s \frac{\partial}{\partial \lambda_1}
(\lambda_1^2 -1) \frac{\partial P_s}{\partial \lambda_1}.
\end{align}
Similarly, part of the third term inside the square bracket of \eqref{eqE.43}
becomes
\begin{align}
\frac{\partial}{\partial \mu_1} (1 - \mu_1^2) \frac{\partial (G_s
P_s)}{\partial \mu_1} &= \frac{\partial}{\partial \mu_1}
(1-\mu_1^2) \left\{P_s \frac{\partial G_s}{\partial \mu_1} + G_s
\frac{\partial P_s}{\partial \mu_1} \right\}\nonumber \\
\label{eqE.45}
&= P_s \frac{\partial}{\partial \mu_1} (1 - \mu_1^2)
\frac{\partial G_s}{\partial \mu_1} + 2 (1 - \mu_1^2)
\frac{\partial G_s}{\partial \mu_1} \frac{\partial P_s}{\partial
\mu_1} + G_s \frac{\partial}{\partial \mu_1} (1 - \mu_1^2)
\frac{\partial P_s}{\partial \mu_1}.
\end{align}

We can rewrite the Laplacian to take into account the separable
functional dependence of the three parts of the wave function, that
is,
\begin{equation}\label{eqE.46}
\frac{1}{\Phi_s(1)} \nabla^2_1 \Phi_s(1) = \frac{\nabla_1^2
F_s}{F_s} + \frac{\nabla_1^2 G_s}{G_s} + \frac{\nabla_1^2
P_s}{P_s} + \frac{2 \nabla_1 F_s \cdot \nabla_1 P_s}{F_s P_s} +
\frac{2 \nabla_1 G_s \cdot \nabla_1 P_s}{G_s P_s}\,,
\end{equation}
since
\begin{align}
\nabla_1 \Phi_s(1)& = \frac{2 (\lambda_1^2-1)^{1/2} }{
R(\lambda_1^2-\mu_1^2)^{1/2} } \frac{\partial \Phi_s(1)}{\partial
\lambda_1} \pmb{e}_{\lambda_1} + \frac{2 (1-\mu_1^2)^{1/2} }{
R(\lambda_1^2-\mu_1^2)^{1/2} } \frac{\partial \Phi_s(1)}{\partial
\mu_1} \pmb{e}_{\mu_1}\nonumber\\
\label{eqE.47}
&\quad + \frac{2}{R (\lambda_1^2-1)^{1/2}
(1-\mu_1^2)^{1/2} } \frac{\partial \Phi_s(1)}{\partial \phi_1}
\pmb{e}_{\phi_1},
\end{align}
where, $\pmb{e}_{\lambda_1}, \pmb{e}_{\mu_1}$, and $\pmb{e}_{\phi_1}$ are the
unit vectors pointing to the respective directions.

Now, we use the actual functional forms of various parts of the
wave function to complete the evaluation of the expectation value
of the Laplacian. Setting $F_s(\lambda_1) = e^{-\alpha\lambda_1}
\lambda_1^{m_s}$, we obtain
\begin{align*}
\nabla_1^2 F_s &= \frac{4}{R^2(\lambda_1^2 - \mu_1^2)}
\frac{\partial}{\partial \lambda_1} ( \lambda_1^2 - 1)
\frac{\partial}{\partial \lambda_1} F_s \\
&=
\frac{4}{R^2(\lambda_1^2 - \mu_1^2)} \frac{\partial}{\partial
\lambda_1} ( \lambda_1^2 - 1) \left\{ -\alpha e^{-\alpha\lambda_1}
\lambda_1^{m_s} + m_s e^{-\alpha\lambda_1} \lambda_1^{m_s-1} \right\} \\
&= \frac{4}{R^2(\lambda_1^2 - \mu_1^2)}
\frac{\partial}{\partial \lambda_1} \left\{-\alpha
e^{-\alpha\lambda_1} ( \lambda_1^{m_s + 2} - \lambda_1^{m_s}) +
m_s e^{-\alpha\lambda_1} ( \lambda_1^{m_s + 1} - \lambda_1^{m_s-1}
) \right\} \\
 &= \frac{4}{R^2(\lambda_1^2 - \mu_1^2)}
\left\{ \alpha^2 e^{-\alpha\lambda_1} ( \lambda_1^{m_s + 2} -
\lambda_1^{m_s} ) - \alpha e^{-\alpha\lambda_1} (
(m_s + 2)\lambda_1^{m_s + 1} - m_s \lambda_1^{m_s - 1}) \right. \\
&\quad \left.  + m_s e^{-\alpha\lambda_1}
\left( (m_s + 1) \lambda_1^{m_s} - (m_s - 1) \lambda_1^{m_s - 2}
\right) - m_s \alpha e^{-\alpha\lambda_1} ( \lambda_1^{m_s + 1} -
\lambda_1^{m_s - 1} ) \right\} \\
 &=
\frac{4\,F_s}{R^2(\lambda_1^2 - \mu_1^2)} \left\{ \alpha^2 (
\lambda_1^2 - 1 ) - 2 \alpha \left( ( m_s + 1) \lambda_1 - m_s
\frac{1}{\lambda_1} \right) + m_s \left( (m_s + 1) - (m_s -
1)\frac{1}{\lambda_1^2} \right) \right\}\,.
\end{align*}

Similarly, setting $G_s(\mu_1) = \mu_1^{j_s}$, we obtain
\begin{align}
\nabla_1^2 G_s &= \frac{4}{R^2(\lambda_1^2 - \mu_1^2)}
\frac{\partial}{\partial \mu_1} ( 1 - \mu_1^2 )
\frac{\partial}{\partial \mu_1} G_s  \nonumber\\
 &=
\frac{4}{R^2(\lambda_1^2 - \mu_1^2)} \frac{\partial}{\partial
\mu_1} ( 1 - \mu_1^2 ) j_s \mu_1^{j_s - 1} \nonumber\\
\label{eqE.48}
&= \frac{4\,G_s}{R^2(\lambda_1^2 - \mu_1^2)} \left( j_s(j_s - 1)
\frac{1}{\mu_1^2} - j_s(j_s + 1) \right)\,.
\end{align}

Since the effect of the Laplacian on a given function is coordinate-free,  we
can consider
the evaluation of $\nabla^2_1\,P_s(r_{12})$, through the spherical
coordinates for a general function $f(r_{12})$ of $r_{12}$
to obtain
\begin{equation}\label{eqE.49}
\nabla^2_1 f(r_{12})  =\nabla^2_{r_{12}} f(r_{12}) =
\frac{1}{r_{12}^2} \frac{d}{d r_{12}} r_{12}^2 \frac{d
f(r_{12})}{d r_{12}}\,.
\end{equation}

Thus, with $P_s(r_{12}) = r_{12}^\ell$,
\begin{equation}\label{eqE.50}
\nabla^2_1 P_s = \frac{1}{r_{12}} \left( r_{12}^2 l(l-1)
r_{12}^{\ell-2} + 2 r_{12} \ell r_{12}^{\ell-1} \right) =
\frac{\ell(\ell+1)}{r_{12}} P_s\,.
\end{equation}

The other terms involved are
\begin{align}
\frac{\partial P_s}{\partial\lambda_1} &= l r_{12}^{\ell-1}
\frac{\partial r_{12}}{\partial \lambda_1}  \nonumber\\
 &=
\frac{R^2 l P_s} {8 r_{12}^2} \left[ 2 \lambda_1 - 2 \lambda_2
\mu_1 \mu_2 - \frac{M}{\lambda_1^2-1} \cos(\phi_1 - \phi_2)
2\lambda_1 \right]  \nonumber \\
&= \frac{R^2 l P_s}{4 r_{12}^2}
\left[\lambda_1 - \lambda_2 \mu_1 \mu_2 -
\frac{M\lambda_1}{(\lambda_1^2-1)} \cos(\phi_1 - \phi_2) \right]
 \nonumber\\
  &= \frac{R^2 l P_s}{4 r_{12}^2} \frac{\lambda_1}{(\lambda_1^2
-1)} \left[ (\lambda_1^2-1) - \lambda_2 \mu_1 \mu_2
\frac{(\lambda_1^2-1)}{\lambda_1} - M \cos(\phi_1 - \phi_2)
\right]\nonumber \\
\label{eqE.51}
&= \frac{R^2 l P_s}{8r_{12}^2} \frac{\lambda_1}{(\lambda_1^2 -1)}
\left[ \frac{4}{R^2} r_{12}^2 + \lambda_1^2 - \lambda_2^2 -
\mu_1^2 - \mu_2^2 + \frac{ 2\lambda_2 \mu_1 \mu_2 }{\lambda_1}
\right]\,,
\end{align}
and
\begin{align}
\frac{\partial P_s}{\partial \mu_1} &= \frac{R^2 l P_s}{8
r_{12}^2} \left[ 2 \mu_1 - 2 \lambda_1 \lambda_2 \mu_2 -
\frac{M}{(1-\mu_1^2)} \cos(\phi_1 - \phi_2) (-2\mu_1) \right]
\nonumber \\
&= \frac{R^2 l P_s}{4 r_{12}^2} \left[ \mu_1 -
\lambda_1 \lambda_2 \mu_2 + \frac{M\mu_1}{(1-\mu_1^2)} \cos(\phi_1
- \phi_2) \right] \nonumber\\
 &= - \frac{R^2 l P_s}{4 r_{12}^2}
\frac{\mu_1}{(1-\mu_1^2)} \left[ -(1 - \mu_1^2) + \lambda_1
\lambda_2 \mu_2 \frac{(1-\mu_1^2)}{\mu_1} -
M \cos(\phi_1 - \phi_2) \right]\nonumber \\
\label{eqE.52}
&= - \frac{R^2 l P_s}{8 r_{12}^2} \frac{\mu_1}{(1-\mu_1^2)}
\left[ \frac{4}{R^2} r_{12}^2 - \lambda_1^2 - \lambda_2^2 +
\mu_1^2 - \mu_2^2 + \frac{ 2 \lambda_1 \lambda_2 \mu_2 }{\mu_1}
\right]\,.
\end{align}

Upon putting the above together, the Laplacian operation part of the
Hamiltonian takes the form
\begin{align}
& \frac{1}{\Phi_s(1)} \nabla^2_1 \Phi_s(1) = \left\{
\frac{\nabla_1^2 F_s}{F_s} + \frac{\nabla_1^2 G_s}{G_s} +
\frac{\nabla_1^2 P_s}{P_s} + \frac{2 \nabla_1 F_s \cdot \nabla_1
P_s}{F_s P_s} + \frac{2 \nabla_1 G_s \cdot \nabla_1 P_s}{G_s P_s}
\right\} \nonumber\\
 =~ &\frac{4}{R^2(\lambda_1^2 - \mu_1^2)}
\left\{ \alpha^2 ( \lambda_1^2 - 1 ) - 2 \alpha \left( (m_s + 1
)\lambda_1 - m_s \frac{1}{\lambda_1} \right) + m_s \left( ( m_s +
1 )  - ( m_s - 1 ) \frac{1}{\lambda_1^2} \right) \right\}
\nonumber\\
& + \frac{4}{R^2 (\lambda_1^2 - \mu_1^2)} \left( j_s (
j_s - 1) \frac{1}{\mu_1^2} - j_s(j_s + 1) \right) + \frac{\ell(\ell+1)}{
r_{12}^2}  \nonumber\\
 & + \frac{1}{(\lambda_1^2-\mu_1^2) }
\left( -\alpha \lambda_1 + m_s \right) \frac{\ell}{r_{12}^2}
\left[\frac{4}{R^2} r_{12}^2 + \lambda_1^2 - \lambda_2^2 - \mu_1^2
- \mu_2^2 + \frac{ 2\lambda_2 \mu_1 \mu_2 }{\lambda_1} \right]\nonumber \\
\label{eqE.53}
& - \frac{1}{ (\lambda_1^2-\mu_1^2) } j_s \frac{\ell}{ r_{12}^2}
\left[ \frac{4}{R^2} r_{12}^2 - \lambda_1^2 - \lambda_2^2 +
\mu_1^2 - \mu_2^2 + \frac{ 2 \lambda_1 \lambda_2 \mu_2 }{\mu_1}
\right]\,.
\end{align}
As before we can construct the inter-term expectation value
integral for the Laplacian using the above relation.
Introducing the function
\begin{equation}\label{eq7a.14}
 X^\nu(m,n,j,k;\ell) = Z^\nu(m,n+2,j,k;\ell) - Z^\nu(m,n,j,k+2;\ell),
\end{equation}
and defining
\begin{equation}\label{eq7a.14a}
(m,n,j,k;\ell) = (m_r,n_r,j_r,k_r;\ell_r) + (m_s,n_s,j_s,k_s;\ell_s),
\end{equation}
we obtain
\begin{align*}
\left\langle \nabla_1^2 \right\rangle_{r,s} &= \left\langle
\lambda_1^{m_r} \lambda_2^{n_r} \mu_1^{j_r} \mu_2^{k_r}
r_{12}^{\ell_r} \mid \nabla_1^2 \mid \lambda_1^{m_s} \lambda_2^{n_s}
\mu_1^{j_s}
\mu_2^{k_s} r_{12}^{\ell_s} \right\rangle\nonumber \\
&= \frac{4}{R^2} \left[ \alpha^2 X^\nu(m+2,n,j,k;\ell) + \left\{
-\alpha^2 + ( m_s - j_s) ( m_s + j_s + 1 + \ell_s ) \right\}
X^\nu(m,n,j,k;\ell) \right. \\
&\quad - 2 \alpha (n_s +1) X^\nu(m+1,n,j,k;\ell) + 2 \alpha n_s
X^\nu(m-1,n,j,k;\ell) \\
&\quad + m_s ( m_s - 1) X^\nu(m-2,n,j,k;\ell) + j_s (j_s - 1)
X^\nu(m,n,j-2,k;\ell) \\
&\quad \left. + \ell_s (\ell_s + 1) \left\{ X^\nu(m,n+2,j,k;\ell-2) -
X^\nu(m,n,j,k+2;\ell-2)
\right\} - \ell_s \alpha X^\nu(m+1,n,j,k;\ell) \right] \\
\intertext{\newpage}
&\quad - \ell_s \alpha
X^\nu(m+3,n,j,k;\ell-2) + \ell_s \alpha X^\nu(m+1,n+2,j,k;\ell-2) \\
&\quad + \ell_s \alpha X^\nu(m+1,n,j+2,k;\ell-2) + \ell_s \alpha
X^\nu(m+1,n,j,k+2;
\ell-2) \\
&\quad - 2 \ell_s \alpha
X^\nu(m,n+1,j+1,k+1;\ell-2)\\
&\quad + \ell_s ( m_s + j_s)  X^\nu(m+2,n,j,k;\ell-2) - \ell_s ( m_s - j_s)
X^\nu(m,n+2,j,k,\ell-2) \\
&\quad - \ell_s ( m_s + j_s) X^\nu(m,n,j+2,k;\ell-2) - \ell_s ( m_s - j_s)
X^\nu(m,n,j,k+2;\ell-2) \\
&\quad - 2 \ell_s m_s X^\nu(m-1,n+1,j+1,k+1;\ell-2) - 2 \ell_s j_s
X^\nu(m+1,n+1,j-1,k+1;\ell-2)\,.
\end{align*}
Thus we have furnished complete details of the electronic kinetic
energy calculations.

\section{Recursion relations and their derivations for Subsection \ref{sec5}.D}
\label{recrel}

In this appendix we provide simple proofs of the recursion
relations which are needed in the analytical calculations.

\begin{center}
{\bf 1.} $\pmb{A(m;\alpha)}$
\end{center}

\begin{align}\label{eqF.55}
A(m;\alpha) &\equiv \int^\infty_1 \lambda^m e^{-\alpha \lambda} \, d\lambda \\
 &= \left. \lambda^m \frac{e^{-\alpha \lambda}
}{-\alpha} \right|_1^\infty + \frac{m}{\alpha}
\int_1^\infty \lambda^{m-1} e^{-\alpha \lambda} \, d\lambda
\nonumber\\
 &= \frac{1}{\alpha} [ e^{-\alpha } + m A(m-1;\alpha) ].\nonumber
\end{align}
When $m = 0$,
\begin{equation}\label{eqF.56}
A(0;\alpha) = \int^\infty_1 e^{-\alpha \lambda} \, d\lambda =
\frac{e^{-\alpha}}{\alpha}.
\end{equation}

The recurrence relation can be used in succession to give
\begin{equation}\label{eqF.57}
A(m;\alpha) = \frac{e^{-\alpha}}{\alpha} \sum^m_{\nu=0}
\frac{m!}{(m-\nu)!} \frac{1}{\alpha^\nu}.
\end{equation}

\begin{center}
{\bf 2.} $\pmb{F(m;\alpha)}$
\end{center}

The definition is
\begin{equation}\label{eqF.58}
F(m;\alpha) \equiv \int^\infty_1 \lambda^m e^{-\alpha \lambda}
Q_0(\lambda) d\lambda\,.
\end{equation}
To prove the recurrence relation for $F(m;\alpha)$,
\begin{align*}
& m F(m-1;\alpha) - (m-2) F(m-3;\alpha) = \int^\infty_1 ( m
\lambda^{m-1} - (m-2) \lambda^{m-3}) e^{-\alpha \lambda} Q_0 (\lambda) d\lambda
\\
= ~&\left. (\lambda^m - \lambda^{m-2}) e^{-\alpha \lambda}
Q_0(\lambda) \right|^\infty_0 + \alpha \int^\infty_1 d\lambda \,
(\lambda^m - \lambda^{m-2}) e^{-\alpha \lambda} Q_0 (\lambda) \\
& - \int^\infty_1 d\lambda \, (\lambda^m -
\lambda^{m-2}) e^{-\alpha \lambda} \left(\frac{1}{2}\right) \left(
\frac{1}{\lambda +1} - \frac{1}{\lambda -1} \right) \\
= ~&\alpha [ F(m;\alpha) - F(m-2;\alpha) ] + \int^\infty_1
d\lambda \, \lambda^{m-2} e^{-\alpha \lambda}\,.
\end{align*}
So, the recurrence relation is
\begin{equation}\label{eqF.59}
F(m;\alpha) = F(m-2;\alpha) + \frac{1}{\alpha} \left[ m
F(m-1;\alpha) - (m-2) F(m-3;\alpha) - A(m-2;\alpha) \right].
\end{equation}

The initial conditions for $F(m;\alpha)$ are already given in \eqref{eq7a.28}
and \eqref{eq7a.29}.

\begin{center}
{\bf 3.} $\pmb{S(m,n;\alpha)}$
\end{center}


$S(m,n;\alpha)$ is defined by
\begin{equation}\label{eqF.64}
S(m,n;\alpha) \equiv \int^\infty_1 d\lambda_1 \int^{\lambda_1}_1
d\lambda_2 \lambda_1^m \lambda_2^n e^{-\alpha (\lambda_1 +
\lambda_2)}
\end{equation}
The recurrence relation is
\begin{equation}\label{eqF.65}
S(m,n;\alpha) = \frac{1}{\alpha} \left[ m S(m-1,n;\alpha) +
A(m+n;2\alpha) \right]\,,
\end{equation}
with
\begin{equation}\label{eqF.66}
S(0,n;\alpha) = \frac{1}{\alpha} A(n;2\alpha)\,.
\end{equation}

By using  $S(m,n;\alpha)$, the following integrals can be
represented in terms of $A(m;\alpha)$ and $S(m,n;\alpha)$:
\begin{align}
& \int^\infty_1 d\lambda_1 \int^{\lambda_1}_1 d\lambda_2
\lambda_1^m \lambda_2^n e^{-\alpha (\lambda_1 + \lambda_2)} =
\int^\infty_1 d\lambda_1 \lambda_1^m e^{-\alpha \lambda_1}
\int^{\lambda_1}_1 d\lambda_2 \lambda_2^n e^{-\alpha \lambda_2}
\nonumber\\
 =~ &\int^\infty_1 d\lambda_1 \lambda_1^m e^{-\alpha
\lambda_1} \left( \left. \frac{\lambda_2^n}{-\alpha} e^{-\alpha
\lambda_2} \right|^{\lambda_1}_1 + \frac{n}{\alpha}
\int^{\lambda_1}_1
d\lambda_2 \lambda_2^{n-1} e^{-\alpha \lambda_2} \right)
\nonumber\\
= ~& - \frac{A(m+n;2\alpha)}{\alpha} +
\frac{e^{-\alpha}}{\alpha} A(m;\alpha) + \frac{n}{\alpha}
\int^\infty_1 d\lambda_1 \int^{\lambda_1}_1 d\lambda_2 \lambda_1^m
\lambda_2^{n-1} e^{-\alpha (\lambda_1 + \lambda_2)}  \nonumber\\
= ~& -\frac{1}{\alpha} \sum^n_{\nu=0} \frac{1}{\alpha^\nu}
A(m+n-\nu;2\alpha) \frac{n!}{(n-\nu)!} +
\frac{e^{-\alpha}}{\alpha} \sum^n_{\nu=0} \frac{n!}{\alpha^\nu
(n-\nu)!} A(m;\alpha)  \nonumber\\
=~ & - \frac{1}{\alpha}
\sum^n_{s=0} \frac{\alpha^s}{\alpha^n} A(m+s;2\alpha)
\frac{n!}{s!} + \frac{e^{-\alpha}}{\alpha} \sum^n_{\nu=0}
\frac{n!}{\alpha^\nu
(n-\nu)!} A(m;\alpha)\nonumber \\
\label{eqF.67}
=~ & - S(n,m;\alpha) + A(n;\alpha) A(m;\alpha).
\end{align}
Furthermore,
\begin{align}
& \int^\infty_1 d\lambda_1 \int^\infty_1 d\lambda_2 \lambda_1^m
\lambda_2^n e^{-\alpha (\lambda_1 + \lambda_2)} = A(m;\alpha)
A(n;\alpha)  \nonumber\\
=~ & \int^\infty_1 d\lambda_1
\int^{\lambda_1}_1 d\lambda_2 \lambda_1^m \lambda_2^n e^{-\alpha
(\lambda_1 + \lambda_2)} + \int^\infty_1 d\lambda_1
\int^\infty_{\lambda_1} d\lambda_2
\lambda_1^m \lambda_2^n e^{-\alpha (\lambda_1 + \lambda_2)}
\nonumber\\
=~ & \int^\infty_1 d\lambda_1 \int^{\lambda_1}_1
d\lambda_2 \lambda_1^m \lambda_2^n e^{-\alpha (\lambda_1 +
\lambda_2)} + \int^\infty_1 d\lambda_1 \int^{\lambda_1}_1
d\lambda_2 \lambda_1^n \lambda_2^m e^{-\alpha (\lambda_1 +
\lambda_2)}\nonumber \\
\label{eqF.68}
=~ & - S(n,m;\alpha) + A(n;\alpha)A(m;\alpha) - S(m,n;\alpha) +
A(m;\alpha)A(n;\alpha)
\end{align}
So
\begin{equation}\label{eqF.69}
S(m,n;\alpha) + S(n,m;\alpha) = A(m;\alpha) A(n;\alpha).
\end{equation}

\begin{center}
{\bf 4.} $\pmb{T(m,n;\alpha)}$
\end{center}

The definition is
\begin{equation}\label{eqF.70}
T(m,n;\alpha) \equiv \frac{m!}{\alpha^{m+1}} \sum^m_{\nu=0}
\frac{\alpha^\nu}{\nu!} F(n+\nu; 2 \alpha).
\end{equation}

The recurrence relation and $T(0,n;\alpha)$ are
\begin{equation}\label{eqF.71}
T(m,n;\alpha) = \frac{1}{\alpha} \left[ m T(m-1,n;\alpha) +
F(m+n;2\alpha) \right]
\end{equation}
and
\begin{equation}\label{eqF.72}
T(0,n;\alpha) = \frac{1}{\alpha} F(n;2\alpha)
\end{equation}

\begin{center}
{\bf 5.} $\pmb{H_0(m,n;\alpha)}$
\end{center}

By definition,
\begin{align}
H_0(m,n,\alpha) &\equiv \int^\infty_1 d\lambda_1 \int^\infty_1
d\lambda_2 \lambda_1^m \lambda_2^n e^{-\alpha (\lambda_1 +
\lambda_2)} Q_0(\lambda_>)  \nonumber\\
\label{eqF.73}
 &= \int^\infty_1
d\lambda_1 \int^{\lambda_1}_1 d\lambda_2 \lambda_1^m \lambda_2^n
e^{-\alpha (\lambda_1 + \lambda_2)} Q_0(\lambda_1) + \int^\infty_1
d\lambda_1 \int^{\lambda_1}_1 d\lambda_2 \lambda_1^n \lambda_2^m
e^{-\alpha (\lambda_1 + \lambda_2)} Q_0(\lambda_1).
\end{align}

The first term on the right-hand side above yields
\begin{align}
& \int^\infty_1 d\lambda_1 \int^{\lambda_1}_1 d\lambda_2
\lambda_1^m \lambda_2^n e^{-\alpha (\lambda_1 + \lambda_2)}
Q_0(\lambda_1)  \nonumber\\
 =~ &\int^\infty_1 d\lambda_1
\lambda_1^m e^{-\alpha \lambda_1} Q_0(\lambda_1)
\int^{\lambda_1}_1 d\lambda_2 \lambda_2^n e^{-\alpha \lambda_2}
\nonumber\\
=~ &\int^\infty_1 d\lambda_1 \lambda_1^m e^{-\alpha
\lambda_1} Q_0(\lambda_1) \left( \left.
\frac{\lambda_2^n}{-\alpha} e^{-\alpha \lambda_2}
\right|^{\lambda_1}_1 + \frac{n}{\alpha} \int^{\lambda_1}_1
d\lambda_2 \lambda_2^{n-1} e^{-\alpha \lambda_2} \right)
\nonumber\\
=~ & - \frac{F(m+n;2\alpha)}{\alpha} +
\frac{e^{-\alpha}}{\alpha} F(m;\alpha) + \frac{n}{\alpha}
H_0(m,n-1;\alpha)  \nonumber\\
 =~ & - \frac{F(m+n;2\alpha)}{\alpha}
+ \frac{e^{-\alpha}}{\alpha} F(m;\alpha)  \nonumber\\
&\quad  +
\frac{n}\alpha \left( - \frac{F(m+n-1;2\alpha)}{\alpha} +
\frac{e^{-\alpha}}{\alpha} F(m;\alpha) + \frac{n-1}{\alpha}
H_0(m,n-2;\alpha) \right)  \nonumber\\
=~ & -\frac{1}{\alpha}
\sum^n_{\nu=0} \frac{1}{\alpha^\nu} F(m+n-\nu;2\alpha)
\frac{n!}{(n-\nu)!} + \frac{e^{-\alpha}}{\alpha} \sum^n_{\nu=0}
\frac{n!}{\alpha^\nu (n-\nu)!} F(m;\alpha)  \nonumber \\
=~ & -
\frac{1}{\alpha} \sum^n_{s=0} \frac{\alpha^s}{\alpha^n}
F(m+s;2\alpha) \frac{n!}{s!} + \frac{e^{-\alpha}}{\alpha}
\sum^n_{\nu=0} \frac{n!}{\alpha^\nu
(n-\nu)!} F(m;\alpha) \nonumber\\
\label{eqF.74}
=~ & - T(n,m;\alpha) + A(n;\alpha) F(m;\alpha),
\end{align}
while the second term in \eqref{eqF.73} is the same as the first term if we
interchange $m$ and $n$. Therefore,
\begin{equation}\label{eqF.75}
H_0(m,n;\alpha) = - T(n,m;\alpha) + A(n;\alpha) F(m;\alpha) -
T(m,n;\alpha) + A(m;\alpha) F(n;\alpha).
\end{equation}

\begin{center}
{\bf 6.} $\pmb{H_1(m,n;\alpha)}$
\end{center}

By definition,
\begin{align}
H_1(m,n;\alpha) &\equiv \int^\infty_1 d\lambda_1 \int^\infty_1
d\lambda_2 \lambda_1^m \lambda_2^n e^{-\alpha (\lambda_1 +
\lambda_2)} P_1(\lambda_<) Q_1(\lambda_>)\nonumber\\
 \label{eqF.76}
 &= \int^\infty_1 d\lambda_1 \int^{\lambda_1}_1
d\lambda_2 \lambda_1^m \lambda_2^{n+1} e^{-\alpha (\lambda_1 +
\lambda_2)} Q_1(\lambda_1) + (\text{same as left, with } m \leftrightarrow n).
\end{align}

The first term in \eqref{eqF.76} is evaluated as
\begin{align}
& \int^\infty_1 d\lambda_1 \int^{\lambda_1}_1 d\lambda_2
\lambda_1^m \lambda_2^{n+1} e^{-\alpha (\lambda_1 + \lambda_2)}
Q_1(\lambda_1)  \nonumber\\
 =~ & \int^\infty_1 d\lambda_1
\int^{\lambda_1}_1 d\lambda_2 \lambda_1^{m+1} \lambda_2^{n+1}
e^{-\alpha (\lambda_1 + \lambda_2)} Q_0(\lambda_1) - \int^\infty_1
d\lambda_1 \int^{\lambda_1}_1 d\lambda_2 \lambda_1^m
\lambda_2^{n+1}
e^{-\alpha (\lambda_1 + \lambda_2)} \nonumber\\
\label{eqF.77}
=~ & \int^\infty_1 d\lambda_1 \int^{\lambda_1}_1 d\lambda_2
\lambda_1^{m+1} \lambda_2^{n+1} e^{-\alpha (\lambda_1 +
\lambda_2)} Q_0(\lambda_1) - S(m,n+1;\alpha).
\end{align}
By combining the two terms in  \eqref{eqF.76}, we obtain
\begin{equation}\label{eqF.78}
H_1(m,n;\alpha) = H_0(m+1,n+1;\alpha) - S(n,m+1;\alpha) -
S(m,n+1;\alpha).
\end{equation}

\begin{center}
{\bf 7.} $\pmb{H_\tau(m,n;\alpha)}$
\end{center}

The recurrence relations for the Legendre polynomials are
\begin{align}\label{eqF.79}
(\tau+1) P_{\tau+1} &= (2\tau + 1) x P_\tau - \tau P_{\tau-1}, \\
\label{eqF.80}
(\tau+1) Q_{\tau+1} &= (2\tau + 1) x Q_\tau - \tau Q_{\tau-1}.
\end{align}
We then have
\begin{align}
H_\tau(m,n;\alpha) &= \int\int \lambda_1^m \lambda_2^n
P_\tau(\lambda_<) Q_\tau(\lambda_>) e^{-\alpha(\lambda_1 +
\lambda_2)} \, d\lambda_1 d\lambda_2  \nonumber\\
&=
\frac{1}{\tau^2} \int\int \lambda_1^m \lambda_2^n \left[
(2\tau - 1) \lambda_< P_{\tau-1} - (\tau-1) P_{\tau-2} \right]\nonumber \\
\label{eqF.81}
&\quad \left[ (2\tau - 1) \lambda_> Q_{\tau-1} - (\tau-1)
Q_{\tau-2} \right] e^{-\alpha(\lambda_1 + \lambda_2)} \,
d\lambda_1 d\lambda_2;
\end{align}
\begin{align}
& \tau^2 H_\tau(m,n;\alpha)  \nonumber\\
 =~ & (2\tau-1)^2
H_{\tau-1}(m+1,n+1;\alpha) + (\tau-1)^2 H_{\tau-2} (m,n;\alpha)\nonumber \\
\label{eqF.82}
& - (2\tau - 1)(\tau - 1) \int\int
\lambda_1^m \lambda_2^n ( \lambda_< P_{\tau-1} Q_{\tau-2} +
\lambda_> P_{\tau-2} Q_{\tau-1}
) e^{-\alpha(\lambda_1 + \lambda_2)} \, d\lambda_1 d\lambda_2 \\
 =~ & (2\tau-1)^2
H_{\tau-1}(m+1,n+1;\alpha) + (\tau-1)^2 H_{\tau-2} (m,n;\alpha)
\nonumber \\
& - (2\tau - 1) \int\int \lambda_1^m \lambda_2^n (
\lambda_< ((2\tau-3) \lambda_< P_{\tau-2} - (\tau-2) P_{\tau-3} )
Q_{\tau-2}
\nonumber\\
&  + \lambda_> P_{\tau-2} ((2\tau-3) \lambda_> Q_{\tau-2} -
(\tau-2) Q_{\tau-3} ) ) e^{-\alpha(\lambda_1 + \lambda_2)} \,
d\lambda_1 d\lambda_2  \nonumber \\
=~  & (2\tau-1)^2
H_{\tau-1}(m+1,n+1;\alpha) + (\tau-1)^2 H_{\tau-2} (m,n;\alpha)
\nonumber\\
& - (2\tau - 1) (2\tau-3) \int\int \lambda_1^m
\lambda_2^n ( \lambda_<^2 + \lambda_>^2) P_{\tau-2} Q_{\tau-2}
e^{-\alpha(\lambda_1 + \lambda_2)} \, d\lambda_1 d\lambda_2\nonumber\\
\label{eqF.83}
& + (2\tau - 1) (\tau-2) \int\int
\lambda_1^m \lambda_2^n ( \lambda_< P_{\tau-3} Q_{\tau-2} +
\lambda_> P_{\tau-2} Q_{\tau-3}
) e^{-\alpha(\lambda_1 + \lambda_2)} \, d\lambda_1 d\lambda_2
\\
=~ & (2\tau-1)^2
H_{\tau-1}(m+1,n+1;\alpha) + (\tau-1)^2 H_{\tau-2} (m,n;\alpha)
\nonumber\\
& - (2\tau - 1) (2\tau-3) ( H_{\tau-2} (m+2,n;\alpha) +
H_{\tau-2} (m,n+2;\alpha) ) \nonumber \\
& + (2\tau - 1) \int\int \lambda_1^m \lambda_2^n (
\lambda_< P_{\tau-3} ( ( 2\tau-5) \lambda_> Q_{\tau-3} - (\tau-3)
Q_{\tau-4} )  \nonumber\\
& + \lambda_> ( ( 2\tau-5) \lambda_<
P_{\tau-3} - (\tau-3) P_{\tau-4} ) Q_{\tau-3} )
e^{-\alpha(\lambda_1 + \lambda_2)}
\, d\lambda_1 d\lambda_2 \nonumber\\
=~ & (2\tau-1)^2
H_{\tau-1}(m+1,n+1;\alpha) + (\tau-1)^2 H_{\tau-2} (m,n;\alpha)
\nonumber\\
& - (2\tau - 1) (2\tau-3) ( H_{\tau-2} (m+2,n;\alpha) +
H_{\tau-2} (m,n+2;\alpha) ) \nonumber \\
& + 2 (2\tau - 1) (2\tau-5) \int\int \lambda_1^m
\lambda_2^n \lambda_< \lambda_> P_{\tau-3} Q_{\tau-3}
e^{-\alpha(\lambda_1 + \lambda_2)} \, d\lambda_1 d\lambda_2 \nonumber\\
\label{eqF.84}
& - (2\tau - 1) (\tau-3) \int\int
\lambda_1^m \lambda_2^n ( \lambda_< P_{\tau-3} Q_{\tau-4} +
 \lambda_> P_{\tau-4} Q_{\tau-3})
e^{-\alpha(\lambda_1 + \lambda_2)} \, d\lambda_1 d\lambda_2.
\end{align}
As shown in the equations
\eqref{eqF.84}--\eqref{eqF.84} above, the same patterns
are repeated until $\tau$ is reduced to 0. So, let's consider the last
term. If $\tau$ is even, the last term is
\begin{align}
& - (2\tau-1) \int \int \lambda_1^m \lambda_2^n ( \lambda_< P_1
Q_0 + \lambda_> P_0 Q_1) e^{-\alpha(\lambda_1 + \lambda_2)} \,
d\lambda_1 d\lambda_2  \nonumber\\
=~ & - (2\tau-1) \int \int
\lambda_1^m \lambda_2^n ( \lambda_<^2 Q_0 + \lambda_>^2 Q_0 -
\lambda_> )
e^{-\alpha(\lambda_1 + \lambda_2)} \, d\lambda_1 d\lambda_2 \nonumber\\
=~ & - (2\tau-1) ( H_0(m+2,n;\alpha) + H_0(m,n+2;\alpha) )
\nonumber\\
\intertext{\newpage}
& + (2\tau-1) \int^\infty_1 d\lambda_1
\int^{\lambda_1}_1 d\lambda_2 \, \lambda_1^{m+1} \lambda_2^n
e^{-\alpha(\lambda_1 + \lambda_2)} \nonumber \\
& + (2\tau-1) \int^\infty_1 d\lambda_1
\int^\infty_{\lambda_1} d\lambda_2 \, \lambda_1^m \lambda_2^{n+1}
e^{-\alpha(\lambda_1 +
\lambda_2)} \nonumber\\
=~ & - (2\tau-1) [ H_0(m+2,n;\alpha) + H_0(m,n+2;\alpha)\nonumber\\
\label{eqF.85}
& -
S(m+1,n;\alpha) - S(n+1,m;\alpha) ]
\end{align}
If $\tau$ is odd, the last term then is
\begin{align}
& (2\tau-1) \int \int \lambda_1^m \lambda_2^n ( \lambda_< P_0 Q_1
+ \lambda_> P_1 Q_0) e^{-\alpha(\lambda_1 + \lambda_2)} \,
d\lambda_1 d\lambda_2  \nonumber\\
=~ & (2\tau-1) \int \int
\lambda_1^m \lambda_2^n ( \lambda_< \lambda_> Q_0 - \lambda_< +
\lambda_> \lambda_< Q_0 )
e^{-\alpha(\lambda_1 + \lambda_2)} \, d\lambda_1 d\lambda_2 \nonumber\\
=~ & 2 (2\tau-1) H_0(m+1,n+1;\alpha)
\nonumber \\
& - (2\tau-1) \int^\infty_1 d\lambda_1
\int^{\lambda_1}_1 d\lambda_2 \, \lambda_1^m \lambda_2^{n+1}
e^{-\alpha(\lambda_1 + \lambda_2)}
 \nonumber\\
 & - (2\tau-1) \int^\infty_1 d\lambda_1
\int^\infty_{\lambda_1} d\lambda_2 \, \lambda_1^{m+1} \lambda_2^n
e^{-\alpha(\lambda_1 + \lambda_2)} \nonumber\\
\label{eqF.86}
=~ & (2\tau-1) [ 2 H_0(m+1,n+1;\alpha) - S(m,n+1;\alpha) -
S(n,m+1;\alpha) ]
\end{align}

\begin{center}
{\bf 8.} $\pmb{H^{(1)}_\tau(m,n;\alpha)}$
\end{center}

The recurrence relations for the associated Legendre polynomial
are
\begin{align}
\label{eqF.87}
(x^2-1)^{1/2} P^{\nu+1}_\tau(x) &= (\tau-\nu) x P^\nu_\tau(x) -
(\tau+\nu)P^\nu_{\tau-1}(x), \\
\label{eqF.88}
(x^2-1)^{1/2} Q^{\nu+1}_\tau(x) &= (\tau-\nu) x Q^\nu_\tau(x) -
(\tau+\nu) Q^\nu_{\tau-1}(x);
\end{align}
\begin{align}\label{eqF.89}
&H^{(1)}_\tau(m,n;\alpha) = \tau^2 \int \int \lambda_1^m
\lambda_2^n ( \lambda_< P_\tau - P_{\tau-1} ) ( \lambda_> Q_\tau -
Q_{\tau-1} ) e^{-\alpha(\lambda_1 + \lambda_2)} \, d\lambda_1
d\lambda_2  \\
 =~ & \tau^2 [ H_\tau(m+1,n+1;\alpha) +
H_{\tau-1}(m,n;\alpha) ]  \nonumber\\
&~ - \tau^2 \int \int
\lambda_1^m \lambda_2^n ( \lambda_< P_\tau Q_{\tau-1} + \lambda_>
P_{\tau-1} Q_\tau ) e^{-\alpha(\lambda_1 + \lambda_2)} \,
d\lambda_1 d\lambda_2  \nonumber\\
=~ & \tau^2 [
H_\tau(m+1,n+1;\alpha) + H_{\tau-1}(m,n;\alpha) ]  \nonumber\\
&~
- \frac{\tau^2}{(2\tau+1)\tau} \left[ (2\tau+1)^2
H_\tau(m+1,n+1;\alpha) + \tau^2 H_{\tau-1}(m,n;\alpha) -
(\tau+1)^2 H_{\tau+1}(m,n;\alpha) \right]  \nonumber \\
=~ &
\frac{\tau(\tau+1)^2}{2\tau+1} H_{\tau+1}(m,n;\alpha) - \tau
(\tau+1) H_\tau(m+1,n+1;\alpha) + \frac{\tau^2(\tau+1)}{2\tau+1}
H_{\tau-1}(m,n;\alpha).\nonumber
\end{align}
So
\begin{equation}\label{eqF.90}
\frac{2\tau+1}{\tau(\tau+1)}H^{(1)}_\tau(m,n;\alpha) = (\tau+1)
H_{\tau+1}(m,n;\alpha) - (2\tau+1) H_\tau(m+1,n+1;\alpha) + \tau
H_{\tau-1}(m,n;\alpha)
\end{equation}

\begin{center}
{\bf 9.} $\pmb{H^{(2)}_\tau(m,n;\alpha)}$
\end{center}

Similarly to the preceding paragraph,
\begin{align}
& H^{(2)}_\tau(m,n;\alpha) \nonumber\\
=~ & \int \int \lambda_1^m
\lambda_2^n [(\lambda_1^2-1)(\lambda_2^2-1)]^{1/2}
((\tau-1) \lambda_< P^1_\tau - (\tau+1) P^1_{\tau-1} )
\nonumber\\
  &\quad \times ( (\tau-1) \lambda_> Q^1_\tau -
(\tau+1) Q^1_{\tau-1}
) e^{-\alpha(\lambda_1 + \lambda_2)} \, d\lambda_1 d\lambda_2
\nonumber\\
 =~ & [ (\tau-1)^2 H^{(1)}_\tau(m+1,n+1;\alpha) +
(\tau+1)^2 H^{(1)}_{\tau-1}(m,n;\alpha) ]
 \nonumber\\
 \label{eqF.91}
 & - (\tau^2 -1) \int \int \lambda_1^m \lambda_2^n
[(\lambda_1^2-1)(\lambda_2^2-1)]^{1/2} ( \lambda_< P^1_\tau
Q^1_{\tau-1} + \lambda_> P^1_{\tau-1} Q^1_\tau )
e^{-\alpha(\lambda_1 + \lambda_2)} \, d\lambda_1 d\lambda_2,
\end{align}
and
\begin{align}
& H^{(1)}_{\tau+1}(m,n;\alpha)  \nonumber\\
=~ & \frac{1}{\tau^2}
\int \int \lambda_1^m \lambda_2^n
[(\lambda_1^2-1)(\lambda_2^2-1)]^{1/2} ((2\tau+1) \lambda_<
P^1_\tau - (\tau+1) P^1_{\tau-1} )  \nonumber\\
& \quad
\times ( (2\tau+1) \lambda_> Q^1_\tau - (\tau+1) Q^1_{\tau-1}
) e^{-\alpha(\lambda_1 + \lambda_2)} \, d\lambda_1 d\lambda_2\nonumber \\
=~ & \frac{1}{\tau^2}  [ (2\tau+1)^2
H^{(1)}_\tau(m+1,n+1;\alpha) + (\tau+1)^2
H^{(1)}_{\tau-1}(m,n;\alpha) ]\nonumber\\
\label{eqF.92}
& - \frac{(2\tau+1)(\tau+1)}{\tau^2} \int \int \lambda_1^m \lambda_2^n
[(\lambda_1^2-1)(\lambda_2^2-1)]^{1/2} ( \lambda_< P^1_\tau
Q^1_{\tau-1} + \lambda_> P^1_{\tau-1} Q^1_\tau )
e^{-\alpha(\lambda_1 + \lambda_2)} \, d\lambda_1 d\lambda_2.
\end{align}

Therefore,
\begin{align}
(2\tau+1) H^{(2)}_\tau(m,n;\alpha) &= \tau^2 (\tau-1)
H^{(1)}_{\tau+1}(m,n;\alpha) - (\tau+2)(\tau-1)(2\tau+1)
H^{(1)}_\tau(m+1,n+1;\alpha) \nonumber\\
\label{eqF.93}
&\quad + (\tau+1)^2(\tau+2)
H^{(1)}_{\tau-1}(m,n;\alpha).
\end{align}

\section{Derivations for the 5-term recurrence relations
\eqref{M21.4}}\label{A4}

\indent

First, we cast both equations \eqref{M21.1} and \eqref{M21.2} into the form
\begin{equation}\label{M24.1}
[(1-x^2)\phi']' + \left[-A-2R_jx + p^2x^2 - \frac{m^2}{1-x^2}\right] \phi =
0,\qquad j=1,2,
\end{equation}
where
\begin{align*}
\text{for } j=1,\qquad &x = \lambda, \quad
\phi = \Lambda(\lambda), \quad R_1=R(Z_a+Z_b)/2
;\\
\text{for } j=2,\qquad &x=\mu, \quad \phi = M(\mu), \quad R_2=R(Z_a-Z_b)/2.
\end{align*}
Set
\begin{equation}\label{M24.2}
\phi(x) = \sum^\infty_{k=0} f_kP^m_k(x)
\end{equation}
as in \eqref{M21.3} and substitute \eqref{M24.2} into \eqref{M24.1}:
\begin{align*}
&\sum^\infty_{k=0} f_k\left[\frac{d}{dx} (1-x^2) \frac{dP^m_k(x)}{dx}\right] +
\sum^\infty_{k=0} f_k \left[-A-2R_jx + p^2x^2 - \frac{m^2}{1-x^2}\right]
P^m_k(x) = 0,\\
&\sum^\infty_{k=0} f_k \left[\frac{d}{dx} (1-x^2) \frac{d}{dx}
- \frac{m^2}{1-x^2}-A\right] P^m_k(x) + \sum^\infty_{k=0} f_k
[-2R_jx+p^2x^2]P^m_k(x) = 0,\\
&\sum^\infty_{k=0} f_k[-k(k+1)-A] P^m_k(x) - \sum^\infty_{k=0} f_kR_jxP^m_k(x) +
\sum^\infty_{k=0} f_kp^2x^2P^m_k(x)  = 0,\\
&\sum^\infty_{k=0} f_k[-k(k+1)-A] P^m_k(x) - \sum^\infty_{k=0} f_kR_j
\left\{\frac1{2k+1} [(k+m)P^m_{k-1}(x) + (k-m+1) P^m_{k+1}(x)]\right\}\\
+&~\sum^\infty_{k=0} f_kp^2\cdot \left\{\frac{(k-m+1)(k-m+2)}{(2k+1)(2k+3)}
P^m_{k+2}(x)\right.\\
 +&~\left. \left[\frac{(k-m+1)(k+m+1)}{(2k+1)(2k+3)} +
\frac{(k-m)(k+m)}{(2k-1)(2k+1)}\right] P^m_k(x)
+ \frac{(k+m-1)(k+m)}{(2k-1)(2k+1)} P^m_{k-2}(x)\right\} = 0.
\end{align*}
We an now shift indices to convert $P^m_{k+2}(x)$, $P^m_{k+1}(x)$,
$P^m_{k-1}(x)$ and $P^m_{k-2}(x)$ to $P^m_k(x)$. We obtain
\begin{align*}
&\sum^\infty_{k=0} \left\{f_{k-2} \frac{p^2(k-m-1)(k-m)}{(2k-3)(2k-1)} - f_{k-1}
\frac{2R_j(k-m)}{2k-1} + f_kC_k\right.\\
&\quad \left. - f_{k+1} \frac{2R_j(k+m+1)}{2k+3} + f_{k+2}
\frac{p^2(k+m+1)(k+m+2)}{(2k+3)(2k+5)}\right\} = 0.
\end{align*}
The terms inside the parentheses above are exactly the 5-term recurrence
relations \eqref{M21.4}.
\vspace{.3in}

\section{Dimensional scaling in spherical coordinates}
\label{dimspher}

For description of diatomic molecules cylindrical coordinates provide a natural
way of making a dimensional (D-) scaling transformation. Here we show how to do
the D-scaling transformation in spherical coordinates, which is useful for
description of atoms.
Let us first consider the Laplacian in
the $D$-dimensional hyperspherical  coordinates
\begin{eqnarray} \nonumber
x_1 &=&
r \cos\theta_1 \sin\theta_2 \sin\theta_3 \cdots \sin\theta_{D-1}, \\
\nonumber x_2 &=&
r \sin\theta_1 \sin\theta_2 \sin\theta_3 \cdots \sin\theta_{D-1}, \\
\nonumber x_3 &=&
r \cos\theta_2 \sin\theta_3 \sin\theta_4 \cdots \sin\theta_{D-1}, \\
\nonumber x_4 &=&
r \cos\theta_3 \sin\theta_4 \sin\theta_5 \cdots \sin\theta_{D-1}, \\
\nonumber \vdots && \vdots \\
\nonumber x_j &=&
r \cos\theta_{j-1} \sin\theta_j \sin\theta_{j+1} \cdots \sin\theta_{D-1}, \\
\nonumber \vdots && \vdots \\
\nonumber x_{D-1} &=&
r \cos\theta_{D-2} \sin\theta_{D-1}, \\ \nonumber
x_D &=& r \cos\theta_{D-1},\\
0\le\theta_1\le 2\pi, && \quad 0\le\theta_j\le \pi,\quad
{\text for} \quad j=2,3,\cdots ,D-1,
\label{eq2}
\end{eqnarray}
where $D$ is a positive integer and $D\ge 3$. Define
\begin{equation}\label{eq3}
h = \prod^{D-1}_{j=0} h_j,
\end{equation}
where
\begin{equation}\label{eq4}
h^2_k = \sum^D_{j=1} \left( \frac{\partial x_j}{\partial \theta_k}
\right)^2.
\end{equation}
Then the scaling factors are
\begin{eqnarray} \nonumber
h_0 &=& 1, \\ \nonumber h_1 &=& r \sin\theta_2 \sin\theta_3 \cdots
\sin \theta_{D-1}, \\ \nonumber
h_2 &=& r \sin\theta_3 \sin\theta_4 \cdots \sin \theta_{D-1}, \\
\nonumber \vdots && \vdots \\
\nonumber
h_k &=& r \sin\theta_{k+1} \sin\theta_{k+2} \cdots \sin \theta_{D-1}, \\
\nonumber \vdots && \vdots \\ \nonumber h_{D-2} &=& r
\sin\theta_{D-1}, \\ \nonumber
h_{D-1} &=& r, \\
h &=& r^{D-1} \sin\theta_2 \sin^2\theta_3 \sin^3\theta_4 \cdots
\sin^{k-1} \theta_k \cdots \sin^{D-2} \theta_{D-1}.\label{eq5}
\end{eqnarray}
The $D$-dimensional Laplacian now becomes
\begin{eqnarray} \nonumber
\nabla^2_D &=& \frac{1}{r^{D-1}} \frac{\partial}{\partial r}
r^{D-1} \frac{\partial}{\partial r} \\
\nonumber && +
\frac{1}{r^2} \sum^{D-2}_{k=1} \frac{1}{\sin^2\theta_{k+1}
\sin^2\theta_{k+2} \cdots \sin^2\theta_{D-1}} \left\{
\frac{1}{\sin^{k-1}\theta_k} \frac{\partial}{\partial \theta_k}
\sin^{k-1}\theta_k \frac{\partial}{\partial \theta_k} \right\} \\
&& + \frac{1}{r^2} \left\{ \frac{1}{\sin^{D-2}\theta_{D-1}}
\frac{\partial}{\partial \theta_{D-1}} \sin^{D-2}\theta_{D-1}
\frac{\partial}{\partial \theta_{D-1}} \right\}\label{eq6}
\end{eqnarray}

Define the generalized orbital angular momentum operators by
\begin{eqnarray} \nonumber
L_1^2 &=& - \frac{\partial^2}{\partial \theta_1^2}, \\
\nonumber L_2^2 &=& - \frac{1}{\sin \theta_2}
\frac{\partial}{\partial \theta_2} \sin\theta_2
\frac{\partial}{\partial \theta_2} + \frac{L_1^2}{\sin^2\theta_2},
\\ \nonumber \vdots && \vdots \\
L_k^2 &=& - \frac{1}{\sin^{k-1}\theta_k} \frac{\partial}{\partial
\theta_k} \sin^{k-1}\theta_k \frac{\partial}{\partial \theta_k} +
\frac{L_{k-1}^2}{\sin^2\theta_k}.\label{eq7}
\end{eqnarray}
Then we have
\begin{equation}\label{eq8}
\nabla_D^2 = K_{D-1}(r) - \frac{L_{D-1}^2}{r^2},
\end{equation}
where
\begin{equation}\label{eq9}
K_{D-1}(r) \equiv \frac{1}{r^{D-1}} \frac{\partial}{\partial r}
\left( r^{D-1} \frac{\partial}{\partial r} \right).
\end{equation}

Let us consider Schr\"odinger equation for a particle moving in D-dimensions in
a central potential $V(r)$:
$$
\left(-\frac{\nabla_D^2}{2}+V(r)\right)\Psi_D=E\Psi_D.
$$
To eliminate the angular dependence we separate
the variables by writing
\begin{equation}\label{eq10}
\Psi_D(r,\Omega_{D-1}) = R(r) Y(\Omega_{D-1}).
\end{equation}
Near the origin $r=0$,
\begin{equation}\label{eq11}
\Psi_D \sim r^l Y(\Omega_{D-1}),
\end{equation}
or
\begin{equation}\label{eq12}
\nabla_D^2 r^l Y(\Omega_{D-1}) =[l(l+D-2) - C] r^{l-2}
Y(\Omega_{D-1}) = 0
\end{equation}
with
\begin{equation}\label{eq13}
L^2_{D-1} Y(\Omega_{D-1}) = CY(\Omega_{D-1}).
\end{equation}
The effective Hamiltonian is given by
\begin{equation}
H_D = - \frac{1}{2} K_{D-1}(r) + \frac{l(l+D-2)}{2r^2} + V(r)
\label{Eq:RadialHamiltonai}
\end{equation}
With the following transformation
\begin{equation}\label{eq15}
\Psi_D = r^{-(D-1)/2}\Phi_D
\end{equation}
the corresponding equation for $\Phi_D$ reads
\begin{equation}\label{eq16}
\left[ - \frac{1}{2} \frac{\partial^2}{\partial r^2} +
\frac{\Lambda(\Lambda+1)}{2r^2} + V(r) \right] \Phi_D = E_D
\Phi_D,
\end{equation}
where
\begin{equation}\label{eq17}
\Lambda = l + \frac{1}{2} (D-3).
\end{equation}
Equation \eqref{eq16} is the Schr\"odinger equation in
D-dimensions for the function $\Phi_D$.

As an example, consider the Sch\"odinger equation for the H-atom in
D-dimensions
\begin{equation}\label{eq18}
\left[ - \frac{1}{2} \nabla^2 - \frac{Z}{r} \right] \Psi = E \Psi.
\end{equation}
In the scaled variables
\begin{eqnarray}\label{eq19}
r_s &=& \frac{3}{2} \frac{r}{r_0}, \\
\label{eq20}
E_s &=& \frac{1}{2} \frac{E}{E_0},
\end{eqnarray}
with $r_0 = D(D-1)/4$ and $E_0 = 2/(D-1)^2$, the Schr\"odinger
equation reads
\begin{equation}\label{eq21}
\left[ - \frac{1}{2} \left( \frac{3}{D} \right)^2 \nabla_s^2 -
\left( \frac{3}{D} \right) \left( \frac{D-1}{2} \right)
\frac{Z}{r_s} \right] \Psi = E_s \Psi.
\end{equation}

Now, let us write the Laplacian in spherical coordinates and
transform the wave function $\Psi $ according to \eqref{eq15}. We obtain
\begin{equation}
\label{dd1}\left[ -\frac 12\left( \frac 3D\right) ^2\left( \frac{d^2}{dr_s^2}%
-\frac{(D-1)(D-3)}{4r_s^2}\right) -\frac{3(D-1)}{2D}\frac Z{r_s}\right] \Psi
=E_s\Psi
\end{equation}
In the limit $D\rightarrow \infty $ Eq. \eqref{dd1} reduces to a simple
algebraic problem of minimization the expression
\begin{equation}
\label{dd2}E_s=\frac 9{8r_s^2}-\frac 32\frac Z{r_s},
\end{equation}
which yields $r_s=3/2Z$ and $E_s=-Z^2/2$. This value coincides with the
ground state energy of the hydrogen atom in 3 dimensions.

\section*{Acknowledgements}

We wish to thank Professor D.\ Herschbach for lectures given at Princeton
in Fall 2003 which stimulate our interest in $D$-scaling, and
 Professor C.~Le Sech for the
personal communication cited in \ref{sec5}.E.(1). We also thank Professor
G.\ Hunter, who read the manuscript and offered constructive criticisms.
 This research is partially
supported by grants from  ONR
(N00014-03-1-0693 and N00014-04-1-0336), NSF DMS 0310580, TITF of Texas
A\&M University, and the Robert A.\ Welch
Foundation (\#A1261).


\begin{thebibliography}{99}

\bibitem{Juds92}  R. Judson and H. Rabitz, Phys. Rev. Lett {\bf 68},
1500 (1992).

\bibitem{PNAS}  M.O. Scully, G.W. Kattawar, R.P. Lucht, T. Opatrny, H.
Pilloff, A. Rebane, A.V. Sokolov and M.S. Zubairy, PNAS {\bf 99},
10994 (2002).

\bibitem{Niel99} M. Nielsen and I. Chuang, ``Quantum Computation and Quantum
Information", Cambridge University Press, Cambridge, U.K., 1999.

\bibitem{Svid04}  A.A. Svidzinsky, M.O. Scully, and D.R. Herschbach, Phys.
Rev. Lett., {\bf 95}, 080401 (2005).

\bibitem{Svid05}  A.A. Svidzinsky, M.O. Scully, and D.R. Herschbach, PNAS,
{\bf 102}, 11985 (2005).

\bibitem{Bohr1}  N. Bohr, Phil. Mag. {\bf 26}, 1, 476, 857 (1913).

\bibitem{Le}  C.\ Le Sech, Phys.\ Rev.\ A {\bf 51} (1995), R2668.

\bibitem{Sech96} C. Le Sech, Phys. Rev. A {\bf 53} (1996), 4610.

\bibitem{Kolo60}  W. Kolos, and C.C.J. Roothaan, Rev. Mod. Phys. {\bf 32},
219 (1960).

\bibitem{Witt80}  E. Witten, Phys. Today {\bf 33} (7), 38 (1980).

\bibitem{Hers92}  ``Dimensional Scaling in Chemical Physics'', Eds. D. R.
Herschbach, J. S. Avery and O. Goscinski, Kluwer Academic Publishers,
Dordrecht, 1992.

\bibitem{Kais93}  S. Kais, S.M. Sung and D.R. Herschbach, J. Chem. Phys.
{\bf 99}, 5184 (1993).

\bibitem{ABB}  M.\ Aubert, N.\ Bessis and G.\ Bessis, Phys.\ Rev.\ A Part I
{\bf 10} (1974), 51; Part II {\bf 10} (1974), 61; Part III {\bf 12} (1975),
2298.

\bibitem{A-FL}  M. Aubert--Fr\'econ and C. Le Sech, J.\ Chem.\ Phys. {\bf 74}
(1981), 2931.

\bibitem{SL}  L.D.A.\ Siebbeles and C.\ Le Sech, J.\ Phys.\ B, {\bf 27}
(1994), 4443.

\bibitem{PTT} S.H. Patil, K.T. Tang and J.P. Toennies,
J.\ Chem.\ Phys. {\bf 111} (1999), 7278.

\bibitem{ScullyEtAl}  M.O. Scully, R.E. Allen, Y. Dou, K.T. Kapale, M. Kim,
G. Chen and A.A. Svidzinsky, Chem. Phys. Lett. {\bf 389}, 385-392 (2004).

\bibitem{KW}  W. Kolos, and L. Wolniewicz, J. Chem. Phys. {\bf 41} (1964)
3663; J. Chem. Phys., {\bf 43} (1965) 2429; J. Chem. Phys., {\bf 49} (1968) 404.

\bibitem{KS}  W. Kolos, K. Szalewicz, and H.J. Monkhorst, J. Chem. Phys.
{\bf 84} (1986), 3278.

\bibitem{james}  H.\ James and A.\ Coolidge, J.\ Chem.\ Phys. {\bf 1}
(1933), 823.


\bibitem{chen2} G.\ Chen and J.\ Zhou, {\em Boundary Element Methods}, Academic
Press, London-New York-San Diego, 1992.

\bibitem{Landau}  L.D. Landau, and E.M. Lifshitz, Quantum Mechanics:
Non-Relavistic Theory, vol 3, 4th ed., Nauka, Moscow, 1986.

\bibitem{Bur}  O. Burrau, Kgl. Danske.\ Vidensk.\ Selsk.\ {\bf 7} (1927), 1.

\bibitem{Heit27}  W. Heitler and F. London, Zeit. f. Phys. {\bf 44}, 455
(1927).

\bibitem{Con}  E.U. Condon, Proc.\ Nat.\ Acad.\ Sci.\ USA {\bf 13} (1927), 466.

\bibitem{BO}  M. Born and J.R. Oppenheimer, Ann.\ Physik {\bf 84} (1927), 457.

\bibitem{HM}  F. Hund and R.S. Mulliken, Phys.\ Rev. {\bf 32} (1928), 186.

\bibitem{CN}  F.A. Cotton and D.G. Nocera, Acc.\ Chem.\ Res. {\bf 33} (2000),
483.

\bibitem{Hin} A.\ Hinchliffe, {\em Molecular Modelling for Beginners},
Wiley, New York, 2003.

\bibitem{hylleraas}  E.A.\ Hylleraas, Zeit.\ f\"ur Physik, {\bf 71} (1931),
739.

\bibitem{Ryc}  J. Rychlewski, {\em Advances in Quantum Chemistry\/}, Academic
Press, New York, 1999, 173-199.

\bibitem{teller} E. Teller and H.L. Sahlin, {\it Physical Chemistry: An
Advanced Treatise}, Vol.5, P.1, (Academic, New York, 1970).

\bibitem{Mul}  R.S. Mulliken, J.\ Chem.\ Phys. {\bf 43} (1965), 52.

\bibitem{Klei98}  U. Kleinekath\"ofer, S.H. Patil, K.T. Tang and J.P.
Toennies, Polish J. Chem., {\bf 72} (1998), 1361.

\bibitem{pat} S.H. Patil and K.T. Tang, {\it Asymptotic methods in
quantum mechanics : application to atoms, molecules and nuclei}, Springer,
Berlin ; New York : Springer, (2000).


\bibitem{parr} R.G. Parr, in {\it The quantum theory of molecular
electronic structure; a lecture-note and reprint volume},
Reading, Mass. W. A. Benjamin, 1972.


\bibitem{K}
U. Kleinekath\"ofer, S.H. Patil, K.T. Tang and J.P. Toennies, Phys.\ Rev.\ A
{\bf 54} (1996), 2840.

\bibitem{Pati00}  S.H. Patil and K.T. Tang, J.\ Chem.\ Phys., {\bf 113}
(2000), 676.

\bibitem{Pati03a}  S.H. Patil and K.T. Tang, J.\ Chem.\ Phys., {\bf 118}
(2003), 4905.

\bibitem{Pati03}  S.H. Patil, Phys.\ Rev.\ A {\bf 68} (2003), 044501.

\bibitem{WH}  R.F.\ Wallis and H.M.\ Hulburt, J.\ Chem.\ Phys. {\bf 22}, 5
(1954), 774.

\bibitem{mclean} A. D. Mclean, A. Weiss, and M. Yoshimine,
                 Rev. Mod. Phys. {\bf 32} (1960) 211.

\bibitem{shull} H. Shull, {\it Physical Chemistry: An
Advanced Treatise}, Vol.5, P.125, (Academic, New York, 1970).

\bibitem{Theo84}  G. Theodorakopoulos, S.C. Farantos, R.J. Buenker, and S.D.
Peyerimhoff, J. Phys. B: Mol. Phys. {\bf 17}, 1453 (1984).

\bibitem{Dock72}  K.K. Docken, and J. Hinze, J. Chem. Phys., {\bf 57}, 4928
(1972).

\bibitem{JT1}
H.J.\ Jahn and E.\ Teller, Proc.\ Roy.\ Soc. {\bf A161} (1937), 220.

\bibitem{JT2}
H.J.\ Jahn and E.\ Teller, Proc.\ Roy.\ Soc. {\bf A164} (1938), 117.

\bibitem{Ya1}
D.R.\ Yarkony, Electronic structure aspects of nonadiabetic processes in
polyatomic systems, in {\em Modern Electronic Structure Theory}, Vol.~2, D.R.\
Yarkony, ed., World Scientific, Singapore, 1995, 642--721. See also his webpage
{\tt http://jhunivase.jhu.edu/$\sim$chem/yarkony.html}

\bibitem{Oh1}
Y.\ \"Ohrn, The Quantum Theory Project at the University of Florida,\newline {\tt
http://www.qtp.ufl.edu/$\sim$ohrn}

\bibitem{eckart}
C. Eckart, Phys. Rev. {\bf 36} (1930), 878.

\bibitem{klein} U. Kleinekathofer, S.H. Patil, K.T. Tang,
and J.P. Toennies, Phys. Rev. {\bf A54}, 2840 (1996).

\bibitem{hamm}B.L. Hammond, W.A. Lester, Jr., and P.J. Reynolds,
{\it Monte Carlo Methods in Ab Initio Quantum Chemistry}, World Scientific,
Singapore (1994).

\bibitem{hesech} C. Le Sech, J. Phys. B: At. Mol. Opt. Phys. {\bf 30}
(1997) L47.

\bibitem{GZ}
V. Guillemin and C. Zener, Proc.\ Nat.\ Acad.\ Sci. {\bf 15}, 314 (1929).

\bibitem{inui} T. Inui, Proc. Phys. Math. Soc. Japan {\bf 20} (1938) 770.

\bibitem{nord} A. Nordsieck, Phys. Rev. {\bf 58}, (1940) 310.

\bibitem{P} S.H. Patil, Phys.\ Rev.\ A {\bf 62} (2000), 052515.

\bibitem{Root60}  C.C.J. Roothaan and A.W. Weiss, Rev. Mod. Phys., {\bf 32}
(1960) 194.

\bibitem{SW} T.\ Shibuya and C.E. Wulfman,
Proc.\ Roy.\ Soc.\ Ser.\ A {\bf 286} (1965), 377.

\bibitem{baber}
W.G.\ Baber and H.R.\ Hass\'e, Proc.\ Cambridge Phil.\ Soc. {\bf 31}, 564
(1935).

\bibitem{jaffe} G.\ Jaff\'e, Zeit. f\"ur Physik, {\bf 87} (1934), 535.

\bibitem{bates}
D.\ Bates, K.\ Ledsham and A.S.\ Stewart, Phil. Trans.\ Roy.\ Soc.\ London {\bf
246}, 215 (1953)

\bibitem{Judd}
B.R.\ Judd, {\em Angular Momentum Theory for Diatomic Molecules}, Academic
Press, New York, 1975.

\bibitem{Ka}
T.\ Kato, Comm.\ Pure Appl.\ Math. {\bf 10} (1957), 151.

\bibitem{Hir}
J.O. Hirschfelder, J.\ Chem.\ Phys. {\bf 39} (1963), 3145.

\bibitem{SML} L.D.A. Siebbles, D.P. Marshall and C. Le Sech,
J.\ Phys.\ B {\bf 26} (1993), L321.

\bibitem{AS} M.\ Abramowitz and I.A.\ Stegun, {\em Handbook of Mathematical
Functions}, 9th printing, Dover, New York, 1973.

\bibitem{Slat63}  J.C. Slater, {\it Quantum theory of molecules and solids},
McGraw-Hill, 1963.

\bibitem{Rose31}  N. Rosen, Phys.\ Rev. {\bf 38} (1931), 2099.

\bibitem{Wang28} S.C. Wang, Phys.\ Rev.\ {\bf 31} (1928), 579.

\bibitem{Boy}
S.F.\ Boys, Proc.\ Roy. Soc.\ London, Ser.\ A {\bf 200} (1950), 542.

\bibitem{Kobu96} J. Kobus, L. Laaksonen and D. Sundholm, Comput. Phys. Commun.
{\bf 98} (1996), 346.

\bibitem{lesechetal}
 M. Aubert-Frecon, P. Ceyzeriat, C. Le Sech, and A. M. Jorus,
J. Chem. Phys. {\bf 75}, 5212-13 (1981); C. Le Sech, A.M. Jorus, and M.
Aubert-Frecon, J. Chem. Phy. {\bf 75}, 2932-34 (1981); C. Le Sech, Phys. Rev.
A.\ Rapid Communication, {\bf 51}, R2668 (1995);

\bibitem{lesechetal1}
C. Le Sech, A. Sarsa, Phys. Rev. A, {\bf 63} 022501 (2001).

\bibitem{Ki}
 J. Killingbeck, Rep. Prog. Phys. {\bf 48} (1985a); J. Phys. A: Math. Gen.
{\bf 18} (1985b), 245; J. Phys. A: Math. Gen. {\bf 18} (1985c), L1025;
G. Hardinger, M.
Aubert-Frecon and G. Hardinger, J. Phys. B: AMO Phys. {\bf 22} (1989), 679.

\bibitem{Fran88}  D.D. Frantz and D.R. Herschbach, Chem. Phys. {\bf 126}, 59
(1988).

\bibitem{Lewi16}  G.N. Lewis, J. Am. Chem. Soc. {\bf 38}, 762 (1916).

\bibitem{Dore87}  D.J. Doren and D.R. Herschbach, J. Chem. Phys. {\bf 87},
433 (1987).

\bibitem{Lapi75}  I.R. Lapidus, Am. J. Phys. {\bf 43}, 790 (1975).

\bibitem{Good92}  D.Z. Goodson, M. L\'opez-Cabrera, D.R. Herschbach and J.D.
Morgani III, J. Chem. Phys. {\bf 97}, 8491 (1992); Phys. Rev. Lett. {\bf 68}%
, 1992 (1992).

\bibitem{ff}  To improve convergence of the $1/D$ expansion a variant
scaling involving a factor $(2D+9)/5D$ is used in front of the $\partial
^2/\partial \phi ^2$ term in Eqs. (\ref{d4}), (\ref{d7}).

\bibitem{Vlec22}  J.H. Van Vleck, Phil. Mag. {\bf 44}, 842 (1922).

\bibitem{Drak02}  G.W.F. Drake, M.M. Cassar, and R.A. Nistor, Phys. Rev. A
{\bf 65}, 054501 (2002).

\bibitem{Shi01}  Q. Shi, S. Kais, F. Remacle and R.D. Levine, Chem. Phys.
Chem. {\bf 2}, 434 (2001).



\end{thebibliography}
\end{document}